\documentclass[fleqn,usenatbib]{mnras}

\pdfoutput=1

\usepackage{mathptmx}

\usepackage[T1]{fontenc}

\DeclareRobustCommand{\VAN}[3]{#2}
\let\VANthebibliography\thebibliography
\def\thebibliography{\DeclareRobustCommand{\VAN}[3]{##3}\VANthebibliography}

\usepackage{graphicx}	
\usepackage{amsmath}	
\usepackage{amssymb}	

\usepackage{enumerate}
\usepackage{natbib}
\defcitealias{Varela-Lavin2022}{VL22}

\title[Lopsidedness and early galactic assembly history]{Lopsidedness as a tracer of early galactic assembly history}

\author[A. Dolfi et al.]{
Arianna Dolfi,$^{1,2}$\thanks{E-mail: arianna.dolfi@userena.cl}
Facundo A. G\'omez,$^{1,2}$
Antonela Monachesi,$^{1,2}$
Silvio Varela-Lavin,$^{1,2}$
Patricia B. Tissera,$^{3,4}$
\and Crist\'obal Sif\'on,$^{5}$ 
and Gaspar Galaz$^{3}$\\ 
$^{1}$Departamento de Astronom\'ia, Universidad de La Serena, Av. Juan Cisternas 1200 Norte, La Serena, Chile\\
$^{2}$Instituto Multidisciplinario de Investigaci\'on y Postgrado, Universidad de La Serena, Ra\'ul Bitr\'an 1305, La Serena, Chile\\
$^{3}$Instituto de Astrofísica, Pontificia Universidad Católica de Chile, Av. Vicuña Mackenna 4860, Santiago, Chile\\
$^{4}$Centro de Astro-Ingenieria, Pontificia Universidad Católica de Chile, Av. Vicuña Mackenna 4860, Santiago, Chile\\ 
$^{5}$Instituto de F\'isica, Pontificia Universidad Cat\'olica de Valpara\'iso, Casilla 4059, Valpara\'iso, Chile\\
}

\date{Accepted XXX. Received YYY; in original form ZZZ}

\pubyear{2023}

\begin{document}
\label{firstpage}
\pagerange{\pageref{firstpage}--\pageref{lastpage}}
\maketitle

\begin{abstract}
Large-scale asymmetries (i.e. lopsidedness) are a common feature in the stellar density distribution of nearby disk galaxies both in low- and high-density environments. In this work, we characterize the present-day lopsidedness in a sample of $1435$ disk-like galaxies selected from the TNG50 simulation. We find that the percentage of lopsided galaxies ($10\%$-$30\%$) is in good agreement with observations if we use similar radial ranges to the observations. However, the percentage ($58\%$) significantly increases if we extend our measurement to larger radii. We find a mild or lack of correlation between lopsidedness amplitude and environment at $z=0$ and a strong correlation between lopsidedness and galaxy morphology regardless of the environment. Present-day galaxies with more extended disks, flatter inner galactic regions and lower central stellar mass density (i.e. late-type disk galaxies) are typically more lopsided than galaxies with smaller disks, rounder inner galactic regions and higher central stellar mass density (i.e. early-type disk galaxies). Interestingly, we find that lopsided galaxies have, on average, a very distinct star formation history within the last $10\, \mathrm{Gyr}$, with respect to their symmetric counterparts. Symmetric galaxies have typically assembled at early times ($\sim8$-$6\, \mathrm{Gyr}$ ago) with relatively short and intense bursts of central star formation, while lopsided galaxies have assembled on longer timescales and with milder initial bursts of star formation, continuing building up their mass until $z=0$. Overall, these results indicate that lopsidedness in present-day disk galaxies is connected to the specific evolutionary histories of the galaxies that shaped their distinct internal properties.
\end{abstract}




\section{Introduction}
The existence of asymmetries in the present-day distribution of light and mass of disk galaxies has been known for a long time. In an early work, \citet{Baldwin1980} studied the distribution of the neutral HI gas in a sample of $20$ disk galaxies and found that half of them displayed asymmetries where the gas on one side of the galaxy was more extended than on the opposite side. Specifically, in four of these galaxies, such asymmetries were very significant, with the ratio between the projected HI density of the two sides of the galaxy being greater than 2:1. These strongly asymmetric galaxies were referred to as lopsided galaxies by \citet{Baldwin1980}. To explain the high frequency of these asymmetries, the authors suggested a lopsided pattern of elliptical orbits.

Follow-up studies that included larger samples of galaxies have further detected the lopsided feature not only in the spatial distribution (e.g. \citealt{Richter1994,Haynes1998}) and large-scale kinematics (e.g. \citealt{Swaters1999}) of the HI gas, but also in the old stellar component of disk galaxies observed in the near infrared (e.g. \citealt{Bloch1994,Rix1995,Zaritsky1997}). The lopsidedness in the stellar component was typically quantified by performing an azimuthal decomposition of the galaxy stellar surface density distribution and measuring the mass-weighted mean amplitude of the first Fourier mode, $\mathrm{A_{1}}$, over the radial range between $1.5$-$2.5$ disk scale lengths. 
The mean value of $\mathrm{A_{1}}$ obtained from a sample of $149$ spiral galaxies by \citet{Bournaud2004} was $\mathrm{A_{1}}\simeq0.1$. As such, this value is now widely adopted as a threshold to identify lopsided stellar disks. The behaviour of the lopsidedness was found to be similar in both the HI gas and stellar component with the amplitude of the asymmetry increasing at larger galactocentric radii \citep{Bournaud2005}. 

Overall, the large fraction of late-type disk galaxies with significant HI (i.e. $\sim50\%$ of the galaxies; \citealt{Richter1994,Haynes1998}) and stellar (i.e. $\sim30\%$ of the galaxies; \citealt{Rix1995,Zaritsky1997,Bournaud2005}) asymmetries suggests that lopsidedness is a rather common phenomenon in present-day disk galaxies. The observed high occurrence rate of the lopsidedness in disk galaxies has, thus, raised the question of whether lopsidedness is a long-lived phenomenon or the result of an external, recent, perturbation \citep{Varela-Lavin2022,Ghosh2022}. 

Several mechanisms that have been proposed to explain the origin of lopsidedness typically involve an external perturbation, such as a tidal interaction (e.g. \citealt{Beale1969}) or minor merger (e.g. \citealt{Walker1996,Zaritsky1997,Zaritsky1999}), which is expected to excite a short-lived (i.e. $\sim1$-$2\, \mathrm{Gyr}$) but strong lopsided feature \citep{Bournaud2005}. 
Therefore, this scenario can potentially explain the origin of the lopsidedness in group environments where tidal perturbations and mergers are more frequent. In fact, the observations that early-type disk galaxies tend to be more lopsided than late-type ones in galaxy groups \citep{Angiras2006,Angiras2007} seem consistent with a tidal origin of the lopsidedness as repetitive tidal encounters can significantly heat up the stellar disk and trigger multiple starbursts, thus producing a gas-poor lenticular galaxy with a thick disk and no prominent spiral arm structure \citep{Bekki2011}. 
However, strong lopsided features are also observed in field galaxies where tidal interaction and mergers are less frequent. In fact, late-type disk galaxies are typically found to be more lopsided than early-type ones in the field \citep{Conselice2000,Bournaud2004,Bournaud2005}. Additionally, the fact that the lopsidedness amplitude is expected to drop quickly after the external perturbation \citep{Bournaud2005}, combined with the observed lack of correlation between lopsidedness and the presence of close companions \citep{Bournaud2005,Wilcots2010}, also disfavours the external perturbation scenario for the origin of lopsidedness. Similar conclusions were reached by \citet{Varela-Lavin2022} (hereafter, \citetalias{Varela-Lavin2022}), using fully cosmological models of late-type galaxy formation from the IllustrisTNG simulations \citep{Nelson2019}.

For this reason, other mechanisms have been proposed that suggest the lopsided feature is a long-lived perturbation. Some of these mechanisms include the response of the disk to a tidally perturbed dark matter halo as a result of past interactions (e.g. \citealt{Jog1997,Jog2002}), an off-centred disk with respect to the dark matter halo (e.g. \citealt{Nordermeer2001}), internal dynamical processes within the disk (e.g. \citealt{Saha2007}) and the accretion of misaligned gas into the disk (e.g. \citealt{Bournaud2005}). 
Specifically, the misaligned gas accretion scenario has received much attention in previous works due to the finding that lopsided galaxies are typically characterized by higher star formation rates, bluer colours, lower metallicity and larger gas fractions than symmetric galaxies. This suggests that lopsidedness may originate as a result of asymmetric star formation following the accretion of misaligned low metallicity gas \citep{Lokas2022}.  
However, the fact that lopsidedness is not only observed in the young but also in the old stellar component of disk galaxies, where the lopsided mode is expected to be damped due to differential rotation, indicates that this cannot be the only mechanism that produces lopsidedness. 
Furthermore, \cite{Rudnick1998} studied the lopsidedness in a sample of $54$ early-type disk galaxies and found that $\sim20\%$ showed significant asymmetries, similarly to the results obtained by \citet{Rix1995} and \citet{Zaritsky1997} for late-type disk galaxies. Therefore, the authors suggested that lopsidedness is more likely associated to mass asymmetries rather than to asymmetric star formation.

Other studies have found that lopsidedness correlates more strongly with morphology \citep{Conselice2000}, as well as with other galaxy structural properties (\citealt{Reichard2008};\citetalias{Varela-Lavin2022}). Specifically, \citet{Conselice2000} found that early-type disk galaxies typically show smaller asymmetries than late-type ones, which generally agrees with the results of \citet{Reichard2008}, who showed that lopsided galaxies are, on average, characterized by lower concentration and stellar surface mass density than symmetric ones. Indeed, \citetalias{Varela-Lavin2022} showed that late-type lopsided galaxies in numerical simulations are typically characterized by lower central surface densities and are, thus, less gravitationally cohesive than symmetric galaxies. As a result, lopsided galaxies are more easily perturbed by external interactions and prone to develop stronger lopsided features than symmetric ones. Additionally, \citetalias{Varela-Lavin2022} also found that the lower central surface density of the late-type lopsided galaxies is likely the result of the more significant growth of their stellar half-mass radius, $R_{1/2}$, towards $z=0$. These results suggest a stronger correlation between lopsidedness and the internal galaxy properties rather than a main external perturbation.

In this work, we aim to study the correlation between lopsidedness, galaxy morphology, environment and galaxy internal properties to establish the connection between present-day asymmetric perturbations on galactic disks and the evolution of galaxies in relation with the environment using the IllustrisTNG simulations. 
Firstly, we aim to investigate the dependence of lopsidedness on the environment by selecting a sample of central and satellite disk-like galaxies embedded in a range of different total halo masses. This allows us to better characterize the correlation between lopsidedness and environment by extending the previous work by \citetalias{Varela-Lavin2022}, who mainly focused on central late-type disk galaxies embedded in a restricted range of total halo masses (i.e. $10^{11.5} < \mathrm{M_{200}/M_{\odot}} < 10^{12.5}$). In addition, we extend the results of \citetalias{Varela-Lavin2022} by investigating whether the correlation between lopsidedness and internal galaxy properties holds in different environments.
Finally, we aim to investigate what mechanisms are mainly responsible for shaping the distinct internal galaxy properties that are strongly linked to lopsidedness (i.e. whether it is the environment or the internal evolution of the galaxy).

The paper is structured as follows. In Sec. \ref{sec:data}, we describe the galaxy sample selected from the IllustrisTNG simulations. In Sec. \ref{sec:lopsidedness}, we describe the method used to estimate and quantify the lopsidedness in our galaxies. In Sec. \ref{sec:role_environment} and Appendix \ref{sec:role_environment_tng100}, we study the dependence of lopsidedness on the environment, as well as the correlation between lopsidedness and galaxy morphology as a function of the environment. In Sec. \ref{sec:evolution_history}, we study the star formation histories of the lopsided and symmetric galaxies to identify differences in how these galaxies assembled their stars and built-up their mass, which may possibly correlate with the lopsidedness. Finally, in Sec. \ref{sec:conclusions}, we provide a summary of our results and conclusions. 


\begin{figure}
    \centering
    \includegraphics[width=0.45\textwidth]{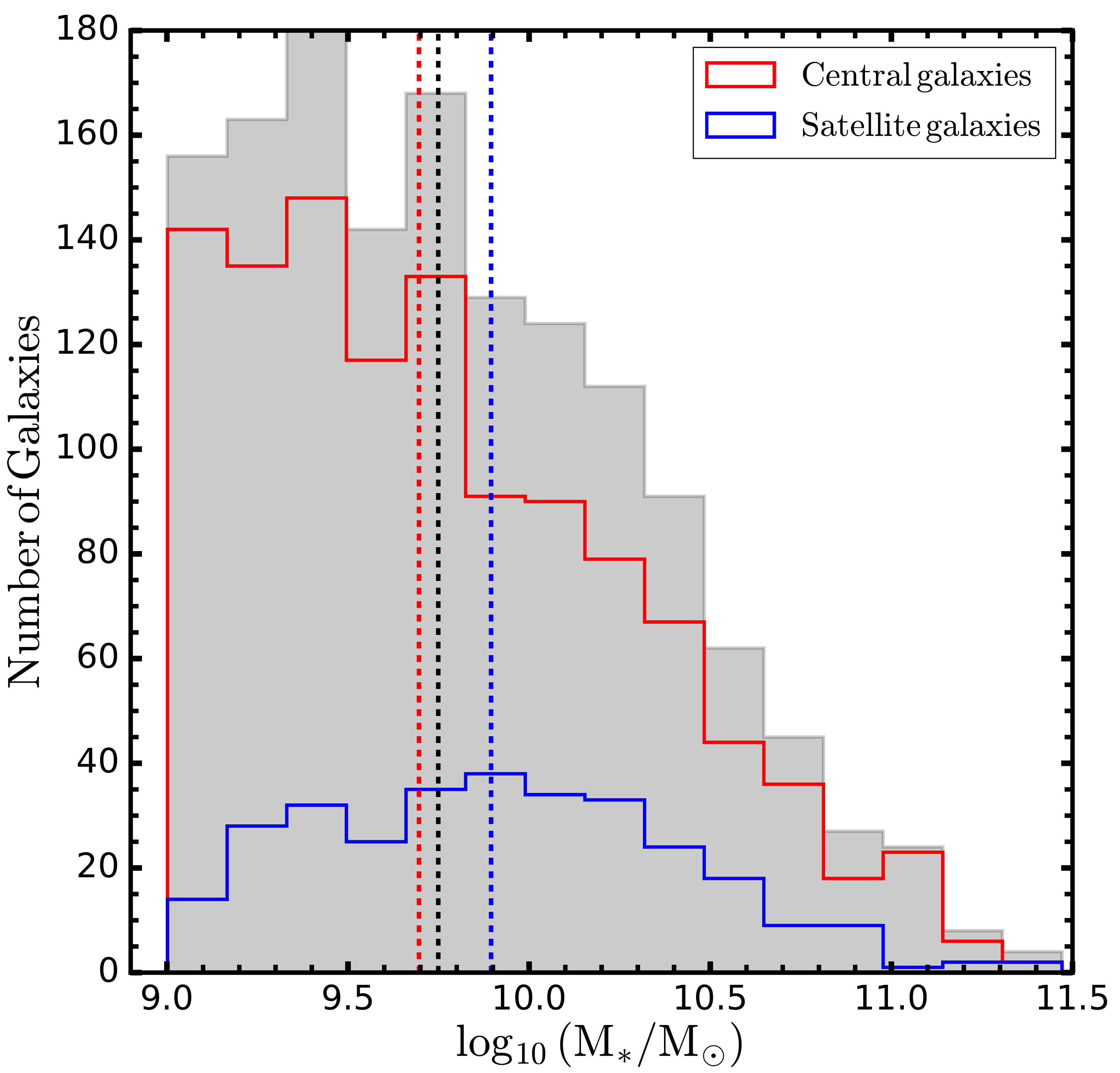}
    \includegraphics[width=0.45\textwidth]{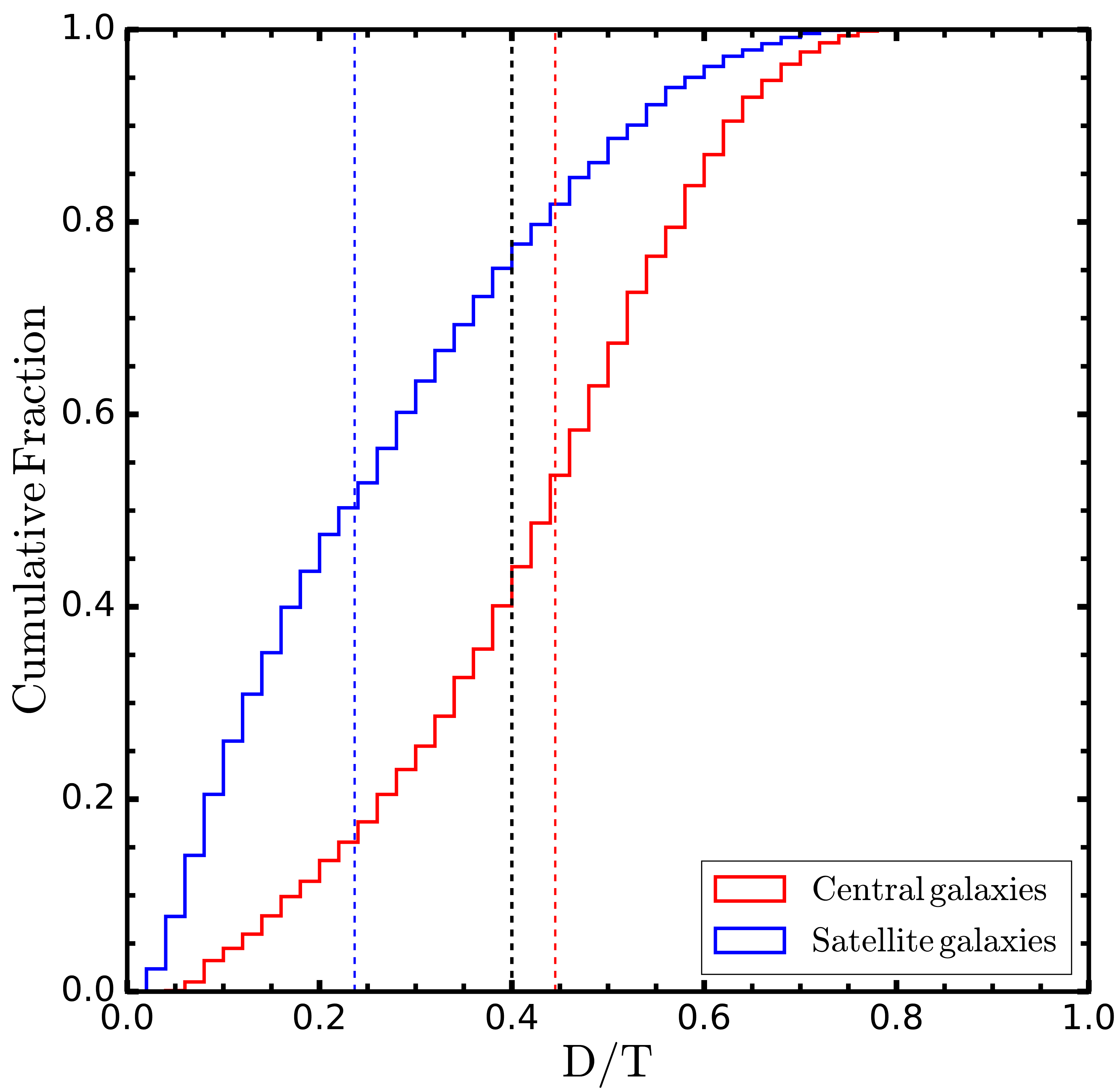}
    \caption{{\bf Top:} Stellar mass distribution of all (grey histogram), central (red histogram) and satellite (blue histogram) galaxies in our galaxy sample selected from the TNG50 simulation as described in Sec. \ref{sec:selection_criteria}. The black, red and blue vertical lines represent the median $\mathrm{M_{*}}$ of all, central and satellite galaxies, respectively. {\bf Bottom:} Cumulative distribution of the $\mathrm{D/T}$ of the central and satellite galaxies in the TNG50 simulation with $\mathrm{N}_{\mathrm{tot,\, stars}} \geq 10^{4}$ and $\mathrm{M}_{*} \geq 10^{9}\, \mathrm{M}_{\odot}$ (i.e. i) and ii) criteria as described in Sec. \ref{sec:selection_criteria}). The red and blue vertical lines represent the median $\mathrm{D/T}$ of the central and satellite galaxies, respectively, while the black vertical line represents the $\mathrm{D/T}\geq0.4$ cut adopted to select the disk galaxies in this work (see Sec. \ref{sec:selection_criteria}). We see that the majority of the satellite galaxies have an early-type morphology in the TNG50 simulation as only $\sim20\%$ have $\mathrm{D/T}\geq0.4$.}
    \label{fig:mass_distribution}
\end{figure}

\section{The data}
\label{sec:data}

\subsection{IllustrisTNG simulations}
\label{sec:simulation}
IllustrisTNG\footnote{\url{https://www.tng-project.org/}}, successor of the original Illustris\footnote{\url{https://www.illustris-project.org/}} project, is a suite of cosmological, gravo-magnetohydrodynamical simulations \citep{Vogelsberger2014,Genel2014,Sijacki2015,Nelson2015}, which includes a comprehensive physical model of galaxy formation designed to realistically trace the formation and evolution of the galaxies across cosmic time \citep{Weinberger2017,Pillepich2018}. A detailed analysis of the updated physical model used in the IllustrisTNG simulations can be found in \citet{Pillepich2018}.

IllustrisTNG comprises three simulations with different cosmological volumes, i.e. $(51.7\, \mathrm{cMpc})^{3}$, $(106.5\, \mathrm{cMpc})^{3}$ and $(302.6\, \mathrm{cMpc})^{3}$, referred to as TNG50, TNG100 and TNG300, respectively. The largest TNG300 simulation enables the study of the massive structures in the Universe (e.g. galaxy clusters) and provides a large statistical galaxy sample, while the smallest TNG50 simulation enables the study of the internal and structural properties of the galaxies thanks to its higher mass resolution \citep{Springel2018,Pillepich2018b,Naiman2018,Nelson2018,Marinacci2018,Nelson2019b,Pillepich2019}. 
On the other hand, the intermediate TNG100 simulation facilitates the comparison between the old Illustris and the updated IllustrisTNG model since it uses the same initial physical conditions, and it offers a good compromise between volume, statistics and resolution, thus making it suitable for several galaxy evolution studies \citep{Nelson2019}. 

All the IllustrisTNG simulations are run with the moving-mesh code \texttt{Arepo} \citep{Springel2010} and were released in \citet{Nelson2019}. A more detailed description of the three IllustrisTNG simulations can be found in the release paper by \citet{Nelson2019} together with a summary of the main changes and improvements between the old Illustris and the updated IllustrisTNG model (see e.g. table 2 of \citealt{Nelson2019}).

In this work, we mainly use the TNG50 simulation \citep{Weinberger2017,Pillepich2018}. 
Specifically, we use the highest resolution version (i.e. TNG50-1; hereafter, TNG50) of the four available runs which is characterized by an initial gas cell mass of $8.5\times 10^{4}\, \mathrm{M_{\odot}}$. This mass resolution enables TNG50 to resolve galaxies down to stellar masses $\mathrm{M}_{*}\sim 10^{7}\, \mathrm{M_{\odot}}$ with more than $100$ stellar particles, thus making it the most suitable of the three IllustrisTNG simulations to study differences in the stellar mass density distribution of central and satellite disk-like galaxies as a function of the environment and their internal and structural properties.
TNG50 also includes one massive galaxy cluster with total halo mass of $\mathrm{M}_{200} \simeq 2\times 10^{14}\, \mathrm{M_{\odot}}$ at redshift $z=0$ (i.e. similar to Coma cluster-like environments), where $\mathrm{M}_{200}$ represents the total mass of the group enclosed in a sphere whose mean density is $\sim200$ times the critical density of the Universe at the time the halo is considered.

On the contrary, the TNG100-1 (hereafter, TNG100) simulation is characterized by an initial gas cell mass of $1.4\times 10^{6}\, \mathrm{M_{\odot}}$, which is roughly a hundred times lower mass resolution than TNG50. Due to their larger volume, we use the TNG100 simulation in Appendix \ref{sec:role_environment_tng100} to complement the study of asymmetries in the stellar mass density distribution of disk-like galaxies as a function of the environment carried out using TNG50. 

Throughout this paper, we assume a flat $\Lambda$CDM cosmological model with parameters from the \citet{Planck2016}, i.e. Hubble constant $\mathrm{H}_{0}=67.8\, \mathrm{kms^{-1}Mpc^{-1}}$, total matter density $\Omega_{\mathrm{m}}=0.3089$, dark energy density $\Omega_{\Lambda}=0.6911$, baryonic matter density $\Omega_{\mathrm{b}}=0.0486$, rms of mass fluctuations at a scale of $8\, h^{-1}\mathrm{Mpc}$, $\sigma_{8}=0.8159$, and primordial spectral index $n_{\mathrm{s}}=0.9667$, which is the same adopted in all the IllustrisTNG simulations.

\subsection{Sample selection criteria}
\label{sec:selection_criteria}
From the TNG50 simulation, we select our sample of disk-like galaxies at $z=0$ based on the following criteria:
\begin{enumerate}[(i)]\itemsep0.2cm
    \item total number of bound stellar particles $\mathrm{N}_{\mathrm{tot,\, stars}} \geq 10^{4}$. 
    This selection criteria is to ensure that each galaxy is well resolved with a large number of stellar particles to more reliably identify and quantify differences in the stellar mass density distribution;
    \item total stellar mass $\mathrm{M}_{*} \geq 10^{9}\, \mathrm{M}_{\odot}$, where $\mathrm{M_{*}}$ represents the sum of the stellar masses of all the stellar particles gravitationally bound to the galaxy at $z=0$;
    \item disk morphology, quantified kinematically by the circularity fraction, $f_{\mathrm{e}}$. The latter is defined as the fractional mass of the stellar particles with circularity, $\epsilon>0.7$. The circularity parameter is defined as $\epsilon = J_{z}/J_{\mathrm{max}}(E)$, where $J_{z}$ is the component of the specific angular momentum of the stellar particle parallel to the angular momentum of the galaxy and $J_{\mathrm{max}}(E)$ is the maximum angular momentum each stellar particle could attain for their given binding energy \citep{Marinacci2014,Joshi2020}.
    In this work, we use the catalog of stellar circularity, angular momenta and axis ratios \citep{Genel2015} provided by the IllustrisTNG project database\footnote{\url{https://www.tng-project.org/data/docs/specifications/}}, which contains kinematic measurements of morphology for all galaxies with $\mathrm{M_{*}}>3.4\times 10^{8}\, \mathrm{M_{\odot}}$ and at least $100$ stellar particles, where $\mathrm{M_{*}}$ is measured within twice the stellar half-mass radius (i.e. $R_{\mathrm{half}}$).
    We include in our sample all galaxies with circularity fraction $\geq 0.4$, which is considered as a proxy of the disk-to-total mass ratio ($\mathrm{D/T}$, hereafter). The circularity values are calculated including all stars within ten times $R_{\mathrm{half}}$. This value of $\mathrm{D/T}$ is chosen so that we have both a variation of morphology within disk galaxies, including S0s.\footnote{We note that we have also tried selecting only galaxies with $\mathrm{D/T}\geq0.5$ and our results do not significantly change.}
\end{enumerate}

We note that the above three selection criteria are similar to those adopted in \citetalias{Varela-Lavin2022}. However, in this work, we do not apply any cut on the total halo mass of the environment where our galaxies reside, and we include both central and satellite galaxies that satisfy the selection criteria described above. As a result, we obtain an extended galaxy sample of $1435$ disk-like galaxies, with $1131$ centrals and $304$ satellites, as compared to the $240$ central late-type disk galaxies of \citetalias{Varela-Lavin2022}. This allow us to study asymmetries in the stellar mass density distribution as a function of environment and galaxy morphology.

The top panel of Fig. \ref{fig:mass_distribution} shows the stellar mass distribution of all our galaxies as well as of the central and satellite galaxies, separately. We see that the central and satellite galaxies have an overall similar stellar mass distribution, but the satellite galaxies are slightly shifted towards higher masses and represent a minority ($21\%$) compared to the central galaxies ($79\%$) in our sample. The low number of satellite galaxies in our sample is the result of the $\mathrm{D/T}$ selection criterion. This is shown from the cumulative distribution of the $\mathrm{D/T}$ of all the central and satellite galaxies in the TNG50 simulation with $\mathrm{N}_{\mathrm{tot,\, stars}} \geq 10^{4}$ and $\mathrm{M}_{*} \geq 10^{9}\, \mathrm{M}_{\odot}$ (i.e. before applying the iii) criterion from Sec. \ref{sec:selection_criteria}) in the bottom panel of Fig. \ref{fig:mass_distribution}. The black vertical line represents the $\mathrm{D/T}\geq0.4$ cut adopted to select the disk-like galaxies in this work (i.e. iii) criterion). We see that only $\sim20\%$ of the satellite galaxies have $\mathrm{D/T}\geq0.4$ as opposed to $60\%$ of the central galaxies, suggesting that the majority of the satellite galaxies have an early-type morphology in the TNG50 simulation. 

\begin{figure}
    \centering
    \includegraphics[width=0.45\textwidth]{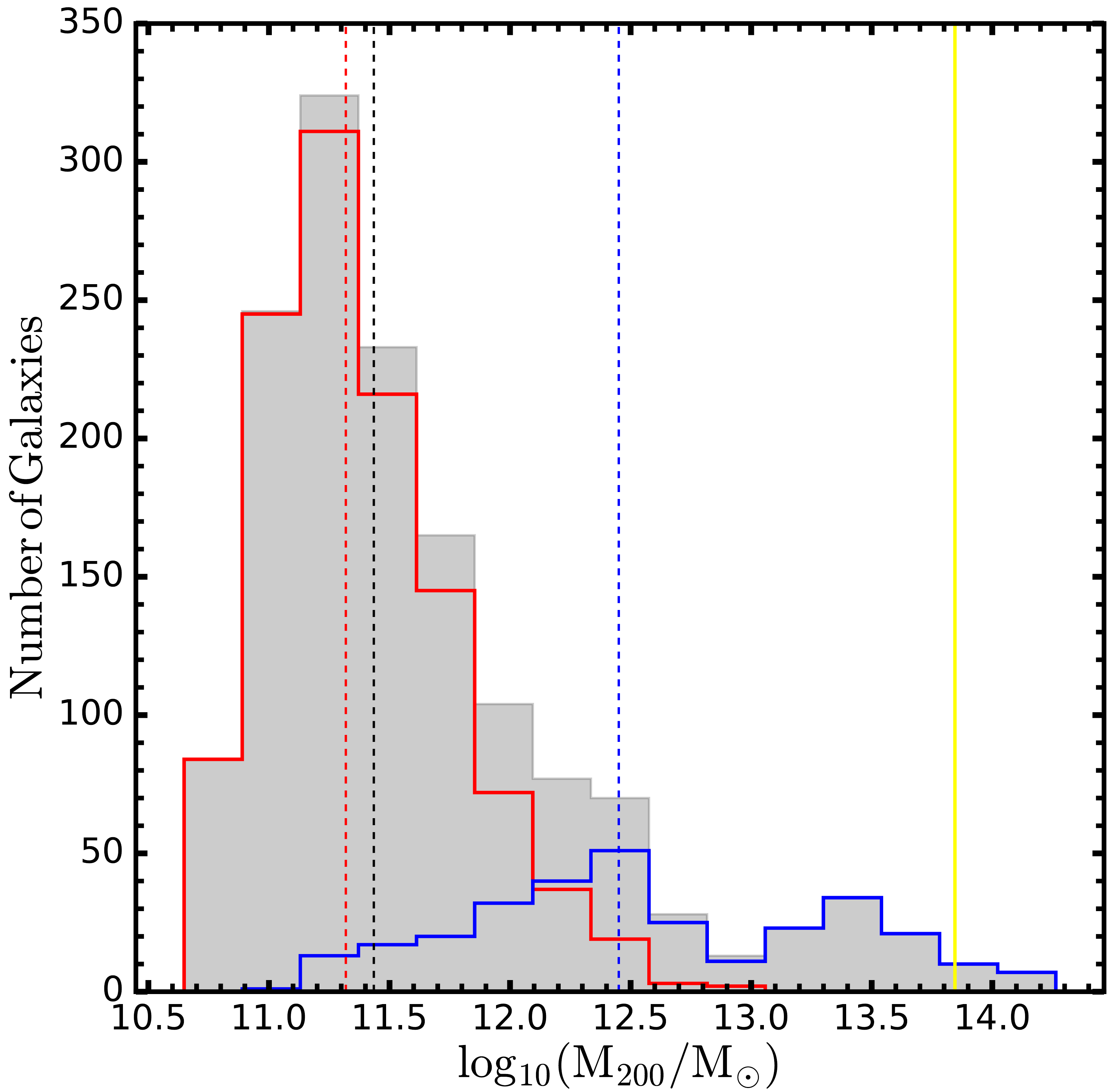}
    \caption{The total halo mass distribution, $\mathrm{M_{200}}$, of the environment where our selected galaxies reside. The grey histogram represents all the galaxies, while the red and blue histograms are for the central and satellite galaxies, respectively. The vertical black, red and blue lines represent the median $\mathrm{M}_{200}$ of the galaxy groups to which all, the central and satellite galaxies belong, respectively, while the vertical yellow line represents the $\mathrm{M}_{200} \sim 7\times 10^{13}\, \mathrm{M}_{\odot}$ of the Fornax cluster. Overall, we see that we are not exploring the highest density environments of rich galaxy clusters (i.e. $\mathrm{M}_{200} \gtrsim 10^{14}\, \mathrm{M}_{\odot}$).}
    \label{fig:halo_mass}
\end{figure}
\begin{figure}
    \centering
    \includegraphics[width=0.45\textwidth]{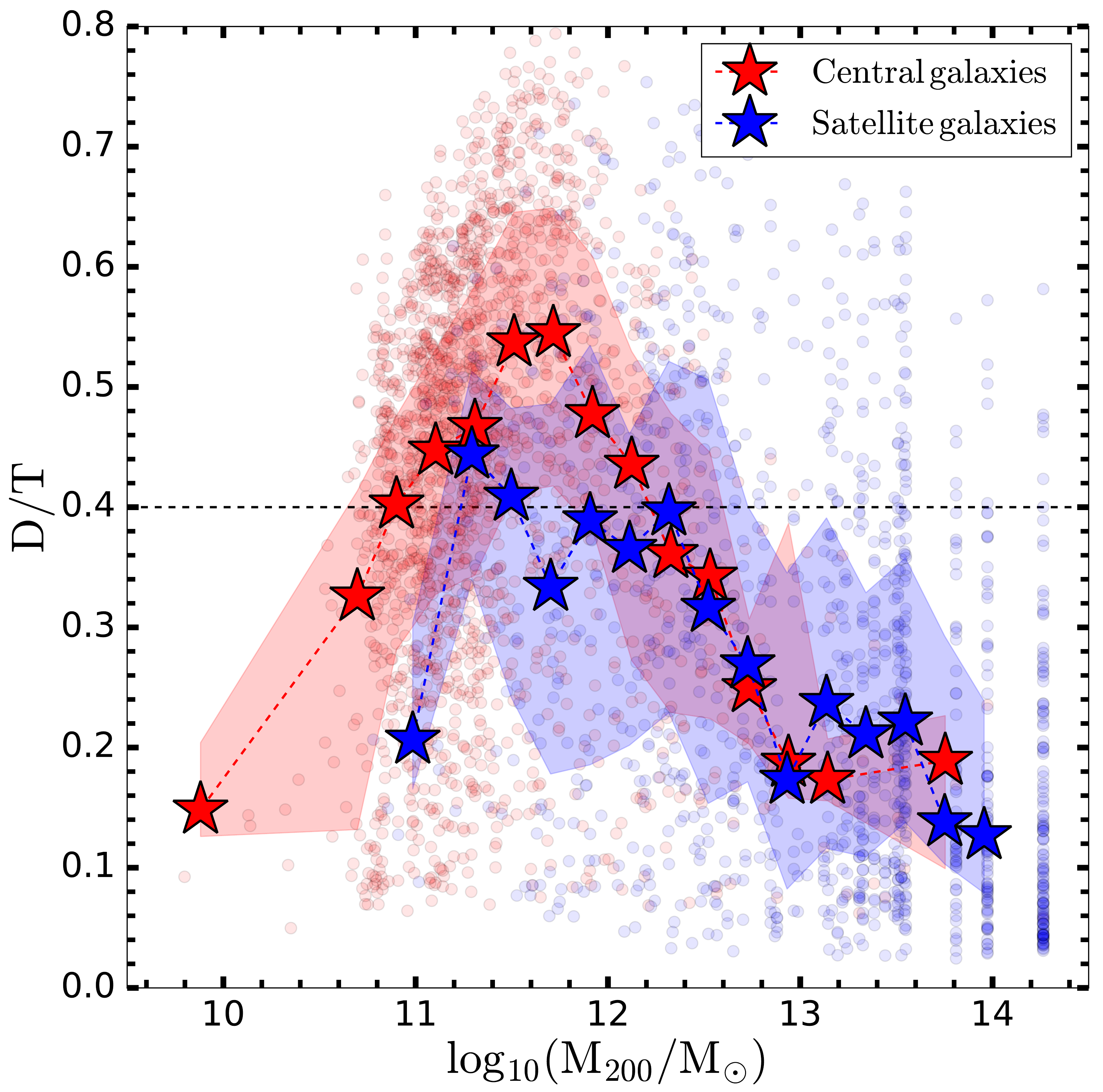}
    \caption{The median $\mathrm{D/T}$ of the central (red stars) and satellite (blue stars) galaxies in the TNG50 simulation with $\mathrm{N}_{\mathrm{tot,\, stars}} \geq 10^{4}$ and $\mathrm{M}_{*} \geq 10^{9}\, \mathrm{M}_{\odot}$ (i.e. i) and ii) criteria as described in Sec. \ref{sec:selection_criteria}) as a function of the total halo mass, $\mathrm{M_{200}}$, of the environment to which they belong. The shaded areas represent the $25\mathrm{th}$-$75\mathrm{th}$ interquartile range of the data in each bin of $\mathrm{M_{200}}$. The background coloured points represent the individual galaxies, while the black horizontal line represents the $\mathrm{D/T}\geq0.4$ cut adopted to select the disk-like galaxies in this work (iii) criteria in Sec. \ref{sec:selection_criteria}). We see that the median $\mathrm{D/T}$ steeply drops to $\mathrm{D/T}<0.4$ for both central and satellite galaxies at large total halo masses, $\mathrm{M}_{200} \gtrsim 10^{12}\, \mathrm{M}_{\odot}$, indicating that galaxies belonging to high-density environments are mainly characterized by an early-type morphology in the TNG50 simulation.}
    \label{fig:disk_fraction-halo_mass}
\end{figure}

\subsection{Environment}
\label{sec:environment}
The total halo mass, $\mathrm{M}_{200}$, can be used as a proxy of the galaxy host environment, with low and high values of $\mathrm{M}_{200}$ typically corresponding to low and high density environments, respectively. 

Fig. \ref{fig:halo_mass} shows the total halo mass distribution, $\mathrm{M}_{200}$, of the environment where our selected galaxies reside. Overall, we see that our galaxy sample reaches up to Fornax cluster-like environments, which have $\mathrm{M}_{200} \sim 7 \times 10^{13}\, \mathrm{M}_{\odot}$ \citep{Drinkwater2001}. In Fig. \ref{fig:halo_mass}, we represent this $\mathrm{M}_{200}$ with a vertical yellow line. We see that there are only a few satellite galaxies belonging to a dense environment with $\mathrm{M}_{200} \sim 10^{14}\, \mathrm{M}_{\odot}$.

Recent observational works have typically adopted a $\mathrm{M}_{200} \simeq 10^{13}\, \mathrm{M}_{\odot}$ division to separate between low- and high-density environments, with field and low-mass galaxy groups with $\mathrm{M}_{200} < 10^{13}\, \mathrm{M}_{\odot}$ classified as low-density environments, and high-mass galaxy groups and clusters with $\mathrm{M}_{200} > 10^{13}\, \mathrm{M}_{\odot}$ classified as intermediate- and high-density environments \citep{Deeley2020,Foster2021}. 
Therefore, based on this classification, we are mainly exploring low-density environments with TNG50.

As mentioned in Sec. \ref{sec:simulation}, the TNG50 simulation contains one massive cluster with $M_{200} \simeq 2 \times 10^{14}\, \mathrm{M}_{\odot}$ at $z=0$ \citep{Nelson2019}, but only few satellite galaxies in our selected galaxy sample belong to galaxy groups with $M_{200} \sim 10^{14}\, \mathrm{M}_{\odot}$ (see Fig. \ref{fig:halo_mass}). 
The reason for the low number of disk galaxies in high-density environments can be likely attributed to the effect of the environment on the galaxy morphology. In fact, the fraction of early-type galaxies is typically found to increase with increasing galaxy density at the expenses of late-type disk galaxies, as it is also suggested from the morphology-density relation \citep{Dressler1980,Dressler1997,Fasano2000}. Therefore, in high-density environments, we expect to be more dominated by galaxies with an earlier type disk morphology. This can be seen in Fig. \ref{fig:disk_fraction-halo_mass}, which shows the median $\mathrm{D/T}$ as a function of the total halo mass, $\mathrm{M_{200}}$, for all the central and satellite galaxies in the TNG50 simulation with $\mathrm{N}_{\mathrm{tot,\, stars}} \geq 10^{4}$ and $\mathrm{M}_{*} \geq 10^{9}\, \mathrm{M}_{\odot}$. The background coloured points represent the individual galaxies, and the black horizontal line represents the $\mathrm{D/T}\geq0.4$ cut adopted to select the disk-like galaxies in this work (see Sec. \ref{sec:selection_criteria}). 
We see that the median $\mathrm{D/T}$ is the largest (i.e. $\mathrm{D/T}\gtrsim0.4$) for total halo masses $10^{11} \lesssim \mathrm{M}_{200} \lesssim 10^{12.5}\, \mathrm{M}_{\odot}$, and it steeply drops to $\mathrm{D/T}<0.4$ for both central and satellite galaxies at large total halo masses, i.e. $\mathrm{M}_{200} \gtrsim 10^{12.5}\, \mathrm{M}_{\odot}$. 
We also see that there are no central galaxies with $\mathrm{D/T}\geq0.4$ for $\mathrm{M}_{200} \geq 10^{13}\, \mathrm{M}_{\odot}$, as also shown in Fig. \ref{fig:halo_mass}. 
This indicates that both central and satellite galaxies located in high-density environments with $\mathrm{M}_{200} \gtrsim 10^{12.5}\, \mathrm{M}_{\odot}$ tend to be characterized by an early-type morphology in the TNG50 simulation.

\begin{figure*}
    \centering
    \includegraphics[width=0.33\textwidth]{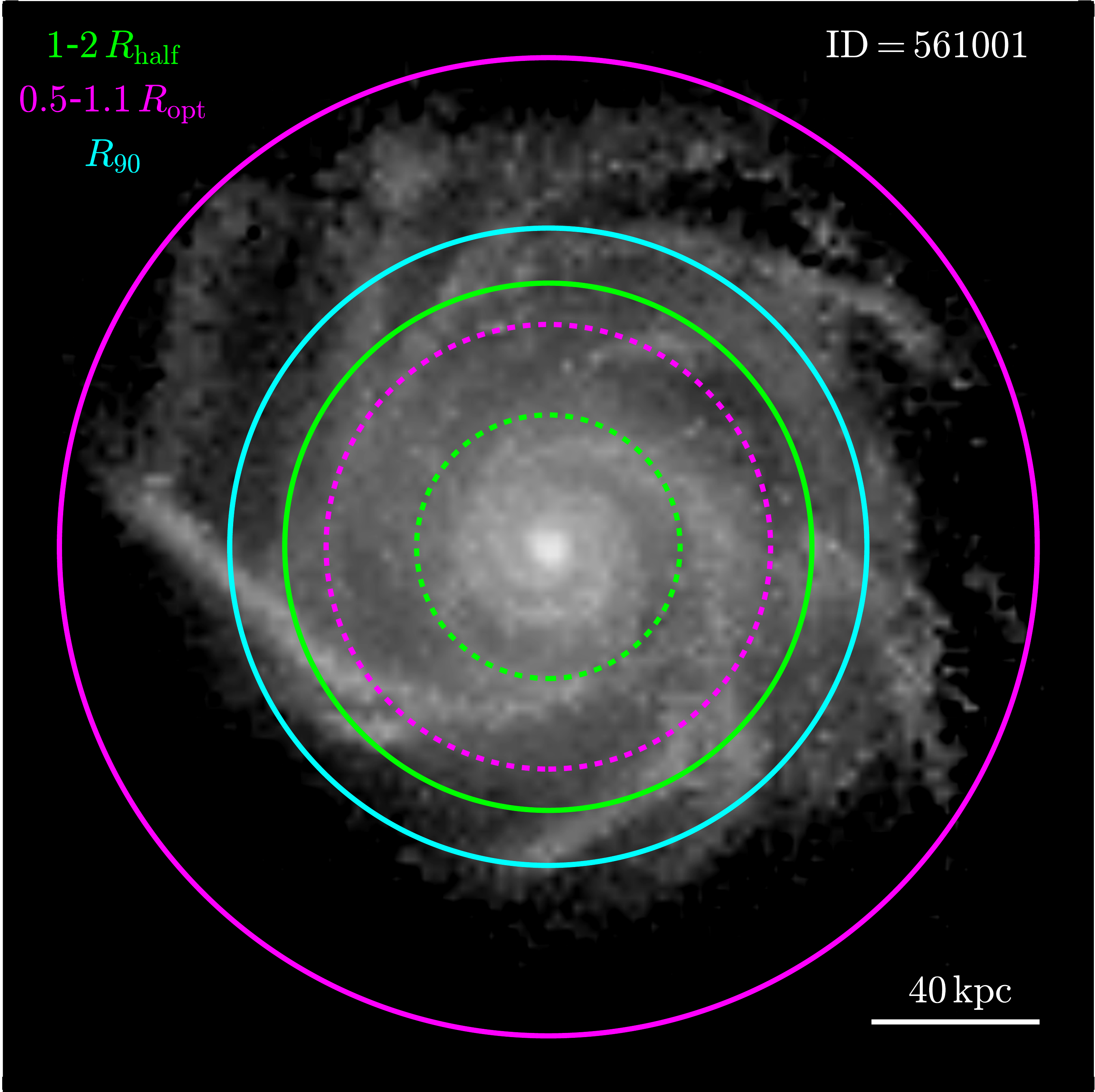}
    \includegraphics[width=0.33\textwidth]{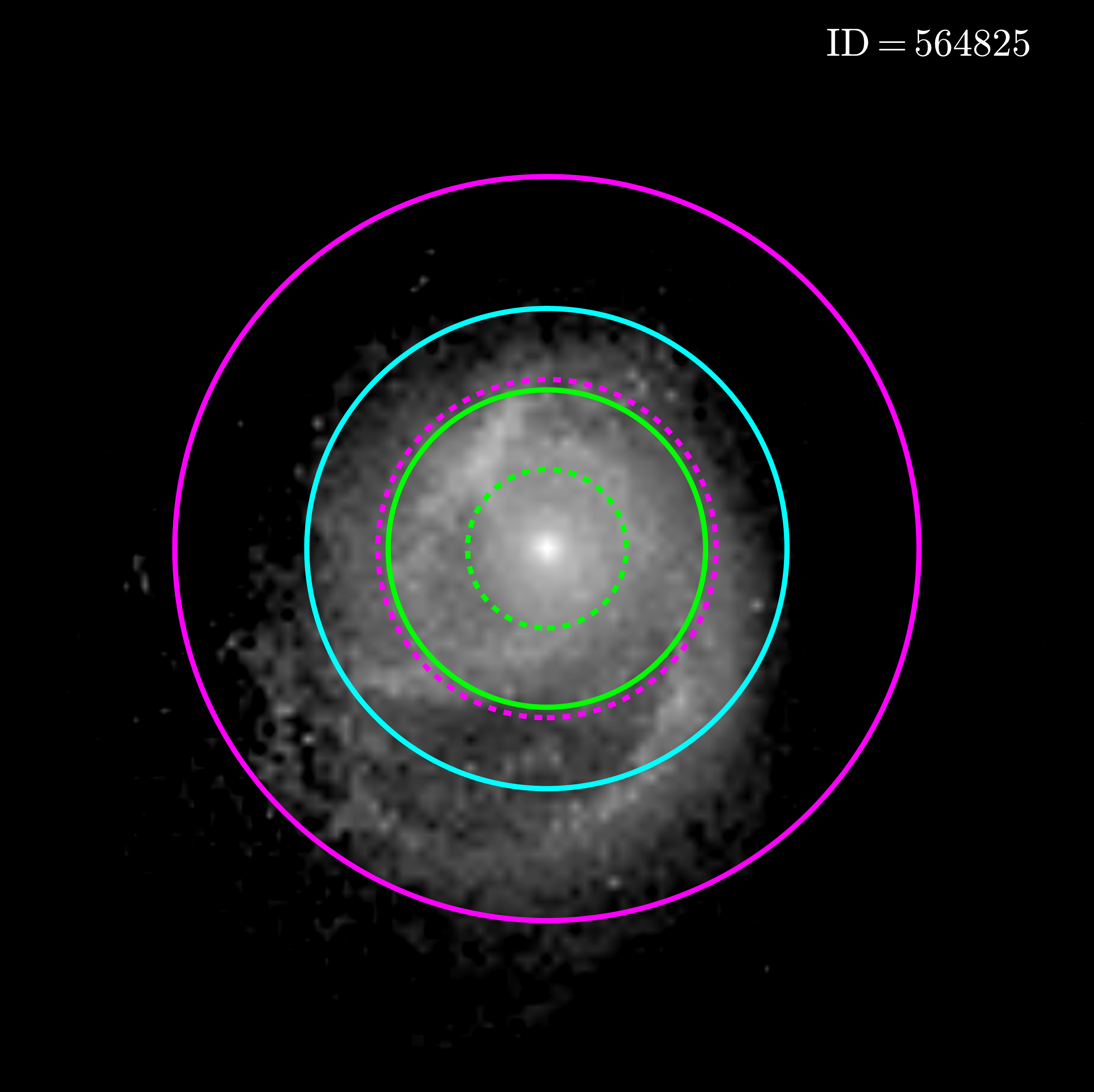}
    \includegraphics[width=0.33\textwidth]{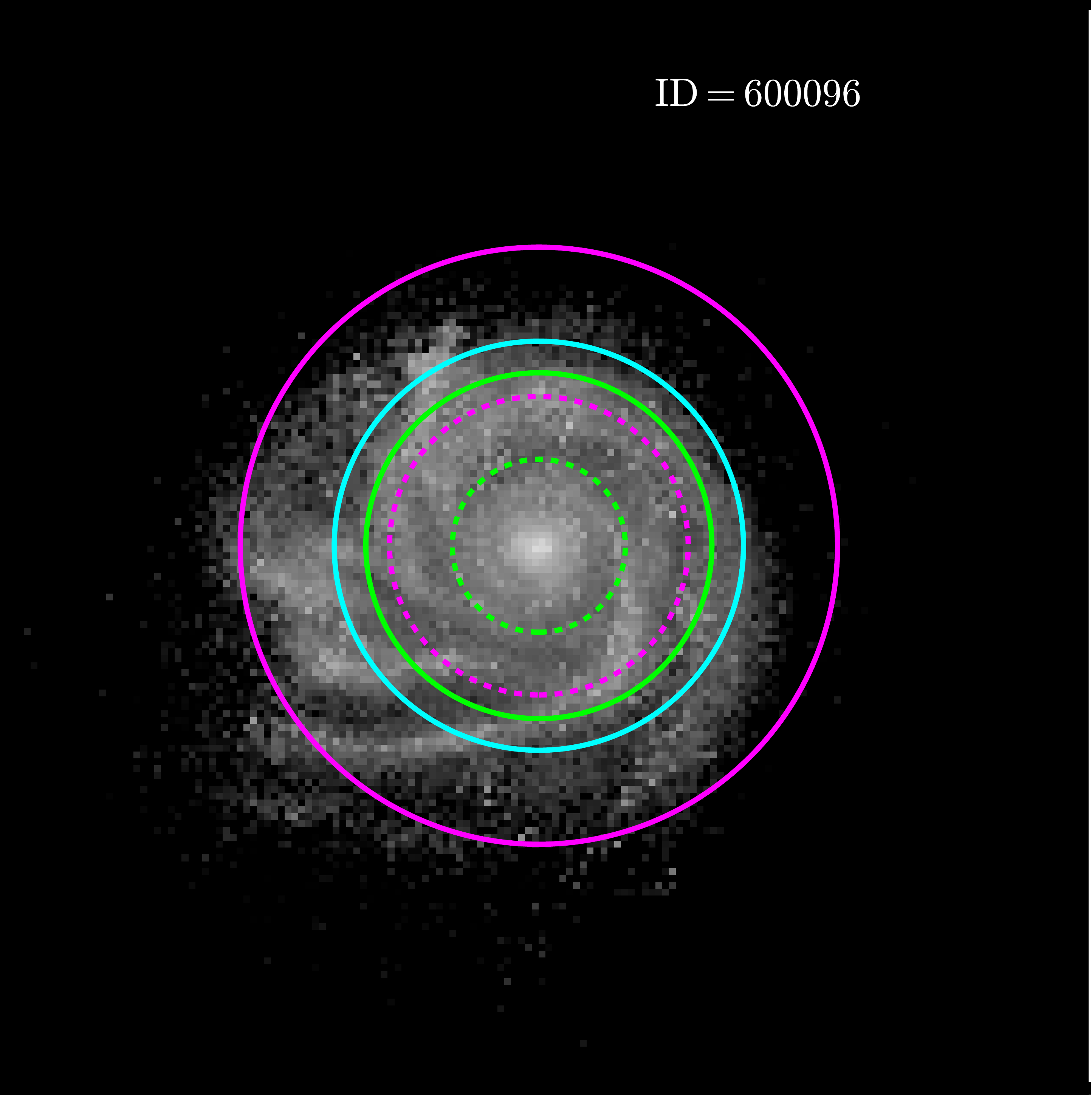}

    \includegraphics[width=0.325\textwidth]{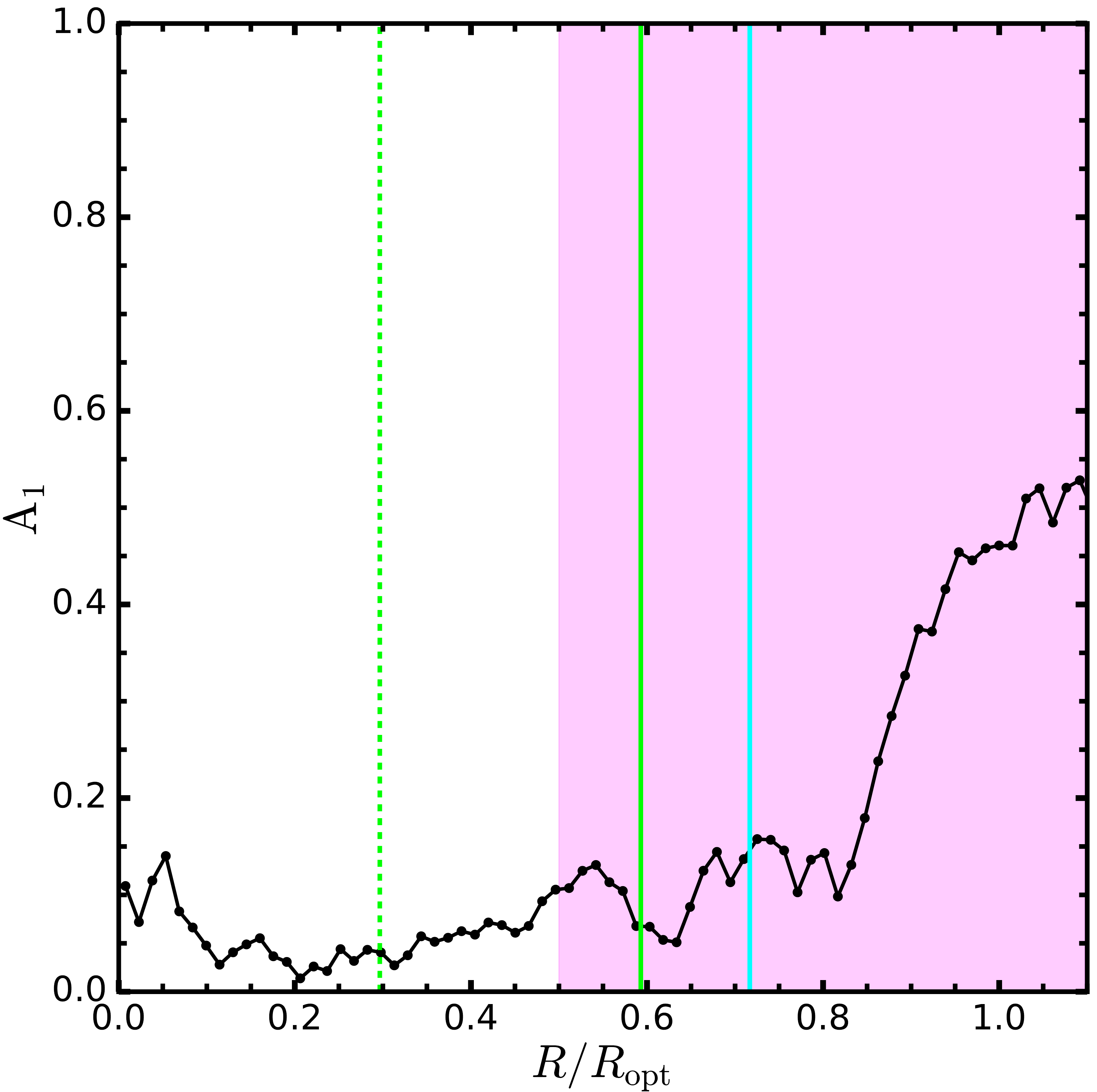}
    \includegraphics[width=0.325\textwidth]{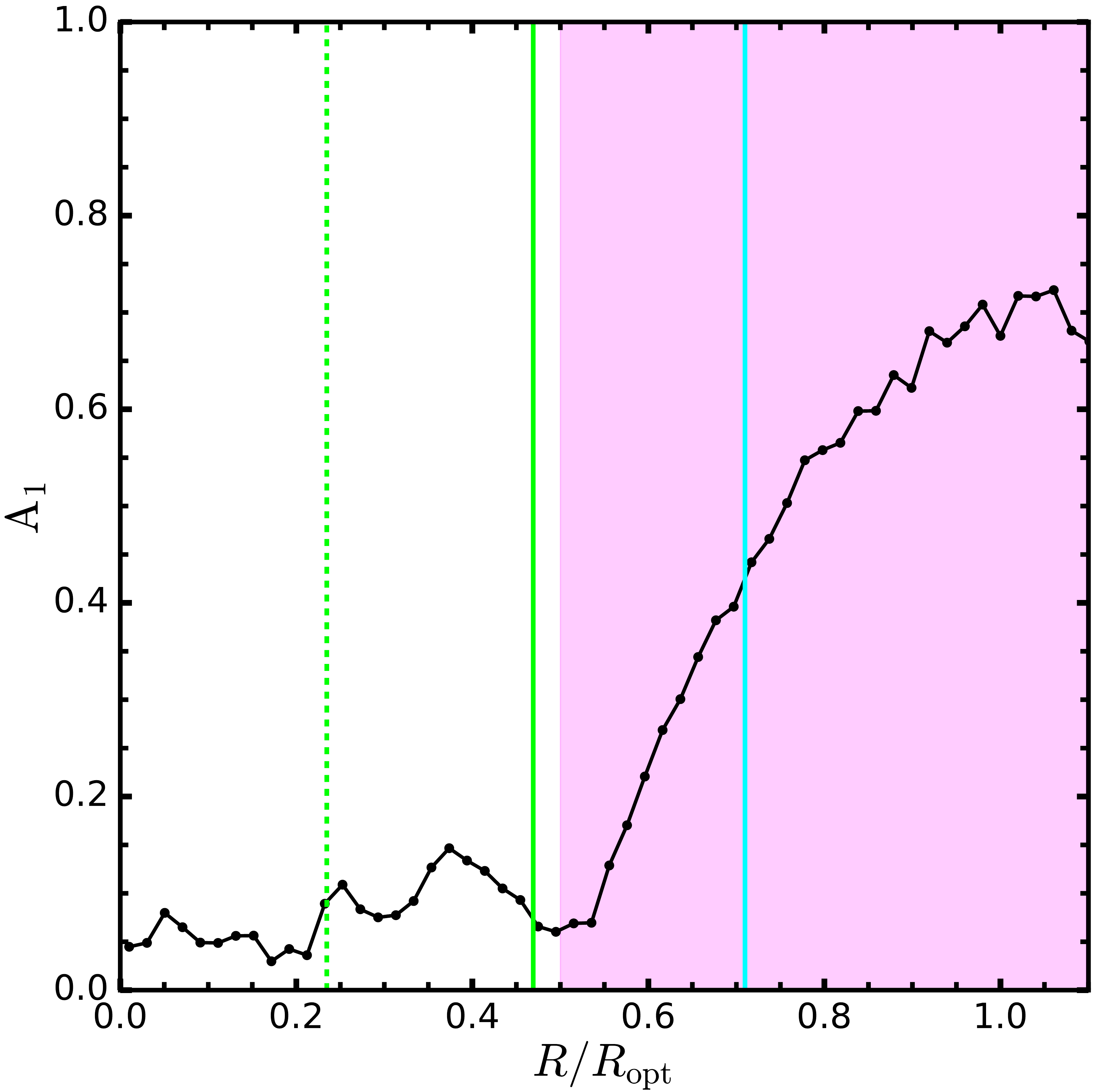}
    \includegraphics[width=0.325\textwidth]{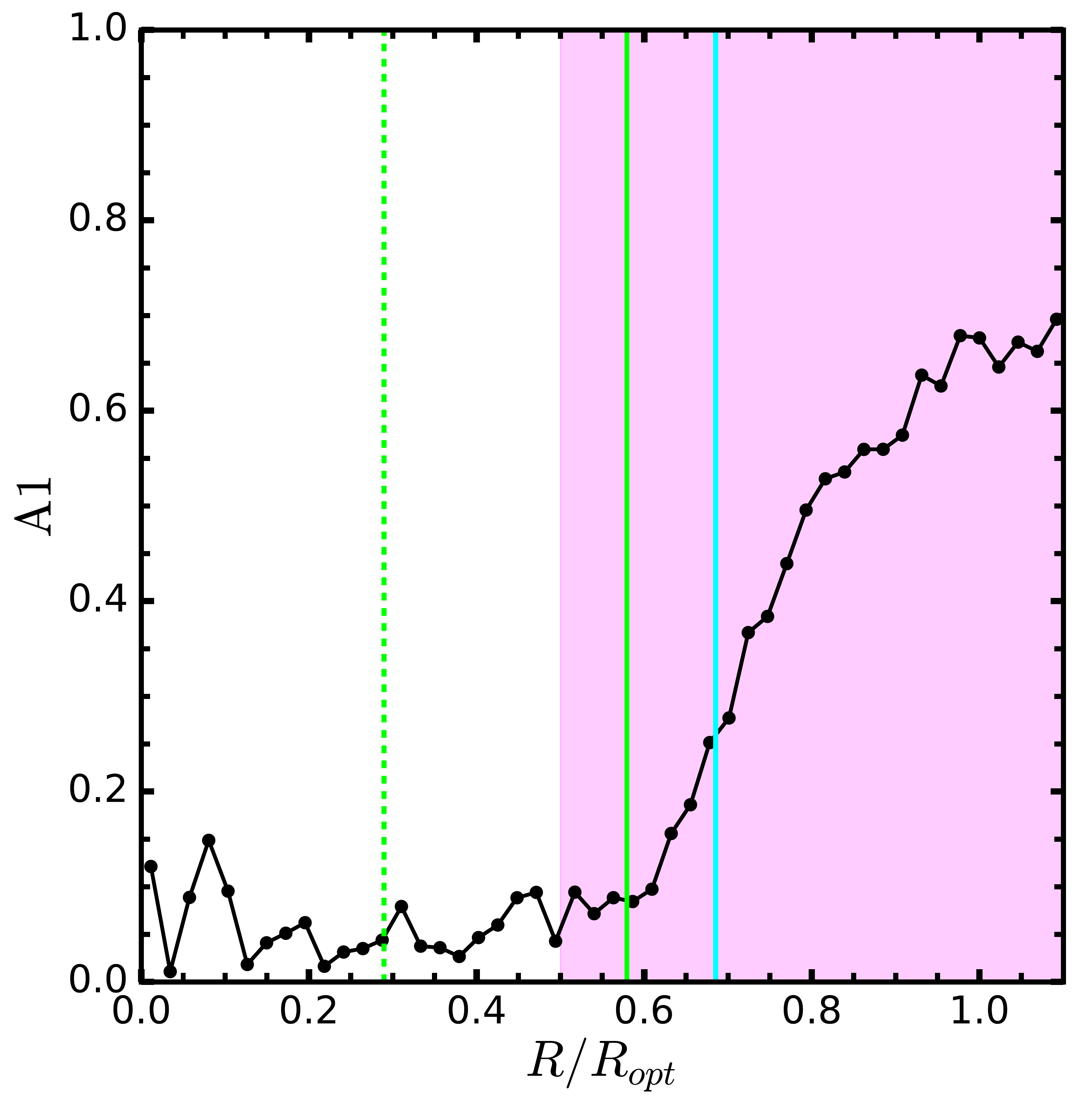}    
    \caption{{\bf Top:} The face-on projection of the V-band stellar surface brightness at $z=0$ of three example galaxies in our selected galaxy sample. The size of each image (i.e. $40\, \mathrm{kpc}$) is indicated in the bottom right corner, while the surface brightness colour mapping ranges from a minimum of $27\, \mathrm{mag\, arcsec^{-2}}$ up to a maximum of $19\, \mathrm{mag\, arcsec^{-2}}$.
    The magenta and green circles represent the radial interval between $0.5$-$1.1\, R_{\mathrm{opt}}$ and $1$-$2\, R_{\mathrm{half}}$, respectively, while the cyan circle represents $R_{90}$. These were the radii adopted to measure the global lopsidedness in previous works (\citealt{Reichard2008,Lokas2022};\citetalias{Varela-Lavin2022}). {\bf Bottom:} The corresponding radial profile of the lopsidedness for the three example galaxies shown in the top panels. The coloured lines correspond to the different radial intervals adopted to measure the lopsidedness in previous works, as described above. We see that all three galaxies are clearly asymmetric, with the lopsidedness amplitude increasing steeply towards large galactocentric radii, and that the radial interval $0.5$-$1.1\, R_{\mathrm{opt}}$ is best at detecting and quantifying the full lopsided feature of the galaxies.}
    \label{fig:faceon_images_examples}
\end{figure*}

\section{Lopsidedness}
\label{sec:lopsidedness}

\subsection{Measuring lopsidedness: Fourier analysis}
\label{sec:lopsidedness_definition}
In this section, we describe the method used to measure asymmetries in the mass distribution of the stellar component of disk-like galaxies, which are typically referred to as lopsidedness.

We calculate the lopsidedness of the galaxies in our selected sample by performing an azimuthal Fourier decomposition of the galaxy stellar mass distribution and measuring the amplitude of the first Fourier mode, $m=1$, as similarly done in previous works (e.g. \citealt{Rix1995,Reichard2008,Lokas2022};\citetalias{Varela-Lavin2022}). The amplitude of the first mode of the Fourier decomposition, i.e. $\mathrm{A_{1}}$, quantifies the lopsidedness of the galaxy which is the large-scale asymmetry in the stellar mass density distribution between one side of the galaxy and the opposite side.

Firstly, we rotate all our galaxies such that the angular momentum vector of the stars is aligned with the $z$-axis (i.e. the stellar disk lies in the $xy$-plane). Then, we bin the galaxy stellar mass in equally-spaced concentric radial annuli of width $0.5\, \mathrm{kpc}$ in the face-on projection. We then apply the Fourier transform to the stellar mass density distribution and we calculate the complex Fourier coefficients within each radial annulus, which are defined as \citep{Grand2016}:

\begin{equation}
    C_{n}(R_{j},t) = \sum_{N} m_{i} e^{-in\theta_{i}},
\label{eq:fourier_decomposition}
\end{equation}

where $m_{i}$ and $\theta_{i}$ is the mass and azimuthal angle of the $i$th stellar particle found within a given radial annulus. The amplitude of the $n$th Fourier mode at a given radial annulus is then given by:

\begin{equation}
    B_{n}(R_{j},t) = \sqrt{ a_{n}(R_{j},t)^{2} + b_{n}(R_{j},t)^{2} },
\end{equation}

with $a_{n}(r)$ and $b_{n}(r)$ being the real and imaginary components of the complex Fourier coefficients in Eq. \ref{eq:fourier_decomposition} within the radial annulus. 
These two components are defined as:

\begin{equation}
\begin{split}
    a_{n}(R_{j},t) = \sum_{i} \mathrm{m}_{i} \cos(n\theta_{i}) \\
    b_{n}(R_{j},t) = \sum_{i} \mathrm{m}_{i} \sin(n\theta_{i}),
\end{split}
\end{equation}

where $\theta_{i}$ is the azimuthal angle and $\mathrm{m}_{i}$ is the mass of the $i$th stellar particle found within a given radial annulus. The sum runs over all stellar particles in the radial annulus under consideration. Finally, the ratio between the amplitude of the first Fourier mode $n=1$ and the zeroth mode $n=0$, measured at each radial annulus is referred to as the lopsidedness parameter, $A_{1}$:

\begin{equation}
    \mathrm{A}_{1}(R_{j},t) = \frac{B_{1}(R_{j},t)}{B_{0}(R_{j},t)},
\label{eq:lopsidedness}
\end{equation}

where the amplitude of the zeroth Fourier mode, $B_{0}(R_{j},t)$, corresponds to the total stellar mass of the stellar particles in the radial annulus. The averaged value of $\mathrm{A_{1}}$ measured over a chosen radial interval (e.g. between $1.5$-$2.5$ disk scale lengths as typically adopted in observations; \citealt{Jog2009}) defines the global lopsidedness amplitude of the galaxy. Then, if $\mathrm{A_{1}}>0.1$, the galaxy is said to be lopsided. Instead, if $\mathrm{A_{1}}<0.1$, the galaxy is said to be symmetric.

\subsection{Global lopsidedness}
\label{sec:global_lopsidedness}
Following \citetalias{Varela-Lavin2022}, to quantify the present-day global lopsidedness of the galaxies in our selected sample, we calculate the average $\mathrm{A}_{1}$ amplitude considering the radial annulus between $0.5$-$1.1\, R_{\mathrm{opt}}$. Here, $R_{\mathrm{opt}}$ is the optical radius of the galaxy defined as the radius where the V-band surface brightness drops down to $\mu_{\mathrm{V}} = 26.5\, \mathrm{mag\, arcsec^{-2}}$. 
Similarly to the lopsidedness, we estimate $R_{\mathrm{opt}}$ by binning the face-on projection of the galaxy stellar light in radial annuli of width $0.5\, \mathrm{kpc}$ and measuring the stellar surface brightness profile normalized by the area of the circular shell in units of $\mathrm{arcsec^{2}}$.  

\begin{figure*}
    \centering
    \includegraphics[width=0.44\textwidth]{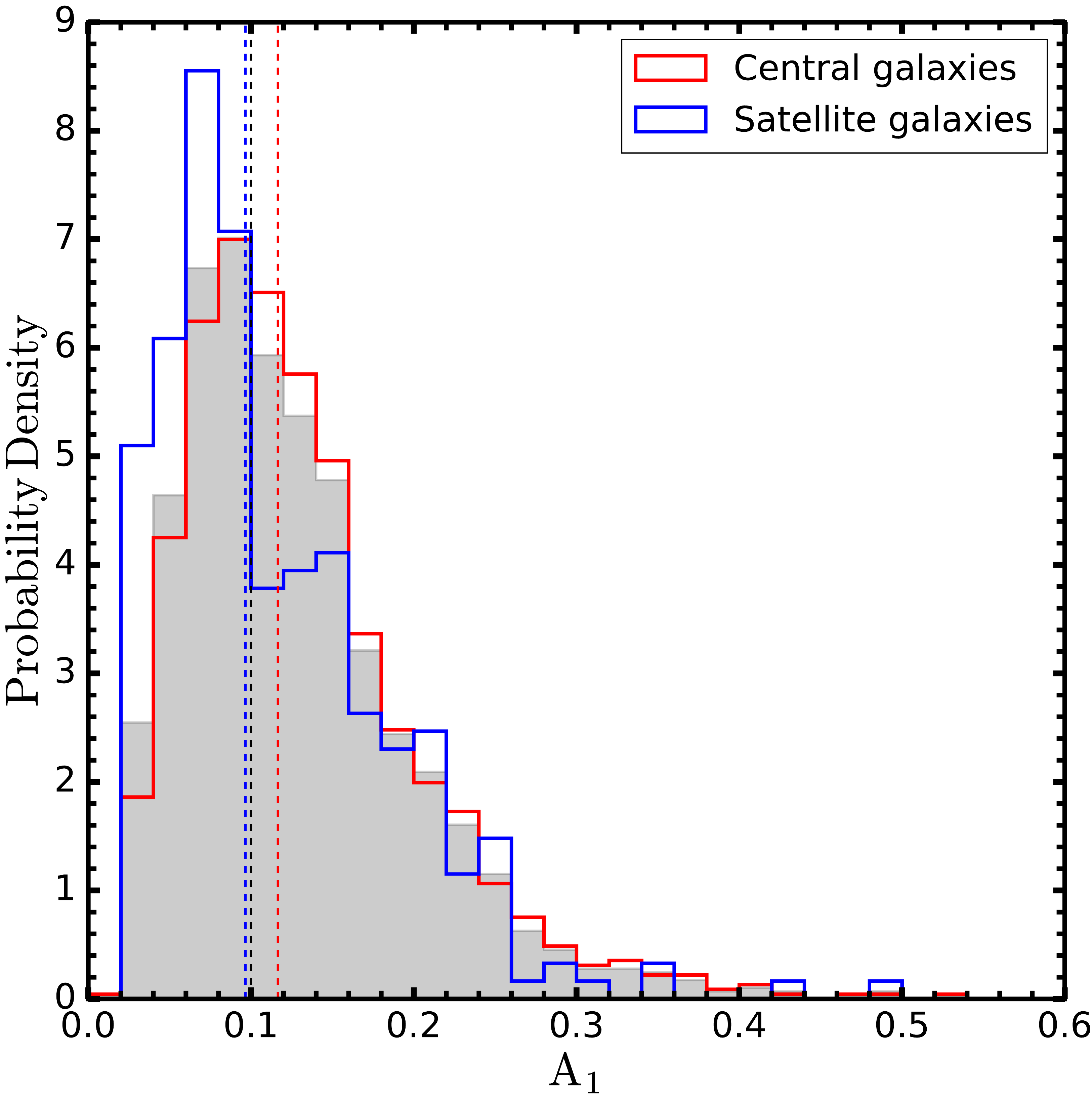}
    \includegraphics[width=0.45\textwidth]{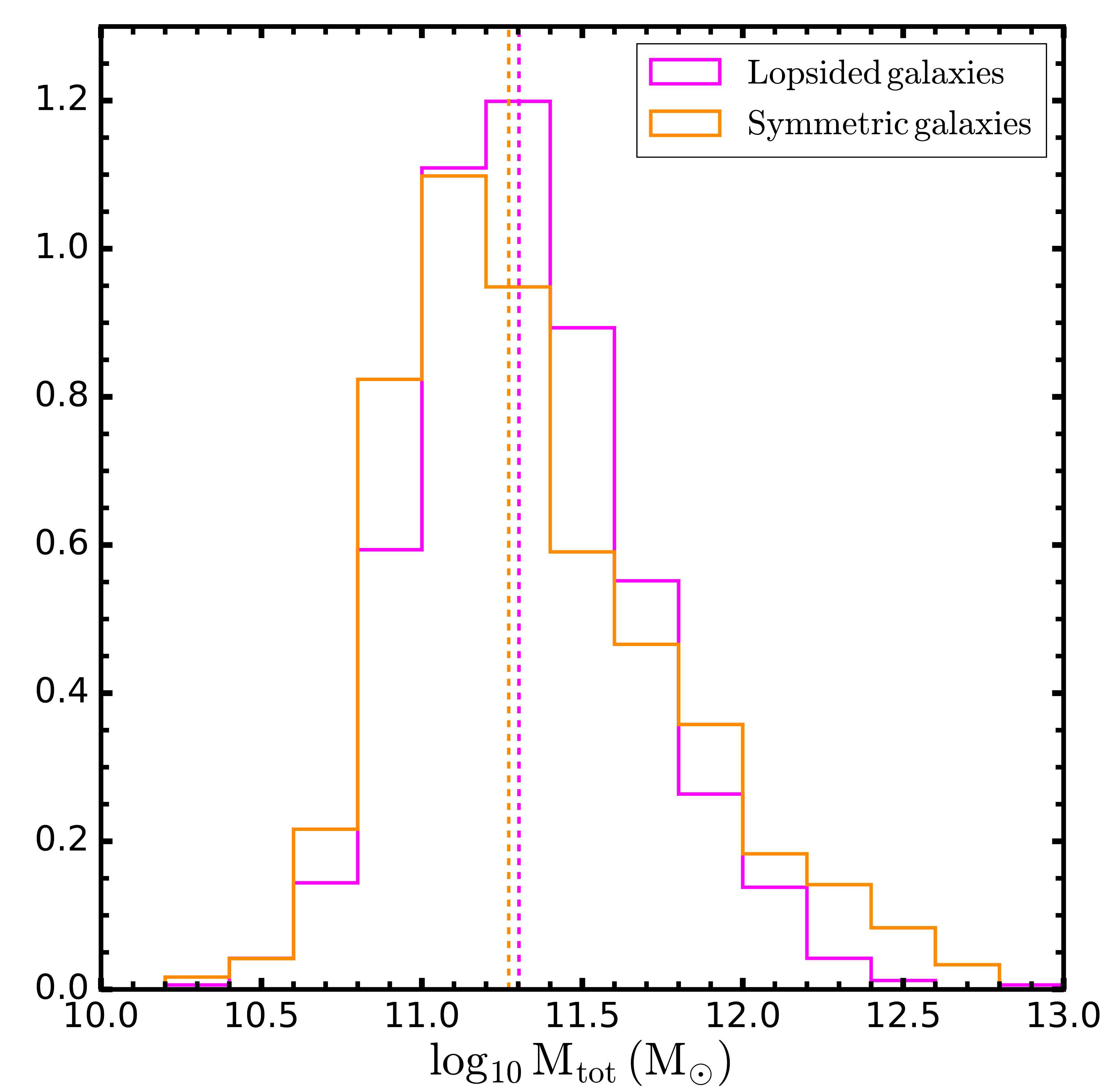}
    \includegraphics[width=0.45\textwidth]{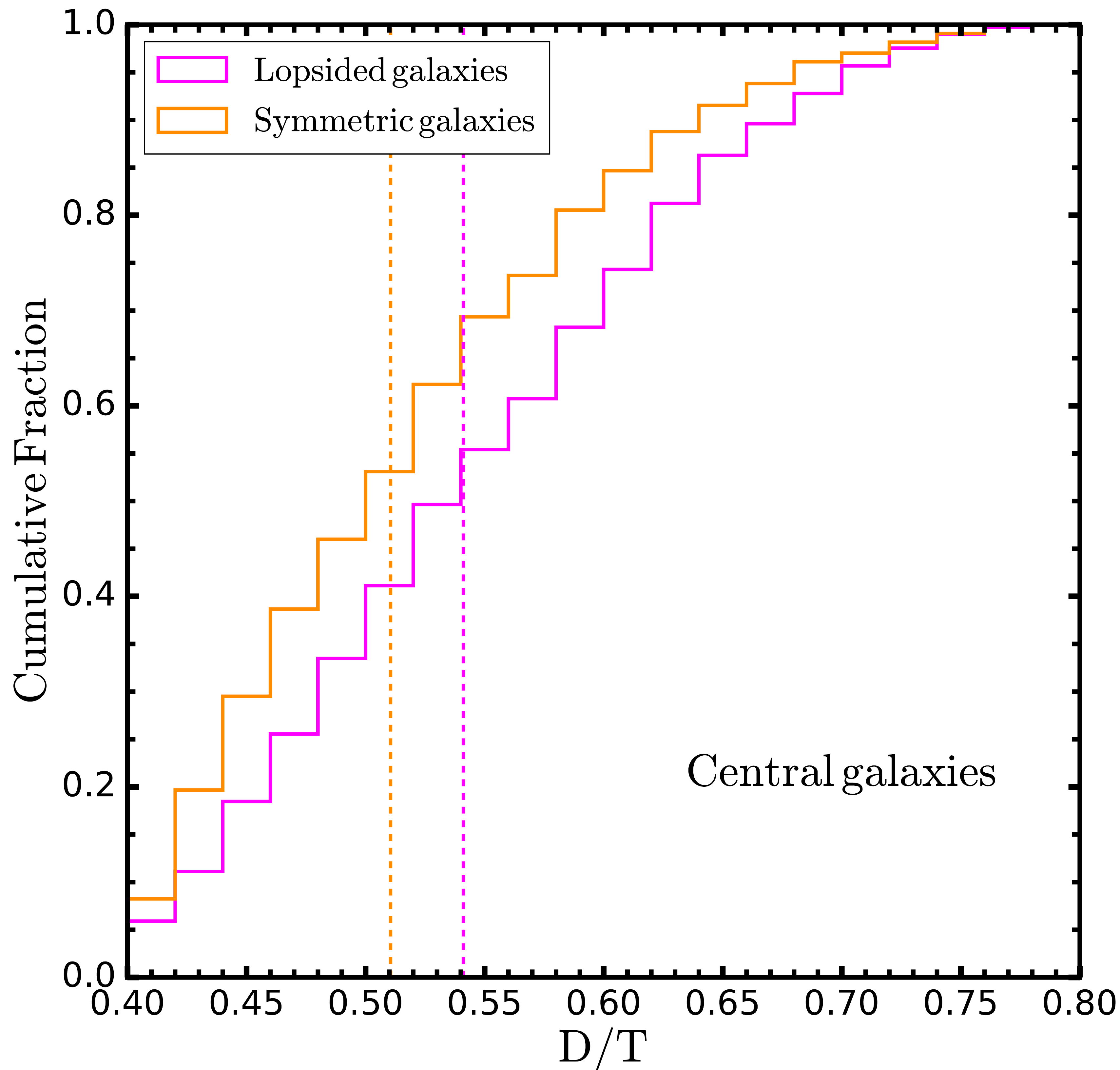}
    \includegraphics[width=0.45\textwidth]{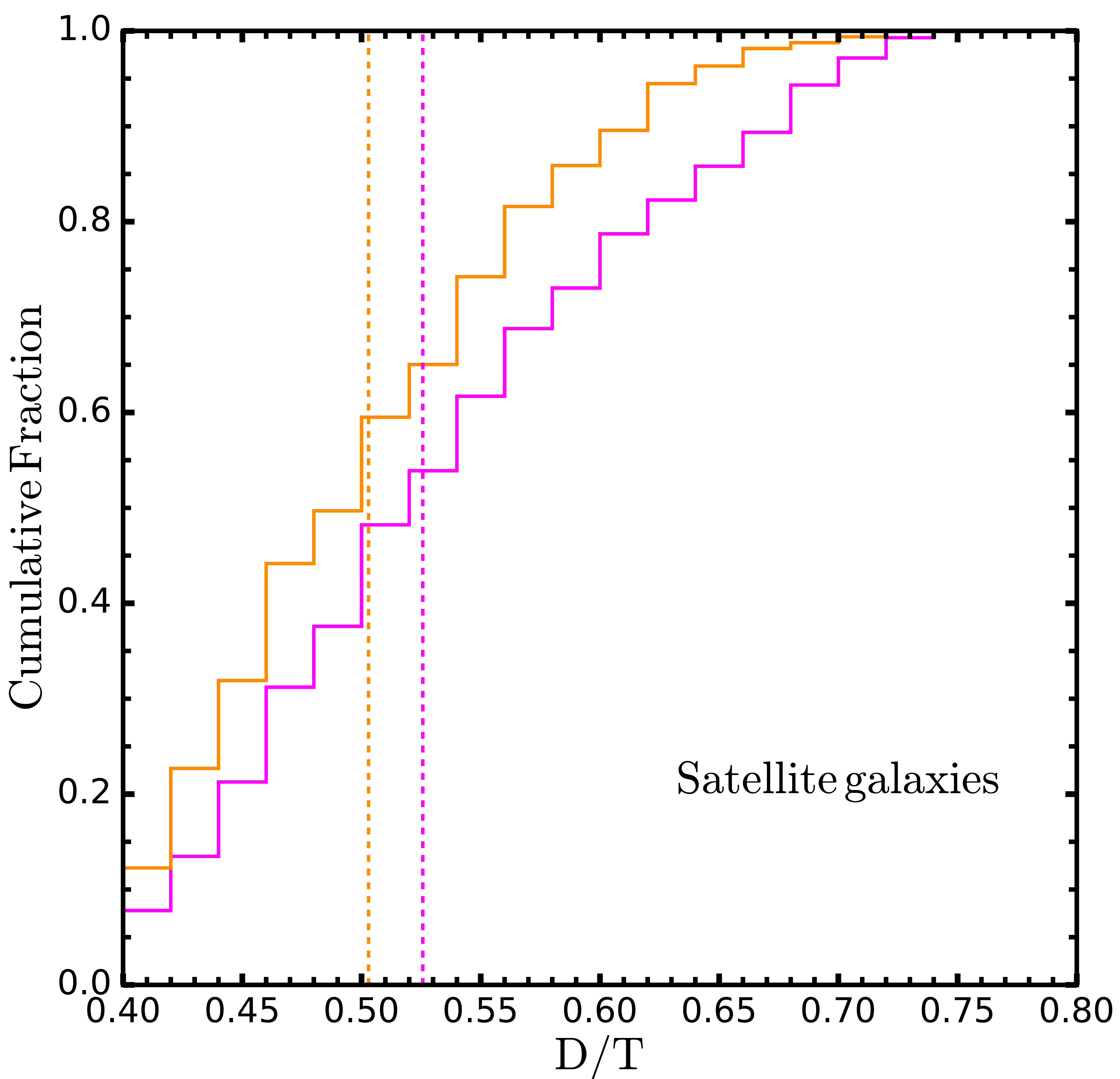}
    \caption{{\bf Top left:} The distribution of the present-day lopsidedness amplitudes of the galaxies in our selected galaxy sample, divided by centrals ($693$ lopsided, $438$ symmetric) and satellites ($141$ lopsided, $163$ symmetric). The global lopsidedness, $\mathrm{A_{1}}$, is averaged over the radial interval between $0.5$-$1.1\, R_{\mathrm{opt}}$. The grey, red and blue histograms represent the lopsidedness distribution of all, central and satellite galaxies, respectively, normalized such that the area underneath integrates to one. The black vertical line represents the threshold $\mathrm{A}_{1}=0.1$ adopted to distinguish between lopsided ($\mathrm{A}_{1}>0.1$) and symmetric ($\mathrm{A}_{1}<0.1$) galaxies, while the red and blue vertical lines represent the median $\mathrm{A}_{1}$ of all the central and satellite galaxies, respectively. {\bf Top right:} The total mass distribution of our selected galaxy sample, divided by lopsided ($A_{1}>0.1$) and symmetric ($A_{1}<0.1$) galaxies. The vertical dashed lines represent the median values of the total mass distributions of the lopsided and symmetric galaxies. {\bf Bottom left \& right:} The cumulative distribution of the $\mathrm{D/T}$ of the lopsided and symmetric galaxies, divided by centrals and satellites. We find that central galaxies are on average more lopsided than satellites galaxies and that symmetric galaxies have on average lower $\mathrm{D/T}$ than lopsided galaxies, indicating a possible correlation between lopsidedness and internal properties of the galaxies.}
    \label{fig:lopsidedness_distribution}
\end{figure*}

In previous works, the global lopsidedness (hereafter, $\mathrm{A}_{1}$) of the galaxy was typically measured between $1.5$-$2.5$ disk scale lengths (see e.g. \citealt{Jog2009} for a review). On the other hand, \citet{Reichard2008} measured $\mathrm{A}_{1}$ between $1\, R_{e}$-$R_{90}$, where $R_{\mathrm{e}}$ and $R_{90}$ are the radii containing $50\%$ and $90\%$ of the total galaxy light in the $z$-band. This outer radial limit of the galaxy is typically set by the sky-background noise of the observations in the near infrared. Using the TNG100 simulation, \citet{Lokas2022} measured $\mathrm{A}_{1}$ between $1$-$2\, R_{\mathrm{half}}$, which is expected to roughly correspond to the radial interval $1.5$-$2.5$ disk scale lengths used in the observations. 

In Fig. \ref{fig:faceon_images_examples}, we compare the measurements of the global $\mathrm{A}_{1}$ estimated between $0.5$-$1.1\, R_{\mathrm{opt}}$ (magenta lines/area) and $1$-$2\, R_{\mathrm{half}}$ (green lines) for three example galaxies in our selected galaxy sample. The top panels of Fig. \ref{fig:faceon_images_examples} show the face-on projection of the V-band stellar surface brightness of the galaxies at $z=0$, which is clearly asymmetric in all three cases. 
The lopsided feature in these three galaxies is also confirmed by the radial $\mathrm{A_{1}}$ profiles shown in the bottom panels of Fig. \ref{fig:faceon_images_examples}, where we see a steep increase in $\mathrm{A}_{1}$ towards large galactocentric radii, starting from $\sim0.5\, R_{\mathrm{opt}}$ and reaching out to the optical limit of the galaxy. This increment of the lopsidedness amplitude with the galactocentric radius is overall consistent with the trend seen in previous works (\citealt{Reichard2008,Jog2009,Lokas2022}\citetalias{Varela-Lavin2022}. 
In Fig. \ref{fig:faceon_images_examples}, we see that we miss almost all of the strong lopsided feature if we consider the radial interval between $1$-$2\, R_{\mathrm{half}}$ (green lines), which is the one adopted by \citet{Lokas2022}.
The radial interval out to $R_{90}$ adopted by \citet{Reichard2008} (cyan line) is better in comparison to $1$-$2\, R_{\mathrm{half}}$ as it reaches out to larger galactocentric radii, but it still underestimates the lopsidedness compared to the value measured within the radial interval between $0.5$-$1.1\, R_{\mathrm{opt}}$. The latter allows us to detect and quantify the full lopsided feature as it reaches out to the optical extension of the galaxies. 

For this reason, we measure the global lopsidedness $\mathrm{A_{1}}$ of our selected galaxies as the average over the radial interval between $0.5$-$1.1\, R_{\mathrm{opt}}$, similarly to \citetalias{Varela-Lavin2022}. The inner $0.5\, R_{\mathrm{opt}}$ threshold is chosen because this is typically the point beyond which the lopsidedness amplitude $\mathrm{A_{1}}$ starts to steadily increase with radius out to the outskirts $1.1\, R_{\mathrm{opt}}$ for the galaxies in our selected galaxy sample. Overall, we find that $834$ out of $1435$ galaxies (i.e. $58\%$) are lopsided (i.e. $\mathrm{A_{1}}>0.1$). This fraction is in good agreement with the percentage (i.e. $50\%$) of lopsided galaxies reported by the early work of \citet{Baldwin1980} from HI studies. We note that the fraction of lopsided galaxies would drop to $10\%$ and $33\%$ if we adopt the radial interval between $1$-$2\, R_{\mathrm{half}}$ and $1\, R_{\mathrm{e}}$-$R_{90}$, respectively, which is in better agreement with the fraction found in observations (i.e. $\sim30\%$). However, the fraction of lopsided galaxies is significantly higher than that in observations if we extend our measurements to the disk optical radius, $R_{\mathrm{opt}}$, suggesting that deep observations of the galaxy outskirts are necessary to detect the lopsided feature.

The top left-hand side panel of Fig. \ref{fig:lopsidedness_distribution} shows the distribution of the global lopsidedness amplitudes of our selected galaxies, divided by centrals ($693$ lopsided, $438$ symmetric) and satellites ($141$ lopsided, $163$ symmetric), while the top right-hand side panel of Fig. \ref{fig:lopsidedness_distribution} shows the total mass distribution of our selected galaxies, divided by lopsided and symmetric.
We see that the lopsidedness distribution of our galaxies is very broad, and it can reach values as large as $\mathrm{A_{1}}\sim0.6$\footnote{We note that two galaxies in our sample also have lopsidedness amplitudes $\mathrm{A_{1}}>0.6$, i.e. $\mathrm{ID}=600096$ with $\mathrm{A}_{1}\simeq0.65$ and $\mathrm{ID}=645391$ with $\mathrm{A}_{1}\simeq0.83$.}. We also see that central galaxies are on average more lopsided than the satellite galaxies, while both lopsided and symmetric galaxies show similar total mass distributions.   
The bottom left- and right-hand side panels of Fig. \ref{fig:lopsidedness_distribution} show the cumulative distribution of the $\mathrm{D/T}$ for the lopsided and symmetric subsamples, divided by centrals and satellites. For both central and satellite galaxies, we see that the symmetric subsample has, on average, lower $\mathrm{D/T}$ than the lopsided one, indicating a possible correlation between lopsidedness, environment and galaxy internal properties, which we will further investigate in the following Sec. \ref{sec:role_environment}. 


\begin{figure*}
    \centering
    \includegraphics[width=0.45\textwidth]{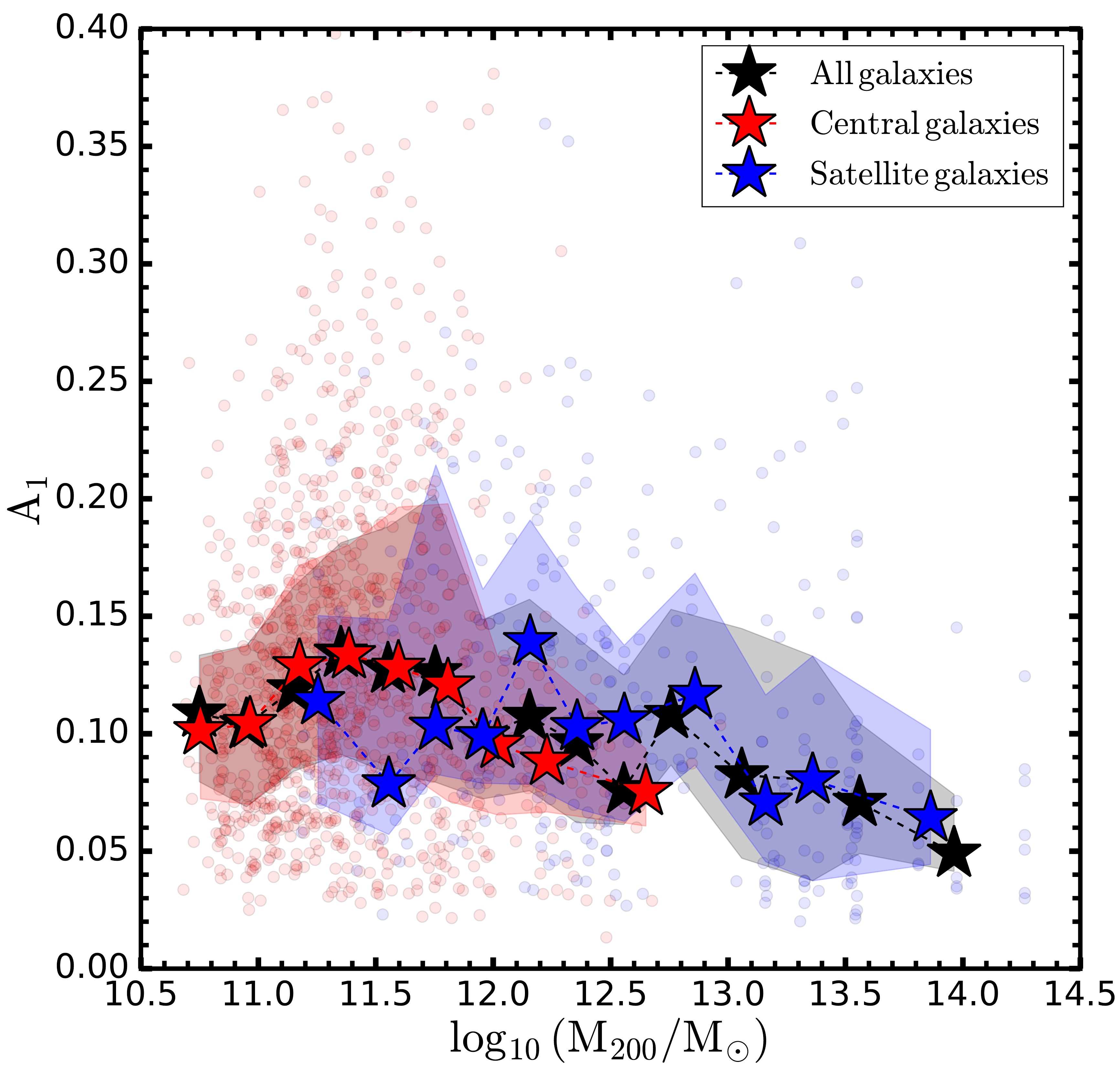}
    \includegraphics[width=0.45\textwidth]{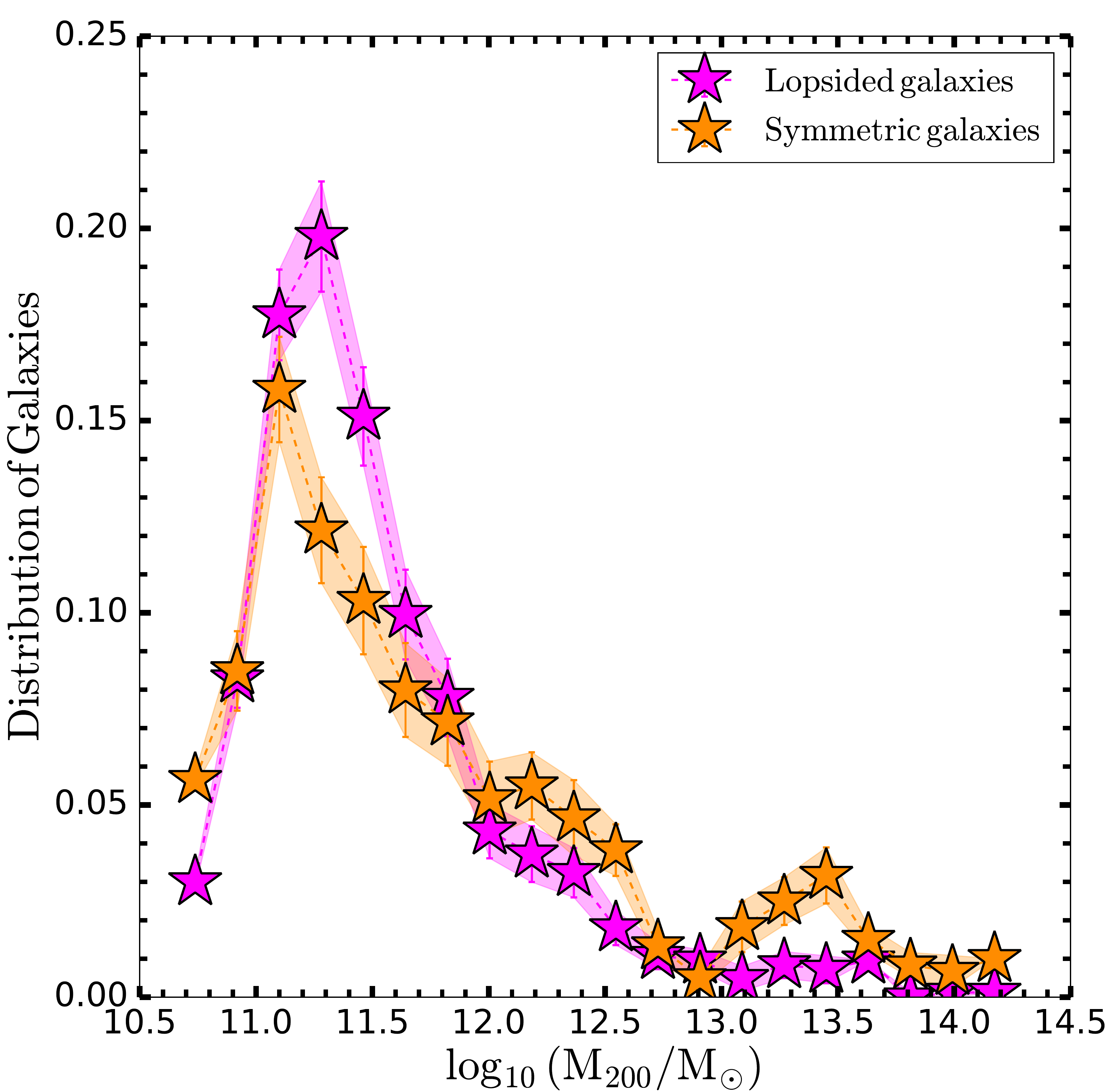}
    \caption{{\bf Left:} The median amplitude of the lopsidedness as a function of the total halo mass, $\mathrm{M_{200}}$, of the environment where our galaxies reside for all the galaxies (black stars) as well as for the central (red stars) and satellite (blue stars) galaxies, separately. The shaded areas represent the $25\mathrm{th}$-$75\mathrm{th}$ interquartile range of the data in each bin, while the background points represent the individual central and satellite galaxies. {\bf Right:} Distribution of the lopsided (magenta stars) and symmetric (orange stars) galaxies as a function of the total halo mass, $\mathrm{M_{200}}$, of the environment where our galaxies reside. The histogram is normalized to unity. The shaded areas represent the corresponding $1\sigma$ errors, calculated from the standard deviation of $500$ bootstrap samples obtained by sampling with replacement our original selected galaxy sample. Overall, we see that the lopsidedness amplitude shows a mild decrease with increasing $\mathrm{M}_{200}$.}
    \label{fig:lopsidedness_environment_dependence_m200}
\end{figure*}

\begin{figure*}
    \centering
    \includegraphics[width=0.33\textwidth]{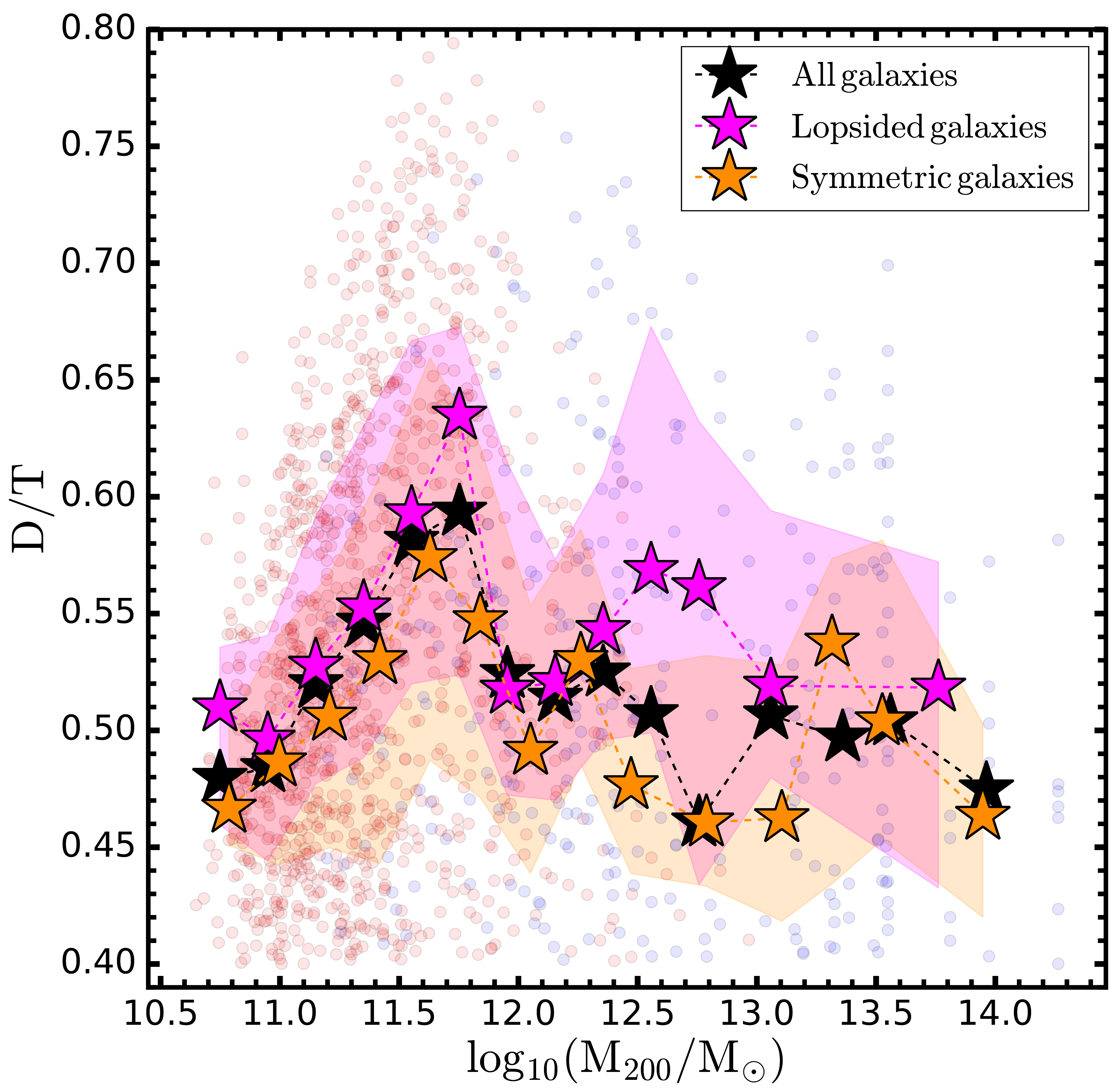}
    \includegraphics[width=0.325\textwidth]{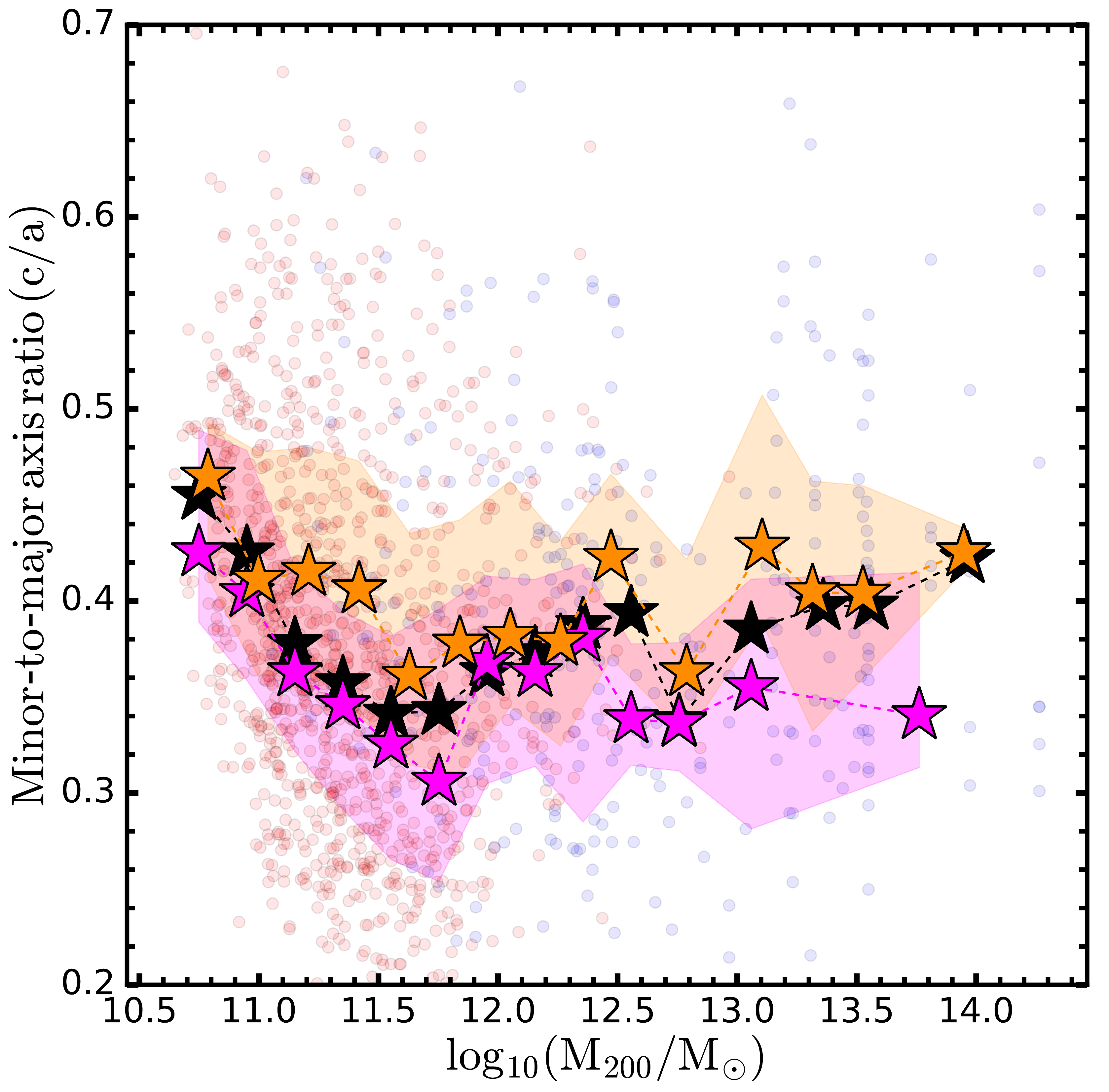}
    \includegraphics[width=0.33\textwidth]{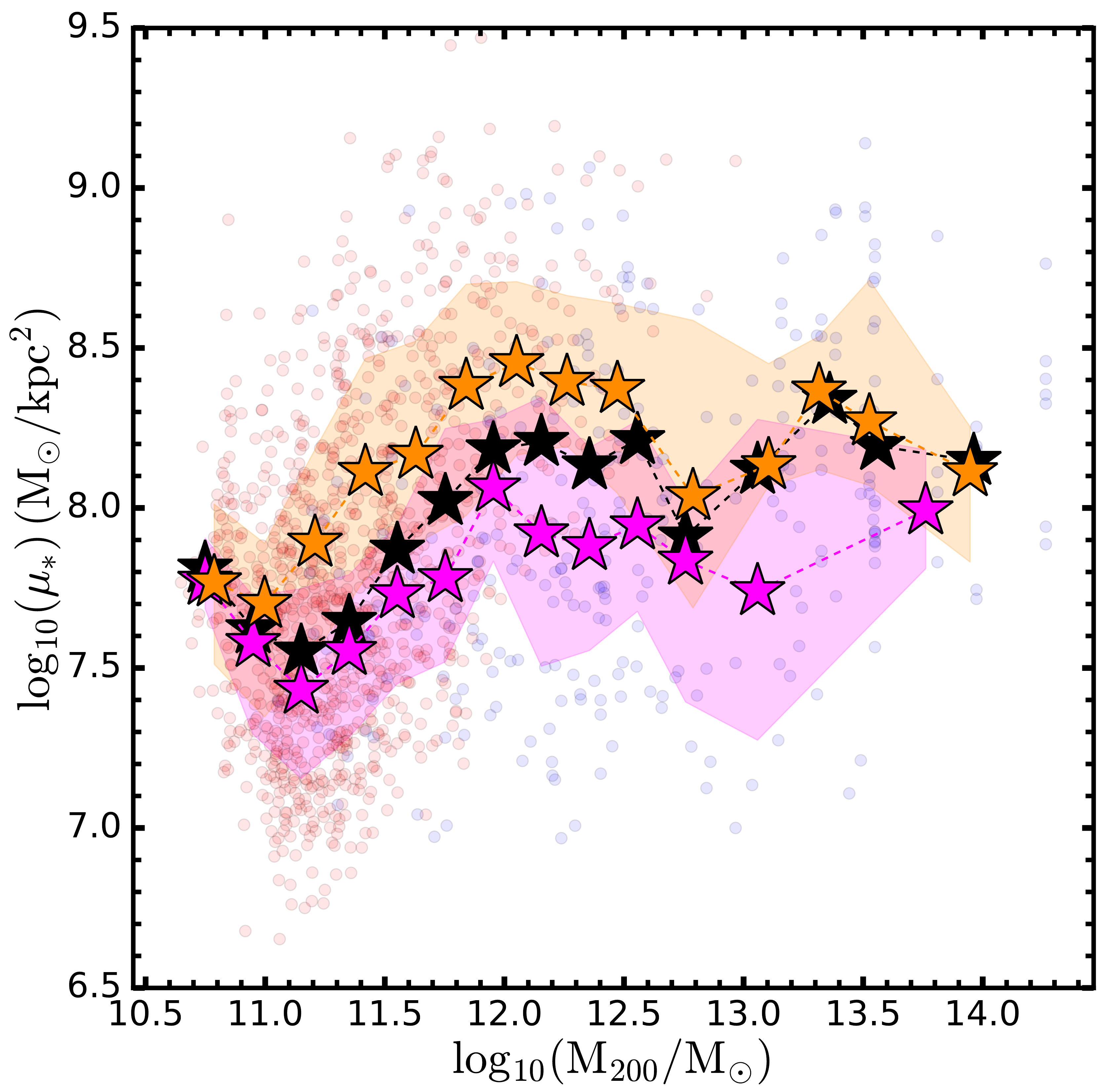}
    \caption{From the left to the right panel, the median disk-to-total ratio ($\mathrm{D/T}$), minor-to-major axis ratio ($\mathrm{c/a}$) and central stellar mass density ($\mu_{*}$) as a function of the total halo mass, $\mathrm{M_{200}}$, respectively, for all (black stars), only the lopsided (magenta stars) and only the symmetric (orange stars) galaxies in our selected galaxy sample. The shaded areas represent the $25\mathrm{th}$-$75\mathrm{th}$ interquartile range of the data in each bin, while the background points represent the individual central and satellite galaxies. Overall, we see that galaxies at low halo masses (i.e. $10^{11} \lesssim \mathrm{M_{200}}/\mathrm{M_{\odot}} \lesssim 10^{12}$), which tend to be more lopsided in Fig. \ref{fig:lopsidedness_environment_dependence_m200}, are characterized by a later type morphology with more extended disks, flatter inner galactic regions and lower central stellar mass density than galaxies at high halo masses (i.e. $\mathrm{M_{200}}/\mathrm{M_{\odot}} \gtrsim 10^{12}$). Additionally, lopsided galaxies are typically characterized by larger $\mathrm{D/T}$, and smaller $\mathrm{c/a}$ and $\mu_{*}$ than symmetric ones regardless of the environment, suggesting a correlation between lopsidedness and the internal properties of the galaxies.}
    \label{fig:environment_dependence_m200}
\end{figure*}

\section{Effect of environment on lopsidedness}
\label{sec:role_environment}

\subsection{Correlations between lopsidedness, environment mass and morphology}
\label{sec:environment_halomass}
As we have seen in Sec. \ref{sec:environment}, our galaxy sample covers from low- ($\mathrm{M}_{200}\lesssim10^{13}\, \mathrm{M}_{\odot}$) to high-density ($\mathrm{M}_{200}\gtrsim10^{13}\, \mathrm{M}_{\odot}$) environments, including Fornax cluster-like environments. However, in high-density environments, we do not have as many galaxies as we have in low-density ones, as described in Sec. \ref{sec:environment}.

In Fig. \ref{fig:lopsidedness_environment_dependence_m200} we study, for central and satellite galaxies, the behaviour of the lopsidedness amplitude, $\mathrm{A}_{1}$, as a function of the total halo mass, $\mathrm{M}_{200}$, of the environment to which each galaxy belongs at $z=0$. To do this, we calculate the median value of $\mathrm{A_{1}}$ in bins of width $\Delta \mathrm{M_{200}} = 0.2\, \mathrm{M_{\odot}}$. For this procedure we require each bin to contain at least ten data points. If this condition is not satisfied, we merge contiguous bins.
We see that the overall median amplitude of $\mathrm{A}_{1}$ shows a mild decrease with increasing $\mathrm{M}_{200}$, indicating a larger fraction of symmetric galaxies than lopsided ones towards increasing $\mathrm{M}_{200}\gtrsim10^{12}\, \mathrm{M}_{\odot}$ (see black stars in the left panel of Fig. \ref{fig:lopsidedness_environment_dependence_m200}). The same decreasing behaviour of $\mathrm{A}_{1}$ with $\mathrm{M}_{200}$ is also seen for the central and satellite galaxies, separately. This result also indicates that satellite galaxies tend to be more symmetric than central ones since the majority of the galaxies are satellites at high halo masses (i.e. $\mathrm{M}_{200}\gtrsim10^{12}\, \mathrm{M}_{\odot}$; see Fig. \ref{fig:halo_mass}), thus explaining the differences in the lopsidedness distributions of central and satellite galaxies described in Fig. \ref{fig:lopsidedness_distribution}.

In Fig. \ref{fig:environment_dependence_m200}, we study the behaviour of the $\mathrm{D/T}$, minor-to-major axis ratio ($c/a$) and central stellar mass density ($\mu_{*}$) as a function of the $\mathrm{M}_{200}$ of the host galaxy group from the left to the right panel, respectively, similarly to Fig. \ref{fig:lopsidedness_environment_dependence_m200}. Here, $c/a$ is given by the eigenvalues of the mass tensor of the stellar component measured within the inner $2\, R_{\mathrm{half}}$, which is a proxy of the shape of the inner galactic regions. Specifically, small values of $c/a$ indicate flatter inner galactic regions, and large values of $c/a$ indicate rounder inner galactic regions. On the other hand, $\mu_{*} = \mathrm{M_{*\, 1/2}}/\pi R_{\mathrm{half}}^{2}$ is defined as the ratio between the stellar mass contained within $R_{\mathrm{half}}$ and the area enclosed within the circular region of radius $R_{\mathrm{half}}$ in the face-on projection of the galaxy.
We see that the mildly decreasing behaviour of $\mathrm{A}_{1}$ with increasing $\mathrm{M}_{200}$ (see Fig. \ref{fig:lopsidedness_environment_dependence_m200}) seems to correlate with the change in the internal properties of the galaxies with $\mathrm{M}_{200}$ (see black stars in Fig. \ref{fig:environment_dependence_m200}). 
Specifically, we see that the overall higher lopsidedness amplitude between $10^{11} \lesssim \mathrm{M_{200}}/\mathrm{M_{\odot}} \lesssim 10^{12}$ in Fig. \ref{fig:lopsidedness_environment_dependence_m200} roughly corresponds to the peak in $\mathrm{D/T}$ and dip in $c/a$ and $\mu_{*}$ in Fig. \ref{fig:environment_dependence_m200}.
Beyond $\mathrm{M_{200}} \sim 10^{11.5}\, \mathrm{M_{\odot}}$, where the overall lopsidedness amplitude starts decreasing, we see that the $\mathrm{D/T}$ also decreases while both $c/a$ and $\mu_{*}$ slightly increase. This also occurs at low $\mathrm{M_{200}} \lesssim 10^{11}\, \mathrm{M_{\odot}}$, where we see that the galaxies are, on average, characterized by low $\mathrm{D/T}$ and high $c/a$ and $\mu_{*}$.

The results in Fig. \ref{fig:lopsidedness_environment_dependence_m200} and \ref{fig:environment_dependence_m200} suggests a correlation between lopsidedness and galaxy morphology, where galaxies with more extended disks\footnote{As shown in Appendix \ref{sec:rhalf}, lopsided galaxies show, on average, larger stellar $R_\mathrm{half}$ and $R_{90}$ than the symmetric counterpart (in addition to the, on average larger $\mathrm{D/T}$ in Fig. \ref{fig:environment_dependence_m200})}, flatter inner galactic regions and lower central stellar mass density (i.e. late-type disk galaxies) are typically more lopsided than galaxies with smaller disks, rounder inner galactic regions and larger central stellar mass density (i.e. early-type disk galaxies). As a result, the mild decrease of $\mathrm{A_{1}}$ towards high $\mathrm{M_{200}}$ is correlated to the change in morphology of the galaxies from late- to early-type disks towards high-density environments. In fact, the change in galaxy morphology with the environment seen in Fig. \ref{fig:environment_dependence_m200}, where galaxies are characterized by, on average, smaller $\mathrm{D/T}$, and larger $c/a$ and $\mu_{*}$ at high $\mathrm{M_{200}}$ than at low ones, is consistent with morphology-density relation according to which the fraction of early-type disk galaxies increases towards high-density environments \citep{Dressler1980,Dressler1997,Fasano2000}.

The result that late-type disk galaxies tend to be more lopsided than early-type ones in low density environments as well as the observed decrease of the median lopsidedness amplitude towards high density environments (i.e. large galaxy groups and clusters; $\mathrm{M_{200}}\gtrsim10^{13}\, \mathrm{M_{\odot}}$) is consistent with the findings from previous works, which showed that lopsided galaxies tend to have lower concentration and stellar surface mass density than symmetric ones (\citealt{Conselice2000,Reichard2008};\citetalias{Varela-Lavin2022}). Thus, \citetalias{Varela-Lavin2022} suggested that lopsidedness is a consequence of galaxies being less gravitationally cohesive and more susceptible to being perturbed compared to compact galaxies, as a result of the same external perturbation.

In Fig. \ref{fig:environment_dependence_m200}, we see that the lopsided and symmetric galaxies show similar $\mathrm{D/T}$, $c/a$ and $\mu_{*}$ behaviours as a function of $\mathrm{M_{200}}$, and they are both characterized by a late-type disk morphology at low halo masses (i.e. $10^{11} \lesssim \mathrm{M_{200}}/\mathrm{M_{\odot}} \lesssim 10^{12}$) and early-type disk morphology at high halo masses ($\mathrm{M_{200}} \sim 10^{12}\, \mathrm{M_{\odot}}$). Nonetheless, we see that the lopsided galaxies always tend to be characterized by, on average, larger $\mathrm{D/T}$, and smaller $c/a$ and $\mu_{*}$ than the symmetric galaxies at all halo masses (i.e. regardless of the environment), confirming the correlation between lopsidedness and galaxy morphology. This is, galaxies characterized by large $\mathrm{D/T}$, and small $c/a$ and $\mu_{*}$ are more prone to develop lopsidedness than galaxies with small $\mathrm{D/T}$, and large $c/a$ and $\mu_{*}$, as a result of a same external perturbation.

Even though our analysis is based on a limiting sampling of high density environments in TNG50, our results suggest a strong correlation between lopsidedness and the internal galaxy properties regardless of the environment.

\begin{figure*}
    \centering
    \includegraphics[width=0.3\textwidth]{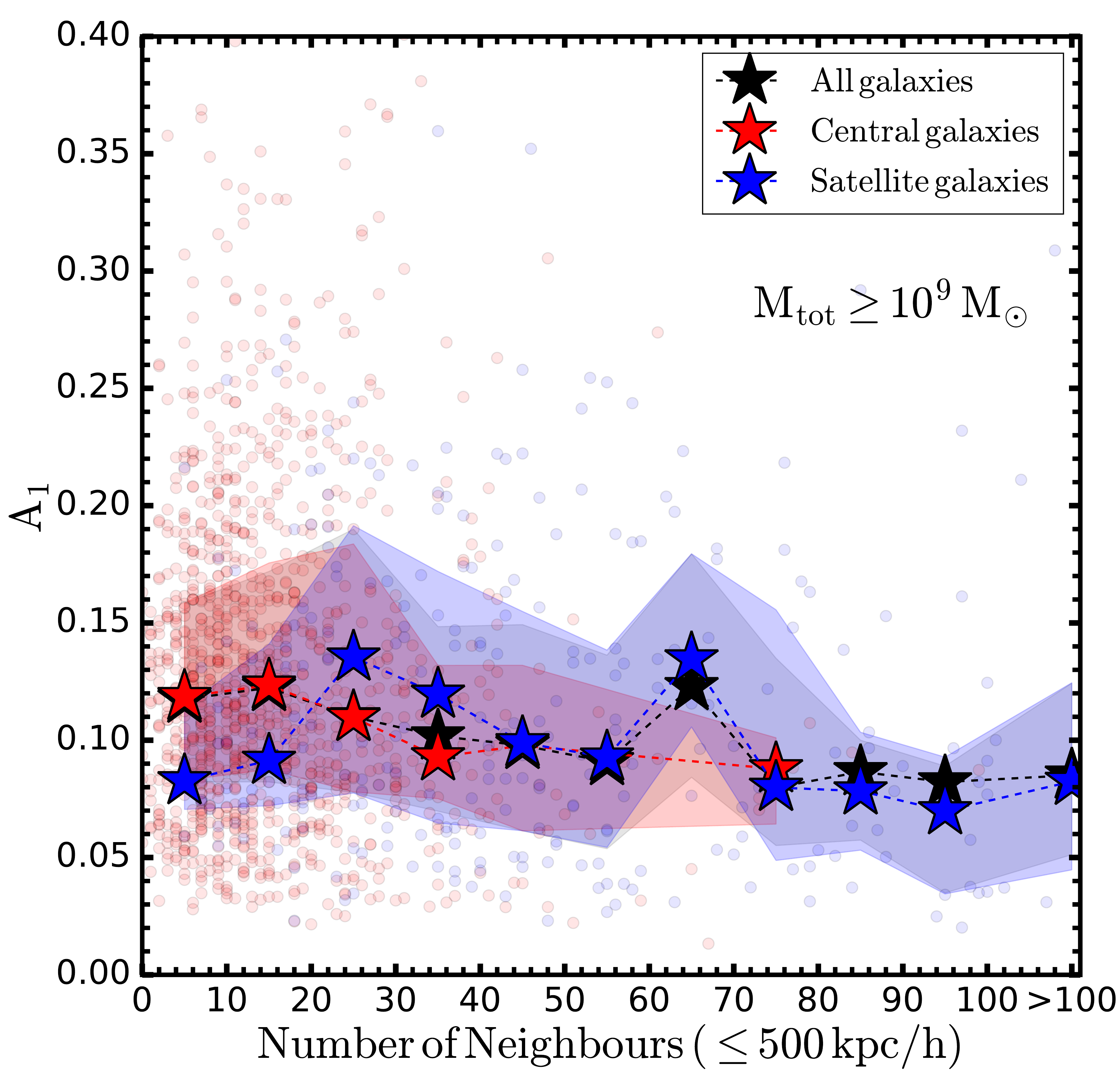}
    \includegraphics[width=0.3\textwidth]{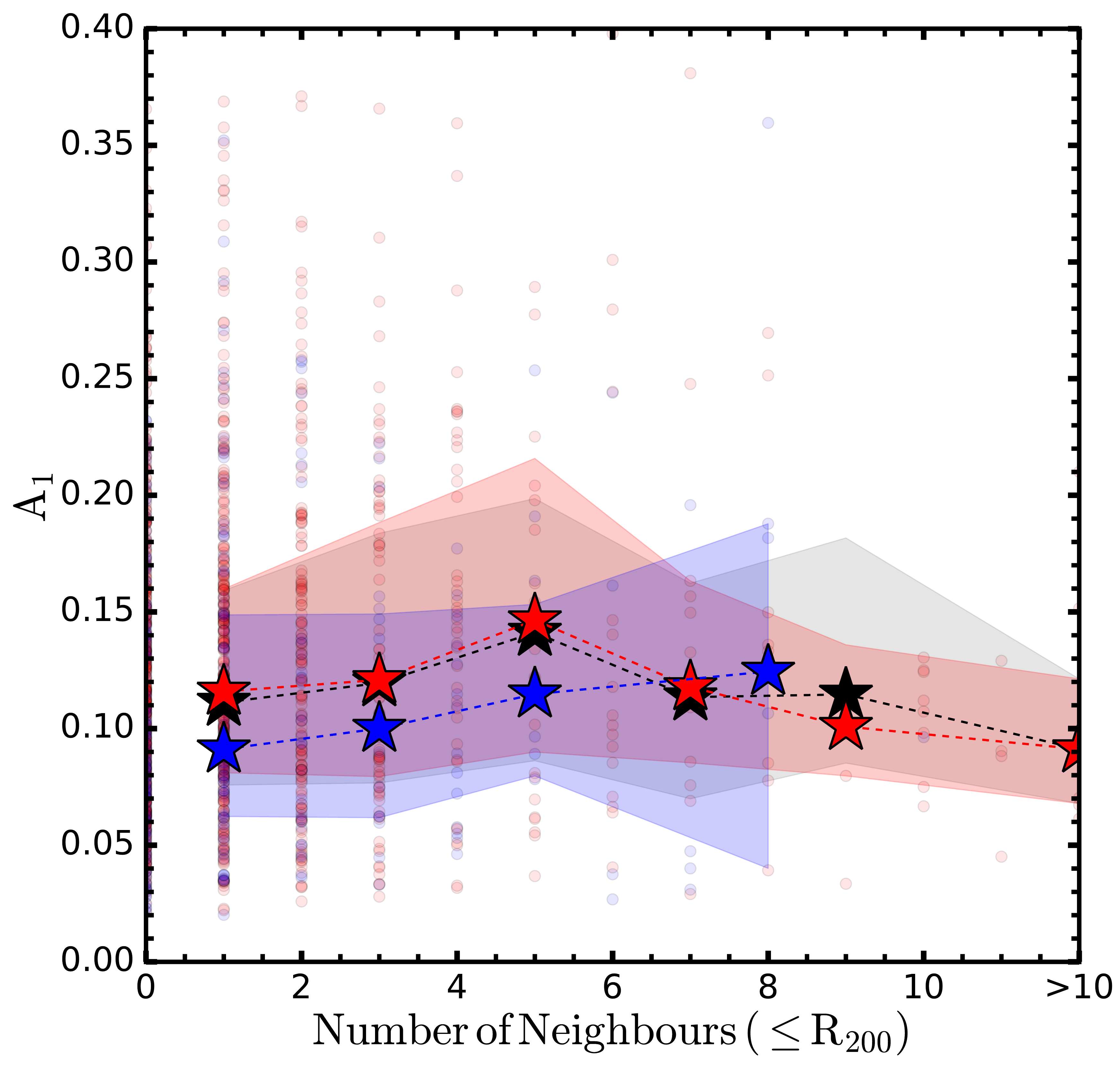}
    \includegraphics[width=0.3\textwidth]{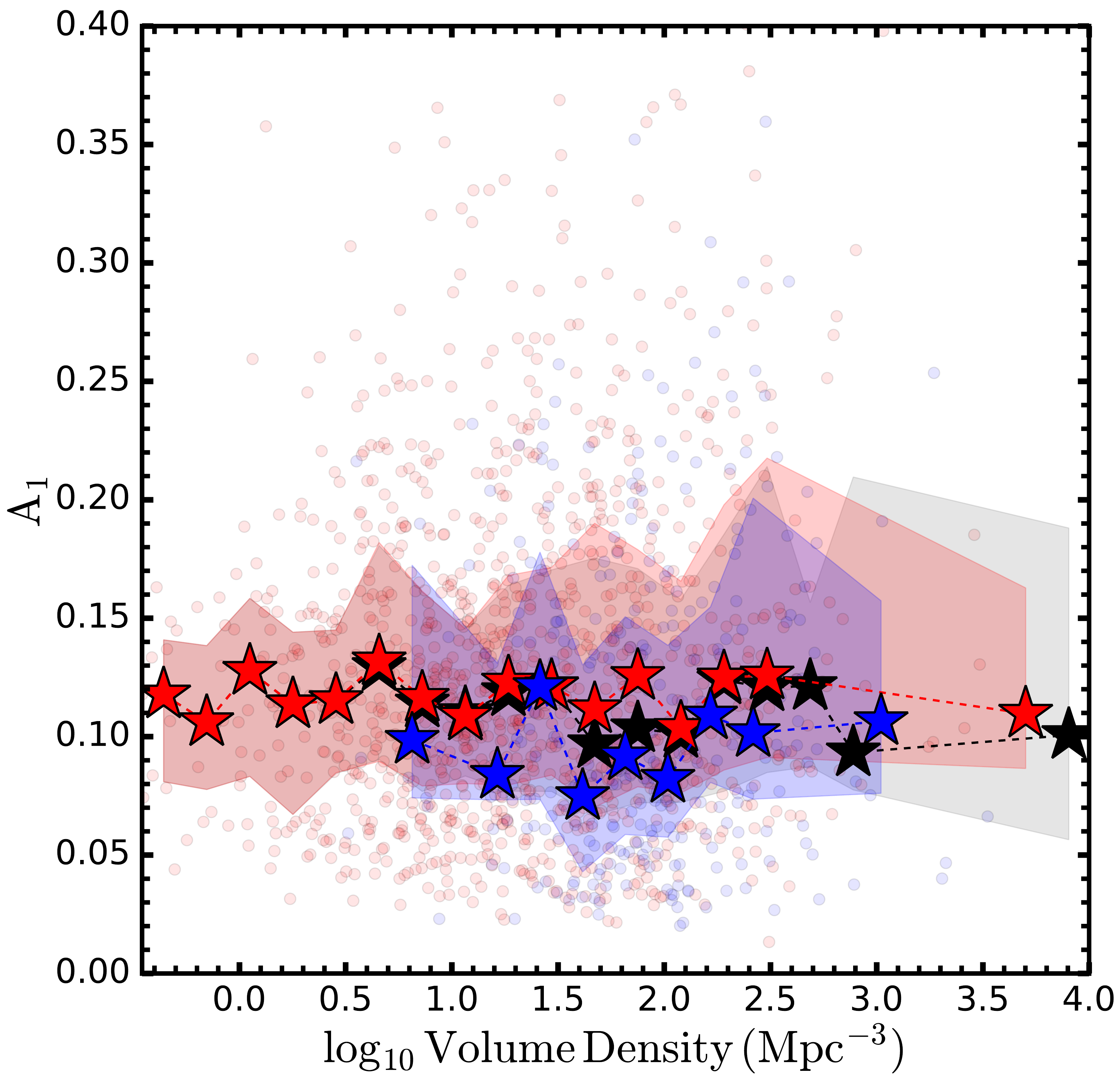}

    \includegraphics[width=0.3\textwidth]{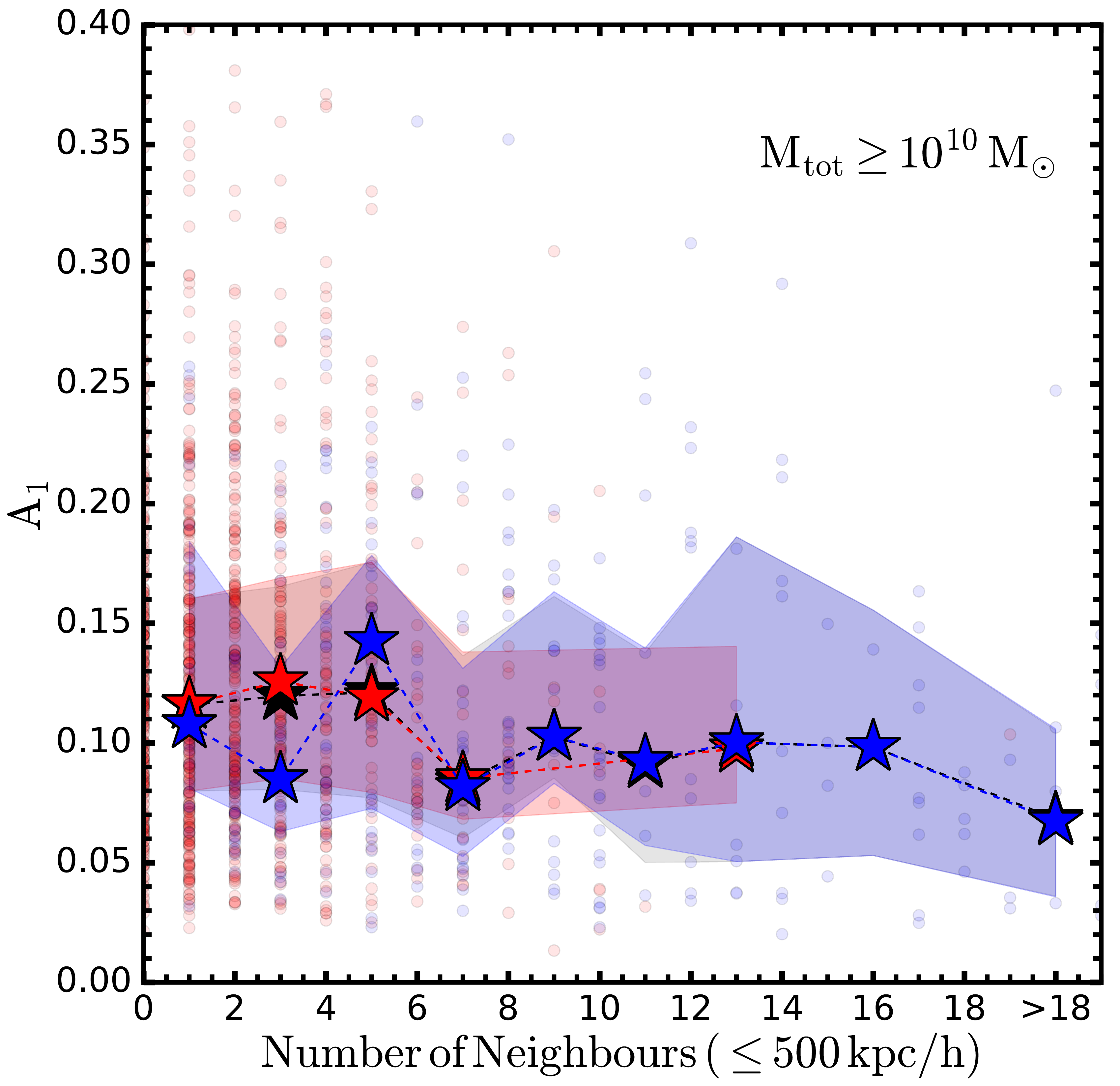}
    \includegraphics[width=0.3\textwidth]{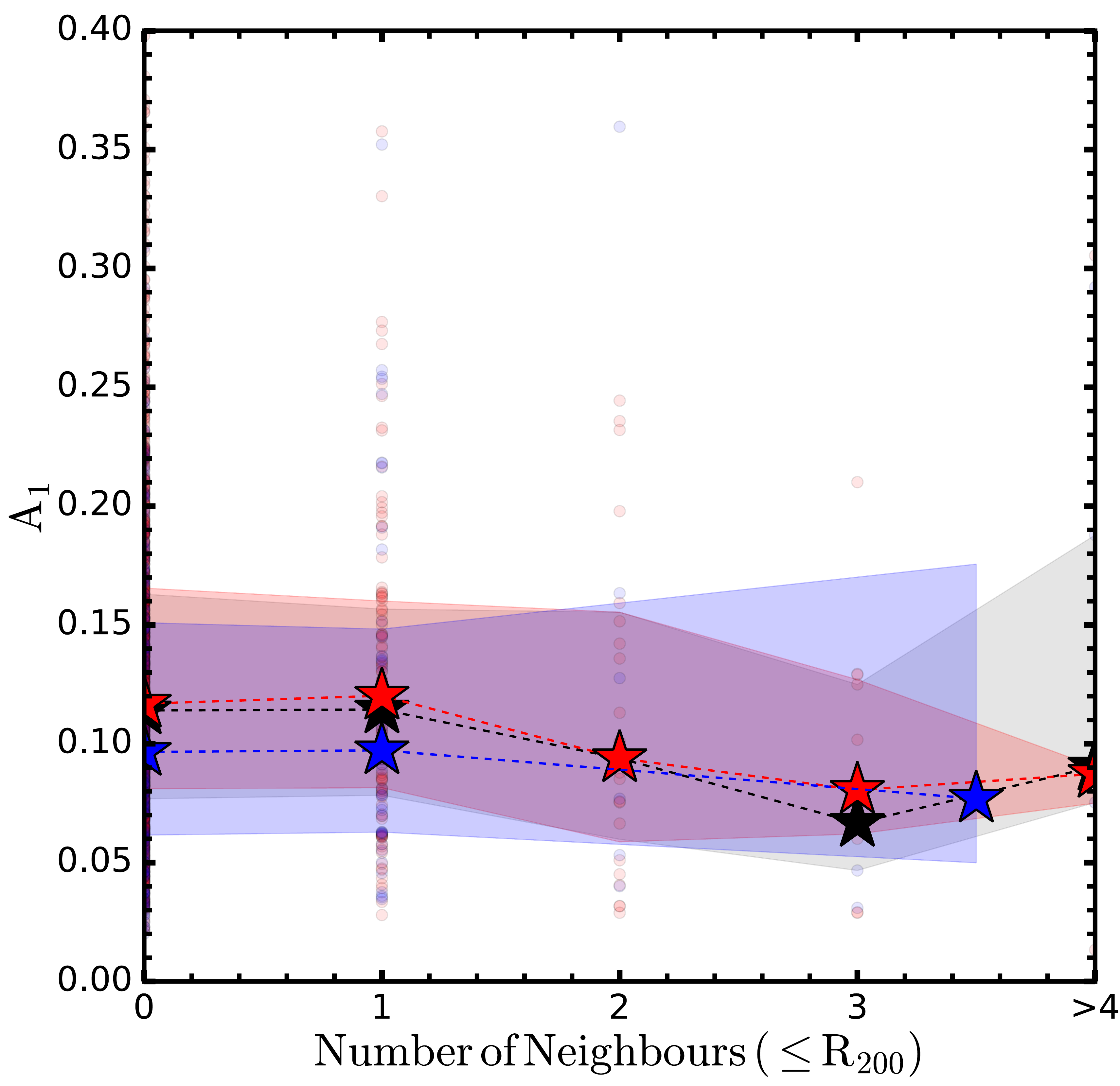}
    \includegraphics[width=0.3\textwidth]{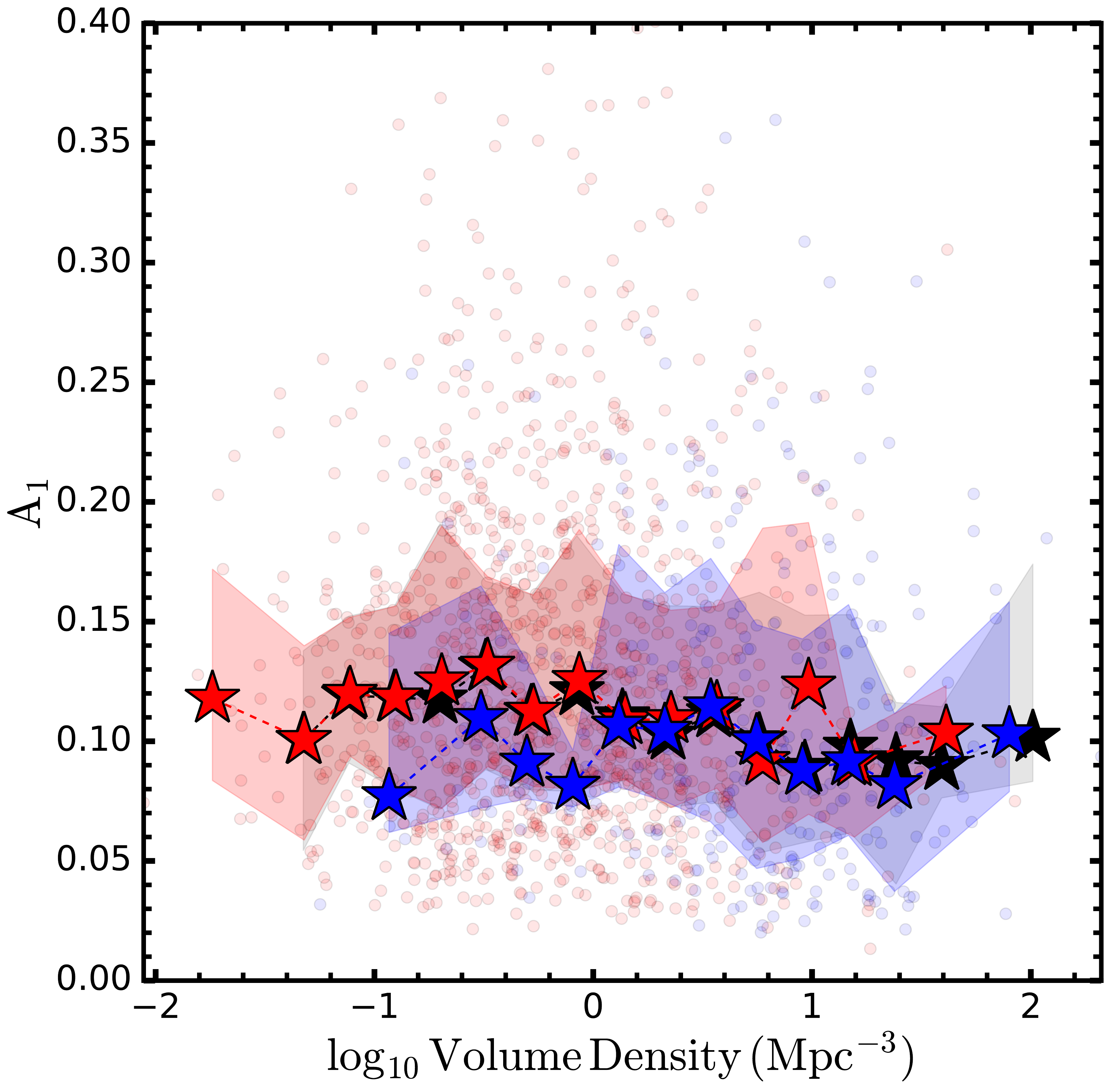}

    \includegraphics[width=0.3\textwidth]{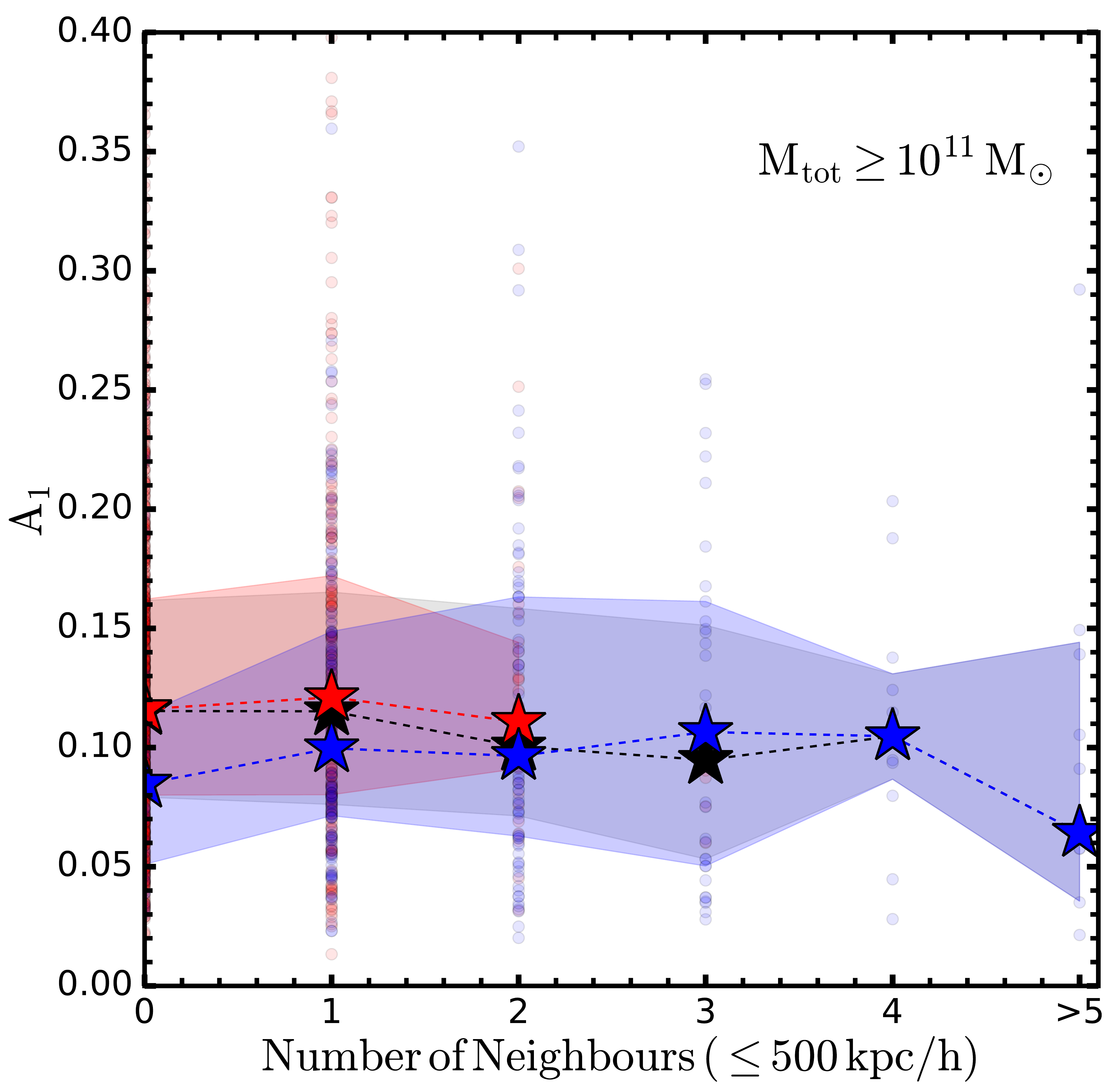}
    \includegraphics[width=0.3\textwidth]{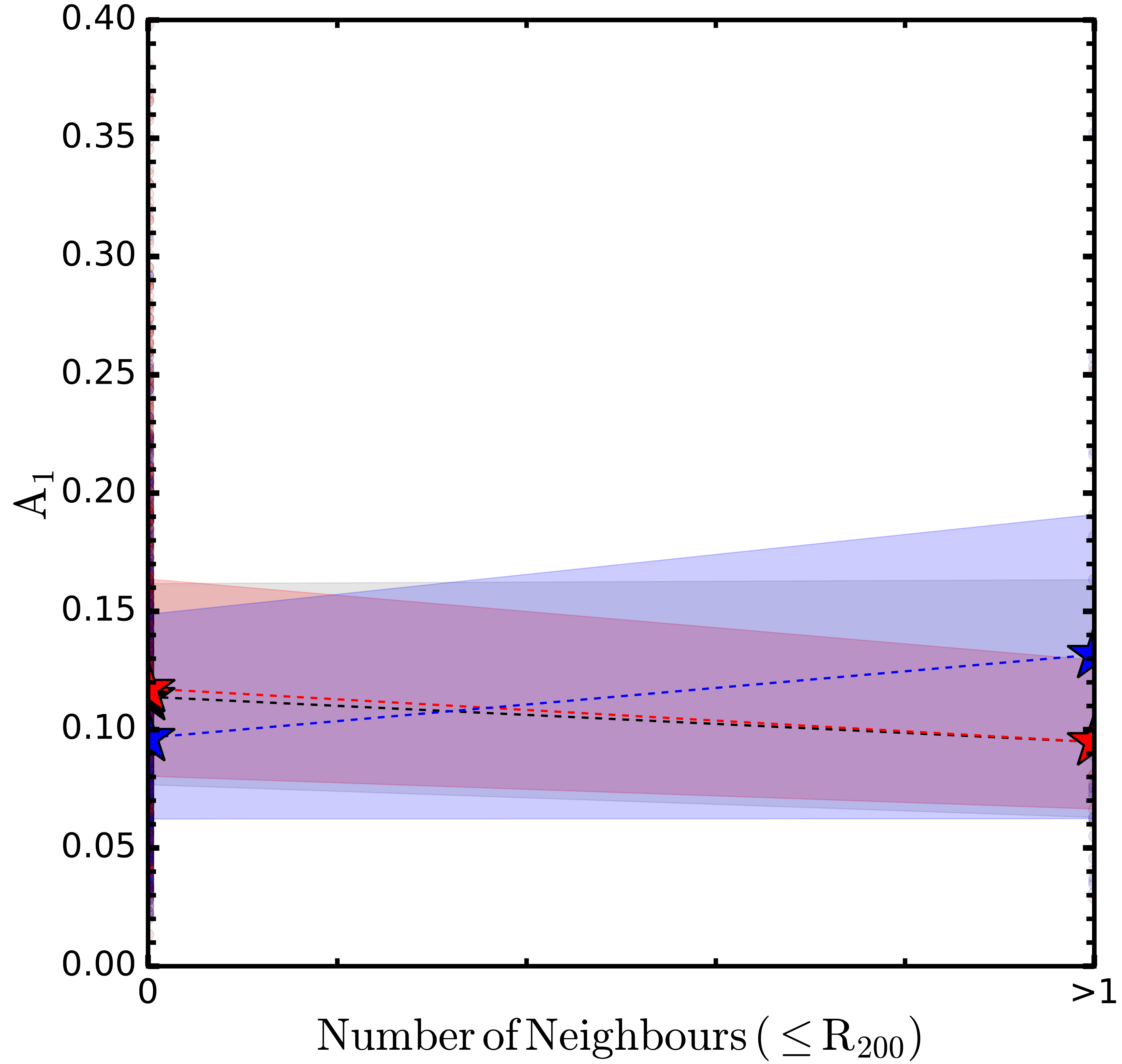}
    \includegraphics[width=0.3\textwidth]{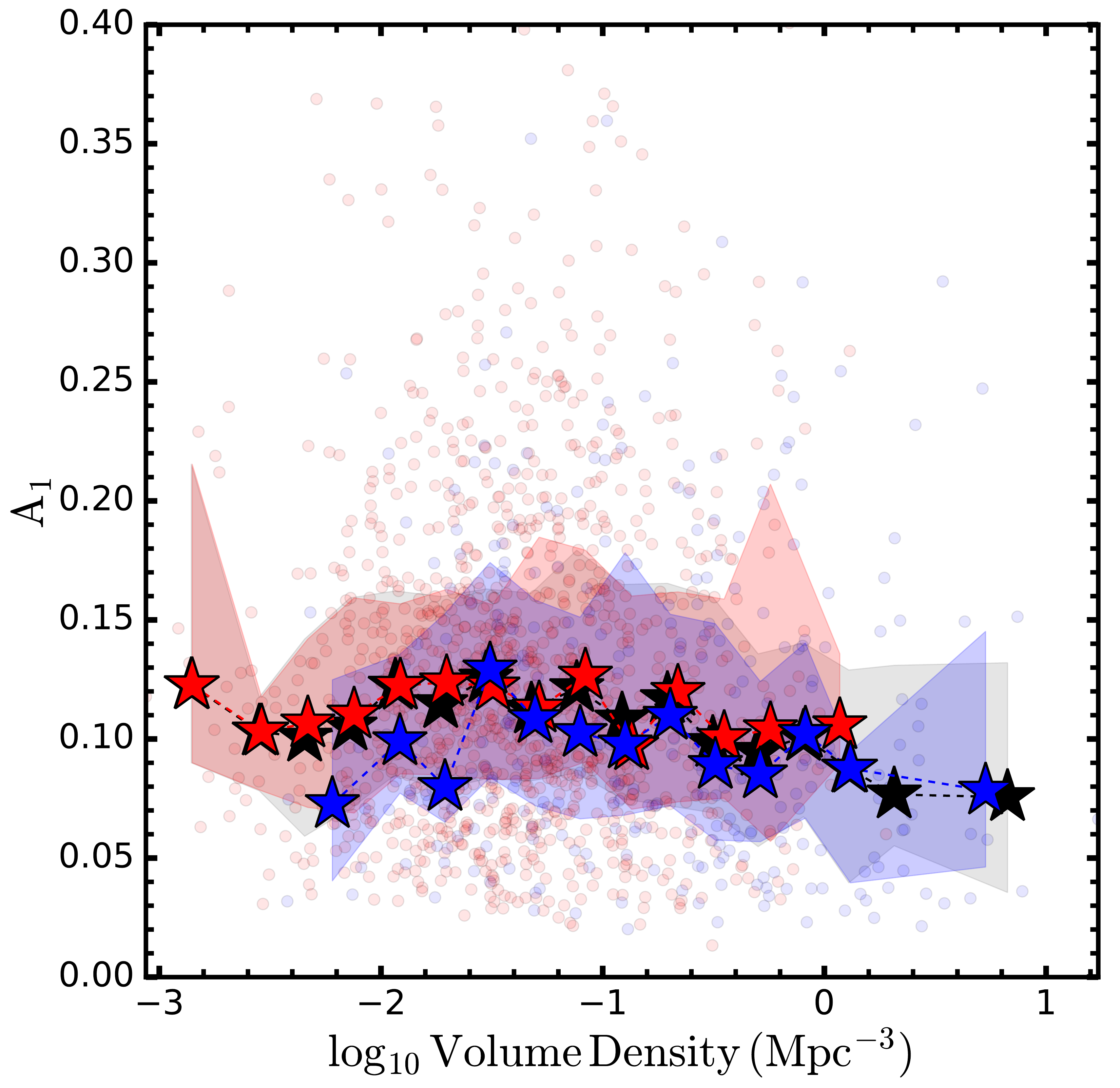}
    \caption{From the left to the right panel, the median amplitude of the lopsidedness, $A_{1}$, as a function of the local environment at $z=0$ defined in i)-iii) as described in Sec. \ref{sec:environment_neighbours} for all the galaxies (black stars) as well as for the central (red stars) and satellite (blue stars) galaxies, separately. The shaded areas represent the $25\mathrm{th}$-$75\mathrm{th}$ interquartile range of the data in each bin. From the top to the bottom panel, the local environment is defined by increasing the minimum total mass of the neighbouring galaxies included in the count, i.e. $\mathrm{M}_{\mathrm{tot}} \geq 10^{9}\, , 10^{10}\, , 10^{11}\, \mathrm{M}_{\odot}$, respectively. The background points represent the individual central (red) and satellite (blue) galaxies. Overall, we see that the results are consistent regardless of the environment definition adopted, and that the median amplitude of the lopsidedness only mildly decreases with the increasing number of close massive neighbours and local volume density of galaxies, suggesting that the lopsided perturbation does not depend on the number of close massive neighbours at $z=0$.} 
    \label{fig:environment_dependence_neighbours}
\end{figure*}

The above results do not exclude that the environment may still have an effect in the origin of the lopsided perturbation. 
By studying the time when each satellite galaxy has fallen into its group environment, we find that lopsided and symmetric galaxies show a similar overall distribution of infall times, but symmetric satellites display a wider spread in infall times than lopsided ones. Some symmetric satellites have fallen into the group environment as early as $\sim9$-$10\, \mathrm{Gyr}$ ago, while there are no lopsided satellites with infall times greater than $\sim8\, \mathrm{Gyr}$ ago, we find that symmetric satellites have fallen into the group $\sim0.5\, \mathrm{Gyr}$ earlier than lopsided ones. If we only consider satellite galaxies in high-density environments (i.e. Fornax cluster-like masses $\mathrm{M_{200}}\gtrsim10^{13}\, \mathrm{M_{\odot}}$), then the infall time difference between the symmetric and lopsided satellites increases to $\sim1\, \mathrm{Gyr}$. On the other hand, if we only consider satellite galaxies in low density environments (i.e. $\mathrm{M_{200}}\lesssim10^{13}\, \mathrm{M_{\odot}}$), then symmetric and lopsided galaxies have similar infall times. 
While this result may indicate that symmetric satellite galaxies have experienced longer interactions or environmental effects (e.g. strangulation, ram-pressure stripping) than lopsided ones, which modified their morphology to that of early-type disks that are typically less lopsided than late-type ones, the overall similarity of the infall time distributions, specifically in low-density environments, indicate that lopsided and symmetric galaxies already had distinct internal properties prior to entering their environment, which allow the development of the lopsided perturbation as a result of an external interaction. Once again, these results suggest a stronger dependence of lopsidedness on galaxy morphology than the environment, which mainly plays a secondary role. The further study of the differences between symmetric and lopsided satellite galaxies, which have fallen at similar times into their group environment, in terms of internal properties and orbits, that may have favoured the development of the lopsided perturbation will be left to a future work.

\subsection{Correlations between lopsidedness, number of close neighbours and morphology}
\label{sec:environment_neighbours}
In the previous Sec. \ref{sec:environment_halomass}, we have seen that the distinct internal galaxy properties, such as their $\mathrm{D/T}$, $c/a$ and $\mu_{*}$, are playing a stronger role in the origin of the lopsidedness than the environment, defined in terms of the total halo mass (i.e. $\mathrm{M_{200}}$) of the environment in which galaxies reside in Sec. \ref{sec:environment_halomass}. 

In this section, we seek to further characterize whether different metrics of the environment do or do not show a correlation with the strength of this asymmetric perturbation. For this purpose, we adopt the following three definitions of the local environment surrounding our galaxies at $z=0$: 
\begin{enumerate}[i)]\itemsep0.2cm
    \item We count the number of galaxies with total mass $\mathrm{M}_{\mathrm{tot}}\geq10^{9},$ or $10^{10},$ or $10^{11}\, \mathrm{M}_{\odot}$ located within a sphere of fixed radius $500\, \mathrm{kpc/h}$ centered on the target galaxy (see e.g. \citealt{Blanton2009,Gargiulo2022});

     \item We count the number of galaxies with total mass $\mathrm{M}_{\mathrm{tot}}\geq10^{9},$ or $10^{10},$ or $10^{11}\, \mathrm{M}_{\odot}$ located within a sphere of radius equals to the virial radius, $R_{200}$, of the target galaxy. For the central galaxies, we use the $R_{200}$ of the galaxy at $z=0$. However, for the satellite galaxies, we use the $R_{200}$ of the galaxy at the simulation snapshot before the galaxy became a satellite (i.e. when it was still a central) as $R_{200}$ is not defined for satellite galaxies. 

    \item We calculate the local volume density, i.e. $\rho_{\mathrm{N_{\mathrm{gal}}}} = \mathrm{N_{\mathrm{gal}}}/(\frac{4}{3}\pi R_{\mathrm{N_{\mathrm{gal}}}}^{3})$, of galaxies with total mass $\mathrm{M}_{\mathrm{tot}}\geq10^{9},\, 10^{10},\, 10^{11}\, \mathrm{M}_{\odot}$ located within a sphere of radius, $R_{\mathrm{N_{\mathrm{gal}}}}$, centered on the target galaxy. Here, $R_{\mathrm{N_{\mathrm{gal}}}}$ is the radius containing the $\mathrm{N_{\mathrm{gal}}}=5$ closest galaxies, similarly to \citet{Cappellari2011}.
    With this definition, we aim to quantify the effect of different distributions of the surrounding galaxies to any target galaxy. 
    We note that we have also repeated this calculation considering the ten closest galaxies, i.e. $\mathrm{N_{\mathrm{gal}}}=10$, however the results do not vary significantly.  
\end{enumerate}

\begin{figure*}
    \centering
    \includegraphics[width=0.3\textwidth]{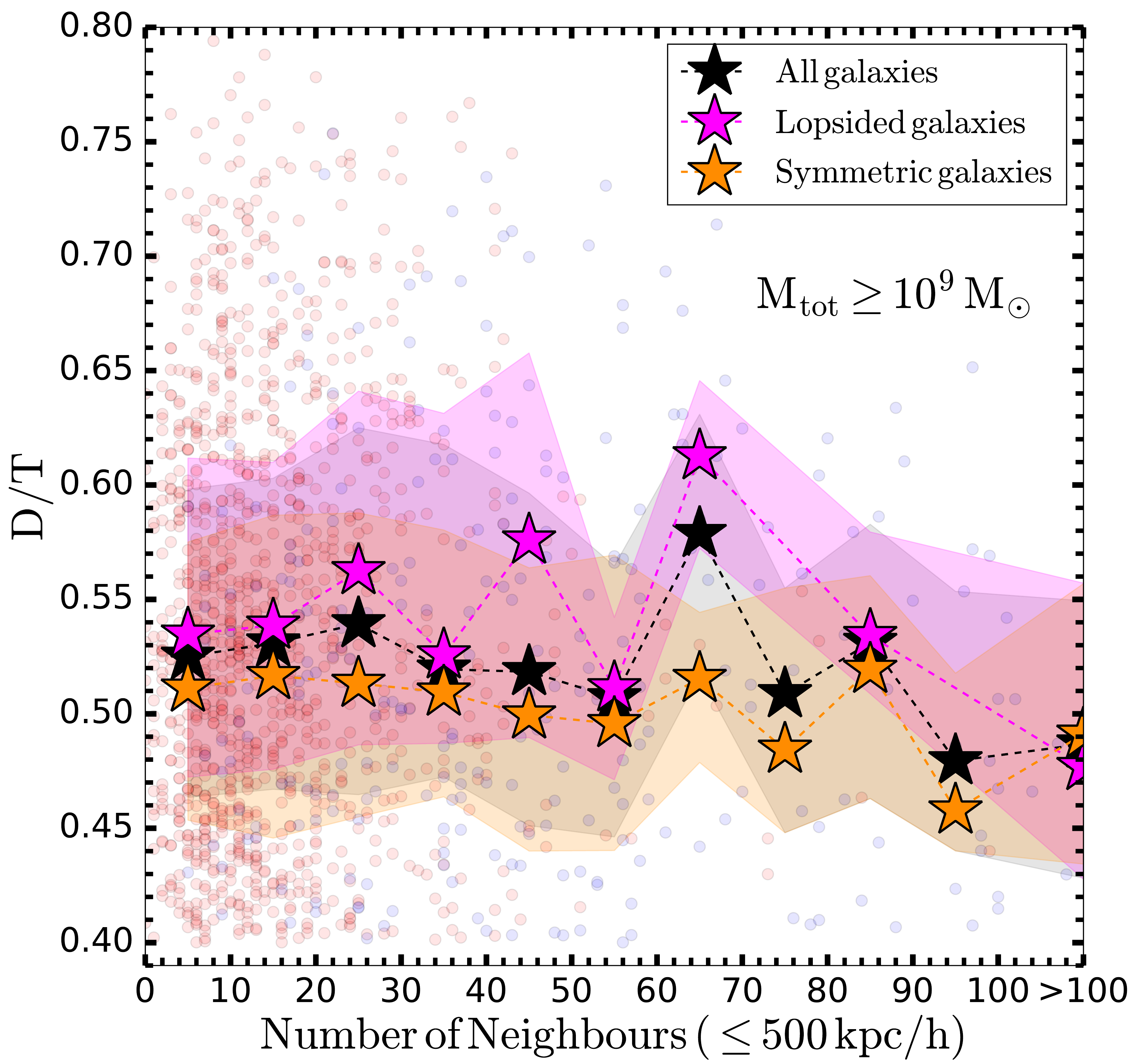}
    \includegraphics[width=0.3\textwidth]{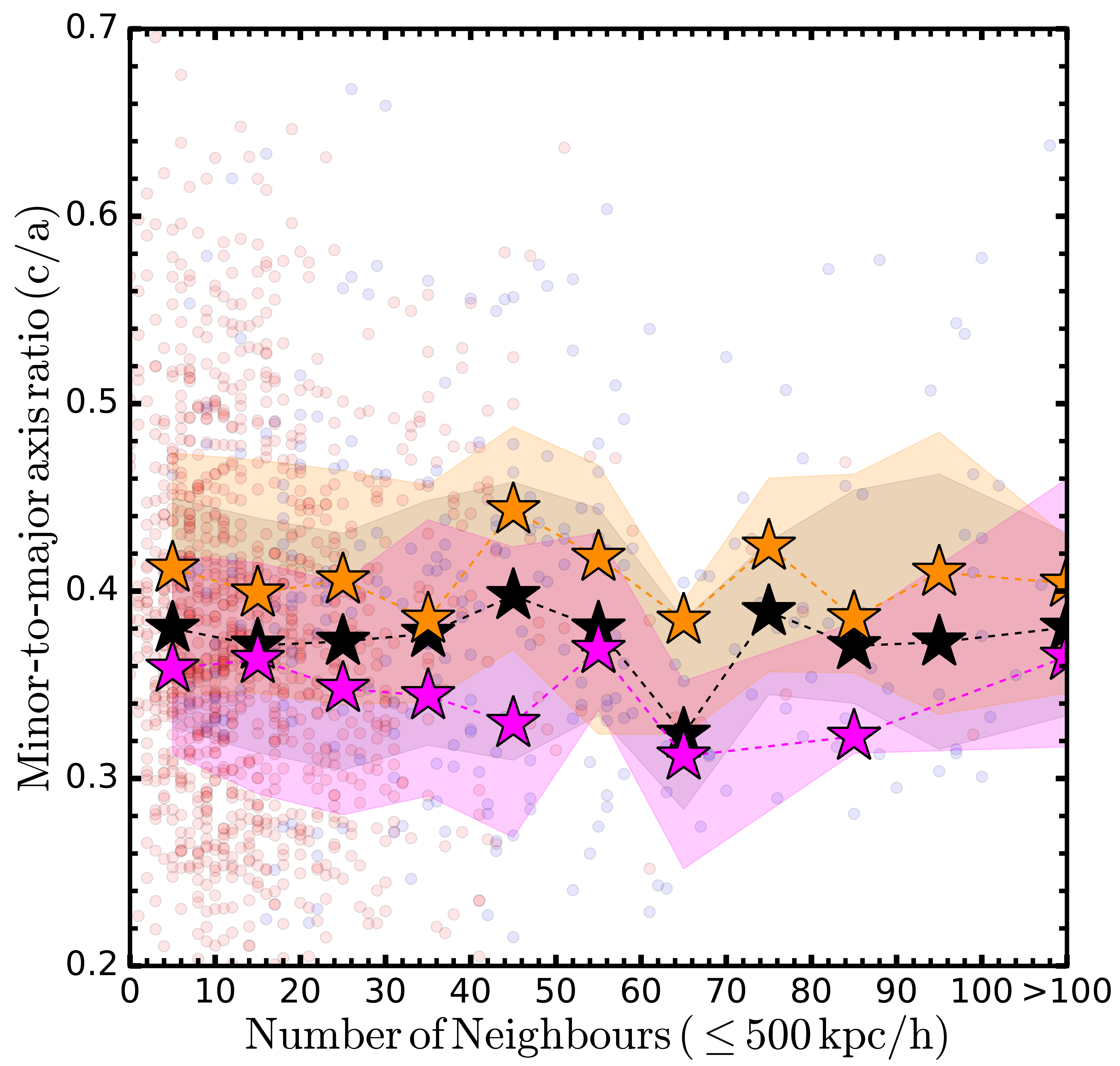}
    \includegraphics[width=0.3\textwidth]{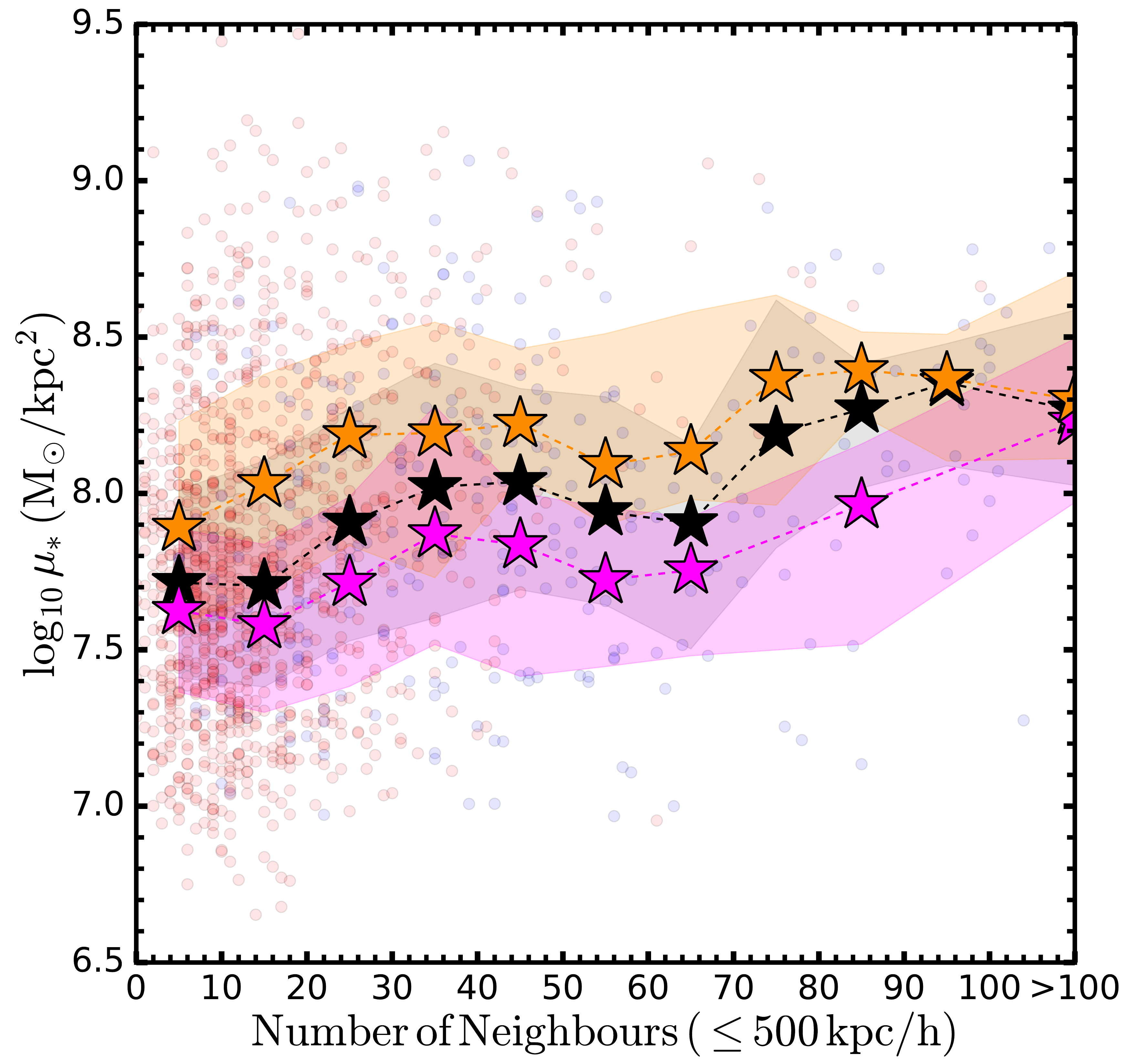}

    \includegraphics[width=0.3\textwidth]{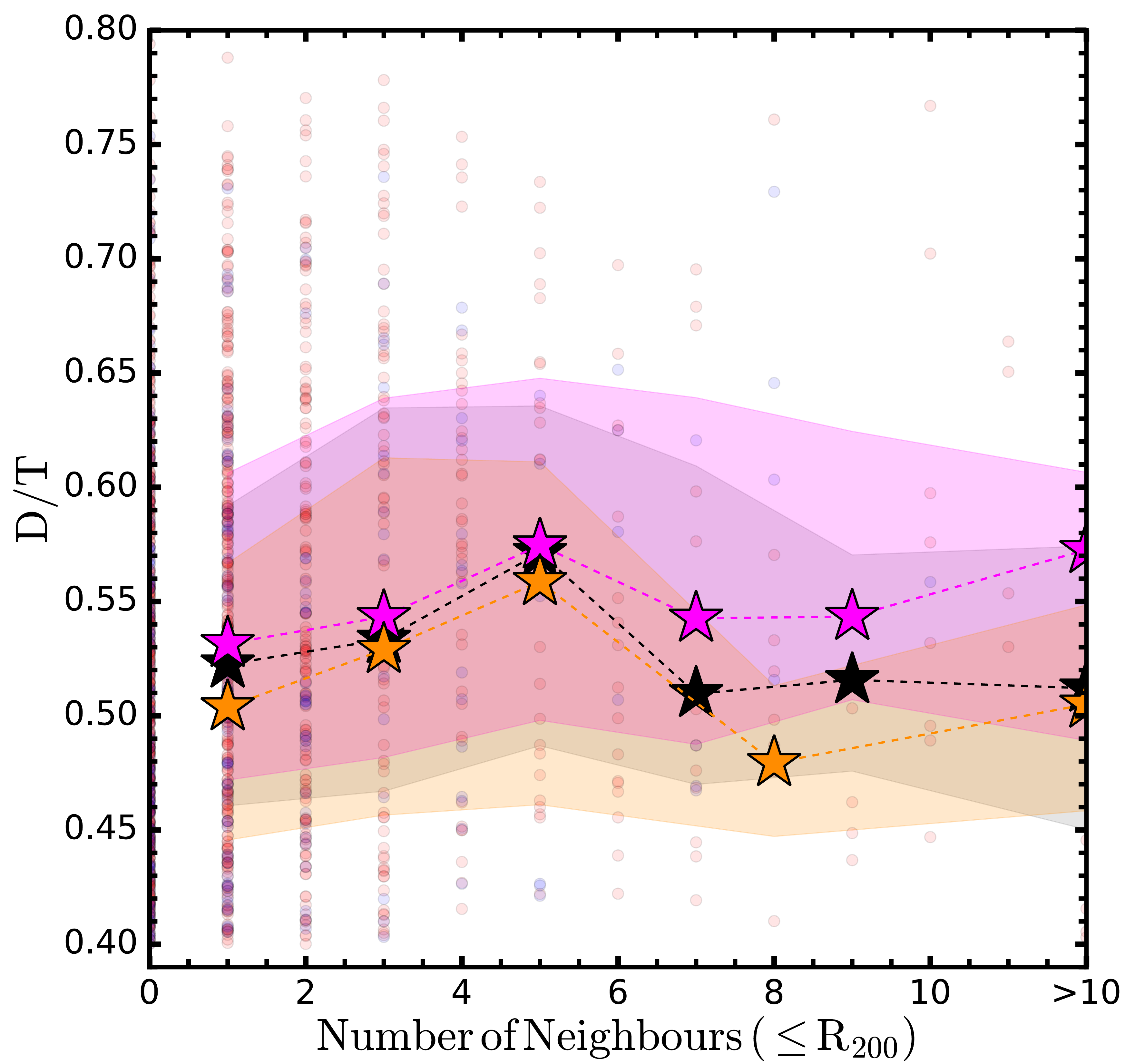}
    \includegraphics[width=0.3\textwidth]{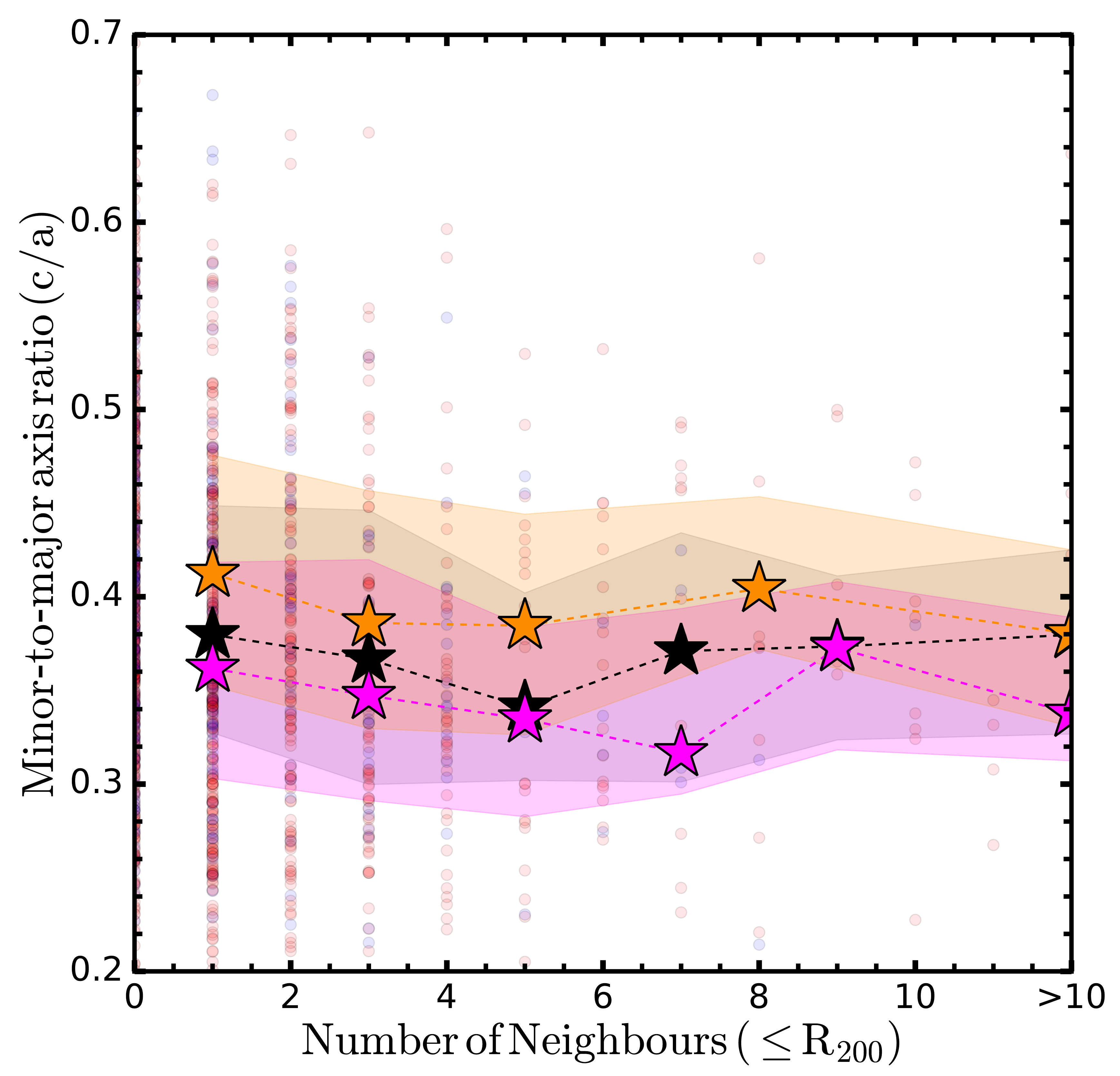}
    \includegraphics[width=0.3\textwidth]{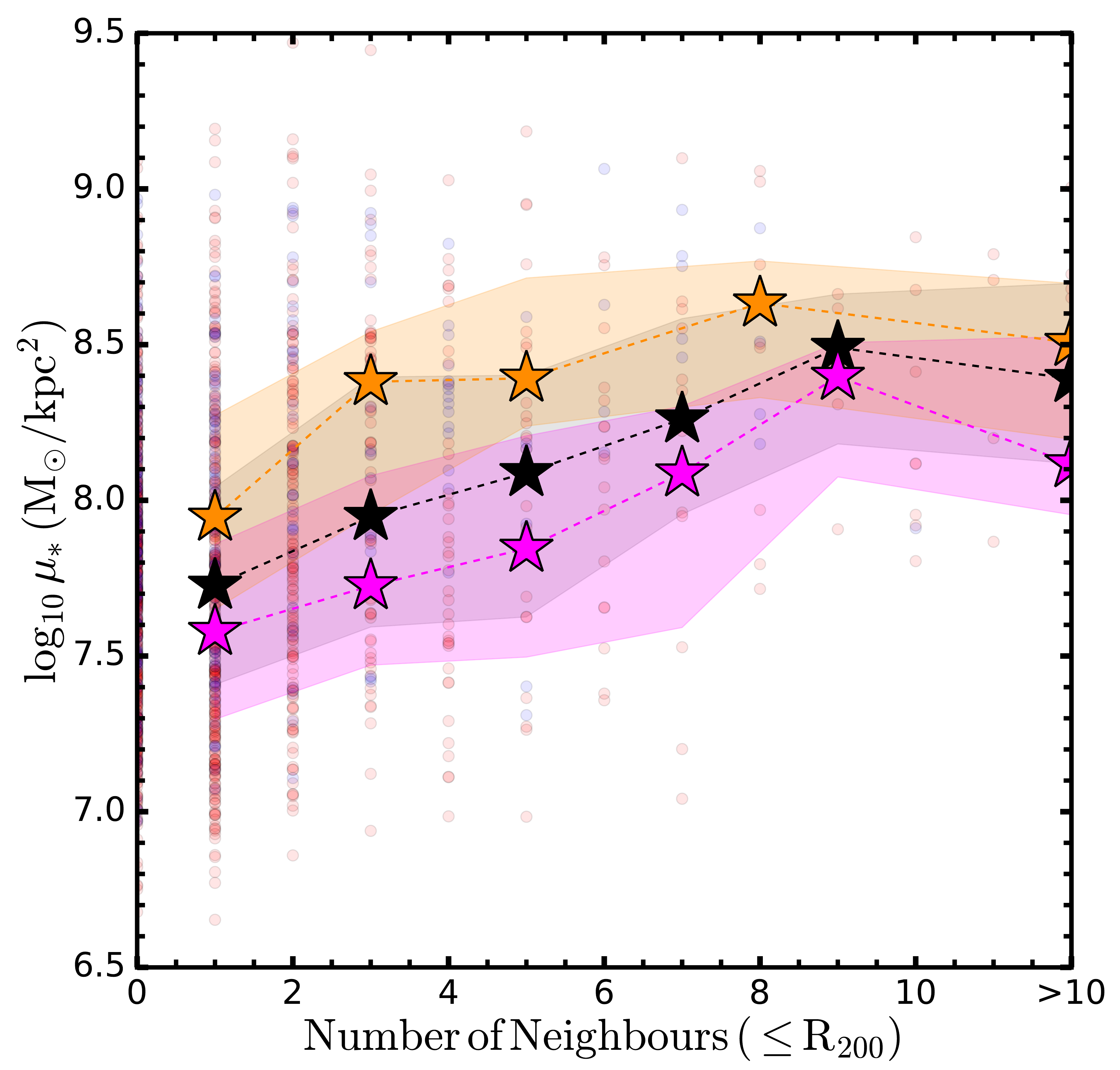}

    \includegraphics[width=0.3\textwidth]{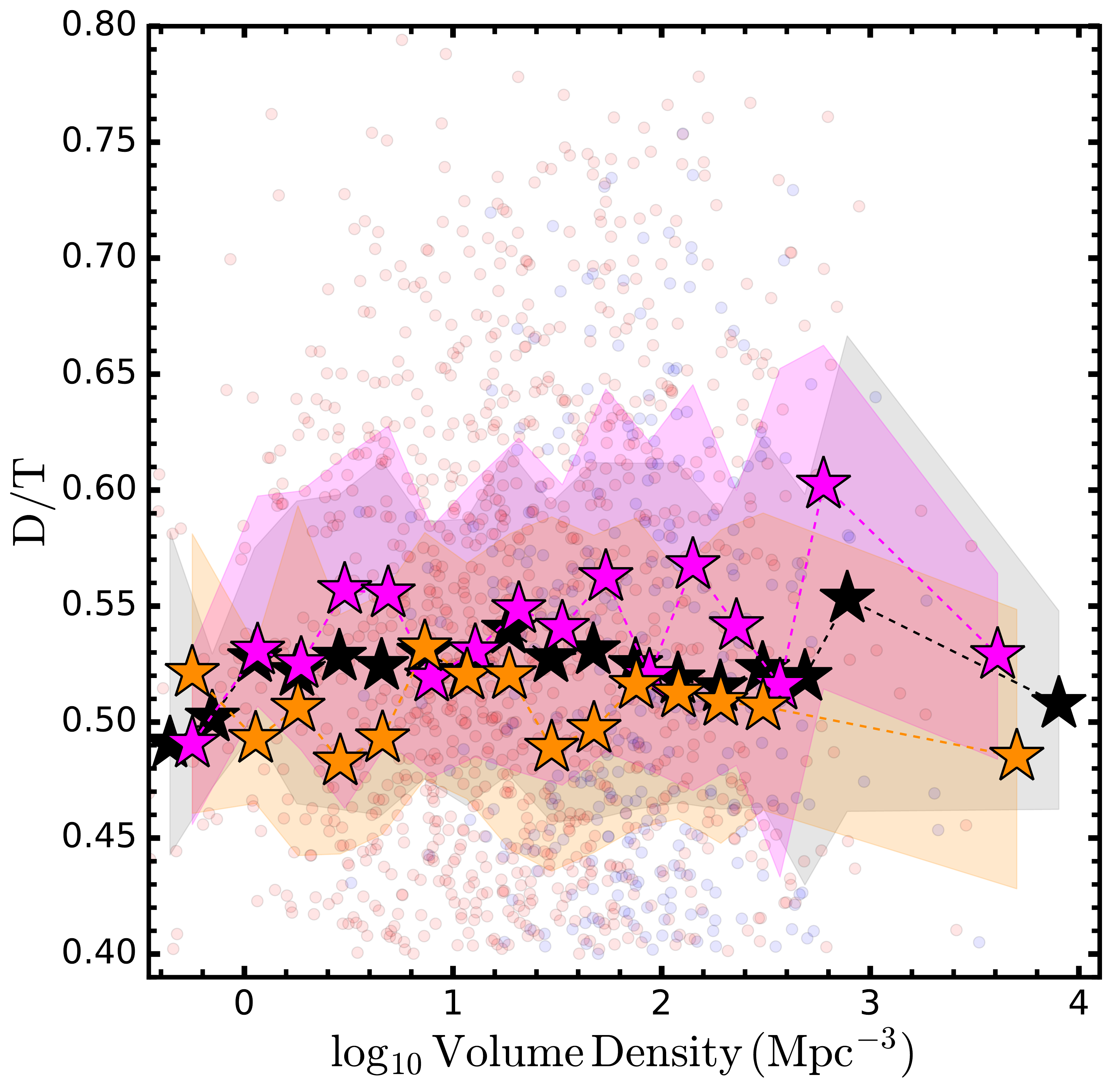}
    \includegraphics[width=0.3\textwidth]{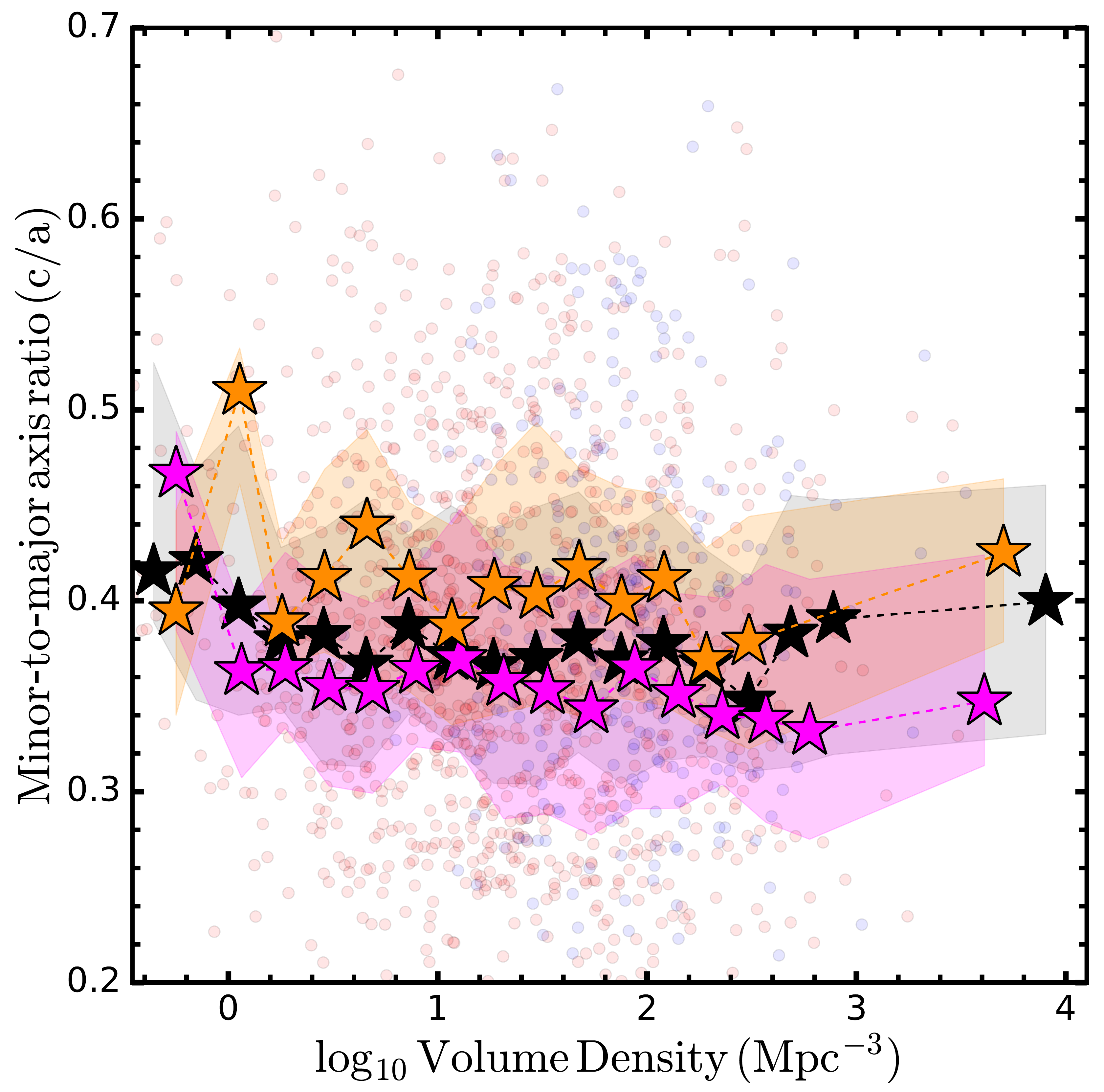}
    \includegraphics[width=0.3\textwidth]{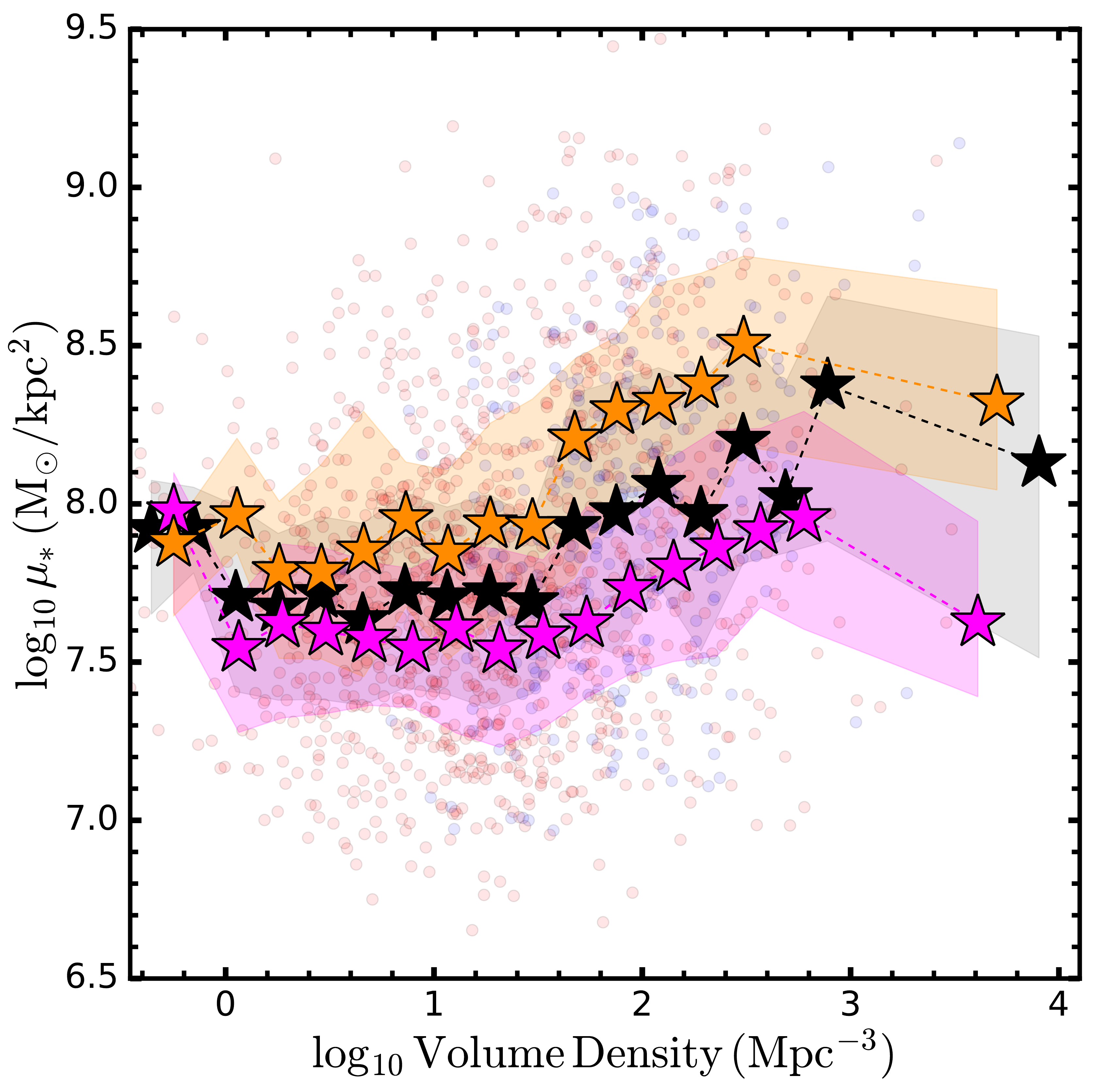}
    \caption{From the left to the right panel, the median disk-to-total ratio ($\mathrm{D/T}$), minor-to-major axis ratio ($\mathrm{c/a}$) and central stellar mass density ($\mu_{*}$) as a function of the local environment at $z=0$ defined in i)-iii), from the top to the bottom panels, for all (black stars), only the lopsided (magenta stars) and only the symmetric (orange stars) galaxies in our selected galaxy sample. The description is as in Fig. \ref{fig:environment_dependence_m200}. Overall, we see that the lopsided galaxies are typically characterized by on average more extended disks, flatter inner galactic regions and lower central stellar mass density than the symmetric galaxies regardless of the number of close massive neighbours and local volume density of galaxies, indicating a stronger correlation between lopsidedness and the internal properties of the galaxies than with the environment.} 
    \label{fig:environment_dependence_neighbours_galaxy_properties}
\end{figure*}

\begin{figure}
    \centering
    \includegraphics[width=0.45\textwidth]{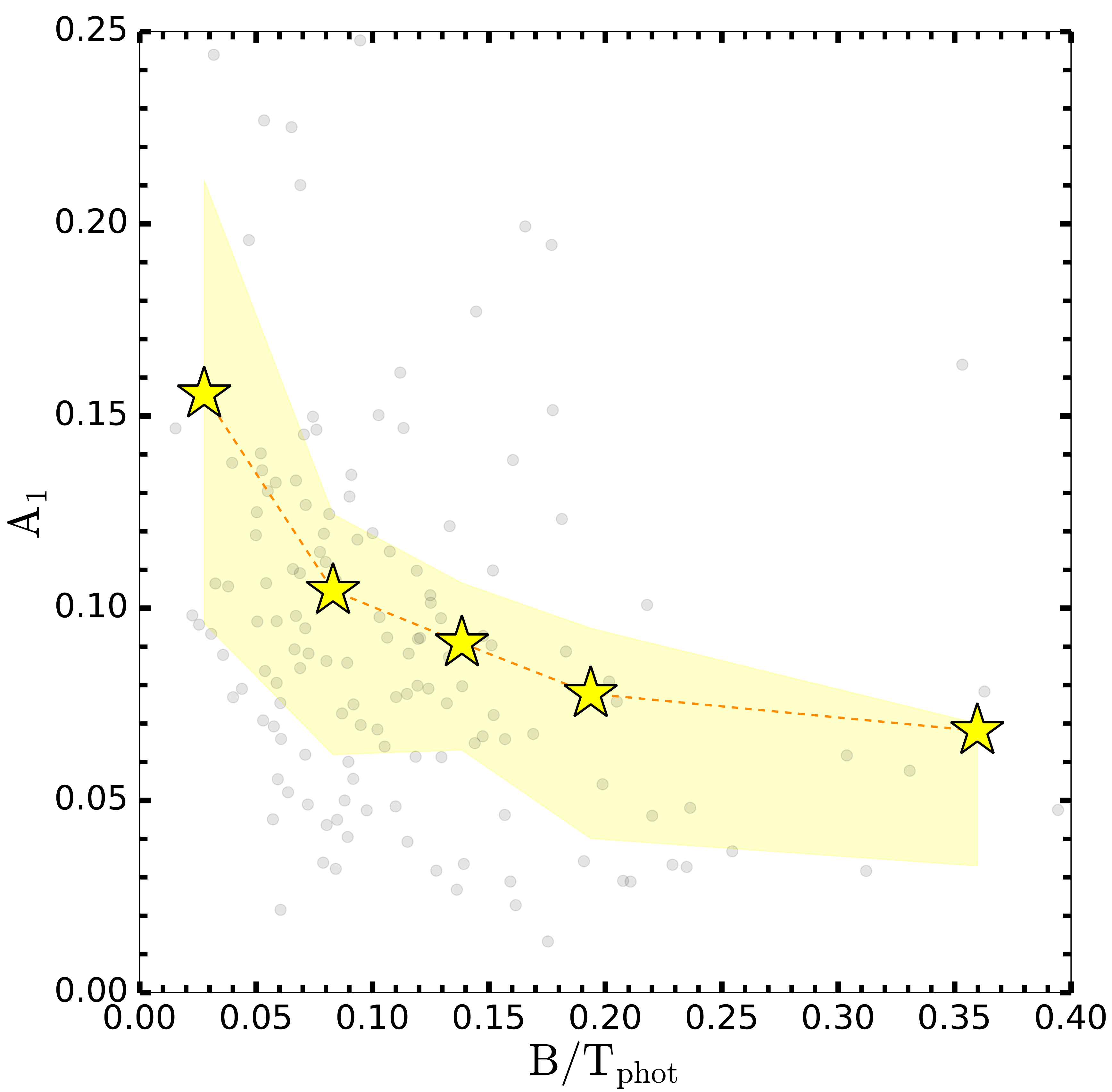}
    \caption{The lopsidedness amplitude as a function of the photometric bulge-to-total ($\mathrm{B/T}$) fractions for a sub-sample of our galaxies in common with the sample of Milky Way/M31-like galaxies selected from TNG50 by \citet{Gargiulo2022}. The photometric $\mathrm{B/T}$ values are taken from \citet{Gargiulo2022}. The background black points represent the individual galaxies, while the yellow stars represent the median $A_{1}$ calculated in each bin. The shaded areas represent the $25\mathrm{th}$-$75\mathrm{th}$ interquartile range of the data in each bin.}
    \label{fig:bt_photometric}
\end{figure}

In Fig. \ref{fig:environment_dependence_neighbours}, we study the behaviour of the lopsidedness amplitude as a function of the local environment at $z=0$ defined in i)-iii) from the left to the right panels, respectively. Here, the median value of $\mathrm{A_{1}}$ is calculated as described in Sec. \ref{sec:environment_halomass}. The different rows show the results obtained considering different minimum galaxy mass threshold when counting galaxies in the corresponding local environments. Firstly, we see that the results are overall consistent regardless of the definition of the environment adopted, as well as for the different galaxy total mass cuts. 
Similarly to Fig. \ref{fig:lopsidedness_environment_dependence_m200}, we see that the overall median lopsidedness amplitude only shows a mild decreasing trend with the increasing number of close massive neighbours and local volume density of galaxies (see black stars in Fig. \ref{fig:environment_dependence_neighbours}). If we separate between central and satellite galaxies, we also find an overall constant to mildly decreasing $\mathrm{A_{1}}$ as a function of the local environment surrounding our galaxies at $z=0$. In the scenario in which a larger number or density of close massive neighbours may indicate a higher probability of interactions, these results suggest that the specific type or frequency of interactions may not be the primary cause to the origin of the lopsided perturbation, in agreement with previous works (e.g. \citealt{Bournaud2005,Wilcots2010,Varela-Lavin2022}).

In Fig. \ref{fig:environment_dependence_neighbours_galaxy_properties}, we show the behaviour of $\mathrm{D/T}$, $c/a$ and $\mu_{*}$, from the left to the right panels, as a function of the local environment at $z=0$ defined in i)-iii), from the top to the bottom panels, respectively, for a minimum total mass cut of the neighbouring galaxies $\mathrm{M}_{\mathrm{tot}}\geq10^{9}\, \mathrm{M}_{\odot}$. For simplicity, we only consider here the case of minimum total mass cut $\mathrm{M}_{\mathrm{tot}}\geq10^{9}\, \mathrm{M}_{\odot}$ for the neighbouring galaxies. 
We see that the central stellar mass density, $\mu_{*}$, mildly increases, while the $\mathrm{D/T}$ mildly decreases with the increasing number of close massive neighbours and local volume density of galaxies, similarly to the trend seen in Fig. \ref{fig:environment_dependence_m200} as a function of $\mathrm{M_{200}}$. The flatness of the inner galactic regions, $c/a$, remains mainly constant as a function of the number of close massive neighbours and local volume density of galaxies.
The change of the $\mathrm{D/T}$ and $\mu_{*}$ as a function of the local environment surrounding our galaxies suggest that galaxies may experience a morphology transformation from a late-type to an early-type disk galaxy, characterized by a more compact disk and prominent bulge. This is likely the result of the more frequent tidal interactions with the higher number of close massive neighbours, which can trigger bursts of star formation close to the galaxy inner regions and produce early-type disk galaxies, such as S0s \citep{Bekki2011,Gargiulo2019}. Additionally, similarly to Fig. \ref{fig:environment_dependence_m200}, we find that lopsided galaxies are always characterized by, on average, larger $\mathrm{D/T}$, smaller $c/a$ and $\mu_{*}$ regardless of the environment. In line with the discussion from Fig. \ref{fig:environment_dependence_neighbours}, where we find an overall constant to mildly decreasing behaviour of $\mathrm{A_{1}}$ with the increasing number and density of close massive neighbours, these results confirm that lopsidedness does not strongly depend on the specific type or frequency of interactions. On the contrary, the higher probability of interactions likely influences the galaxy morphology such that galaxies become more resistant to developing the lopsided perturbation from future external interactions. Thus, these results confirm the strong correlation between lopsidedness and internal galaxy properties, as suggested in previous works (e.g. \citealt{Conselice2000,Bournaud2005,Reichard2008,Wilcots2010}{;\citetalias{Varela-Lavin2022}), regardless of the environment.      

\begin{figure}
    \centering
    \includegraphics[width=0.45\textwidth]{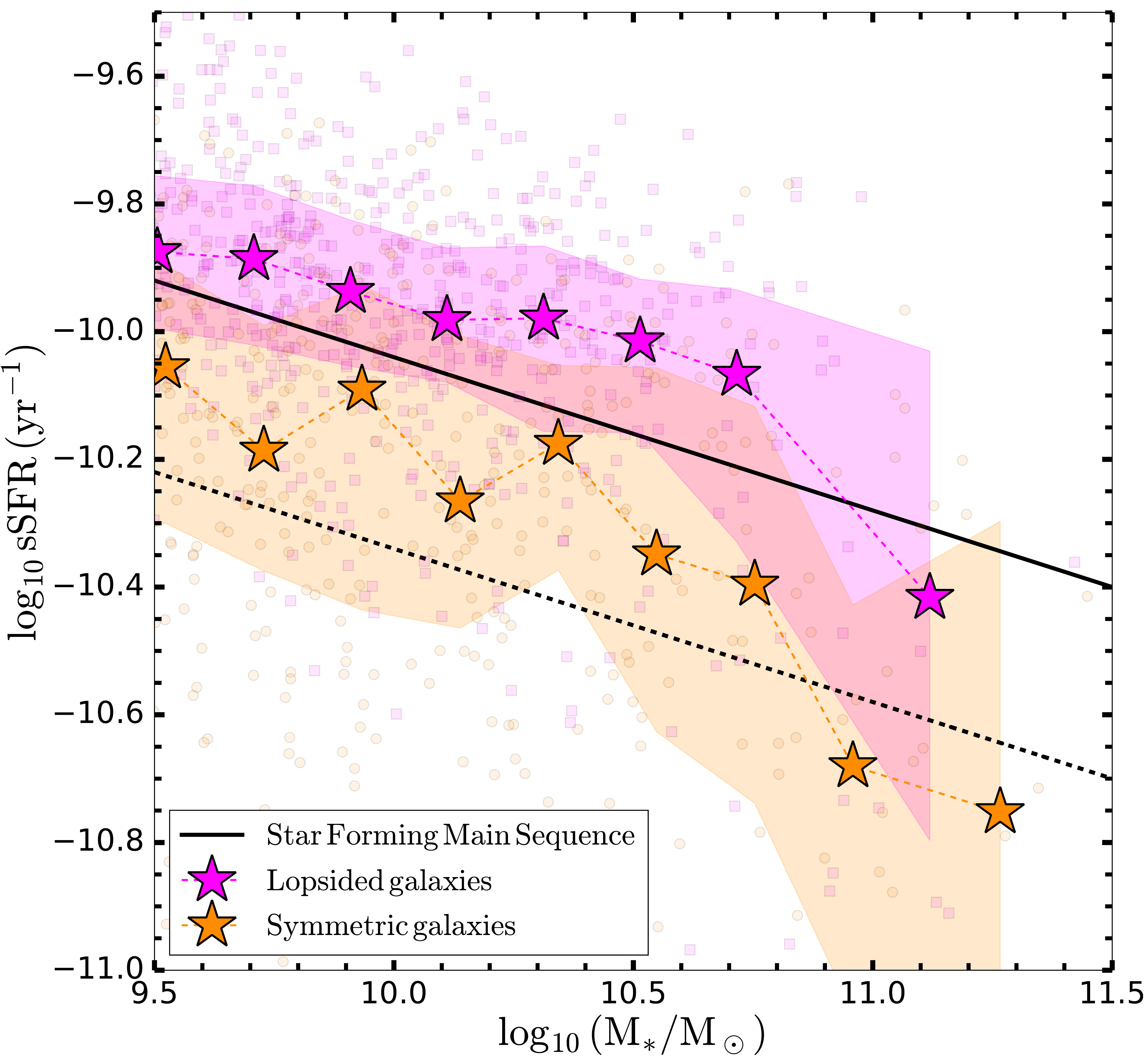}
    \caption{The specific star formation rate (sSFR) as a function of the galaxy total stellar mass, $\mathrm{M_{*}}$, at $z=0$ for the lopsided (magenta) and symmetric (orange) galaxies in our selected galaxy sample. The coloured stars represent the median sSFR of the lopsided and symmetric galaxies, while the shaded areas represent the $25\mathrm{th}$-$75\mathrm{th}$ interquartile range of the data in each stellar mass bin. The background coloured points represent the individual lopsided and symmetric galaxies. The black solid line shows the best-fit relation to the $\mathrm{SFR}$-$\mathrm{M}_{*}$ taken from \citet{Renzini2015}, which defines the star forming main sequence with its $0.3\, \mathrm{dex}$ scatter (black dashed line). We see that the lopsided galaxies are on average more star forming than the symmetric ones, with median sSFR typically above the star forming main sequence.}
    \label{fig:star_formation_rate_present_day}
\end{figure}

\subsection{Bulge-less galaxies}
\label{sec:bulgeless_galaxies}
In light of the correlation between lopsidedness and central stellar mass density, present-day galaxies with low bulge-to-total ($\mathrm{B/T}$) mass fractions represent possible lopsided candidates. In Fig. \ref{fig:bt_photometric}, we show the behaviour of the lopsidedness amplitude, $A_{1}$, as a function of the photometric $\mathrm{B/T}$ for a sub-sample of our galaxies. 
This galaxy sub-sample includes $151$ galaxies, which are in common with the sample of Milky Way/M31-like galaxies selected from TNG50 by \citet{Gargiulo2022}. For these galaxies, the photometric $\mathrm{B/T}$ values are taken from \citet{Gargiulo2022}, who performed a two component decomposition of the galaxy surface brightness profiles. In Fig. \ref{fig:bt_photometric}, we clearly see a strong correlation between lopsidedness and photometric $\mathrm{B/T}$, where the median $A_{1}$\footnote{Here, the median $A_{1}$ is calculated as described in Sec. \ref{sec:environment_halomass}.} decreases with increasing $\mathrm{B/T}$, suggesting that a strong lopsided perturbation is more likely to develop in bulge-less galaxies, as it is observed in the M101 galaxy (e.g. \citealt{Beale1969}).

\begin{figure*}
    \centering
    \includegraphics[width=0.32\textwidth]{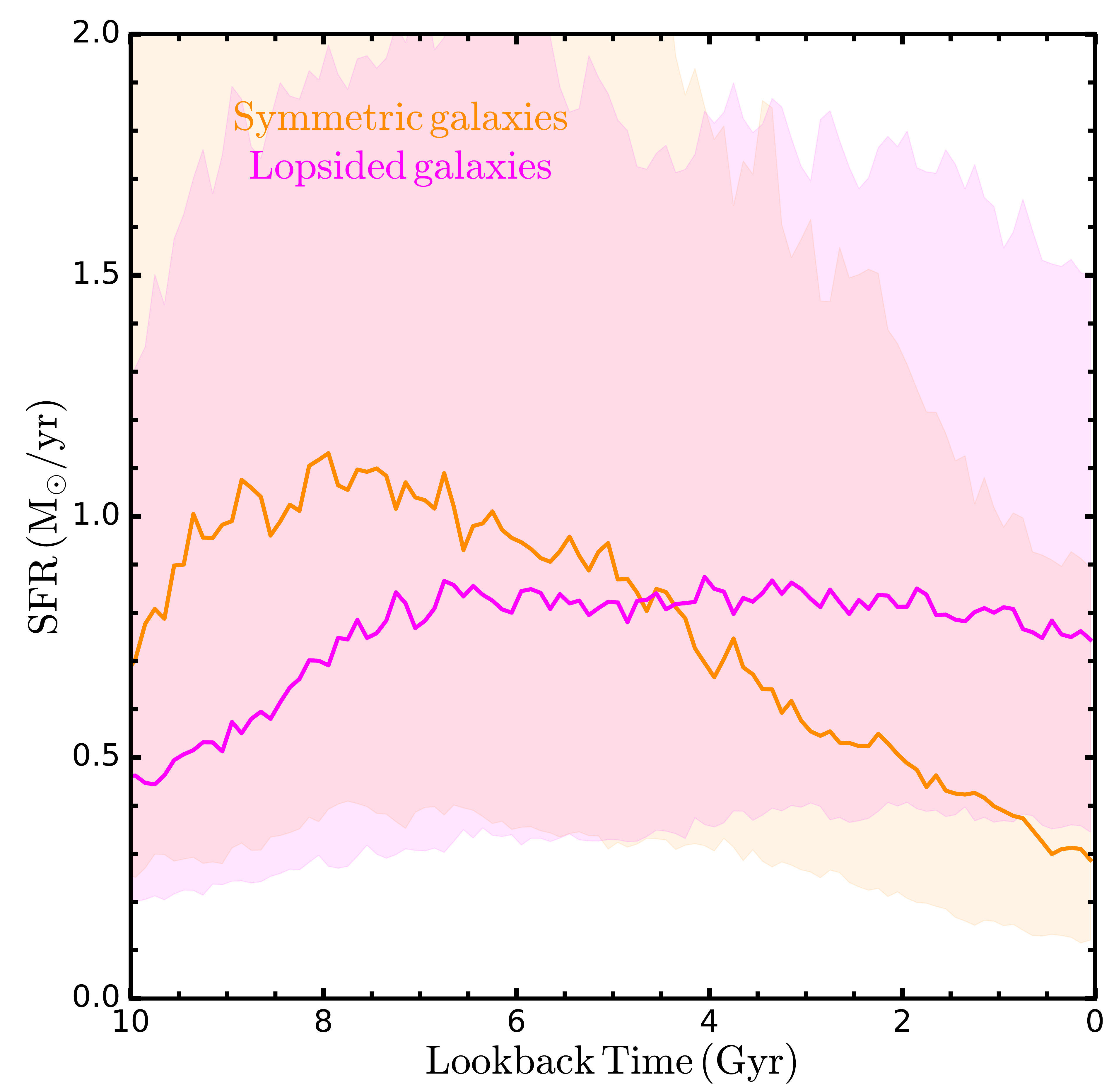}
    \includegraphics[width=0.33\textwidth]{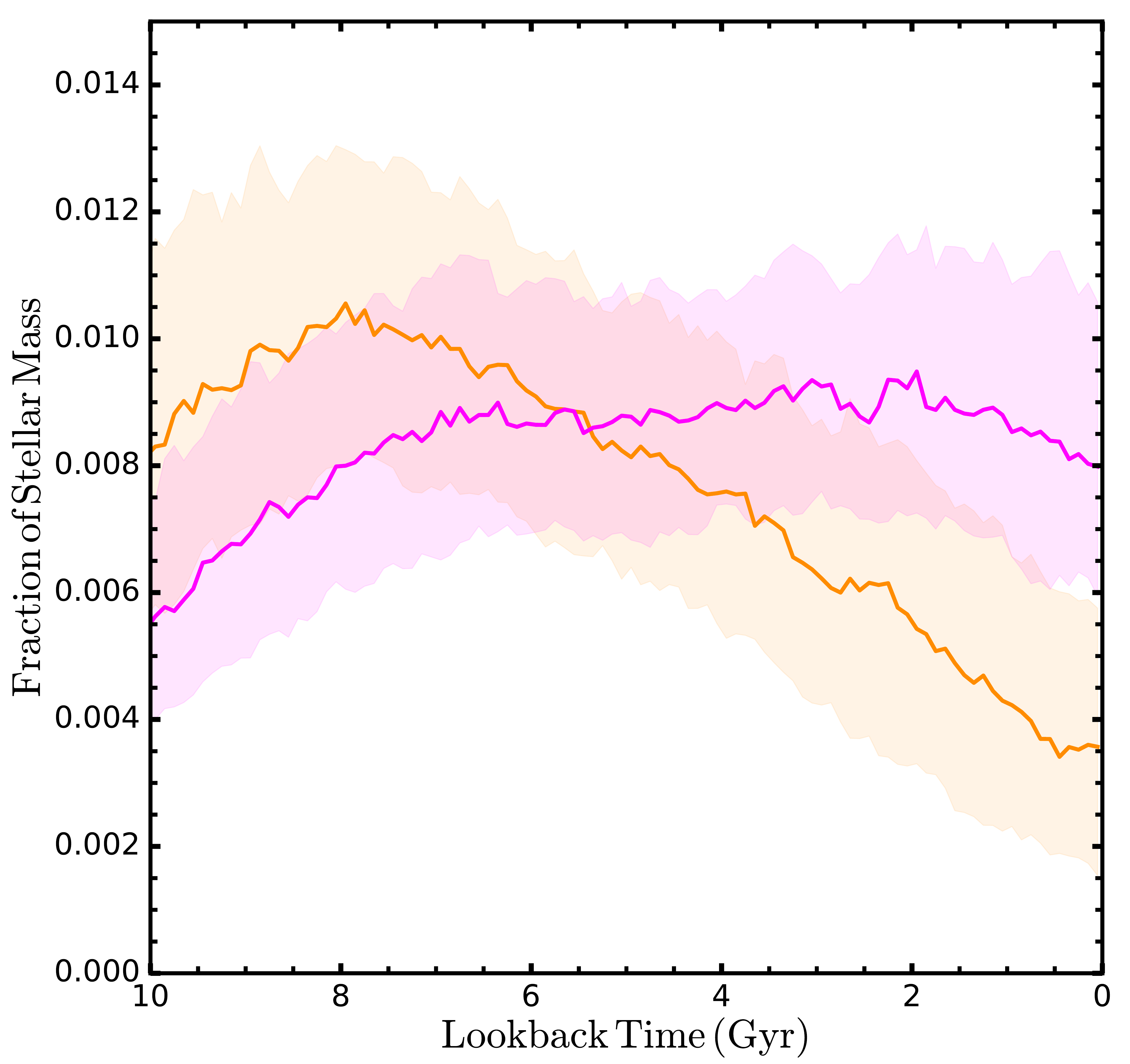}
    \includegraphics[width=0.32\textwidth]{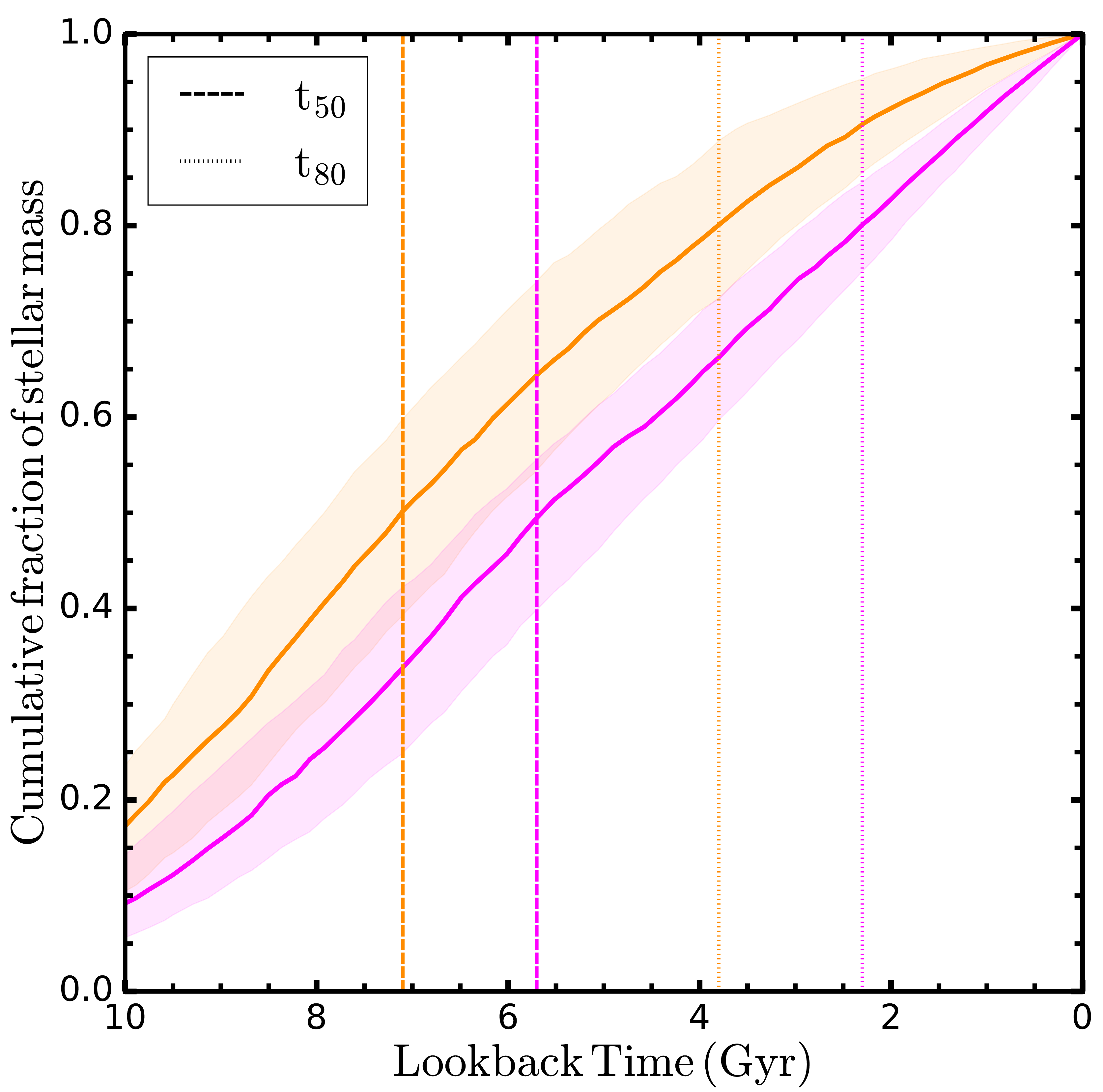}
    \caption{{\bf Left:} The stacked star formation history (SFH) within the last $10\, \mathrm{Gyr}$ of the lopsided (magenta) and symmetric (orange) galaxies in our selected galaxy sample, calculated in bins of $100\, \mathrm{Myr}$ as described in Sec. \ref{sec:star_formation_history}. The solid lines represent the stacked SFH of the strongly lopsided and strongly symmetric sub-samples (see Sec. \ref{sec:star_formation_history}). The shaded areas represent the $25\mathrm{th}$-$75\mathrm{th}$ interquartile range of the data in each time interval. {\bf Middle:} The median fraction of the stellar mass within the last $10\, \mathrm{Gyr}$ of the lopsided and symmetric galaxies in our selected galaxy sample, calculated in bins of $100\, \mathrm{Myr}$. The shaded areas represent the $25\mathrm{th}$-$75\mathrm{th}$ interquartile range of the data in each time interval. {\bf Bottom:} The cumulative fraction of the galaxy stellar mass of the strongly lopsided and strongly symmetric sub-samples as a function of time. The vertical dashed and dotted lines show the time at which the galaxy built up $50\%$ (i.e. $t_{50}$) and $80\%$ (i.e. $t_{80}$) of the total stellar mass, respectively. The description is as in the left panel. We see that the symmetric galaxies are characterized by a peaked SFR at early times that steeply decreases until $z=0$, while the lopsided galaxies maintain an overall constant SFR until $z=0$, suggesting an earlier assembly for the symmetric galaxies than the lopsided ones.}
    \label{fig:star_formation_history_all}
\end{figure*}

\subsection{Correlations between lopsidedness, environment and morphology in the TNG100 simulation}
\label{sec:tng100}
To explore the effect of the environment using a more statistically representative sample of galaxies towards higher density environments (i.e. $\mathrm{M_{200}}\gtrsim10^{12}\, \mathrm{M_{\odot}}$) at $z=0$, we analyze the TNG100 simulation. 
We repeat the analysis performed in the previous Sec. \ref{sec:environment_halomass} and \ref{sec:environment_neighbours} for TNG50 and we find the following results in TNG100, which we briefly summarize below (see Appendix \ref{sec:role_environment_tng100} for more details):
\begin{itemize}
    \item We find that the behaviour of the lopsidedness amplitude also shows an overall constant to mildly decreasing trend as a function of the environment, typically independent of the metric adopted, similarly to the TNG50 simulation. However, we find that the lopsidedness amplitude tend to decrease more steeply with the increasing number of massive neighbours within the galaxy virial radius, $R_{200}$.
    \item We find that galaxies tend to be more disk-like in TNG50 than in TNG100 at a given total halo mass, with larger $\mathrm{D/T}$, and smaller $c/a$ and $\mu_{*}$, on average. However, we still find evidence in TNG100 of a mild trend of decreasing $\mathrm{D/T}$ and increasing $\mu_{*}$ towards high-density environments which correlates with the change in the lopsidedness amplitude. Again, this indicates a dependence of lopsidedness with the galaxy morphology as also seen in TNG50 in Fig. \ref{fig:lopsidedness_environment_dependence_m200} and \ref{fig:environment_dependence_m200}.
    \item We find that lopsided galaxies tend to be characterized by, on average, larger $\mathrm{D/T}$ and smaller $c/a$ and $\mu_{*}$ than the symmetric galaxies also in TNG100, independently of the environment.
\end{itemize}

Therefore, the TNG100 simulation confirms the results from the TNG50 simulation that lopsidedness does not significantly depend on the environment but it more strongly correlates
with the distinct internal properties of the galaxies, which may originate either as a result of the environmental evolution or the different specific evolutionary histories of the galaxies.

\begin{figure*}
    \centering
    \includegraphics[width=0.34\textwidth]{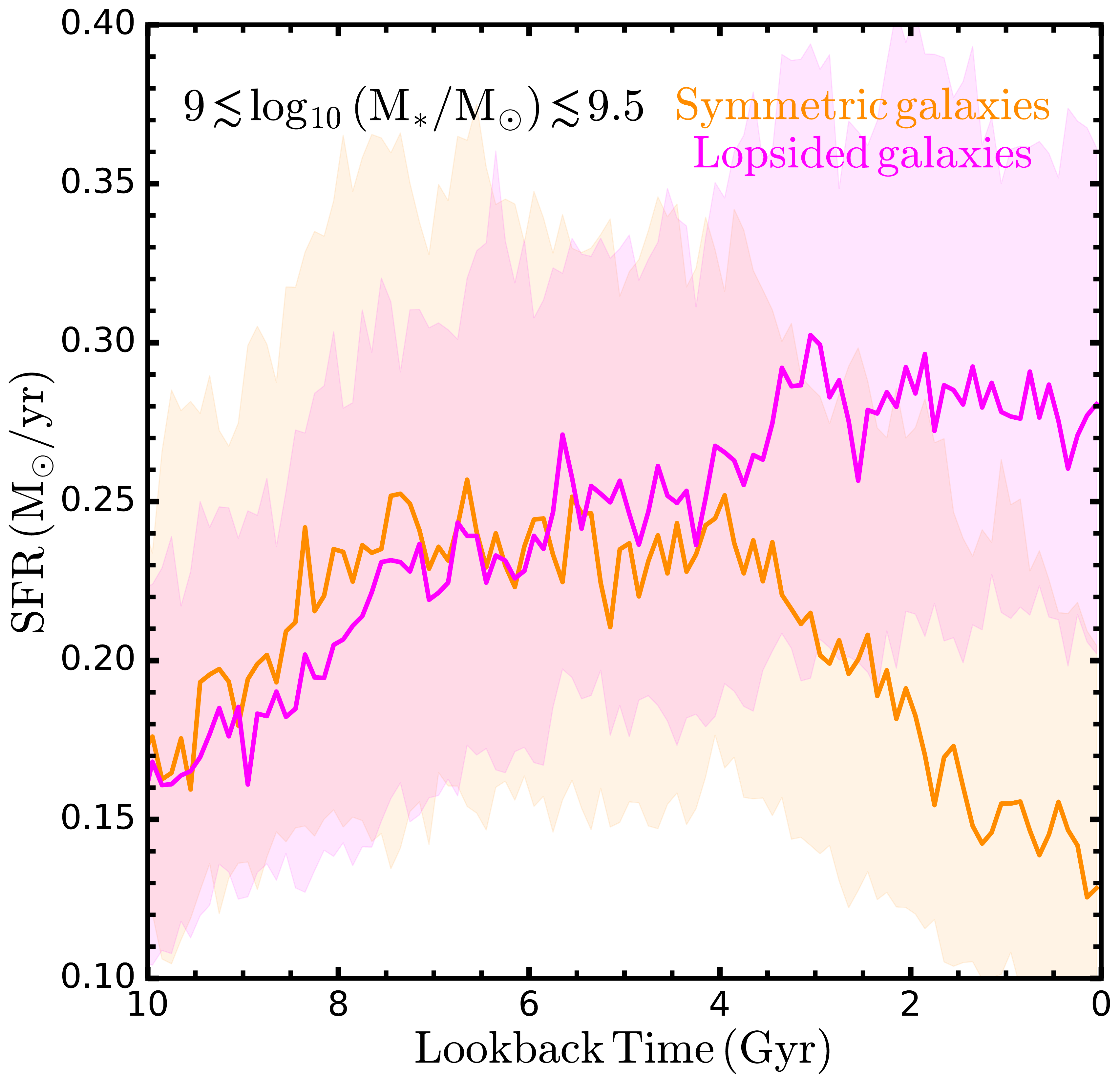}
    \includegraphics[width=0.33\textwidth]{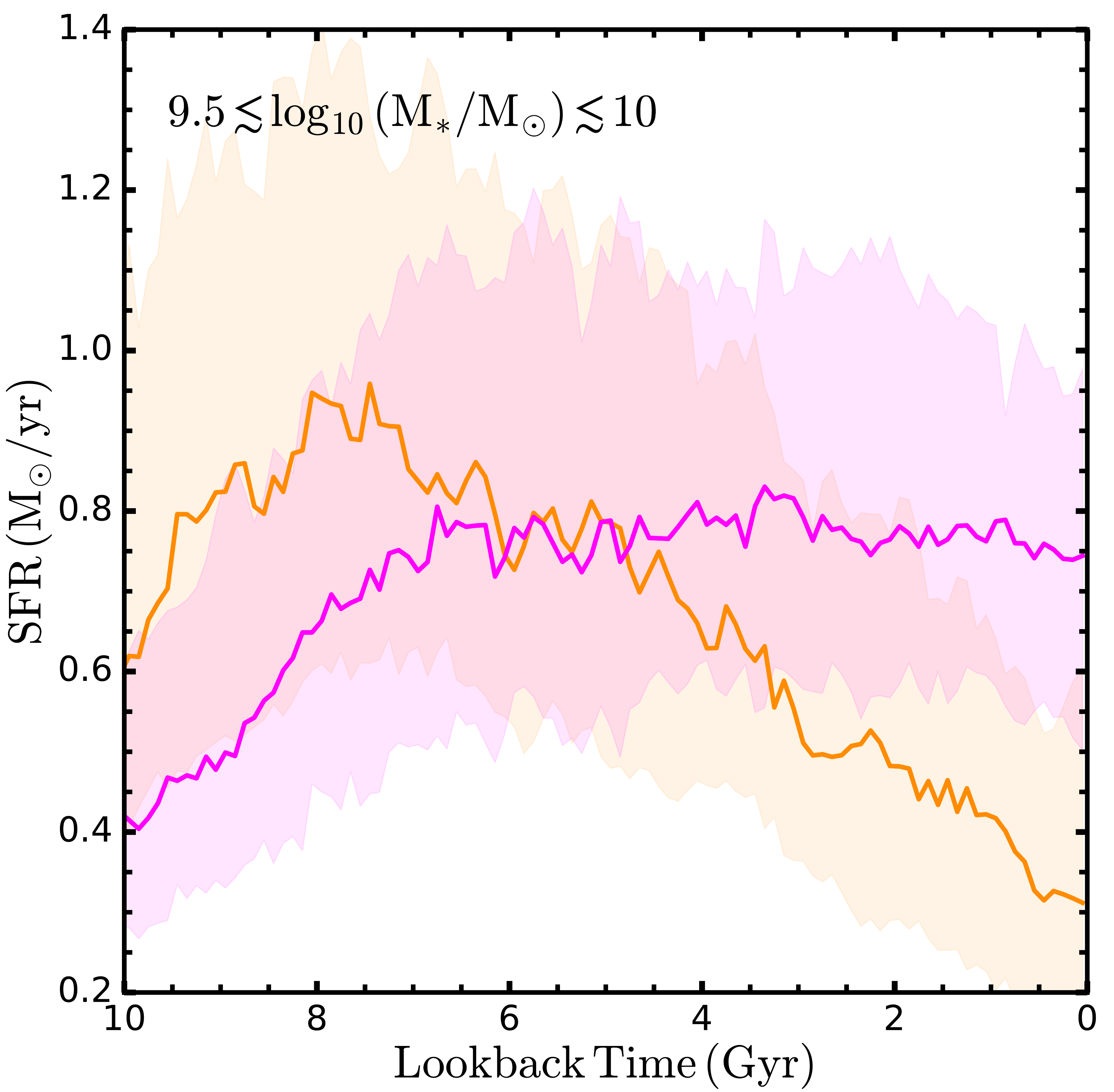}
    \includegraphics[width=0.32\textwidth]{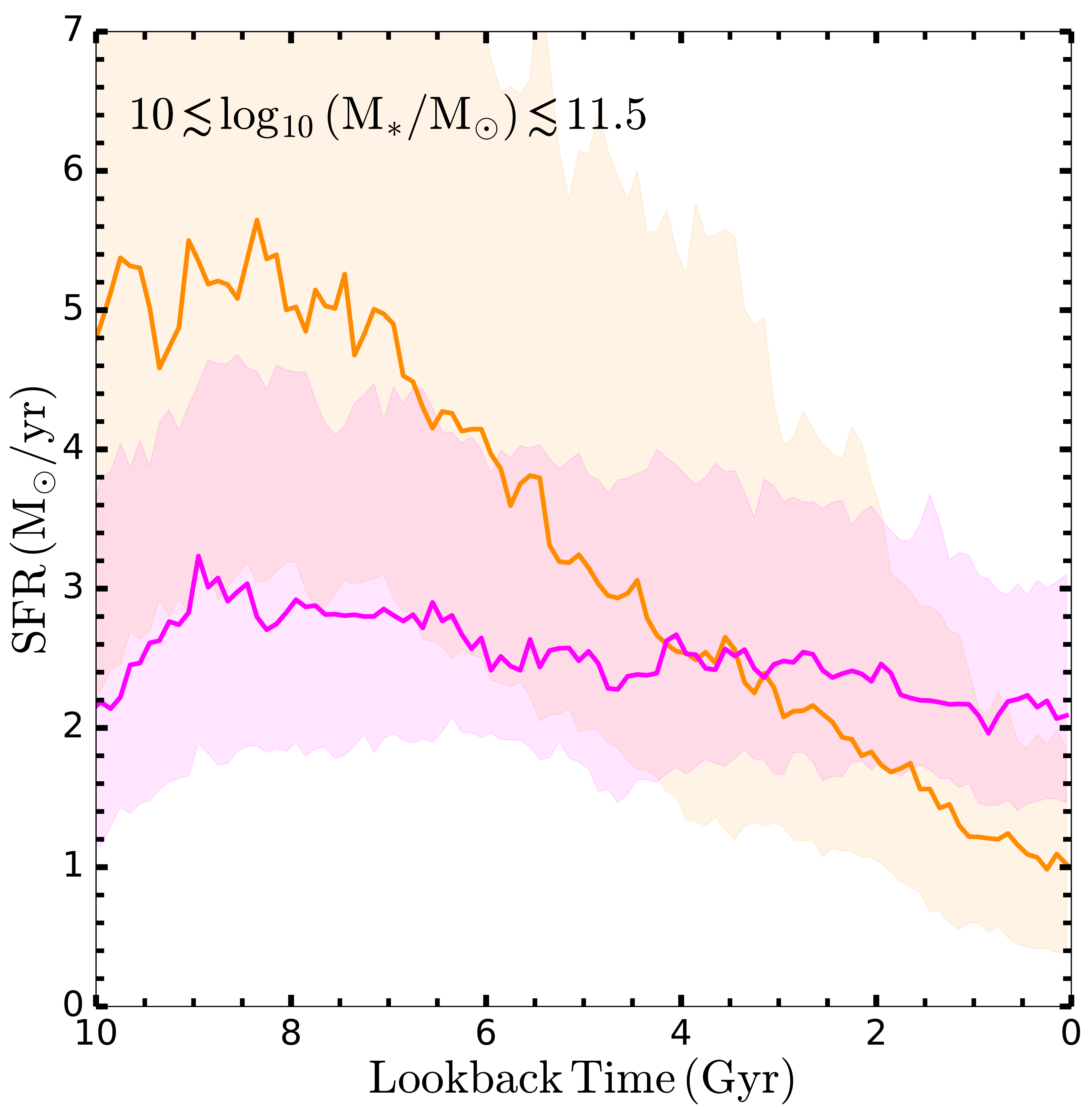}    

    \includegraphics[width=0.33\textwidth]{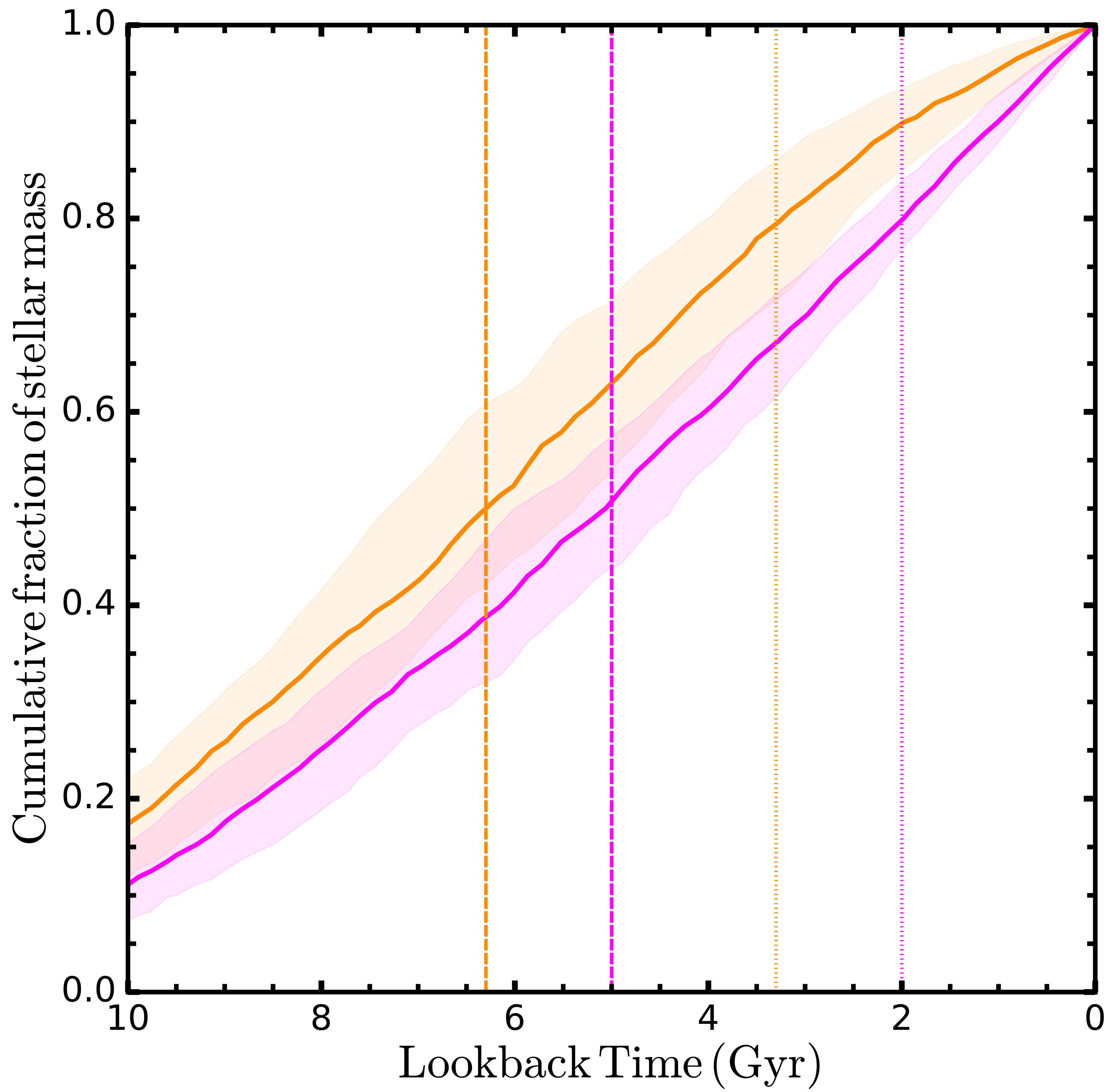}
    \includegraphics[width=0.33\textwidth]{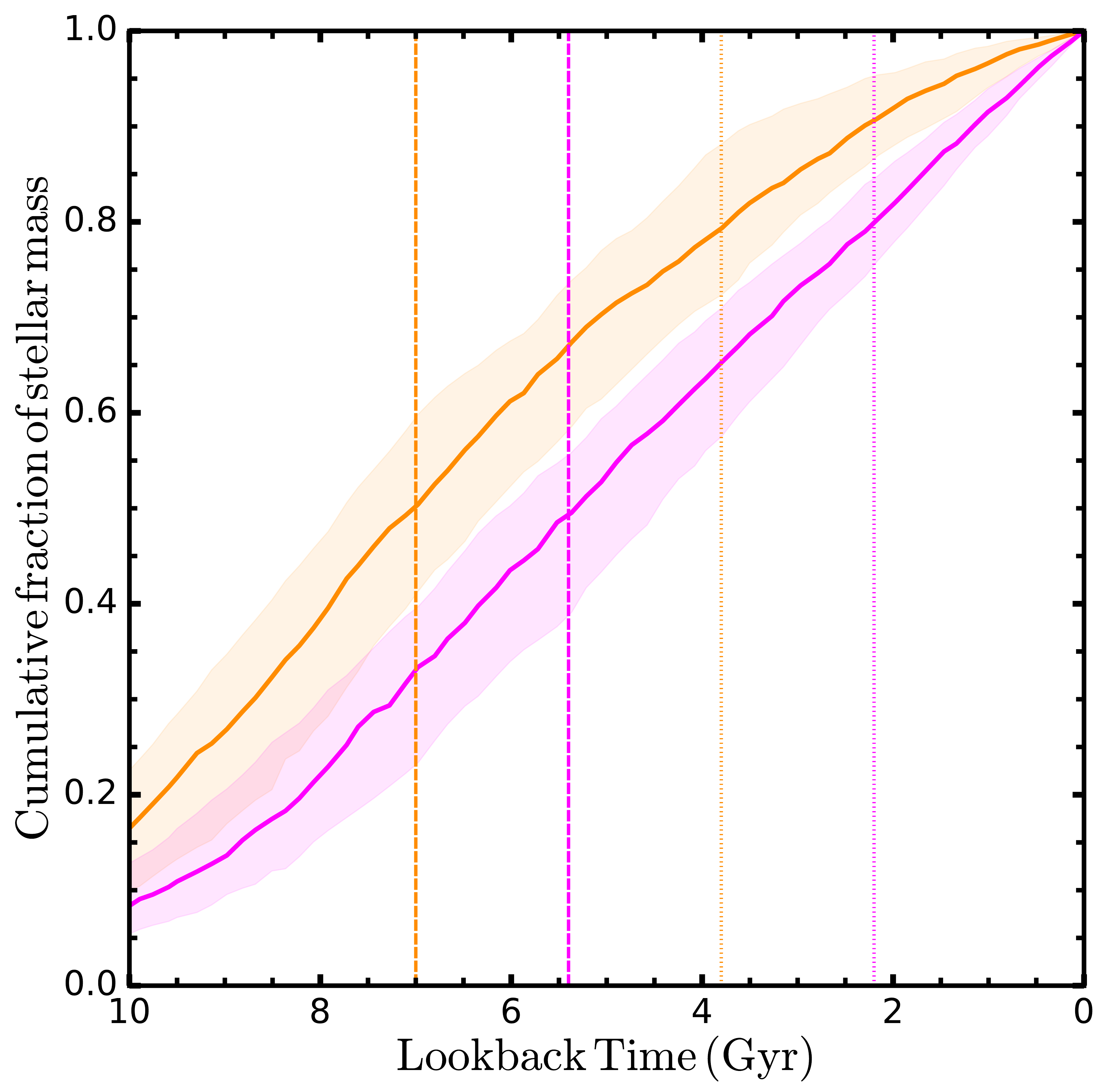}
    \includegraphics[width=0.33\textwidth]{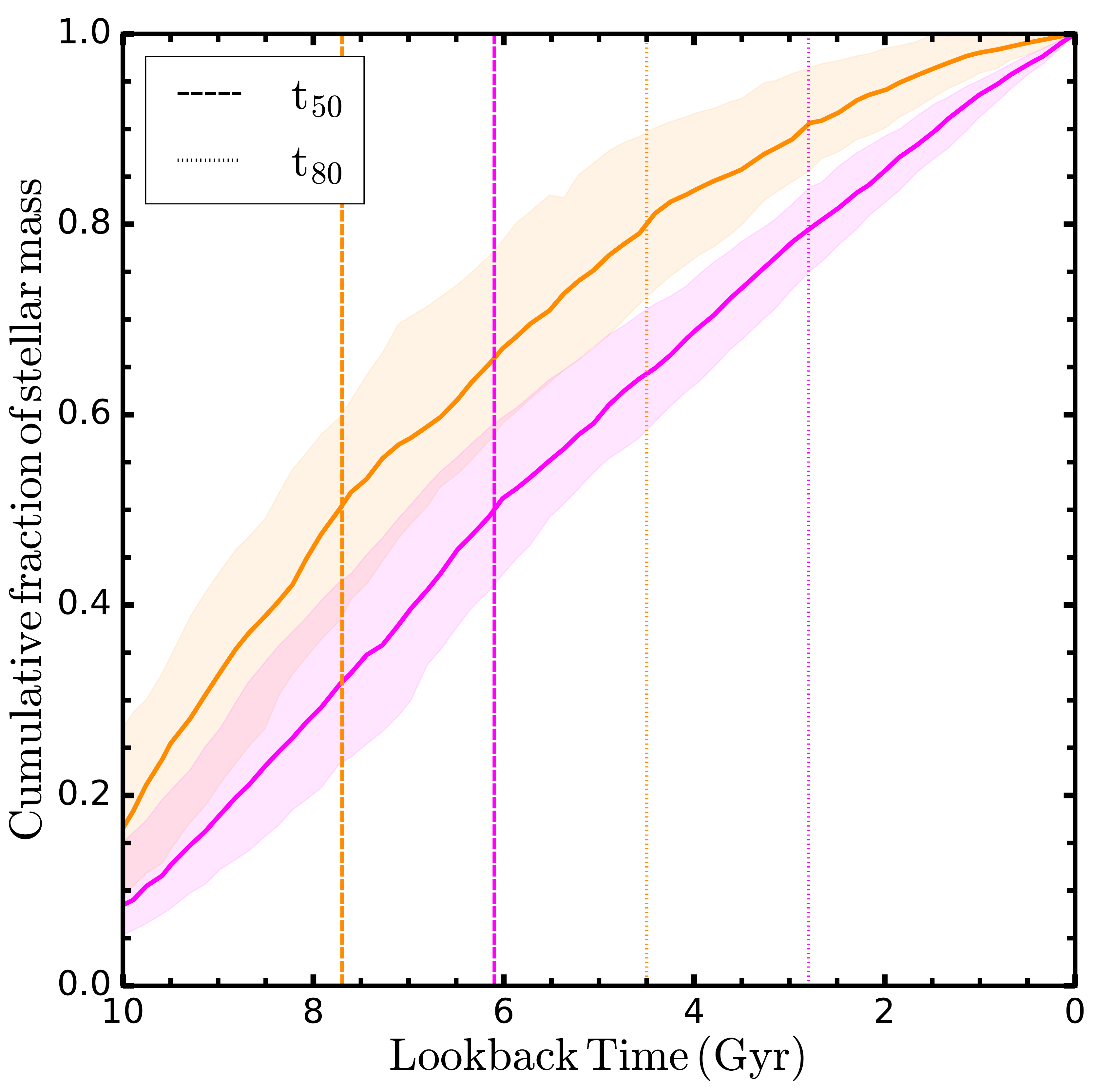}
    \caption{{\bf Top:} The stacked star formation history (SFH) of the lopsided and symmetric galaxies in our selected galaxy sample within the last $10\, \mathrm{Gyr}$ divided in bins of galaxy stellar mass, defined as described in Sec. \ref{sec:star_formation_history}. {\bf Bottom:} The cumulative fraction of the galaxy stellar mass of the strongly lopsided and strongly symmetric sub-samples as a function of time, divided in bins of galaxy stellar mass. The description is as in Fig. \ref{fig:star_formation_history_all}. We see that the lopsided and symmetric galaxies show the two distinct SFH behaviours from the previous Fig. \ref{fig:star_formation_history_all} regardless of the galaxy stellar mass, which reinforces the possibility that these galaxies have assembled through different evolutionary pathways.}
    \label{fig:star_formation_history_massbins}
\end{figure*} 

\section{The evolution histories of the lopsided and symmetric galaxies}
\label{sec:evolution_history}

\subsection{The star formation history}
\label{sec:star_formation_history}
In the previous Sec. \ref{sec:role_environment}, we have studied the correlation between lopsidedness, environment and galaxy morphology at $z=0$. We have seen that the lopsided galaxies are typically characterized by distinct internal properties (i.e. larger $\mathrm{D/T}$, and smaller $c/a$ and $\mu_{*}$) than the symmetric galaxies, not only in high-density environments but also in low-density ones where galaxy interactions are less common.
In this section, we aim to further investigate the nature of the distinct internal properties of the lopsided and symmetric galaxies by studying their evolution histories.

In Fig. \ref{fig:star_formation_rate_present_day}, we show the present-day specific star formation rate (sSFR) as a function of galaxy total stellar mass at $z=0$ for the lopsided and symmetric galaxies in our selected sample. 
We see that the lopsided galaxies are, on average, more star forming than the symmetric ones with median sSFR typically above the star forming main sequence (SFMS; \citealt{Renzini2015}).
This is generally consistent with the results by \citet{Lokas2022}, who studied lopsided galaxies in the TNG100 simulation and found that they typically show larger gas fractions, higher star formation rates (SFR), bluer colours and lower metallicity at the present-day than the symmetric counterparts within the inner $2\, R_{\mathrm{half}}$. Since the authors did not identify any particular type of past interaction in the evolution histories of the galaxies that could have led to their present-day lopsidedness, they suggested asymmetric star formation as mechanism for the origin of lopsidedness due to the large present-day gas fractions and recent increase in SFR seen in their lopsided sample. 
They also suggest an accreted origin for the gas due to its low metallicity nature (compared to the metallicity of the gas in their symmetric sample). Indeed, they find that, while only $9\%$ of their lopsided galaxies experienced a significant gas-rich merger in recent times (i.e. $z<0.1$), almost all of them continued to accrete small satellite galaxies with $\mathrm{M_{\mathrm{tot}}} \sim 10^{8}\, \mathrm{M_{\odot}}$ until the present-day. However, further studies would be necessary to understand how the in-situ or early/late accreted nature of the gas and its evolution until the present-day may be linked to the origin of lopsidedness.

\begin{figure}
    \centering
    \includegraphics[width=0.45\textwidth]{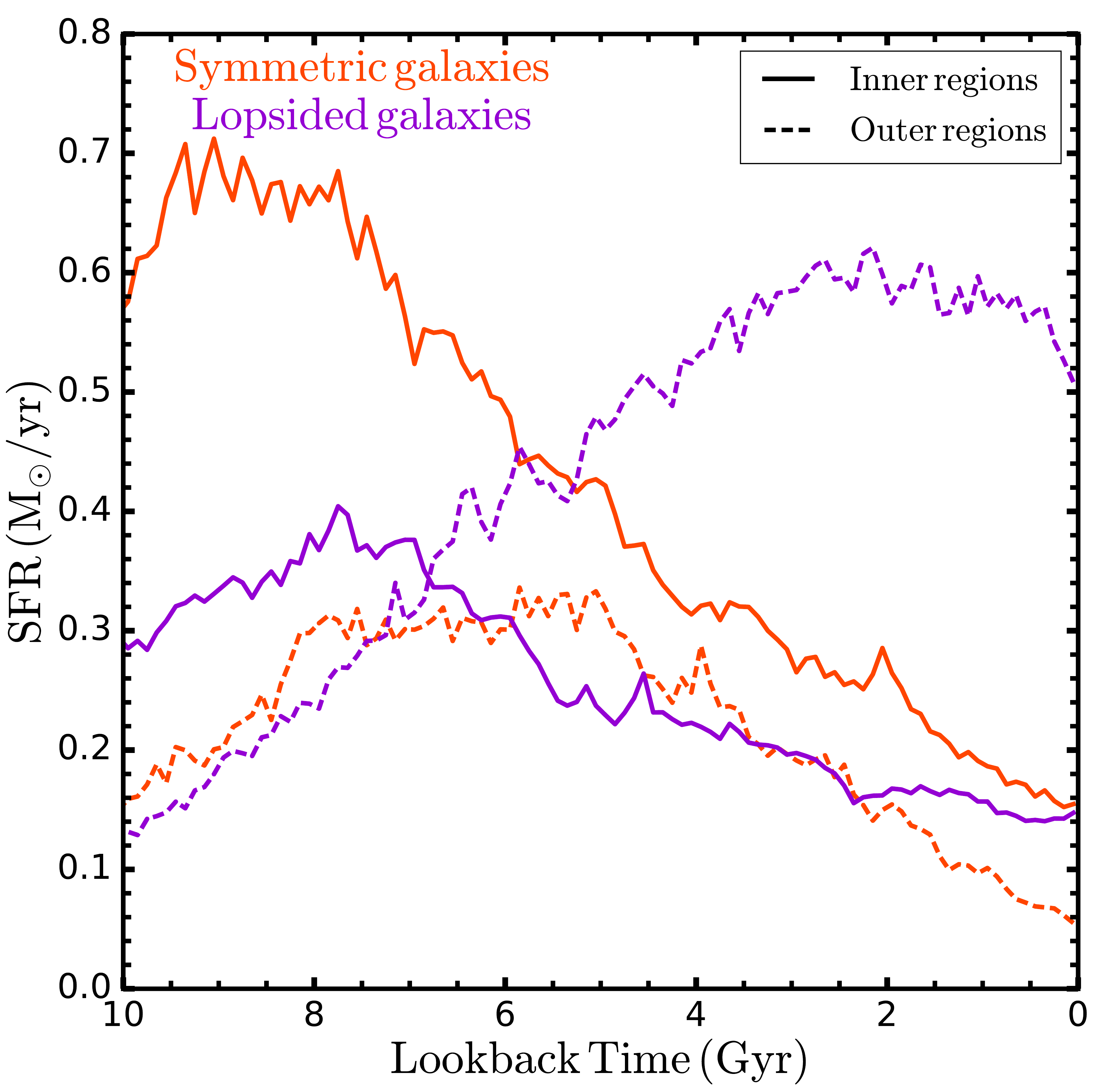}
    \caption{The stacked star formation history (SFH) of the strongly lopsided and strongly symmetric sub-samples within the last $10\, \mathrm{Gyr}$ for the inner (i.e. $<3.75\, \mathrm{kpc}$; solid lines) and outer (i.e. $>3.75\, \mathrm{kpc}$; dashed lines) components of the galaxies, separately. Here, we omit showing the $25\mathrm{th}$-$75\mathrm{th}$ interquartile range as the scatter of the data in each bin to avoid overcrowding. We see that the symmetric galaxies show similar SFH in the inner and outer components, characterized by two peaks in the SFR at early times (i.e. $\sim8\, \mathrm{Gyr}$ and $\sim6\, \mathrm{Gyr}$ ago), respectively, beyond which the SFR steadily decreases until $z=0$. On the other hand, the outer regions of the lopsided galaxies show a steadily increasing SFH until $z=0$ with the peak in the SFR occurring very recently at $\sim2\, \mathrm{Gyr}$ ago, suggesting that they experienced a more secular evolution where the star formation was slowly quenched in the inner regions, but it continued forming new stars in the outer regions of the galaxies until $z=0$.}
    \label{fig:star_formation_history_core_all}
\end{figure}

To further explore this, in Fig. \ref{fig:star_formation_history_all}, we study the star formation histories (SFH) of the lopsided and symmetric galaxies, i.e. the SFR as a function of the lookback time. Our goal is to investigate whether there are any differences in how these galaxies formed their stars and built up their mass over time, thus explaining their distinct present-day internal properties. 
Specifically, we consider the sub-samples of the most strongly lopsided and most strongly symmetric galaxies, which are obtained by selecting the two outermost quartiles in the $\mathrm{A_{1}}$ distribution in Fig. \ref{fig:lopsidedness_distribution}. These two sub-samples contain $359$ galaxies each. We then derive the stacked SFH of our lopsided and symmetric sub-samples within the last $10\, \mathrm{Gyr}$ (see solid lines in the top panel of Fig. \ref{fig:star_formation_history_all}). For this purpose, we compute the median SFR of each sub-sample at different times. We note that we also derive the stacked SFR of the full lopsided and symmetric samples, using the same method described above. However, considering the consistency of the results, we will focus the discussion on the strongly lopsided and strongly symmetric sub-samples throughout the rest of the section. 
Here, the individual SFH of a galaxy is calculated by summing up the mass\footnote{Here, mass refers to the initial mass of the star particle when it was formed.} of all the star particles born in $100\, \mathrm{Myr}$ time intervals between now (i.e. lookback time, $t=0\, \mathrm{Gyr}$) and $t=10\, \mathrm{Gyr}$ ago, and normalizing it by the time interval (i.e. $100\, \mathrm{Myr}$) to obtain the SFR in units of $\mathrm{M}_{\odot}/\mathrm{yr}$. For this calculation, we consider only the star particles located within the optical radius of the galaxy in the face-on projection. 

Fig. \ref{fig:star_formation_history_all} reveals that the lopsided and symmetric galaxies show two distinct SFH behaviours. While the symmetric galaxies are characterized by a peaked SFR at early times (i.e. $\sim8\, \mathrm{Gyr}$ ago) that steeply decreases until $z=0$, lopsided galaxies are, instead, characterized by a mildly increasing SFR until $\sim8\, \mathrm{Gyr}$ ago which, subsequently, remains overall constant until $z=0$. 
We also see that the lopsided galaxies show on average lower SFR than the symmetric ones at early times (i.e. more than $\sim4\, \mathrm{Gyr}$ ago), while the opposite is true at recent times (i.e. within the last $\sim4\, \mathrm{Gyr}$), consistent with the finding that the lopsided galaxies are more star forming than the symmetric ones at $z=0$ (see Fig. \ref{fig:star_formation_rate_present_day}).

In the middle and right panels of Fig. \ref{fig:star_formation_history_all}, we show the stellar mass fraction as a function of time, highlighting how these two sub-samples assemble their mass over time.
Specifically, we see that the symmetric galaxies have typically built up their total stellar mass earlier and more rapidly than the lopsided ones. In fact, the symmetric galaxies already reached $50\%$ and $80\%$ of their total stellar mass at $\mathrm{t_{50}}\sim7.1\, \mathrm{Gyr}$ and $\mathrm{t_{80}}\sim3.8\, \mathrm{Gyr}$ ago, while this occurs at $\mathrm{t_{50}}\sim5.7\, \mathrm{Gyr}$ and $\mathrm{t_{80}}\sim2.3\, \mathrm{Gyr}$ ago for the lopsided galaxies, respectively. Thus, lopsided galaxies assemble their mass $\sim1.5\, \mathrm{Gyr}$ later, on average, than the symmetric ones.

In Fig. \ref{fig:star_formation_history_massbins}, we see that the results shown in Fig. \ref{fig:star_formation_history_all} continue to hold regardless of the galaxy stellar mass. 
Here, the stellar mass bins are defined by dividing the galaxies into three equally-sized galaxy groups, containing the same number of galaxies: low ($\mathrm{M_{*}}\lesssim10^{9.5}\, \mathrm{M_{\odot}}$), intermediate ($10^{9.5}\, \mathrm{M_{\odot}}\lesssim \mathrm{M_{*}}\lesssim10^{10}\, \mathrm{M_{\odot}}$) and high ($\mathrm{M_{*}}\gtrsim10^{10}\, \mathrm{M_{\odot}}$) stellar mass bins.
In the top panels of Fig. \ref{fig:star_formation_history_massbins}, we see that the symmetric galaxies are always characterized by a peaked SFR at early times that steeply decreases until $z=0$, as seen in Fig. \ref{fig:star_formation_history_all}, independently of the mass bin considered. 
The duration and occurrence of the peak in SFR of the symmetric galaxies shows some differences between each mass bin. For example, in the low mass bin, the peak in SFR is less pronounced and lasts longer (i.e. between $\sim8$-$4\, \mathrm{Gyr}$ ago), while the peak in SFR is stronger and shorter in time (occurring $\sim8\, \mathrm{Gyr}$ ago) in the intermediate and high mass bins, consistent with the "downsizing" formation scenario. The peak in SFR is also typically stronger in the high mass bin than in the intermediate one.
The lopsided galaxies also show an overall similar behaviour in all mass bins without any prominent peak in SFR, consistent with the typically constant SFR as a function of time shown in Fig. \ref{fig:star_formation_history_all}.
Interestingly, we note some differences between the SFH of the lopsided and symmetric galaxies in each mass bin. In the low mass bin, we see that the lopsided and symmetric galaxies have similar early SFR (i.e. $\gtrsim4\, \mathrm{Gyr}$ ago), but they significantly differ at later times. In fact, the SFR of the lopsided galaxies shows a steadily increasing behaviour towards $z=0$, as opposed to the peaked and decreasing one of the symmetric galaxies. In the intermediate mass bin, the SFR of the lopsided and symmetric galaxies significantly differs both at early and late times and, as in the low mass bin, we distinguish between an increasing to constant SFR behaviour for the lopsided galaxies and a peaked and decreasing one for the symmetric counterpart towards $z=0$. Finally, in the high mass bin, both the lopsided and symmetric galaxies show a peaked and decreasing SFR behaviour towards $z=0$, but the peak in SFR significantly differs at early times. In fact, the lopsided galaxies show a much flatter peak in SFR $\sim8\, \mathrm{Gyr}$ than the symmetric counterpart.
In the bottom panels of Fig. \ref{fig:star_formation_history_massbins}, we see, once again, that the two distinct SFH behaviours are driving the different stellar mass growth of the lopsided and symmetric galaxies over time in all mass bin, with the symmetric galaxies typically assembling $50\%$ and $80\%$ of their total stellar mass $\sim1.5\, \mathrm{Gyr}$ earlier, on average, than the lopsided ones.

Based on the results shown in Fig. \ref{fig:star_formation_history_all} and \ref{fig:star_formation_history_massbins}, it is clear that the lopsided and symmetric galaxies have followed different evolution pathways that contributed to shaping their distinct present-day internal properties.
The symmetric galaxies have assembled at early times with relatively short and intense bursts of central star formation, possibly triggered by gas-rich major mergers. Since galaxies that form at early times assemble their stars at a time when the Universe was denser and richer in gas, this scenario is generally expected to produce compact galaxies with small sizes (e.g. \citealt{Wellons2015}) which is consistent with the, on average, higher central stellar mass density, smaller disk size and larger thickness of the inner galactic regions of the symmetric galaxies at $z=0$ seen in Fig. \ref{fig:environment_dependence_m200} and \ref{fig:environment_dependence_neighbours_galaxy_properties}.
On the other hand, lopsided galaxies have assembled on longer timescales and with milder initial bursts of star formation. They continued building up their mass until recent times, possibly through the continuous accretion of gas that kept the SFR roughly constant until $z=0$ \citep{Combes2004}. This scenario may have contributed to producing the lower central stellar mass density and larger disk size, on average, of the lopsided galaxies at $z=0$ seen in Fig. \ref{fig:environment_dependence_m200} and \ref{fig:environment_dependence_neighbours_galaxy_properties}, since they are characterized by a more secular evolution that continued growing the galaxy stellar mass until the present-day without significant initial starbursts as in the symmetric galaxies. 

\begin{figure*}
    \centering
    \includegraphics[width=0.33\textwidth]{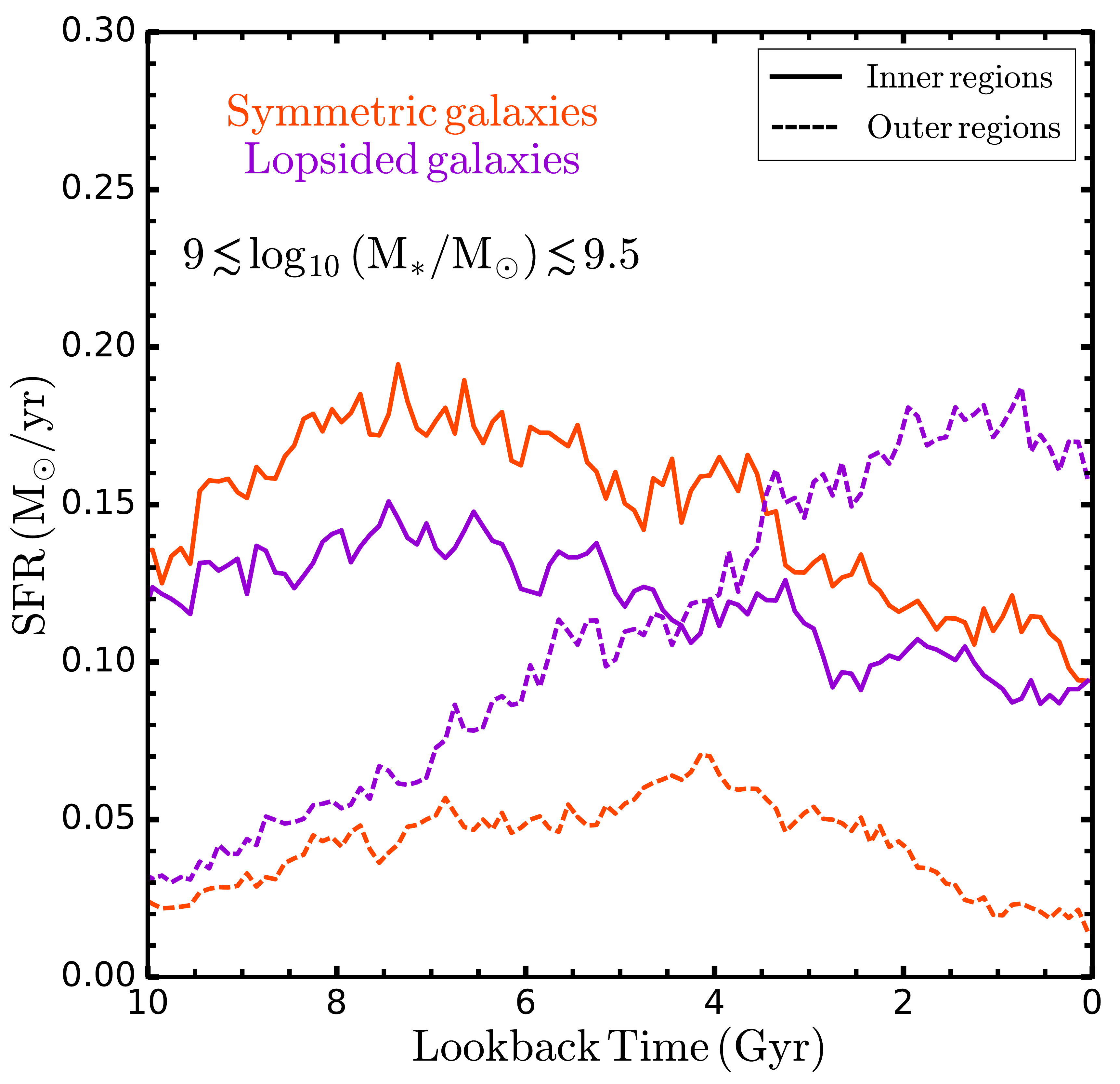}
    \includegraphics[width=0.33\textwidth]{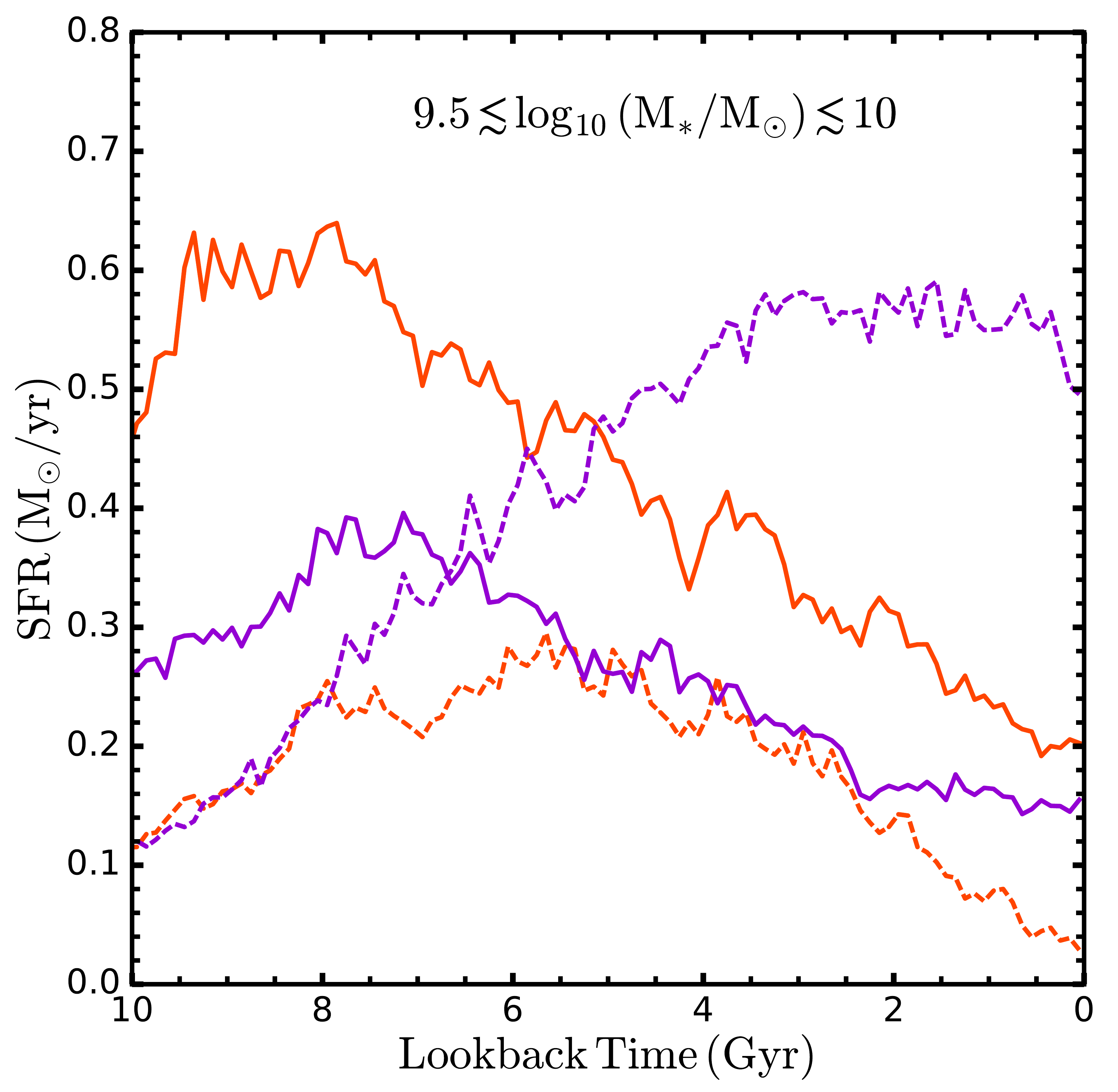}
    \includegraphics[width=0.33\textwidth]{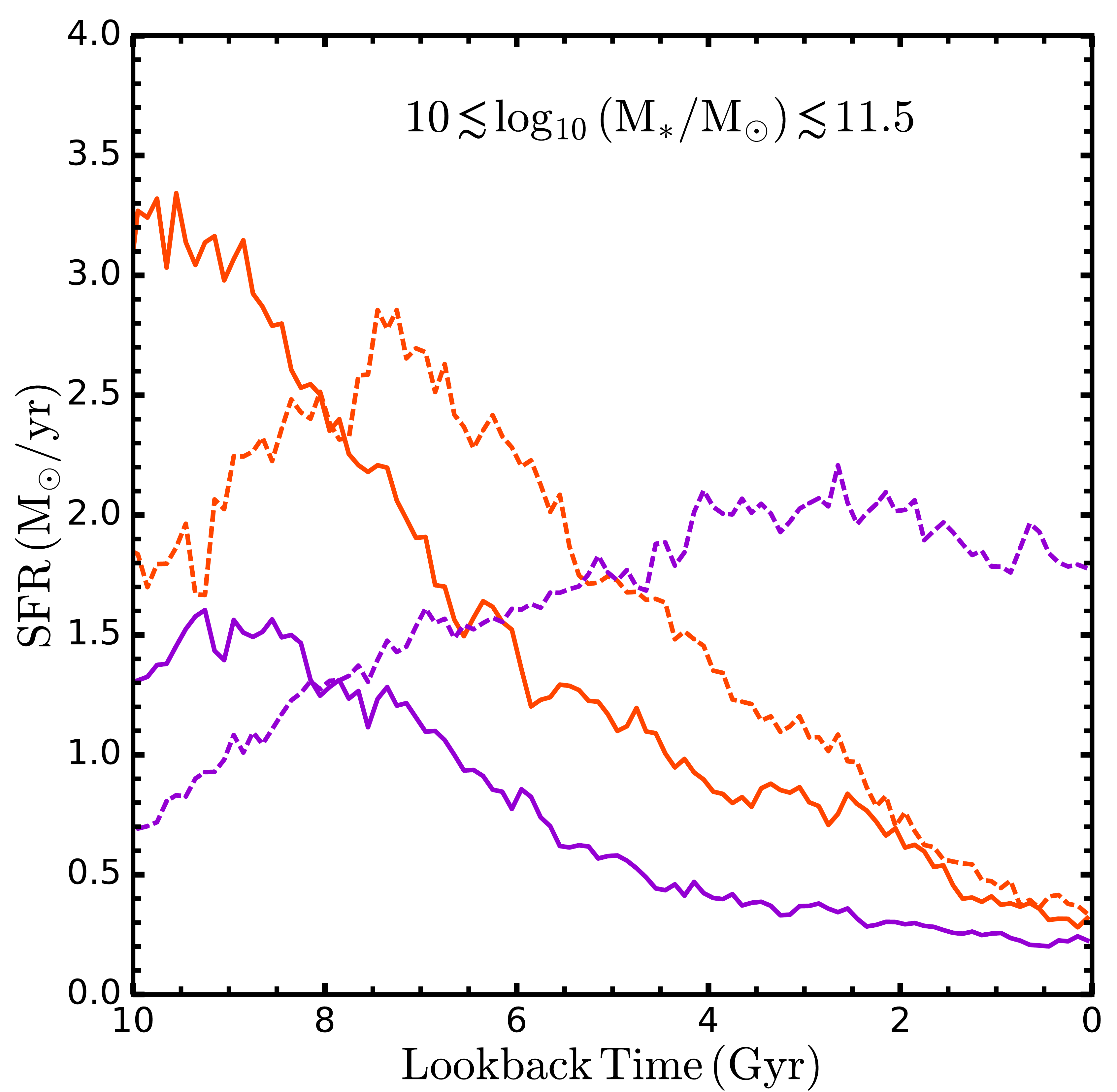}
    \caption{The stacked star formation history (SFH) of the strongly lopsided and strongly symmetric sub-samples within the last $10\, \mathrm{Gyr}$ for the inner (i.e. $<3.75\, \mathrm{kpc}$; solid lines) and outer ($>3.75\, \mathrm{kpc}$; dashed lines) components of the galaxies, separately, divided in bins of galaxy stellar mass, defined as described in Sec. \ref{sec:star_formation_history}. We see that the SFH of the inner and outer regions of the lopsided and symmetric galaxies are still consistent with those shown in Fig. \ref{fig:star_formation_history_core_all} regardless of the stellar mass, suggesting an early assembly history for the symmetric galaxies and a more extended and secular evolution for the lopsided galaxies.}
    \label{fig:star_formation_history_core_subsamples}
\end{figure*}

In order to further explore the different SFH displayed by these two different galaxy sub-samples, we dissect each galaxy into the inner (i.e. $<3.75\, \mathrm{kpc}$; solid lines) and outer (i.e. $>3.75\, \mathrm{kpc}$; dashed lines) components in Fig. \ref{fig:star_formation_history_core_all}. Here, the value to separate between the inner and outer galaxy components corresponds to the median of the $R_{\rm half}$ distribution of all the galaxies in our sample at $z=0$\footnote{We have also tried using the $R_{\rm half}$ of each galaxy to dissect the galaxy into the inner and outer components and we obtain similar results.}. For each component, we show the stacked SFH of the strongly lopsided and strongly symmetric sub-samples within the last $10\, \mathrm{Gyr}$. Overall, we see that the symmetric galaxies have typically assembled both their inner and outer components in two early bursts of star formation $\sim8$ and $\sim6\, \mathrm{Gyr}$ ago. However, the inner components of the symmetric galaxies have typically assembled with a significantly more intense star formation burst than the outer components. 
The lopsided galaxies have also assembled their inner components at similar early times as the symmetric counterpart, but with a significantly milder star formation burst. 
Interestingly, the lopsided galaxies have typically assembled their outer components very recently (i.e. $\sim2\, \mathrm{Gyr}$ ago) with, on average, a more intense peak in SFR than that of their inner components. The peak in SFR of the outer components of the lopsided galaxies is similar in intensity to that of the inner components of the symmetric galaxies.

In Fig. \ref{fig:star_formation_history_core_subsamples}, we see that the results shown in Fig. \ref{fig:star_formation_history_core_all} continue to hold independently of the galaxy stellar mass. However, in each mass bin, we find some interesting differences in terms of the amplitude in the peak of SFR. The symmetric galaxies are characterized by very low SFR in their outer components in the low and intermediate mass bins. Instead, in the high mass bin, the symmetric galaxies have assembled their outer components at similar times as the inner ones at early times (i.e. $\gtrsim6\, \mathrm{Gyr}$ ago), in a second relatively short and intense peak in SFR. In all mass bins, the symmetric galaxies continued evolving mostly passively at later times (i.e. $\lesssim5\, \mathrm{Gyr}$ ago) until $z=0$. 
As previously mentioned, this is consistent with the earlier type disk morphology (i.e. small disk size, rounder shape of the inner galactic regions and high central stellar mass density) of the symmetric galaxies shown in Sec. \ref{sec:environment_halomass} and \ref{sec:environment_neighbours}. Lopsided galaxies also typically show two peaks in SFR in all mass bins, one of which is associated to the inner component and the other to the outer one. However, while the peak in SFR of the inner component occurs at similar early times (i.e. $\sim8\, \mathrm{Gyr}$ ago) as that of the symmetric counterpart, the peak in SFR of the outer components occurs at significantly later times (i.e. $\sim2\, \mathrm{Gyr}$ ago) and with a stronger or similar amplitude than that of the inner components in all mass bins.
This shows that the lopsided galaxies have also likely assembled their inner components at early times, possibly with a short burst of central star formation. However, the peak in SFR of the inner components of the lopsided galaxies is milder than that of the symmetric counterpart, thus explaining the lower central stellar mass density, on average, of the lopsided galaxies with respect to the symmetric counterpart. 
Then, at later times, lopsided galaxies continued growing their outer components until $z=0$ as a result of the continuous accretion of gas that maintained the SFR overall constant. This explains their larger disk size and higher flatness of the inner galactic regions compared to the symmetric counterpart. Overall, this scenario is also consistent with the results found by \citetalias{Varela-Lavin2022}, where it is shown that the stellar half-mass radius (i.e. $R_{\mathrm{half}}$) of the lopsided galaxies increases more significantly than that of the symmetric counterpart towards $z=0$. 

\section{Summary and Conclusions}
\label{sec:conclusions}
In this work, we have studied the presence of asymmetries (i.e. lopsidedness) in the mass distribution of the stellar component of disk-like galaxies selected from the TNG50 simulation at $z=0$, as described in Sec. \ref{sec:data}. Our sample contains both central and satellite galaxies located in environments characterized by a range of different total halo masses, from $\mathrm{M_{200}}/\mathrm{M_{\odot}} \sim 10^{10.5}$ (i.e. low-density environments) to $\mathrm{M_{200}}/\mathrm{M_{\odot}} \sim 10^{14}$ (i.e. Fornax cluster-like environments).
As in previous works, we have quantified the global lopsidedness by performing an azimuthal Fourier decomposition of the galaxy stellar mass in the face-on view and measuring the amplitude of the first Fourier mode, $\mathrm{A}_{1}$, in the radial interval between $0.5$-$1.1\, R_{\mathrm{opt}}$, where $R_{\mathrm{opt}}$ is the optical radius of the galaxy (see Sec. \ref{sec:lopsidedness}). 
Overall, we identify $834$ lopsided galaxies out of the $1435$ in our sample (i.e. $58\%$)). As previously mentioned in Sec. \ref{sec:lopsidedness}, this fraction of lopsided galaxies is significantly larger that that obtained using the observational radial interval, suggesting that deep observations extending out to the galaxy outskirts (i.e. disk optical radius, $R_{\mathrm{opt}}$) are necessary for detecting the full lopsided perturbation.

In Sec. \ref{sec:role_environment}, we have investigated the dependence of the lopsidedness with both the environment and galaxy morphology at $z=0$. We list here our main results from this analysis:
\begin{enumerate}[(i)]\itemsep0.2cm
    \item we find a mild correlation between the lopsidedness amplitude and the environment, with $\mathrm{A_{1}}$ only weakly decreasing with increasing total halo masses as well as number and local volume density of close massive neighbours, i.e. high-density environments;
    \item we find a correlation between lopsidedness and galaxy morphology, where galaxies with more extended disks, flatter inner galactic regions and lower central stellar mass density (i.e. late-type disk galaxies) are typically more lopsided than galaxies with smaller disks, rounder inner galactic regions and larger central stellar mass density. The former (late-type) galaxies are also typically located in lower density environments than the latter (early-type) galaxies, suggesting a change in galaxy morphology with the environment consistent with the observational results from the morphology-density relation;
    \item lopsided galaxies tend to be characterized by, on average, more extended disks, flatter inner galactic regions and lower central stellar mass density than the symmetric counterpart, independently of the environment. 
\end{enumerate}

Overall, these results show evidence of a stronger correlation between lopsidedness and the internal galaxy properties than with the environment, which mainly plays a secondary role in the origin of lopsidedness. The role played by the environment can be characterized by: 1) shaping the evolution of galaxies and their key internal properties in terms of lopsidedness, such as their central surface density and 2) inducing the onset of lopsided perturbations in galaxies with favourable internal properties as a result of an interaction \citepalias{Varela-Lavin2022}. The lack of correlation between lopsidedness and environment, and its stronger dependence on galaxy morphology is confirmed using a more statistically representative sample of galaxies in the TNG100 simulation towards higher density environments (i.e. $\mathrm{M_{200}}\gtrsim10^{12}\, \mathrm{M_{\odot}}$) at $z=0$ (see Sec. \ref{sec:tng100}). 

The fact that lopsided galaxies are characterized by, on average, more extended disks, flatter inner galactic regions and lower central stellar mass density than the symmetric counterpart independently of the environment suggests that other mechanisms, such as internal evolution, may be responsible for shaping the distinct present-day galaxy properties. 
Indeed, in Sec. \ref{sec:star_formation_history}, we find two distinct star formation history behaviours for the lopsided and symmetric galaxies. The symmetric galaxies are characterized by a peaked SFR at early times (i.e. $\sim8\, \mathrm{Gyr}$ ago) that steeply decreases until $z=0$, while the lopsided galaxies are characterized by an overall constant SFR until $z=0$. The lopsided galaxies also show, on average, lower SFR than the symmetric ones at early times (i.e. more than $\sim4\, \mathrm{Gyr}$ ago), while the opposite is true at recent times (i.e. within the last $\sim4\, \mathrm{Gyr}$). 
These results suggest that the lopsided and symmetric galaxies have followed different evolution pathways that contributed to shaping their distinct present-day internal properties.
The symmetric galaxies have assembled at early times with relatively short and intense bursts of central star formation, possibly triggered by gas-rich major mergers, that shaped their, on average, higher central stellar mass density, smaller disk size and larger thickness of the inner galactic regions at $z=0$ (e.g. \citealt{Wellons2015}). 
On the other hand, lopsided galaxies have assembled on longer timescales and with milder central bursts of star formation than the symmetric counterpart. They continued building up their outer regions until recent times, possibly through the continuous accretion of gas that kept the SFR roughly constant until $z=0$ \citep{Combes2004}. This secular evolution scenario contributed to producing the lower central stellar mass density, larger disk size and higher flatness of the inner galactic regions of the lopsided galaxies with respect to the symmetric counterparts at $z=0$.

To conclude, within this work, we find evidence of a clear connection between lopsidedness and the evolution of galaxies, which highlight the potential of probing the likely evolution history of present-day observed galaxies based on the presence of asymmetric features in their disks. For example, the correlation between lopsidedness and photometric $\mathrm{B/T}$ (see Sec. \ref{sec:bulgeless_galaxies}) indicates that a strong lopsided perturbation is more likely to develop in bulge-less galaxies, as it is observed in M101 \citep{Beale1969}, which may have evolved through a more secular evolution scenario based on our results.
In a follow-up work, we aim to explore the origin of the distinct SFH of the lopsided and symmetric galaxies by characterizing the nature of the central prominent peak in SFR at early times of the symmetric galaxies, in comparison to the early mild peak in SFR of the lopsided galaxies. To achieve this goal, we aim to characterize the environment of the progenitors of present-day lopsided and symmetric galaxies at high-redshifts (i.e. lookback time $\sim6$-$10\, \mathrm{Gyr}$ ago) to establish whether the intense burst in star formation in the central regions of the symmetric galaxies may be triggered by early interactions with external galaxies or gas-rich mergers. Overall, this will allow us not only to establish how the distinct SFH can shape the evolution of galaxies and their key internal properties in terms of lopsidedness through cosmic time, but also to probe the role of the environment and internal processes in shaping the SFH and evolution of the galaxies to their present-day properties and morphology.  

\section*{Acknowledgements}
The authors gratefully acknowledge support by the ANID BASAL project FB210003. FAG and AM acknowledge support by FONDECYT Regular grant 1211370 and 1212046, respectively, and funding from the Max Planck Society through a “PartnerGroup” grant. PBT acknowledges partial funding from FONDECYT 1200703 and the LACEGAL Network (Horizon 2030). CS acknowledges support from the Agencia Nacional de Investigaci\'on y Desarrollo (ANID) through FONDECYT grant no. 11191125. SVL acknowledges financial support from ANID "Beca de Doctorado Nacional" 21221776.

\section*{Data Availability}
The data used in this work are publicly available from the IllustrisTNG simulations \citep{Nelson2019}.



\bibliographystyle{mnras}
\bibliography{biblio} 

\begin{thebibliography}{}
\makeatletter
\relax
\def\mn@urlcharsother{\let\do\@makeother \do\$\do\&\do\#\do\^\do\_\do\%\do\~}
\def\mn@doi{\begingroup\mn@urlcharsother \@ifnextchar [ {\mn@doi@}
  {\mn@doi@[]}}
\def\mn@doi@[#1]#2{\def\@tempa{#1}\ifx\@tempa\@empty \href
  {http://dx.doi.org/#2} {doi:#2}\else \href {http://dx.doi.org/#2} {#1}\fi
  \endgroup}
\def\mn@eprint#1#2{\mn@eprint@#1:#2::\@nil}
\def\mn@eprint@arXiv#1{\href {http://arxiv.org/abs/#1} {{\tt arXiv:#1}}}
\def\mn@eprint@dblp#1{\href {http://dblp.uni-trier.de/rec/bibtex/#1.xml}
  {dblp:#1}}
\def\mn@eprint@#1:#2:#3:#4\@nil{\def\@tempa {#1}\def\@tempb {#2}\def\@tempc
  {#3}\ifx \@tempc \@empty \let \@tempc \@tempb \let \@tempb \@tempa \fi \ifx
  \@tempb \@empty \def\@tempb {arXiv}\fi \@ifundefined
  {mn@eprint@\@tempb}{\@tempb:\@tempc}{\expandafter \expandafter \csname
  mn@eprint@\@tempb\endcsname \expandafter{\@tempc}}}

\bibitem[\protect\citeauthoryear{{Angiras}, {Jog}, {Omar}  \&
  {Dwarakanath}}{{Angiras} et~al.}{2006}]{Angiras2006}
{Angiras} R.~A.,  {Jog} C.~J.,  {Omar} A.,   {Dwarakanath} K.~S.,  2006,
  \mn@doi [\mnras] {10.1111/j.1365-2966.2006.10418.x}, \href
  {https://ui.adsabs.harvard.edu/abs/2006MNRAS.369.1849A} {369, 1849}

\bibitem[\protect\citeauthoryear{{Angiras}, {Jog}, {Dwarakanath}  \&
  {Verheijen}}{{Angiras} et~al.}{2007}]{Angiras2007}
{Angiras} R.~A.,  {Jog} C.~J.,  {Dwarakanath} K.~S.,   {Verheijen} M.~A.~W.,
  2007, \mn@doi [\mnras] {10.1111/j.1365-2966.2007.11779.x}, \href
  {https://ui.adsabs.harvard.edu/abs/2007MNRAS.378..276A} {378, 276}

\bibitem[\protect\citeauthoryear{{Baldwin}, {Lynden-Bell}  \&
  {Sancisi}}{{Baldwin} et~al.}{1980}]{Baldwin1980}
{Baldwin} J.~E.,  {Lynden-Bell} D.,   {Sancisi} R.,  1980, \mn@doi [\mnras]
  {10.1093/mnras/193.2.313}, \href
  {https://ui.adsabs.harvard.edu/abs/1980MNRAS.193..313B} {193, 313}

\bibitem[\protect\citeauthoryear{{Beale} \& {Davies}}{{Beale} \&
  {Davies}}{1969}]{Beale1969}
{Beale} J.~S.,  {Davies} R.~D.,  1969, \mn@doi [\nat] {10.1038/221531a0}, \href
  {https://ui.adsabs.harvard.edu/abs/1969Natur.221..531B} {221, 531}

\bibitem[\protect\citeauthoryear{{Bekki} \& {Couch}}{{Bekki} \&
  {Couch}}{2011}]{Bekki2011}
{Bekki} K.,  {Couch} W.~J.,  2011, \mn@doi [\mnras]
  {10.1111/j.1365-2966.2011.18821.x}, \href
  {https://ui.adsabs.harvard.edu/abs/2011MNRAS.415.1783B} {415, 1783}

\bibitem[\protect\citeauthoryear{{Blanton} \& {Moustakas}}{{Blanton} \&
  {Moustakas}}{2009}]{Blanton2009}
{Blanton} M.~R.,  {Moustakas} J.,  2009, \mn@doi [\araa]
  {10.1146/annurev-astro-082708-101734}, \href
  {https://ui.adsabs.harvard.edu/abs/2009ARA&A..47..159B} {47, 159}

\bibitem[\protect\citeauthoryear{{Block}, {Bertin}, {Stockton}, {Grosbol},
  {Moorwood}  \& {Peletier}}{{Block} et~al.}{1994}]{Bloch1994}
{Block} D.~L.,  {Bertin} G.,  {Stockton} A.,  {Grosbol} P.,  {Moorwood}
  A.~F.~M.,   {Peletier} R.~F.,  1994, \aap, \href
  {https://ui.adsabs.harvard.edu/abs/1994A&A...288..365B} {288, 365}

\bibitem[\protect\citeauthoryear{{Bournaud}, {Combes}  \& {Jog}}{{Bournaud}
  et~al.}{2004}]{Bournaud2004}
{Bournaud} F.,  {Combes} F.,   {Jog} C.~J.,  2004, \mn@doi [\aap]
  {10.1051/0004-6361:20040114}, \href
  {https://ui.adsabs.harvard.edu/abs/2004A&A...418L..27B} {418, L27}

\bibitem[\protect\citeauthoryear{{Bournaud}, {Combes}, {Jog}  \&
  {Puerari}}{{Bournaud} et~al.}{2005}]{Bournaud2005}
{Bournaud} F.,  {Combes} F.,  {Jog} C.~J.,   {Puerari} I.,  2005, \mn@doi
  [\aap] {10.1051/0004-6361:20052631}, \href
  {https://ui.adsabs.harvard.edu/abs/2005A&A...438..507B} {438, 507}

\bibitem[\protect\citeauthoryear{{Cappellari} et~al.,}{{Cappellari}
  et~al.}{2011}]{Cappellari2011}
{Cappellari} M.,  et~al., 2011, \mn@doi [\mnras]
  {10.1111/j.1365-2966.2011.18600.x}, \href
  {https://ui.adsabs.harvard.edu/abs/2011MNRAS.416.1680C} {416, 1680}

\bibitem[\protect\citeauthoryear{{Combes}}{{Combes}}{2004}]{Combes2004}
{Combes} F.,  2004, in {Block} D.~L.,  {Puerari} I.,  {Freeman} K.~C.,
  {Groess} R.,   {Block} E.~K.,  eds,  Astrophysics and Space Science Library
  Vol. 319, Penetrating Bars Through Masks of Cosmic Dust. p.~57 (\mn@eprint
  {arXiv} {astro-ph/0406306}), \mn@doi{10.1007/978-1-4020-2862-5_4}

\bibitem[\protect\citeauthoryear{{Conselice}, {Bershady}  \&
  {Jangren}}{{Conselice} et~al.}{2000}]{Conselice2000}
{Conselice} C.~J.,  {Bershady} M.~A.,   {Jangren} A.,  2000, \mn@doi [\apj]
  {10.1086/308300}, \href
  {https://ui.adsabs.harvard.edu/abs/2000ApJ...529..886C} {529, 886}

\bibitem[\protect\citeauthoryear{{Deeley} et~al.,}{{Deeley}
  et~al.}{2020}]{Deeley2020}
{Deeley} S.,  et~al., 2020, \mn@doi [\mnras] {10.1093/mnras/staa2417}, \href
  {https://ui.adsabs.harvard.edu/abs/2020MNRAS.498.2372D} {498, 2372}

\bibitem[\protect\citeauthoryear{{Dressler}}{{Dressler}}{1980}]{Dressler1980}
{Dressler} A.,  1980, \mn@doi [\apj] {10.1086/157753}, \href
  {https://ui.adsabs.harvard.edu/abs/1980ApJ...236..351D} {236, 351}

\bibitem[\protect\citeauthoryear{{Dressler} et~al.,}{{Dressler}
  et~al.}{1997}]{Dressler1997}
{Dressler} A.,  et~al., 1997, \mn@doi [\apj] {10.1086/304890}, \href
  {https://ui.adsabs.harvard.edu/abs/1997ApJ...490..577D} {490, 577}

\bibitem[\protect\citeauthoryear{{Drinkwater}, {Gregg}  \&
  {Colless}}{{Drinkwater} et~al.}{2001}]{Drinkwater2001}
{Drinkwater} M.~J.,  {Gregg} M.~D.,   {Colless} M.,  2001, \mn@doi [\apjl]
  {10.1086/319113}, \href
  {https://ui.adsabs.harvard.edu/abs/2001ApJ...548L.139D} {548, L139}

\bibitem[\protect\citeauthoryear{{Fasano}, {Poggianti}, {Couch}, {Bettoni},
  {Kj{\ae}rgaard}  \& {Moles}}{{Fasano} et~al.}{2000}]{Fasano2000}
{Fasano} G.,  {Poggianti} B.~M.,  {Couch} W.~J.,  {Bettoni} D.,
  {Kj{\ae}rgaard} P.,   {Moles} M.,  2000, \mn@doi [\apj] {10.1086/317047},
  \href {https://ui.adsabs.harvard.edu/abs/2000ApJ...542..673F} {542, 673}

\bibitem[\protect\citeauthoryear{{Foster} et~al.,}{{Foster}
  et~al.}{2021}]{Foster2021}
{Foster} C.,  et~al., 2021, \mn@doi [\pasa] {10.1017/pasa.2021.25}, \href
  {https://ui.adsabs.harvard.edu/abs/2021PASA...38...31F} {38, e031}

\bibitem[\protect\citeauthoryear{{Gargiulo} et~al.,}{{Gargiulo}
  et~al.}{2019}]{Gargiulo2019}
{Gargiulo} I.~D.,  et~al., 2019, \mn@doi [\mnras] {10.1093/mnras/stz2536},
  \href {https://ui.adsabs.harvard.edu/abs/2019MNRAS.489.5742G} {489, 5742}

\bibitem[\protect\citeauthoryear{{Gargiulo} et~al.,}{{Gargiulo}
  et~al.}{2022}]{Gargiulo2022}
{Gargiulo} I.~D.,  et~al., 2022, \mn@doi [\mnras] {10.1093/mnras/stac629},
  \href {https://ui.adsabs.harvard.edu/abs/2022MNRAS.512.2537G} {512, 2537}

\bibitem[\protect\citeauthoryear{{Genel} et~al.,}{{Genel}
  et~al.}{2014}]{Genel2014}
{Genel} S.,  et~al., 2014, \mn@doi [\mnras] {10.1093/mnras/stu1654}, \href
  {https://ui.adsabs.harvard.edu/abs/2014MNRAS.445..175G} {445, 175}

\bibitem[\protect\citeauthoryear{{Genel}, {Fall}, {Hernquist}, {Vogelsberger},
  {Snyder}, {Rodriguez-Gomez}, {Sijacki}  \& {Springel}}{{Genel}
  et~al.}{2015}]{Genel2015}
{Genel} S.,  {Fall} S.~M.,  {Hernquist} L.,  {Vogelsberger} M.,  {Snyder}
  G.~F.,  {Rodriguez-Gomez} V.,  {Sijacki} D.,   {Springel} V.,  2015, \mn@doi
  [\apjl] {10.1088/2041-8205/804/2/L40}, \href
  {https://ui.adsabs.harvard.edu/abs/2015ApJ...804L..40G} {804, L40}

\bibitem[\protect\citeauthoryear{Ghosh, Saha, Jog, Combes  \& Di~Matteo}{Ghosh
  et~al.}{2022}]{Ghosh2022}
Ghosh S.,  Saha K.,  Jog C.~J.,  Combes F.,   Di~Matteo P.,  2022, \mn@doi
  [Monthly Notices of the Royal Astronomical Society] {10.1093/mnras/stac461},
  511, 5878

\bibitem[\protect\citeauthoryear{{Grand}, {Springel}, {G{\'o}mez}, {Marinacci},
  {Pakmor}, {Campbell}  \& {Jenkins}}{{Grand} et~al.}{2016}]{Grand2016}
{Grand} R. J.~J.,  {Springel} V.,  {G{\'o}mez} F.~A.,  {Marinacci} F.,
  {Pakmor} R.,  {Campbell} D. J.~R.,   {Jenkins} A.,  2016, \mn@doi [\mnras]
  {10.1093/mnras/stw601}, \href
  {https://ui.adsabs.harvard.edu/abs/2016MNRAS.459..199G} {459, 199}

\bibitem[\protect\citeauthoryear{Haynes, Hogg, Maddalena, Roberts  \& van
  Zee}{Haynes et~al.}{1998}]{Haynes1998}
Haynes M.~P.,  Hogg D.~E.,  Maddalena R.~J.,  Roberts M.~S.,   van Zee L.,
  1998, \mn@doi [The Astronomical Journal] {10.1086/300166}, 115, 62

\bibitem[\protect\citeauthoryear{Jog}{Jog}{1997}]{Jog1997}
Jog C.~J.,  1997, \mn@doi [The Astrophysical Journal] {10.1086/304721}, 488,
  642

\bibitem[\protect\citeauthoryear{{Jog}}{{Jog}}{2002}]{Jog2002}
{Jog} C.~J.,  2002, \mn@doi [\aap] {10.1051/0004-6361:20020832}, \href
  {https://ui.adsabs.harvard.edu/abs/2002A&A...391..471J} {391, 471}

\bibitem[\protect\citeauthoryear{Jog \& Combes}{Jog \& Combes}{2009}]{Jog2009}
Jog C.~J.,  Combes F.,  2009, \mn@doi [Physics Reports]
  {https://doi.org/10.1016/j.physrep.2008.12.002}, 471, 75

\bibitem[\protect\citeauthoryear{Joshi, Pillepich, Nelson, Marinacci, Springel,
  Rodriguez-Gomez, Vogelsberger  \& Hernquist}{Joshi et~al.}{2020}]{Joshi2020}
Joshi G.~D.,  Pillepich A.,  Nelson D.,  Marinacci F.,  Springel V.,
  Rodriguez-Gomez V.,  Vogelsberger M.,   Hernquist L.,  2020, \mn@doi [Monthly
  Notices of the Royal Astronomical Society] {10.1093/mnras/staa1668}, 496,
  2673

\bibitem[\protect\citeauthoryear{{{\L}okas}}{{{\L}okas}}{2022}]{Lokas2022}
{{\L}okas} E.~L.,  2022, \mn@doi [\aap] {10.1051/0004-6361/202142845}, \href
  {https://ui.adsabs.harvard.edu/abs/2022A&A...662A..53L} {662, A53}

\bibitem[\protect\citeauthoryear{Marinacci, Pakmor  \& Springel}{Marinacci
  et~al.}{2013}]{Marinacci2014}
Marinacci F.,  Pakmor R.,   Springel V.,  2013, \mn@doi [Monthly Notices of the
  Royal Astronomical Society] {10.1093/mnras/stt2003}, 437, 1750

\bibitem[\protect\citeauthoryear{{Marinacci} et~al.,}{{Marinacci}
  et~al.}{2018}]{Marinacci2018}
{Marinacci} F.,  et~al., 2018, \mn@doi [\mnras] {10.1093/mnras/sty2206}, \href
  {https://ui.adsabs.harvard.edu/abs/2018MNRAS.480.5113M} {480, 5113}

\bibitem[\protect\citeauthoryear{{Naiman} et~al.,}{{Naiman}
  et~al.}{2018}]{Naiman2018}
{Naiman} J.~P.,  et~al., 2018, \mn@doi [\mnras] {10.1093/mnras/sty618}, \href
  {https://ui.adsabs.harvard.edu/abs/2018MNRAS.477.1206N} {477, 1206}

\bibitem[\protect\citeauthoryear{{Nelson} et~al.,}{{Nelson}
  et~al.}{2015}]{Nelson2015}
{Nelson} D.,  et~al., 2015, \mn@doi [Astronomy and Computing]
  {10.1016/j.ascom.2015.09.003}, \href
  {https://ui.adsabs.harvard.edu/abs/2015A&C....13...12N} {13, 12}

\bibitem[\protect\citeauthoryear{{Nelson} et~al.,}{{Nelson}
  et~al.}{2018}]{Nelson2018}
{Nelson} D.,  et~al., 2018, \mn@doi [\mnras] {10.1093/mnras/stx3040}, \href
  {https://ui.adsabs.harvard.edu/abs/2018MNRAS.475..624N} {475, 624}

\bibitem[\protect\citeauthoryear{{Nelson} et~al.,}{{Nelson}
  et~al.}{2019a}]{Nelson2019}
{Nelson} D.,  et~al., 2019a, \mn@doi [Computational Astrophysics and Cosmology]
  {10.1186/s40668-019-0028-x}, \href
  {https://ui.adsabs.harvard.edu/abs/2019ComAC...6....2N} {6, 2}

\bibitem[\protect\citeauthoryear{{Nelson} et~al.,}{{Nelson}
  et~al.}{2019b}]{Nelson2019b}
{Nelson} D.,  et~al., 2019b, \mn@doi [\mnras] {10.1093/mnras/stz2306}, \href
  {https://ui.adsabs.harvard.edu/abs/2019MNRAS.490.3234N} {490, 3234}

\bibitem[\protect\citeauthoryear{Noordermeer, Sparke  \& Levine}{Noordermeer
  et~al.}{2001}]{Nordermeer2001}
Noordermeer E.,  Sparke L.~S.,   Levine S.~E.,  2001, \mn@doi [Monthly Notices
  of the Royal Astronomical Society] {10.1046/j.1365-8711.2001.04924.x}, 328,
  1064

\bibitem[\protect\citeauthoryear{{Pillepich} et~al.,}{{Pillepich}
  et~al.}{2018a}]{Pillepich2018}
{Pillepich} A.,  et~al., 2018a, \mn@doi [\mnras] {10.1093/mnras/stx2656}, \href
  {https://ui.adsabs.harvard.edu/abs/2018MNRAS.473.4077P} {473, 4077}

\bibitem[\protect\citeauthoryear{{Pillepich} et~al.,}{{Pillepich}
  et~al.}{2018b}]{Pillepich2018b}
{Pillepich} A.,  et~al., 2018b, \mn@doi [\mnras] {10.1093/mnras/stx3112}, \href
  {https://ui.adsabs.harvard.edu/abs/2018MNRAS.475..648P} {475, 648}

\bibitem[\protect\citeauthoryear{{Pillepich} et~al.,}{{Pillepich}
  et~al.}{2019}]{Pillepich2019}
{Pillepich} A.,  et~al., 2019, \mn@doi [\mnras] {10.1093/mnras/stz2338}, \href
  {https://ui.adsabs.harvard.edu/abs/2019MNRAS.490.3196P} {490, 3196}

\bibitem[\protect\citeauthoryear{{Planck Collaboration} et~al.,}{{Planck
  Collaboration} et~al.}{2016}]{Planck2016}
{Planck Collaboration} et~al., 2016, \mn@doi [\aap]
  {10.1051/0004-6361/201525830}, \href
  {https://ui.adsabs.harvard.edu/abs/2016A&A...594A..13P} {594, A13}

\bibitem[\protect\citeauthoryear{Reichard, Heckman, Rudnick, Brinchmann  \&
  Kauffmann}{Reichard et~al.}{2008}]{Reichard2008}
Reichard T.~A.,  Heckman T.~M.,  Rudnick G.,  Brinchmann J.,   Kauffmann G.,
  2008, \mn@doi [The Astrophysical Journal] {10.1086/526506}, 677, 186

\bibitem[\protect\citeauthoryear{{Renzini} \& {Peng}}{{Renzini} \&
  {Peng}}{2015}]{Renzini2015}
{Renzini} A.,  {Peng} Y.-j.,  2015, \mn@doi [\apjl]
  {10.1088/2041-8205/801/2/L29}, \href
  {https://ui.adsabs.harvard.edu/abs/2015ApJ...801L..29R} {801, L29}

\bibitem[\protect\citeauthoryear{{Richter} \& {Sancisi}}{{Richter} \&
  {Sancisi}}{1994}]{Richter1994}
{Richter} O.~G.,  {Sancisi} R.,  1994, \aap, \href
  {https://ui.adsabs.harvard.edu/abs/1994A&A...290L...9R} {290, L9}

\bibitem[\protect\citeauthoryear{{Rix} \& {Zaritsky}}{{Rix} \&
  {Zaritsky}}{1995}]{Rix1995}
{Rix} H.-W.,  {Zaritsky} D.,  1995, \mn@doi [\apj] {10.1086/175858}, \href
  {https://ui.adsabs.harvard.edu/abs/1995ApJ...447...82R} {447, 82}

\bibitem[\protect\citeauthoryear{{Rudnick} \& {Rix}}{{Rudnick} \&
  {Rix}}{1998}]{Rudnick1998}
{Rudnick} G.,  {Rix} H.-W.,  1998, \mn@doi [\aj] {10.1086/300518}, \href
  {https://ui.adsabs.harvard.edu/abs/1998AJ....116.1163R} {116, 1163}

\bibitem[\protect\citeauthoryear{{Saha}, {Combes}  \& {Jog}}{{Saha}
  et~al.}{2007}]{Saha2007}
{Saha} K.,  {Combes} F.,   {Jog} C.~J.,  2007, \mn@doi [\mnras]
  {10.1111/j.1365-2966.2007.12382.x}, \href
  {https://ui.adsabs.harvard.edu/abs/2007MNRAS.382..419S} {382, 419}

\bibitem[\protect\citeauthoryear{{Sijacki}, {Vogelsberger}, {Genel},
  {Springel}, {Torrey}, {Snyder}, {Nelson}  \& {Hernquist}}{{Sijacki}
  et~al.}{2015}]{Sijacki2015}
{Sijacki} D.,  {Vogelsberger} M.,  {Genel} S.,  {Springel} V.,  {Torrey} P.,
  {Snyder} G.~F.,  {Nelson} D.,   {Hernquist} L.,  2015, \mn@doi [\mnras]
  {10.1093/mnras/stv1340}, \href
  {https://ui.adsabs.harvard.edu/abs/2015MNRAS.452..575S} {452, 575}

\bibitem[\protect\citeauthoryear{{Springel}}{{Springel}}{2010}]{Springel2010}
{Springel} V.,  2010, \mn@doi [\mnras] {10.1111/j.1365-2966.2009.15715.x},
  \href {https://ui.adsabs.harvard.edu/abs/2010MNRAS.401..791S} {401, 791}

\bibitem[\protect\citeauthoryear{{Springel} et~al.,}{{Springel}
  et~al.}{2018}]{Springel2018}
{Springel} V.,  et~al., 2018, \mn@doi [\mnras] {10.1093/mnras/stx3304}, \href
  {https://ui.adsabs.harvard.edu/abs/2018MNRAS.475..676S} {475, 676}

\bibitem[\protect\citeauthoryear{{Swaters}, {Schoenmakers}, {Sancisi}  \& {van
  Albada}}{{Swaters} et~al.}{1999}]{Swaters1999}
{Swaters} R.~A.,  {Schoenmakers} R.~H.~M.,  {Sancisi} R.,   {van Albada} T.~S.,
   1999, \mn@doi [\mnras] {10.1046/j.1365-8711.1999.02332.x}, \href
  {https://ui.adsabs.harvard.edu/abs/1999MNRAS.304..330S} {304, 330}

\bibitem[\protect\citeauthoryear{{Varela-Lavin}, {G{\'o}mez}, {Tissera},
  {Besla}, {Garavito-Camargo}  \& {Marinacci}}{{Varela-Lavin}
  et~al.}{2022}]{Varela-Lavin2022}
{Varela-Lavin} S.,  {G{\'o}mez} F.~A.,  {Tissera} P.~B.,  {Besla} G.,
  {Garavito-Camargo} N.,   {Marinacci} F.,  2022, arXiv e-prints, \href
  {https://ui.adsabs.harvard.edu/abs/2022arXiv221116577V} {p. arXiv:2211.16577}

\bibitem[\protect\citeauthoryear{{Vogelsberger} et~al.,}{{Vogelsberger}
  et~al.}{2014}]{Vogelsberger2014}
{Vogelsberger} M.,  et~al., 2014, \mn@doi [\nat] {10.1038/nature13316}, \href
  {https://ui.adsabs.harvard.edu/abs/2014Natur.509..177V} {509, 177}

\bibitem[\protect\citeauthoryear{{Walker}, {Mihos}  \& {Hernquist}}{{Walker}
  et~al.}{1996}]{Walker1996}
{Walker} I.~R.,  {Mihos} J.~C.,   {Hernquist} L.,  1996, \mn@doi [\apj]
  {10.1086/176956}, \href
  {https://ui.adsabs.harvard.edu/abs/1996ApJ...460..121W} {460, 121}

\bibitem[\protect\citeauthoryear{{Weinberger} et~al.,}{{Weinberger}
  et~al.}{2017}]{Weinberger2017}
{Weinberger} R.,  et~al., 2017, \mn@doi [\mnras] {10.1093/mnras/stw2944}, \href
  {https://ui.adsabs.harvard.edu/abs/2017MNRAS.465.3291W} {465, 3291}

\bibitem[\protect\citeauthoryear{Wellons et~al.,}{Wellons
  et~al.}{2015}]{Wellons2015}
Wellons S.,  et~al., 2015, \mn@doi [Monthly Notices of the Royal Astronomical
  Society] {10.1093/mnras/stv303}, 449, 361

\bibitem[\protect\citeauthoryear{{Wilcots}}{{Wilcots}}{2010}]{Wilcots2010}
{Wilcots} E.~M.,  2010, in {Verdes-Montenegro} L.,  {Del Olmo} A.,   {Sulentic}
  J.,  eds,  Astronomical Society of the Pacific Conference Series Vol. 421,
  Galaxies in Isolation: Exploring Nature Versus Nurture. p.~149

\bibitem[\protect\citeauthoryear{Zaritsky \& Rix}{Zaritsky \&
  Rix}{1997}]{Zaritsky1997}
Zaritsky D.,  Rix H.-W.,  1997, \mn@doi [The Astrophysical Journal]
  {10.1086/303692}, 477, 118

\bibitem[\protect\citeauthoryear{{Zaritsky} \& {Rix}}{{Zaritsky} \&
  {Rix}}{1999}]{Zaritsky1999}
{Zaritsky} D.,  {Rix} H.~W.,  1999, in {Barnes} J.~E.,  {Sanders} D.~B.,  eds,
  Vol. 186, Galaxy Interactions at Low and High Redshift. p.~117

\makeatother
\end{thebibliography}




\appendix

\section{The role of the environment in the TNG100 simulation}
\label{sec:role_environment_tng100}

\begin{figure}
    \centering
    \includegraphics[width=0.45\textwidth]{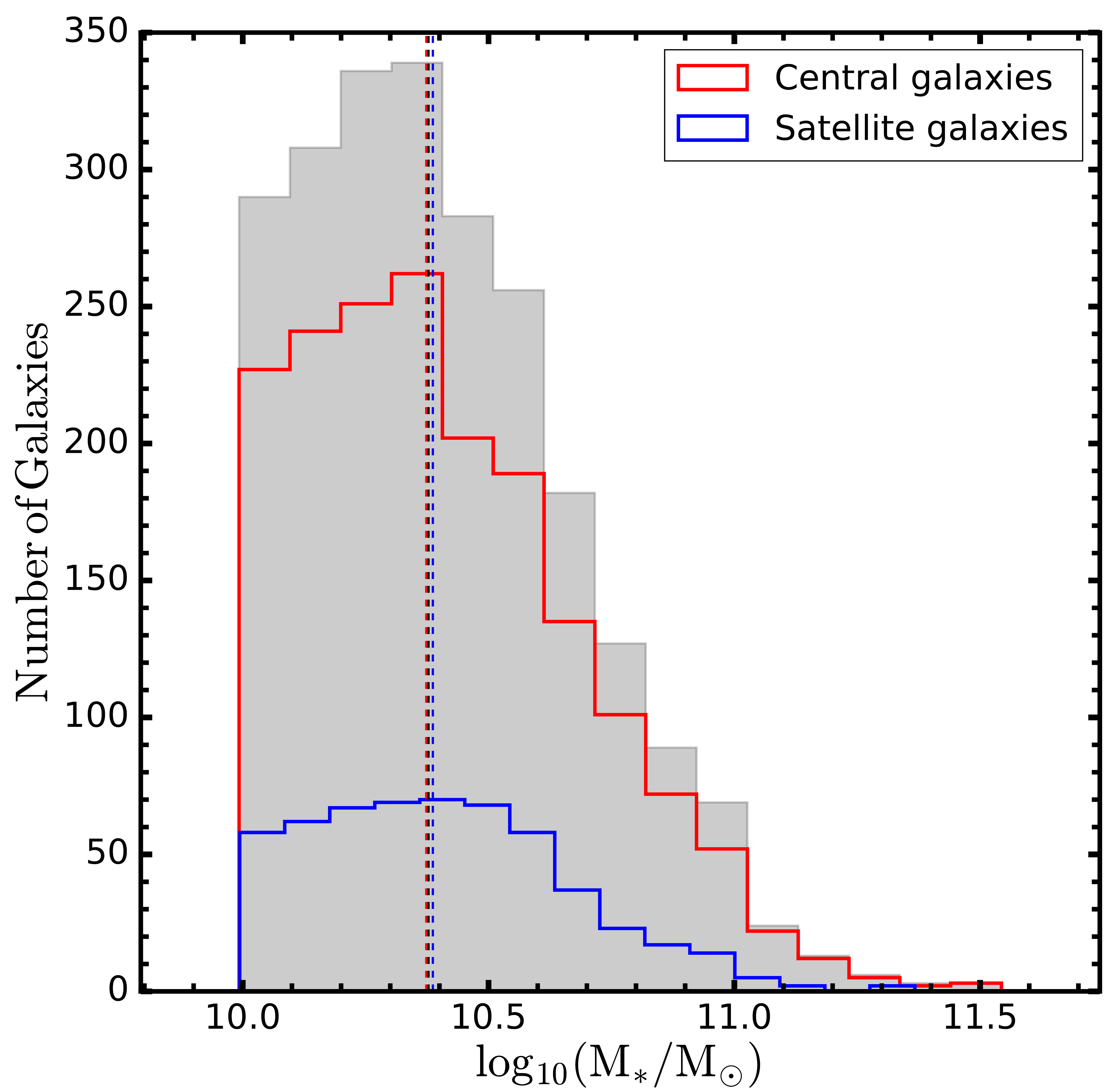}
    \includegraphics[width=0.45\textwidth]{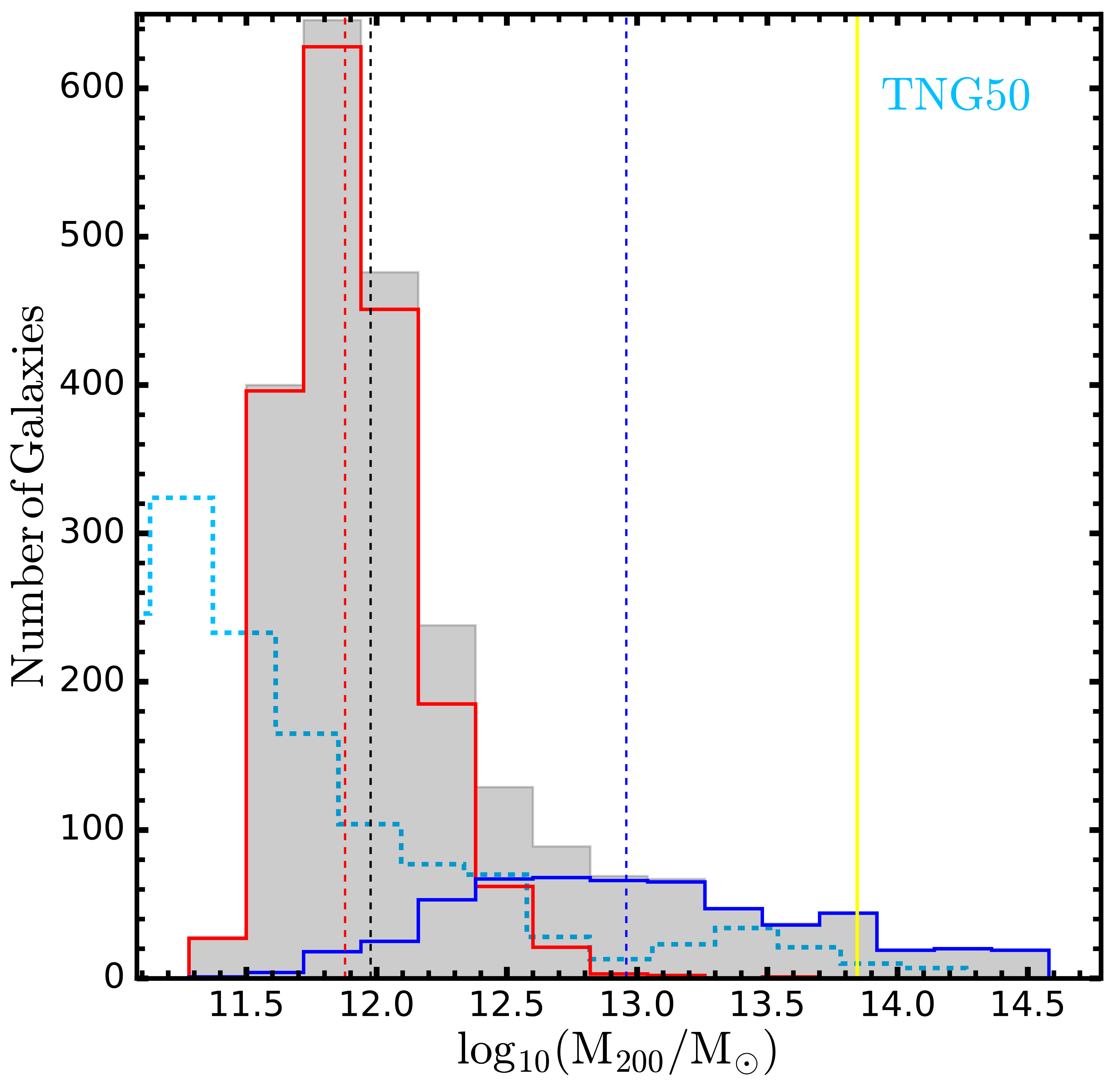}
    \caption{Stellar mass distribution (top) and total halo mass distribution (bottom) of all (black histogram), central (red histogram) and satellite (blue histogram) galaxies in our galaxy sample, which we selected from the TNG100 simulation as described in Sec. \ref{sec:selection_criteria_tng100}. The dashed light blue histogram represents the total halo mass distribution of the galaxies in TNG50 from Fig. \ref{fig:halo_mass}. The description is as in Fig. \ref{fig:mass_distribution} and \ref{fig:halo_mass}. We see that the satellite galaxies with $\mathrm{D/T}\geq0.4$ represent a minority also in the TNG100 simulation, and that TNG100 provides a larger number of galaxies at higher halo masses (i.e. $\mathrm{M_{200}}\gtrsim10^{11.5}\, \mathrm{M_{\odot}}$).}
    \label{fig:mass_distribution_tng100}
\end{figure}

\subsection{The galaxy sample}
\label{sec:selection_criteria_tng100}
We use the highest resolution version, i.e. TNG100-1 (hereafter, TNG100), of the three available simulation runs, and we select our sample of disk-like galaxies in TNG100 at $z=0$ similarly to Sec. \ref{sec:selection_criteria} by assuming the following two selection criteria, i.e.:
\begin{enumerate}[(i)]\itemsep0.2cm
    \item total number of bound stellar particles, $\mathrm{N}_{\mathrm{tot,\, stars}} \geq 10^{4}$;
    \item disk morphology, $\mathrm{D/T}\geq0.4$.
\end{enumerate}

Overall, our selected galaxy sample includes $2328$ galaxies, with $1776$ centrals and $552$ satellites. The top panel of Fig. \ref{fig:mass_distribution_tng100} shows the stellar mass distribution of all our galaxies as well as of the central and satellite galaxies, separately. 
We see that the central and satellite galaxies have a similar stellar mass distribution with median $\mathrm{M_{*}}\sim10^{10.4}\, \mathrm{M_{\odot}}$, which is shifted towards higher stellar masses compared to TNG50. This difference is due to the lower mass resolution of TNG100 than TNG50 and, as a result, low mass galaxies (i.e. $\mathrm{M_{*}}\lesssim10^{10}\, \mathrm{M_{\odot}}$) are not resolved by a sufficiently high number of stellar particles in TNG100 that satisfies our selection criteria described above. However, similarly to TNG50, we also find that only $\sim20\%$ of the satellite galaxies have $\mathrm{D/T}\geq0.4$ as opposed to $\sim45\%$ of the central galaxies in TNG100. For this reason, the central galaxies represent the majority ($76\%$) compared to the satellite galaxies ($23\%$), similarly to what we have seen in Fig. \ref{fig:mass_distribution} of Sec. \ref{sec:selection_criteria} for TNG50. 

On the other hand, the bottom panel of Fig. \ref{fig:mass_distribution_tng100} shows the total halo mass distribution, $\mathrm{M_{200}}$, of all our galaxies as well as of the central and satellite galaxies, separately. Overall, we see that both the TNG100 and TNG50 simulations cover similar halo masses up to Fornax cluster-like environments (i.e. $\mathrm{M_{200}}\sim7\times10^{13}\, \mathrm{M_{\odot}}$), however the TNG100 simulation adds a larger number of galaxies at higher halo masses than TNG50, specifically beyond $\mathrm{M_{200}}\sim10^{11.5}\, \mathrm{M_{\odot}}$ where the number of galaxies steeply declines in TNG50.

\subsection{The lopsidedness distribution}
\label{sec:lopsidedness_tng100}
We calculate the present-day global lopsidedness, $\mathrm{A_{1}}$, of the galaxies in the TNG100 simulation following the procedure described in Sec. \ref{sec:lopsidedness} for TNG50. Therefore, from the face-on projection of the stellar surface density, we estimate the optical radius, $R_{\mathrm{opt}}$, of the galaxy and measure the lopsidedness, $\mathrm{A_{1}}(r)$ (see Eq. \ref{eq:lopsidedness}), in radial annuli of width $0.5\, \mathrm{kpc}$ out to $R_{\mathrm{opt}}$. Them, the global lopsidedness, $\mathrm{A_{1}}$, is calculated by averaging the $\mathrm{A_{1}}(r)$ amplitudes measured in each radial annulus between $0.5$-$1.1\, R_{\mathrm{opt}}$.

Fig. \ref{fig:lopsidedness_distribution_tng100} shows the distribution of the present-day lopsidedness of our selected galaxy sample from the TNG100 simulation. Overall, we find that $1780$ galaxies are lopsided (i.e. $\mathrm{A_{1}}>0.1$), while the remaining $545$ galaxies are symmetric (i.e. $\mathrm{A_{1}}>0.1$). 
Overall, the lopsidedness distribution of the galaxies in TNG100 and TNG50 (see Fig. \ref{fig:lopsidedness_distribution}) are generally consistent with the central galaxies being on average more lopsided than the satellite galaxies, even though this difference is less evident in TNG100 than in TNG50. 

However, we note that we find a much larger number of lopsided galaxies reaching values as large as $\mathrm{A_{1}}\sim0.3$-$0.4$ compared to \citet{Lokas2022}, who also used the TNG100 simulation, although this difference may be due to the different radial interval used in \citet{Lokas2022} (i.e. $1$-$2\, R_{\mathrm{half}}$) and in this work (i.e. $0.5$-$1.1\, R_{\mathrm{opt}}$) to measure the lopsidedness that can significantly affect the estimation of the lopsidedness, as previously discussed in Sec. \ref{sec:global_lopsidedness}.

\begin{figure}
    \centering
    \includegraphics[width=0.45\textwidth]{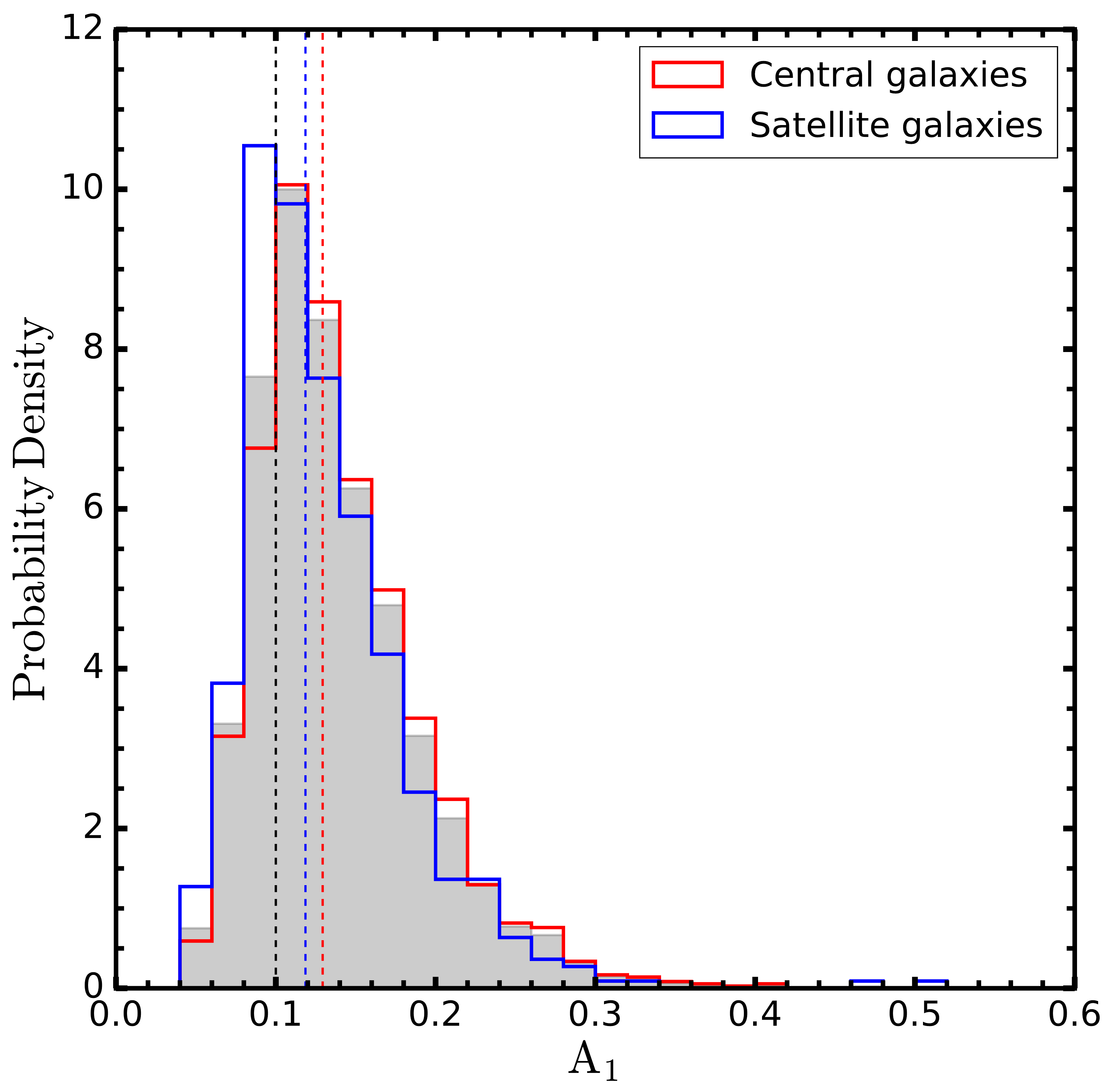}
    \caption{The distribution of the present-day lopsidedness amplitudes of the galaxies in our selected galaxy sample in TNG100, divided by centrals ($1402$ lopsided, $373$ symmetric) and satellites ($378$ lopsided, $172$ symmetric). The description is as in Fig. \ref{fig:lopsidedness_distribution}. We see that the lopsidedness distribution of the galaxies in TNG100 and TNG50 are generally consistent with the central galaxies being on average more lopsided than the satellite galaxies, even though this difference is less evident in TNG100 than in TNG50.}
    \label{fig:lopsidedness_distribution_tng100}
\end{figure}

\begin{figure*}
    \centering
    \includegraphics[width=0.45\textwidth]{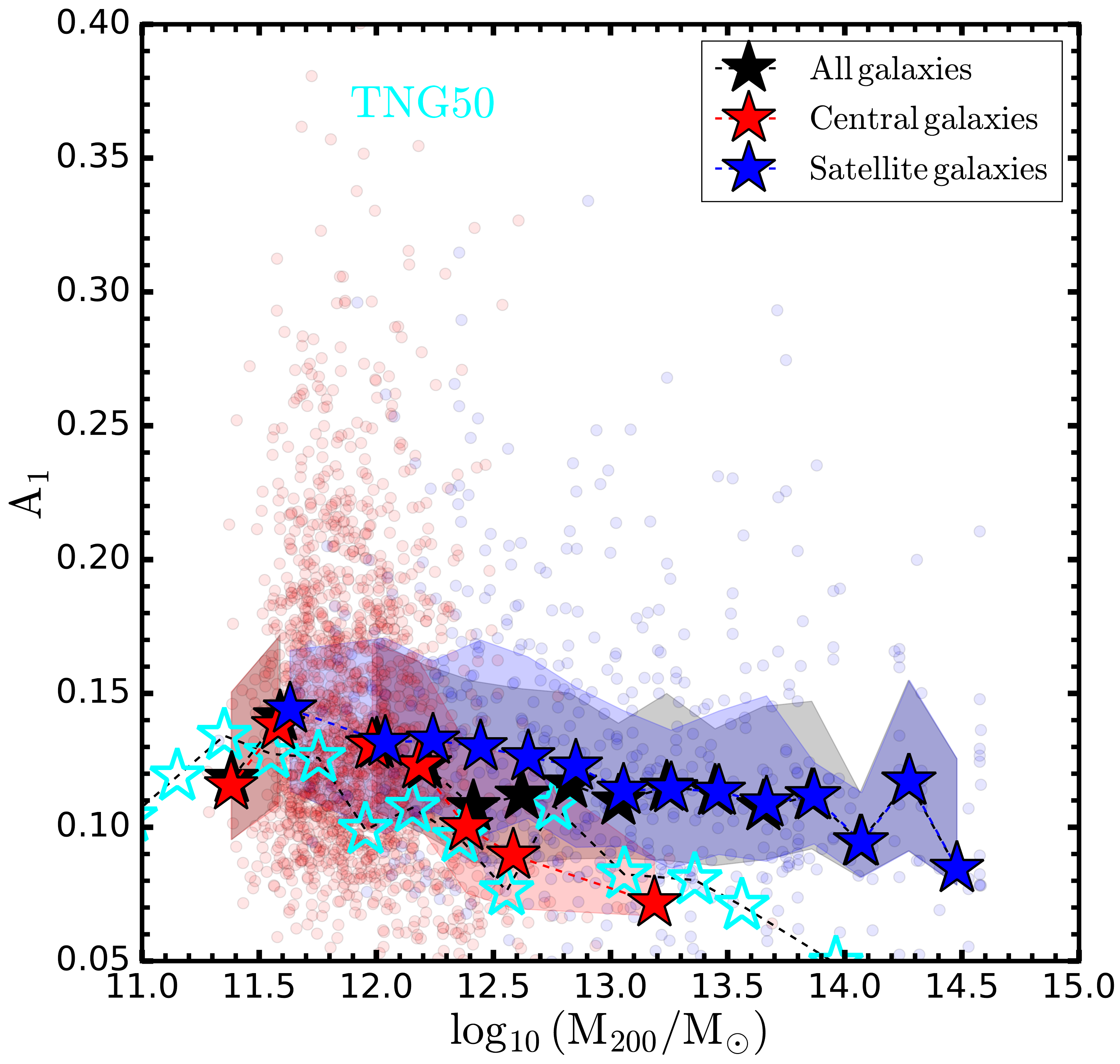}
    \includegraphics[width=0.45\textwidth]{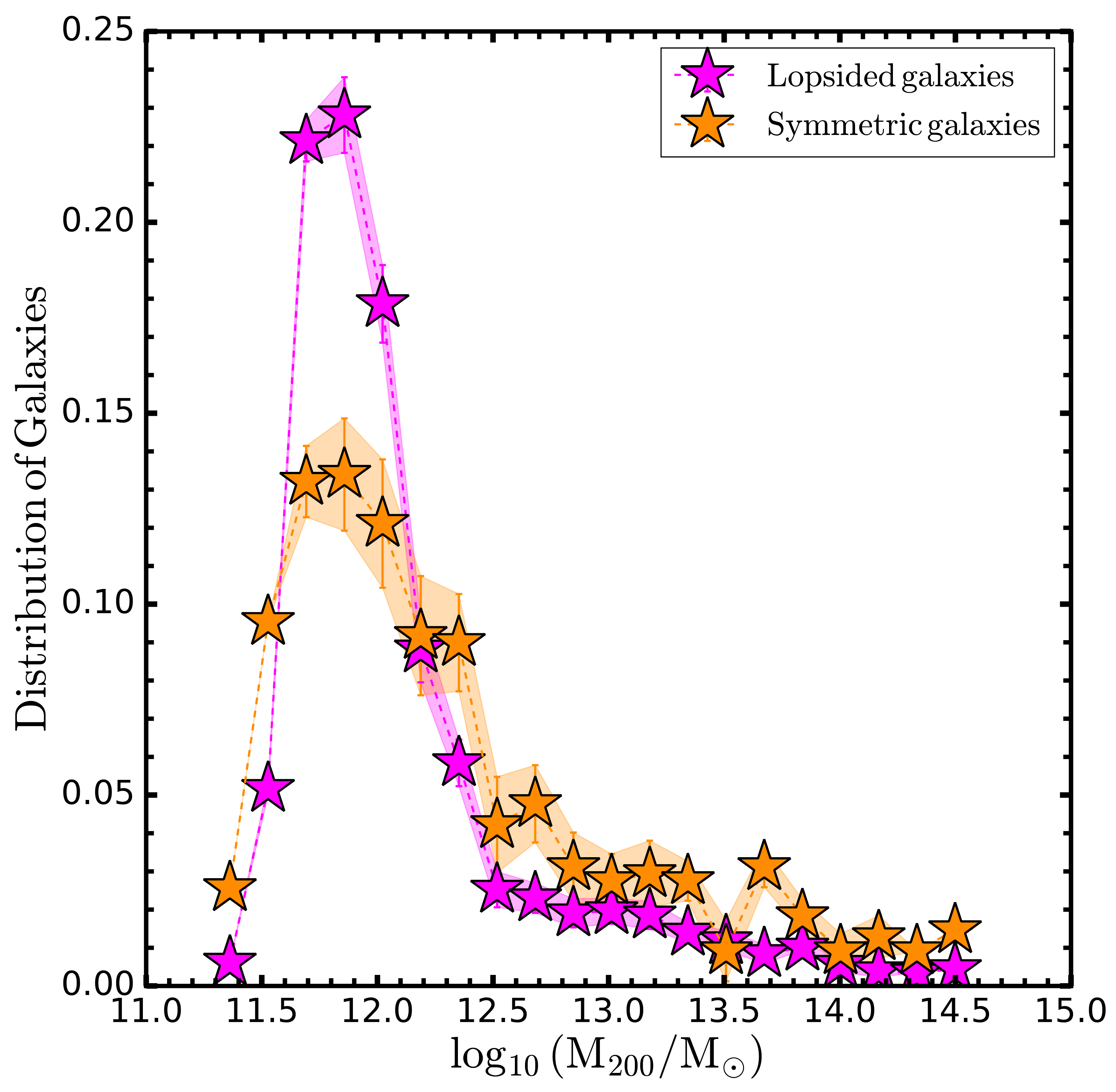}
    \caption{{\bf Left:} The median amplitude of the lopsidedness as a function of the total halo mass, $\mathrm{M_{200}}$, of the environment where our galaxies reside for the full sample (black stars) as well as for the central (red stars) and satellite (blue stars) galaxies, separately, selected from the TNG100 simulation. The light blue stars represent the median $\mathrm{A_{1}}$ as a function of $\mathrm{M_{200}}$ from the TNG50 simulation shown in Fig. \ref{fig:lopsidedness_environment_dependence_m200}. The description is as in Fig. \ref{fig:lopsidedness_environment_dependence_m200}. {\bf Right:} Distribution of the lopsided and symmetric galaxies as a function of $\mathrm{M_{200}}$ in the TNG100 simulation. The histogram is normalized to unity. The shaded areas represent the corresponding $1\sigma$ errors, calculated from the standard deviation of $500$ bootstrap samples obtained by sampling with replacement our original selected galaxy sample. We see that the lopsidedness amplitude shows a milder decrease with increasing $\mathrm{M}_{200}$ in the TNG100 than in the TNG50 simulation, once again suggesting that the environment is not the primary driver of the lopsided perturbation.}
    \label{fig:lopsidedness_environment_dependence_m200_tng100}
\end{figure*}

\subsection{The total halo mass of the host environment of the galaxies}
\label{sec:environment_halomass_tng100}
In Fig. \ref{fig:lopsidedness_environment_dependence_m200_tng100}, we study the behaviour of the lopsidedness amplitude, $\mathrm{A_{1}}$, as a function of the total halo mass, $\mathrm{M_{200}}$, of the environment where our galaxies reside at $z=0$ (black stars). We see that the median amplitude of $\mathrm{A}_{1}$ remains overall constant with increasing $\mathrm{M_{200}}$ in the TNG100 simulation with only a mild drop between $\mathrm{M_{200}}\sim10^{12}$-$10^{12.5}\, \mathrm{M_{\odot}}$, as opposed to the mild, but continuous, decreasing trend seen in TNG50 (light blue stars). 
The flatter behaviour of $\mathrm{A}_{1}$ as a function of $\mathrm{M_{200}}$ in TNG100 than in TNG50 may be due to the larger statistical galaxy sample provided by TNG100 at higher halo masses, specifically beyond $\mathrm{M_{200}}\sim10^{12}\, \mathrm{M_{\odot}}$ (see Fig. \ref{fig:mass_distribution}), where the median lopsidedness amplitude begins decreasing in TNG50. 
Therefore, the overall flatness of $\mathrm{A_{1}}$ as a function of $\mathrm{M_{200}}$ shown in Fig. \ref{fig:lopsidedness_environment_dependence_m200_tng100} using the TNG100 simulation seems to confirm the finding that the lopsidedness amplitude does not significantly depend on the environment, as we have previously suggested in Sec. \ref{sec:environment_halomass} using the TNG50 simulation.

\begin{figure*}
    \centering
    \includegraphics[width=0.33\textwidth]{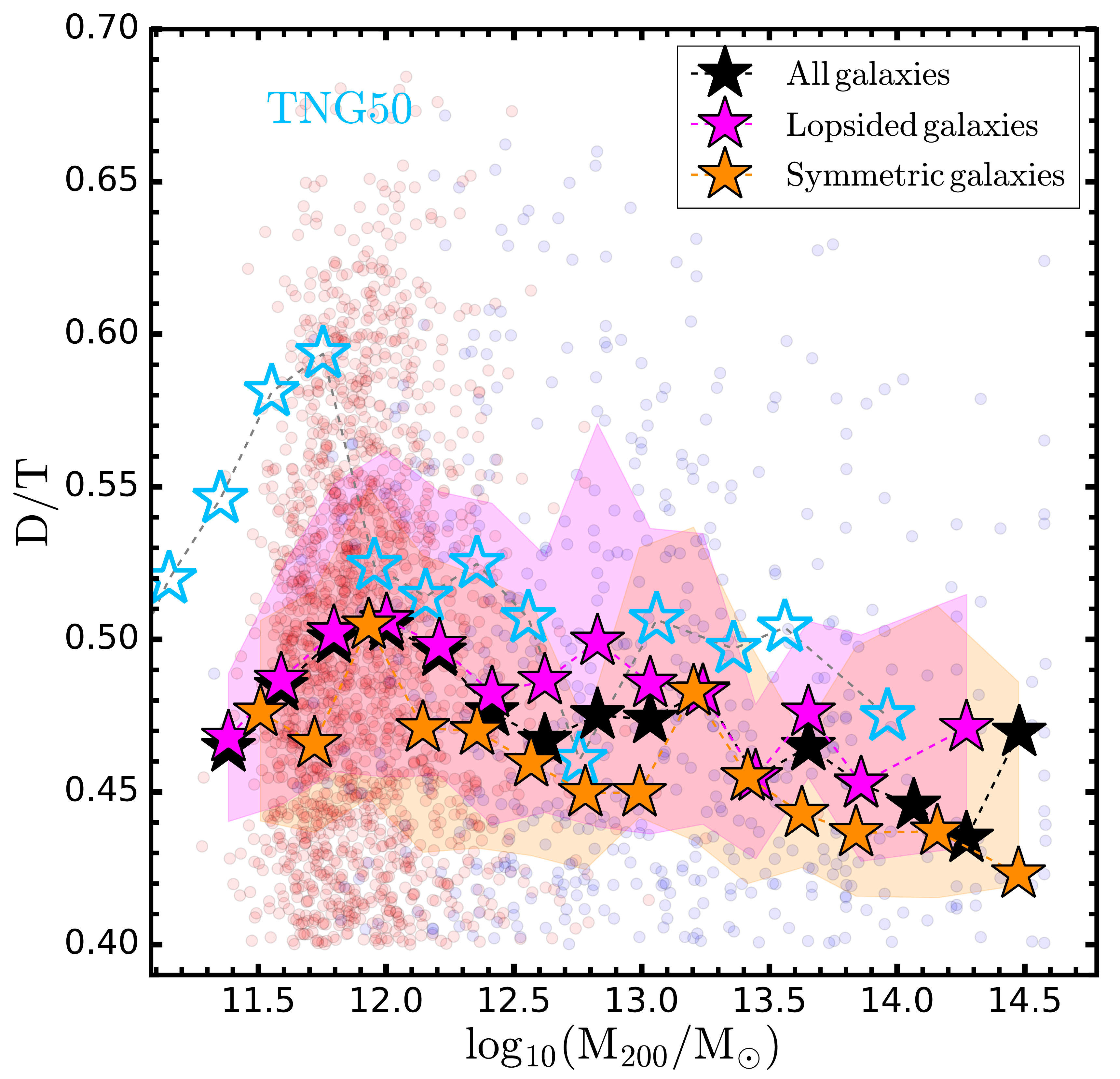}
    \includegraphics[width=0.33\textwidth]{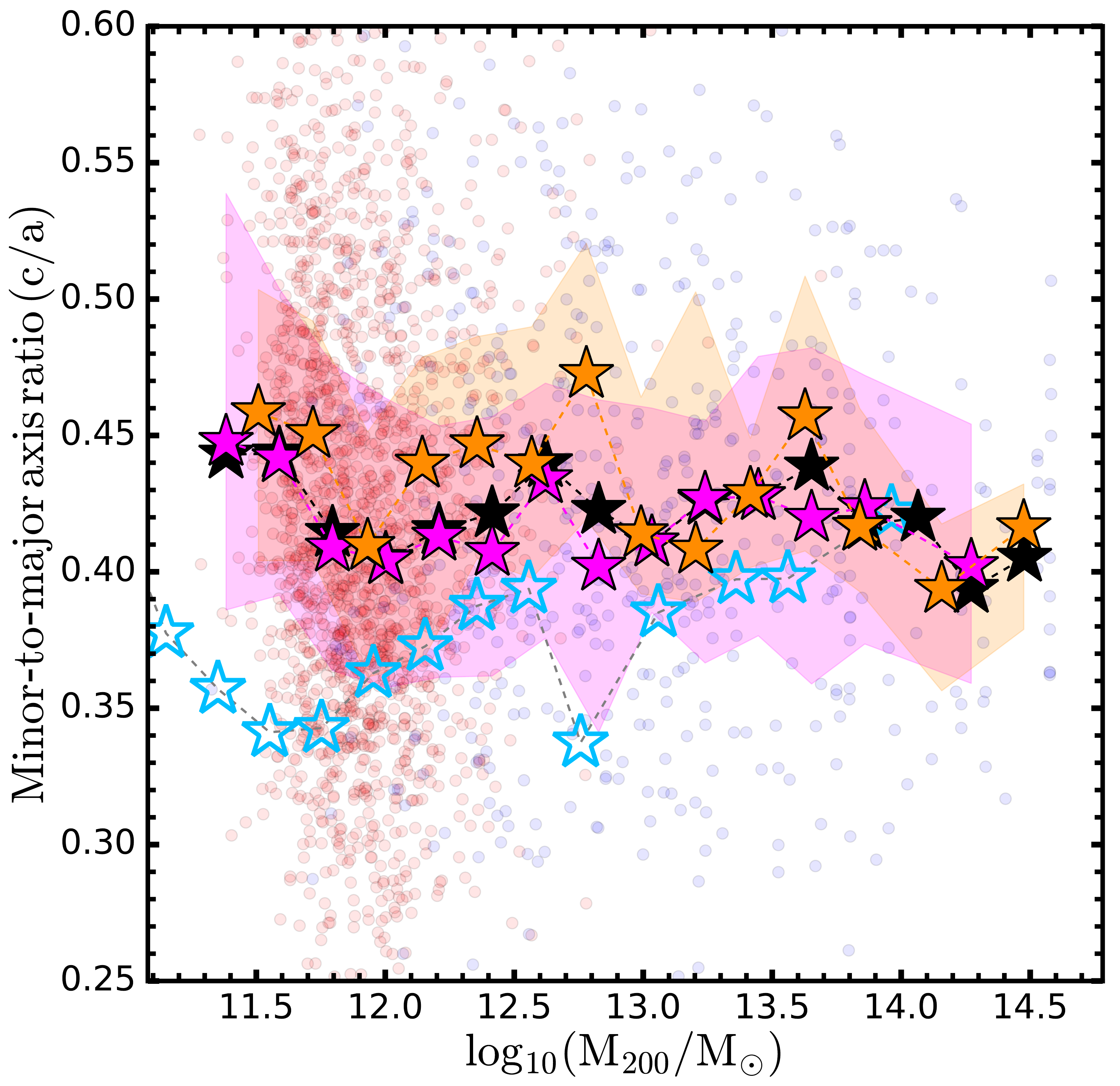}
    \includegraphics[width=0.33\textwidth]{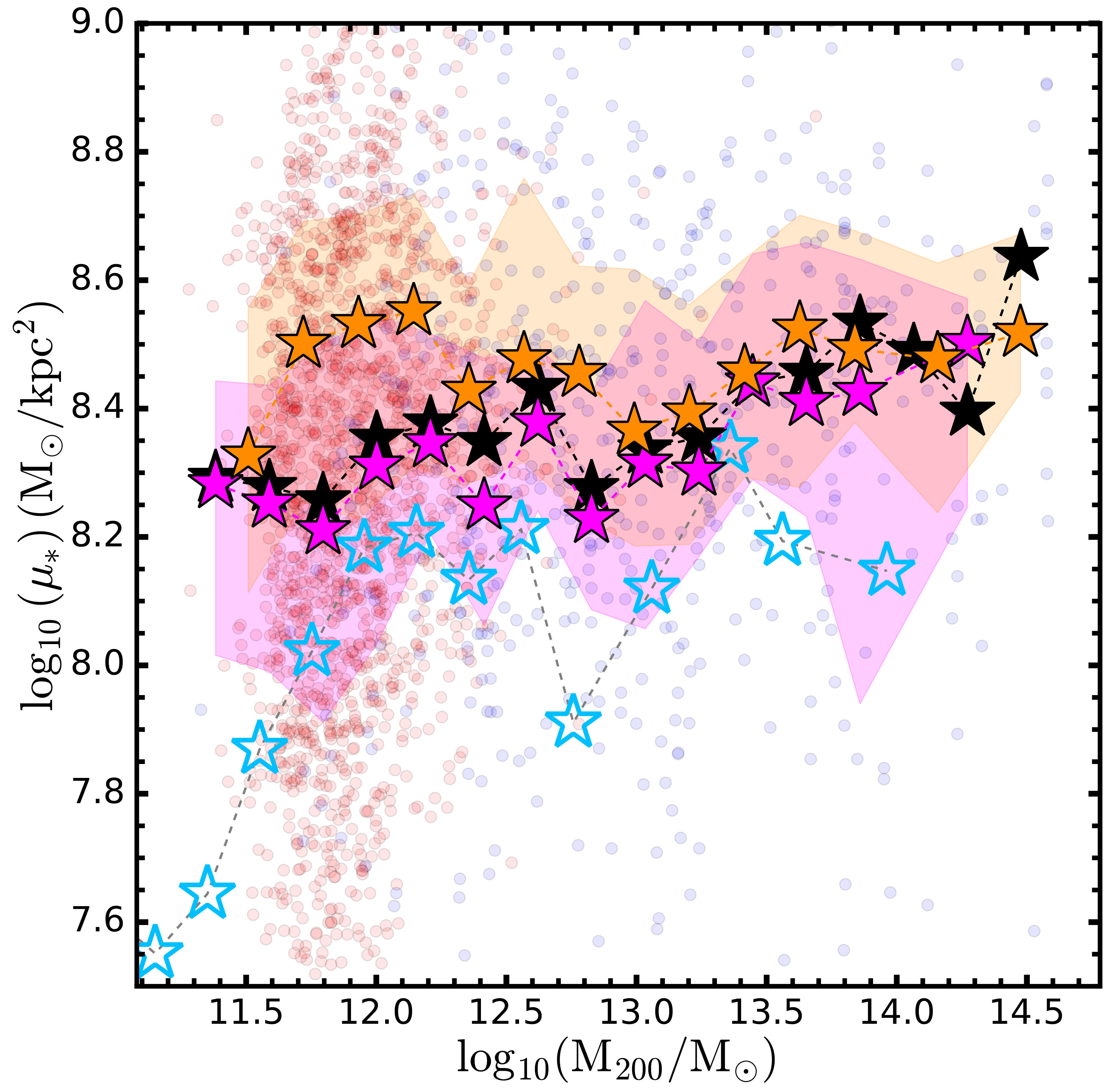}
    \caption{From the left to the right panel, the median disk-to-total ratio ($\mathrm{D/T}$), minor-to-major axis ratio ($\mathrm{c/a}$) and central stellar mass density ($\mu_{*}$) as a function of the total halo mass, $\mathrm{M_{200}}$, of the host environment, respectively, for all (black stars), only the lopsided (magenta stars) and only the symmetric (orange stars) galaxies in our selected galaxy sample from the TNG100 simulation. The blue stars represent the results from the TNG50 simulation shown in Fig. \ref{fig:environment_dependence_m200}. The description is as in Fig. \ref{fig:environment_dependence_m200}. Overall, we see that the galaxies in TNG100 tend to be characterized by on average smaller $\mathrm{D/T}$ and larger $c/a$ and $\mu_{*}$ than in TNG50, which may be a result of the lower mass resolution of TNG100 that does not allow to reliably resolve the different structural features of the galaxies. However, we see that the $\mathrm{D/T}$ and $\mu_{*}$ in TNG100 show a mildly decreasing and increasing trend with $\mathrm{M_{200}}$, respectively, consistent with the results from TNG50 shown in Fig. \ref{fig:environment_dependence_m200}, even though the differences in $\mathrm{D/T}$, $c/a$ and $\mu_{*}$ between the lopsided and symmetric galaxies are less evident in TNG100 than in TNG50.}
    \label{fig:environment_dependence_m200_tng100}
\end{figure*}

Similarly to Fig. \ref{fig:environment_dependence_m200}, in Fig. \ref{fig:environment_dependence_m200_tng100}, we study the behaviour of the $\mathrm{D/T}$, $c/a$ and $\mu_{*}$ as a function of the total halo mass, $\mathrm{M_{200}}$, of the host environment at $z=0$ (black stars) from the left to the right panel, respectively. Firstly, we see that the galaxies tend to be characterized by on average larger $\mathrm{D/T}$ and smaller $c/a$ and $\mu_{*}$ in TNG50 than in TNG100, suggesting that the galaxies in TNG50 are more disk-like than in TNG100. Secondly, we also see that the differences between the internal properties of the lopsided and symmetric galaxies are less evident in TNG100 (see Fig. \ref{fig:environment_dependence_m200_tng100}) than in TNG50 (see Fig. \ref{fig:environment_dependence_m200}).
These differences in the internal properties of the galaxies between TNG50 and TNG100 may be the result of the lower resolution of TNG100, which does not enable to reliably resolve differences in the structural and internal properties of the galaxies and their disks.

Nevertheless, in Fig. \ref{fig:environment_dependence_m200_tng100}, we still see that the overall median $\mathrm{D/T}$ and $\mu_{*}$ are characterized by a mild decreasing and increasing trend with increasing $\mathrm{M_{200}}$, respectively, suggesting a change to earlier type disk morphology towards high halo masses (i.e. high-density environments), consistent with the results from the TNG50 simulation shown in Fig. \ref{fig:environment_dependence_m200}.
Additionally, we also see that the slightly higher lopsidedness amplitude between $10^{11.5}\lesssim\mathrm{M_{200}}/\mathrm{M_{\odot}}\lesssim10^{12}$ in Fig. \ref{fig:lopsidedness_environment_dependence_m200_tng100} roughly corresponds to a small peak in $\mathrm{D/T}$ and small dip in $\mu_{*}$ in Fig. \ref{fig:environment_dependence_m200_tng100}. 
Therefore, these results from the TNG100 simulation also seem to suggest the possibility of a stronger correlation between the lopsidedness and the internal properties of the galaxies rather than the environment, consistent with the results from TNG50 described in Sec. \ref{sec:environment_halomass}. 

\begin{figure*}
    \centering
    \includegraphics[width=0.3\textwidth]{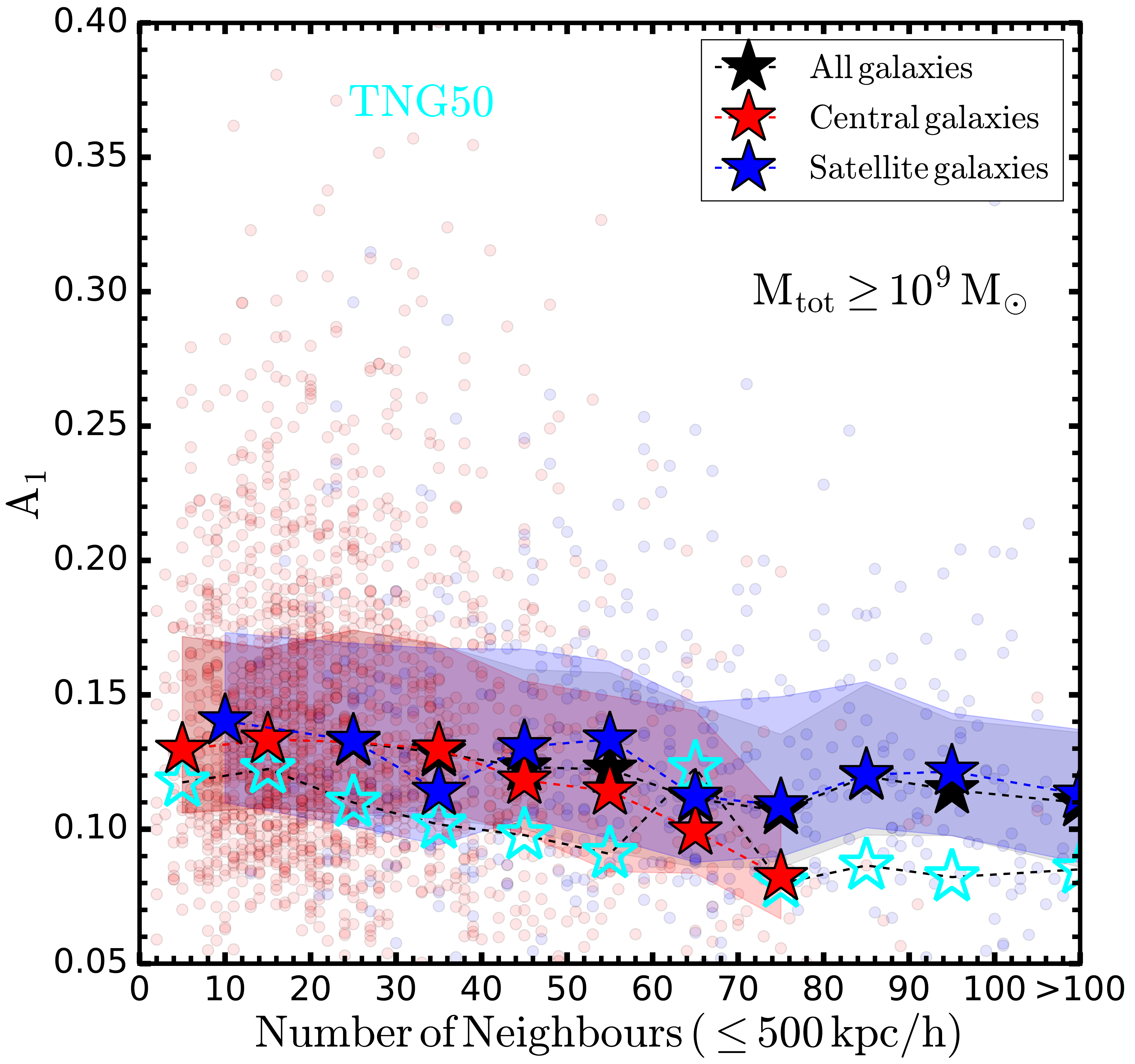}
    \includegraphics[width=0.3\textwidth]{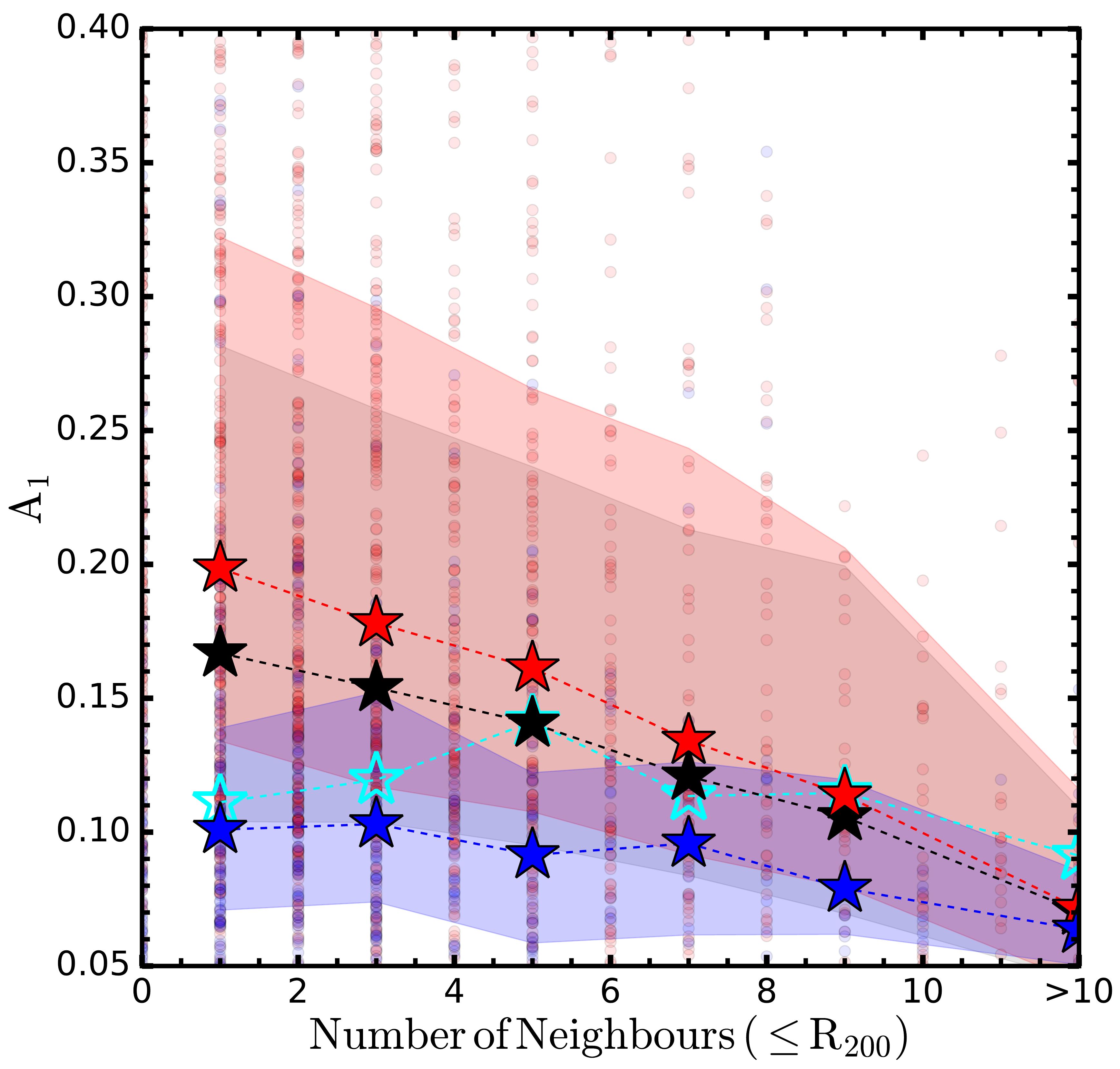}
    \includegraphics[width=0.3\textwidth]{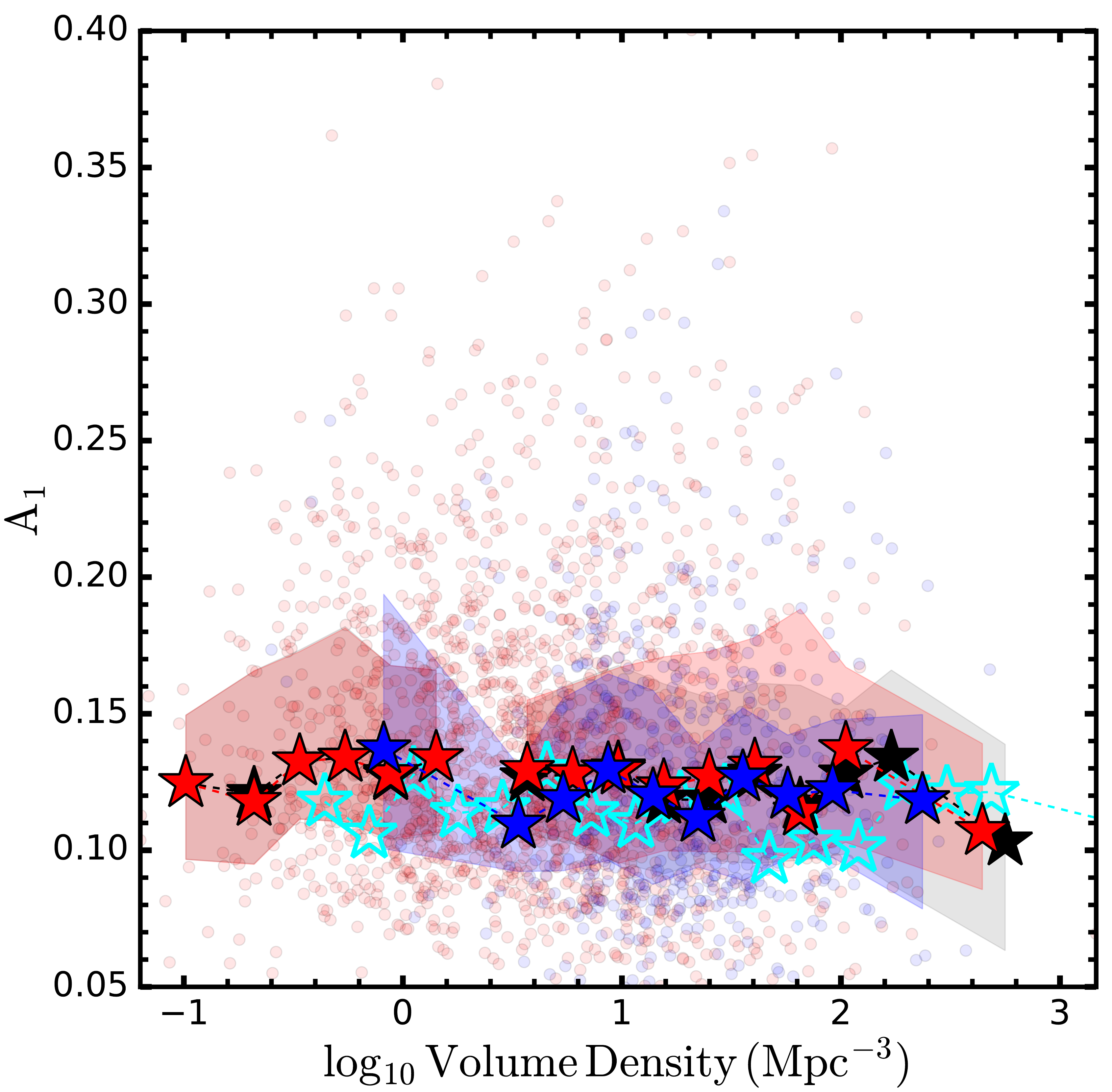}

    \includegraphics[width=0.3\textwidth]{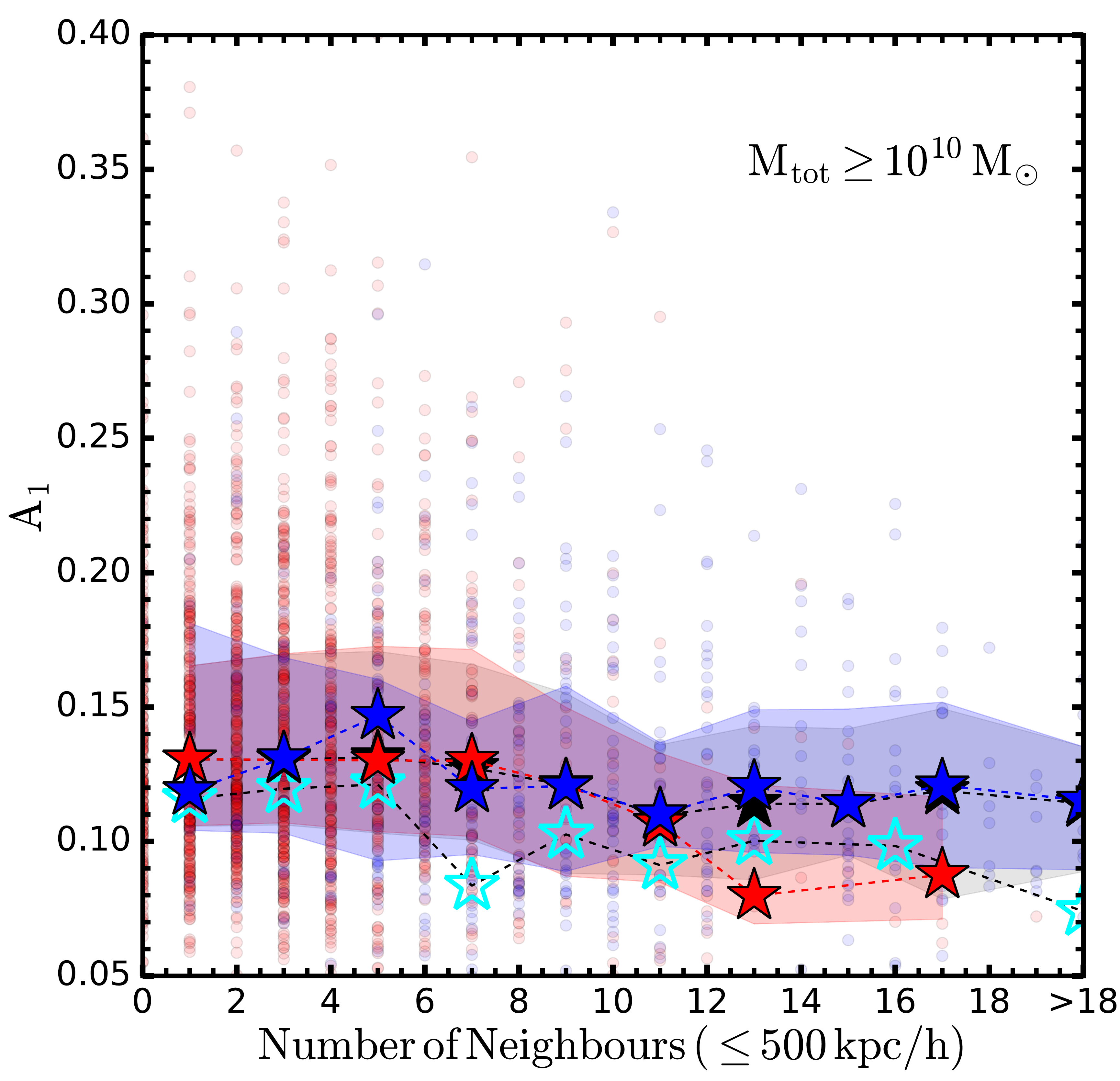}
    \includegraphics[width=0.3\textwidth]{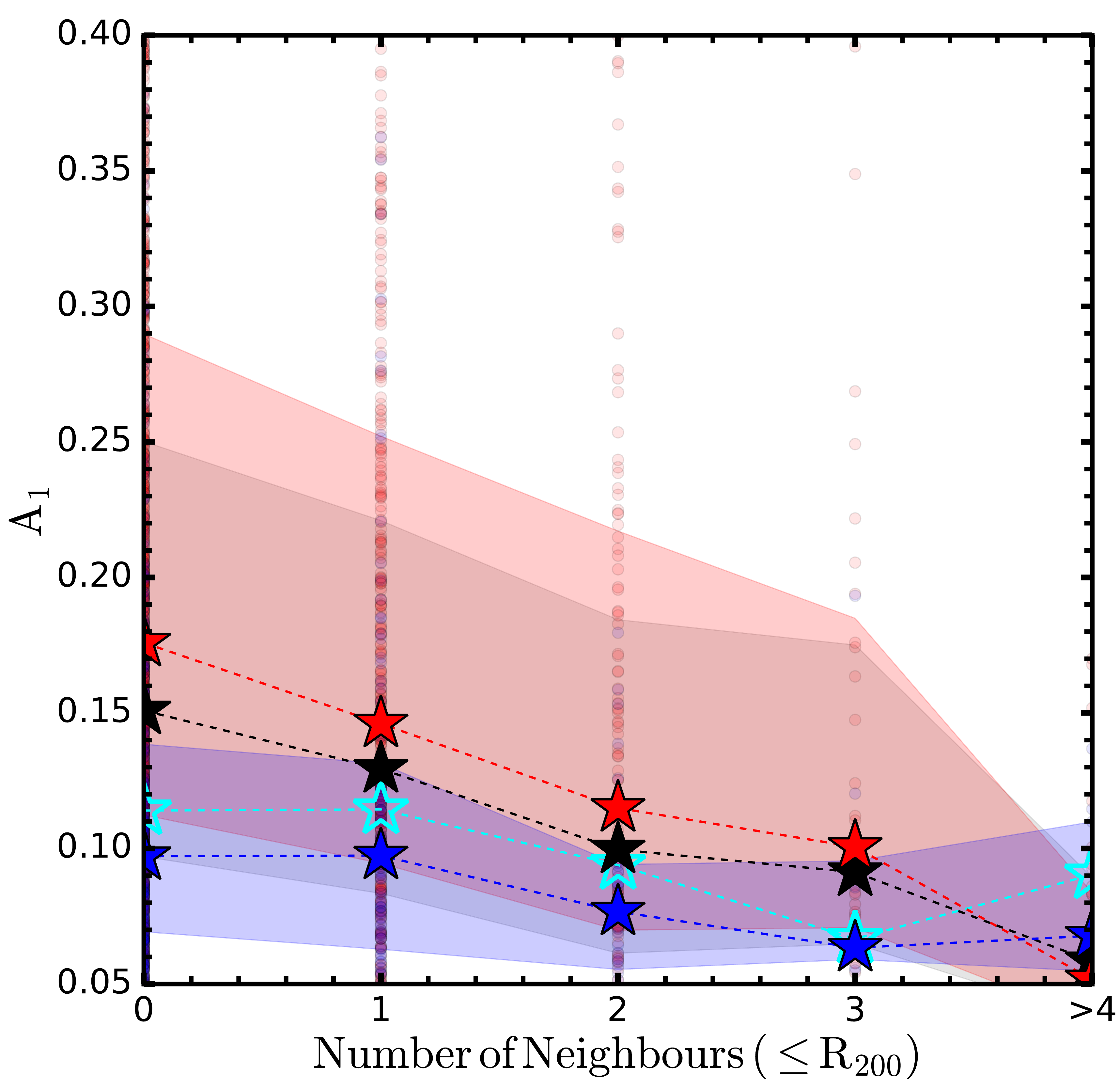}
    \includegraphics[width=0.3\textwidth]{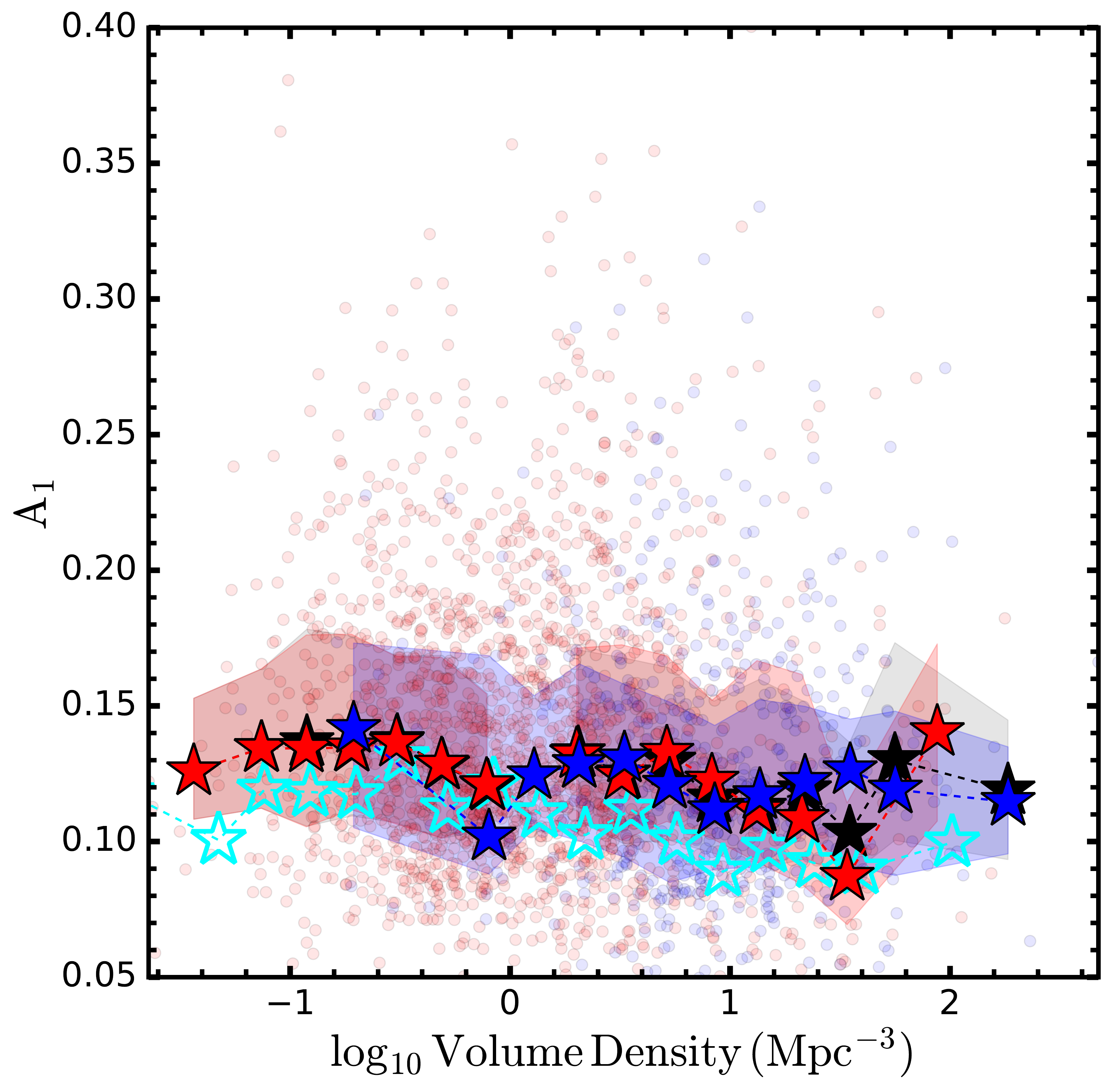}

    \includegraphics[width=0.3\textwidth]{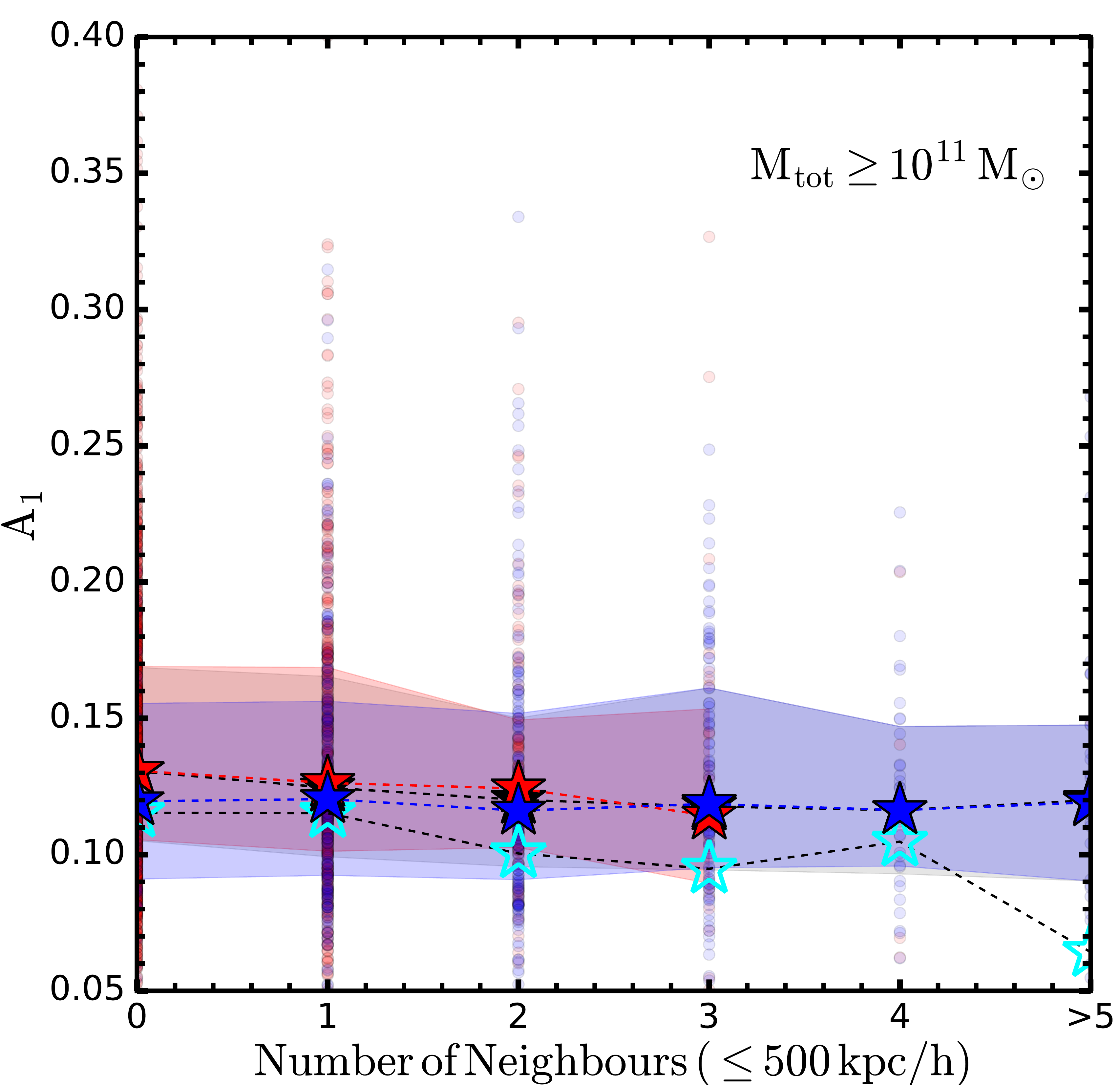}
    \includegraphics[width=0.3\textwidth]{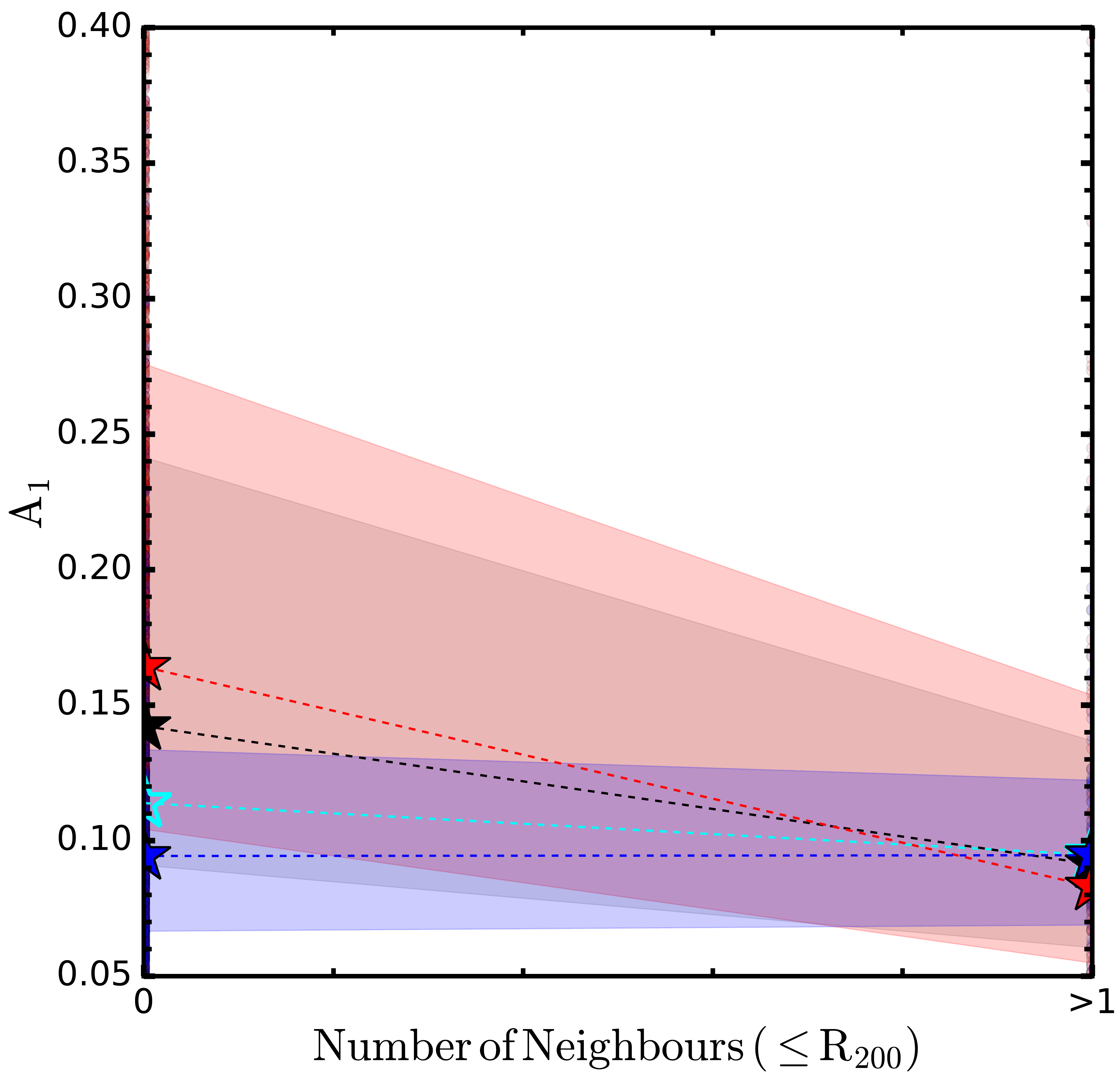}
    \includegraphics[width=0.3\textwidth]{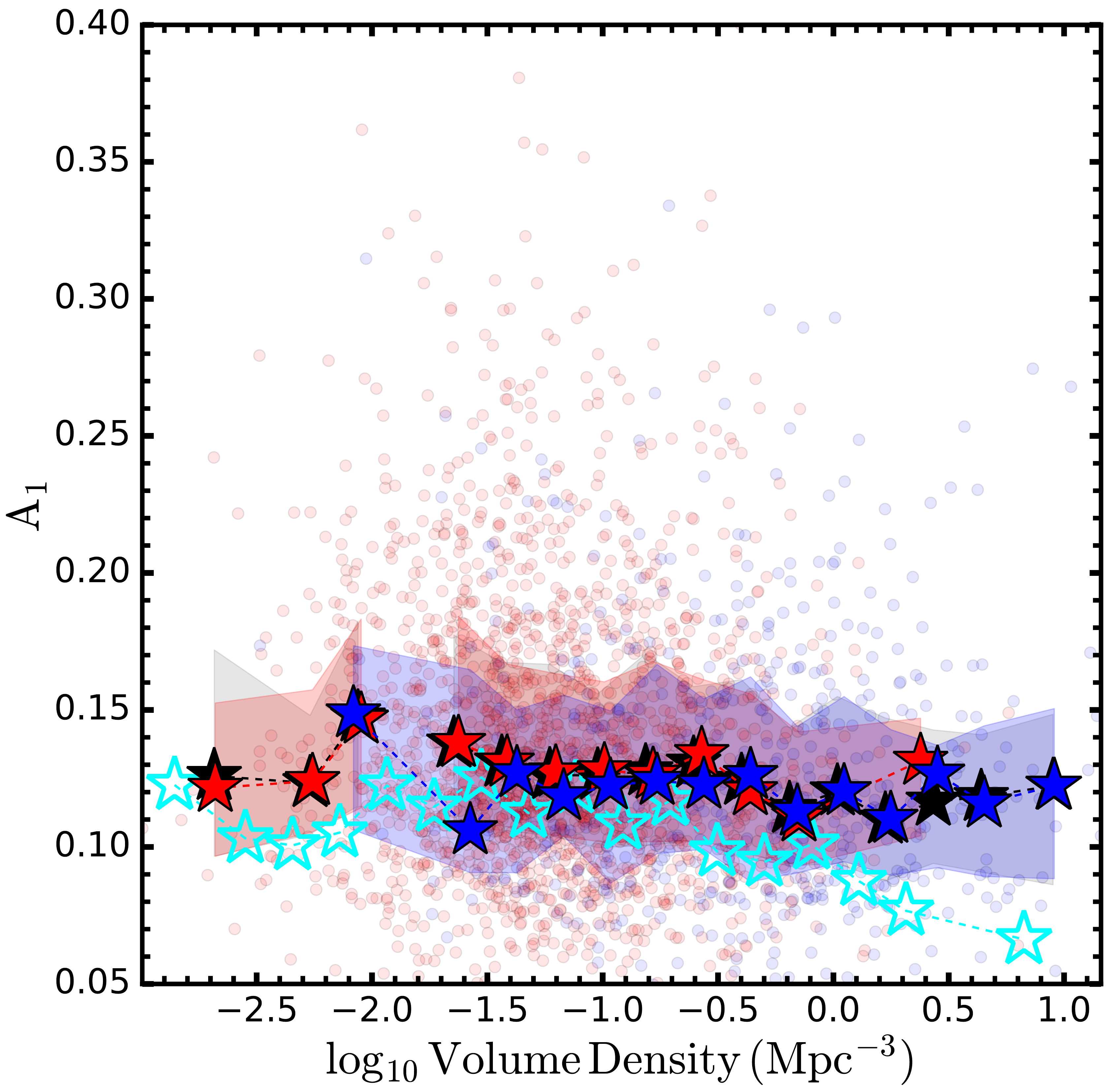}
    \caption{The median amplitude of the lopsidedness, $A_{1}$, as a function of the local environment at $z=0$ defined as described in Sec. \ref{sec:environment_neighbours} for all (black stars), only the central (red stars) and only the satellite (blue stars) galaxies in our selected galaxy sample from the TNG100 simulation. The light blue stars represent the result from the TNG50 simulation shown in Fig. \ref{fig:environment_dependence_neighbours}. The description is as in Fig. \ref{fig:environment_dependence_neighbours}. Overall, we see that the behaviour of the median amplitude of the lopsidedness shows a similar trend in TNG100 and TNG50 as a function of the environment, confirming the hypothesis that the lopsidedness does now significantly depend on the environment as it was suggested using the TNG50 simulation.} 
    \label{fig:environment_dependence_neighbours_tng100}
\end{figure*}

\subsection{The local environment of the galaxies}
\label{sec:environment_neighbours_tng100}
Similarly to Fig. \ref{fig:environment_dependence_neighbours_galaxy_properties}, in Fig. \ref{fig:environment_dependence_neighbours_tng100}, we study the behaviour of the lopsidedness amplitude as a function of the local environment at $z=0$ defined in i)-iii) (see Sec. \ref{sec:environment_neighbours}) from the left to the right panel, respectively, for our selected galaxy sample from the TNG100 simulation. In each panel, we also show the corresponding result from the TNG50 simulation as light blue stars for comparison.
We see that the median amplitude of the lopsidedness also shows an overall constant to mild decreasing trend with the increasing number of close massive neighbours and local volume density of galaxies in TNG100, similarly to TNG50. Here, the exception is represented by the environmental definition based on $R_{200}$, where we see that the median lopsidedness amplitude shows a steeper decrease (middle panels in Fig. \ref{fig:environment_dependence_neighbours_tng100}. 
As already mentioned in the previous Sec. \ref{sec:environment_halomass_tng100}, the mild decreasing trend with the environment indicates that lopsidedness does not significantly depend on the number and density of close massive neighbours (i.e. the frequency and type of interactions), consistent with the results from the TNG50 simulation (see Sec. \ref{sec:environment_neighbours}).

In Fig. \ref{fig:environment_dependence_neighbours_galaxy_properties_tng100}, we show the behaviour of the $\mathrm{D/T}$, $c/a$ and $\mu_{*}$ as a function of the local environment at $z=0$ defined in i)-iii) from the left to the right panel, respectively, for a total mass cut of the neighbouring galaxies $\mathrm{M_{\mathrm{tot}}}\geq10^{9}\, \mathrm{M_{\odot}}$, as done for TNG50 in Fig. \ref{fig:environment_dependence_neighbours_galaxy_properties}). Similarly to Fig. \ref{fig:environment_dependence_m200_tng100}, we see that the galaxies have larger $\mathrm{D/T}$ and smaller $c/a$ and $\mu_{*}$ in TNG50 than in TNG100, possibly suggesting the effect of the different mass resolution between the two simulations, and that the $\mathrm{D/T}$ and $\mu_{*}$ show a mild decreasing and increasing trend with the increasing number of close massive neighbours and local volume density of galaxies, respectively, while the $c/a$ remains overall constant, possibly suggesting a change to earlier type disk morphology towards higher density environments as also seen in TNG50 (see Sec. \ref{sec:role_environment}). However, in Fig. \ref{fig:environment_dependence_neighbours_galaxy_properties_tng100}, the lopsided galaxies still seem to be characterized by somewhat larger $\mathrm{D/T}$ and smaller $c/a$ and $\mu_{*}$ than the symmetric galaxies, similarly to Fig. \ref{fig:environment_dependence_m200_tng100} as a function of $\mathrm{M_{200}}$, even though the differences are less evident than in TNG50 (see Sec. \ref{sec:role_environment}).  

\begin{figure*}
    \centering
    \includegraphics[width=0.3\textwidth]{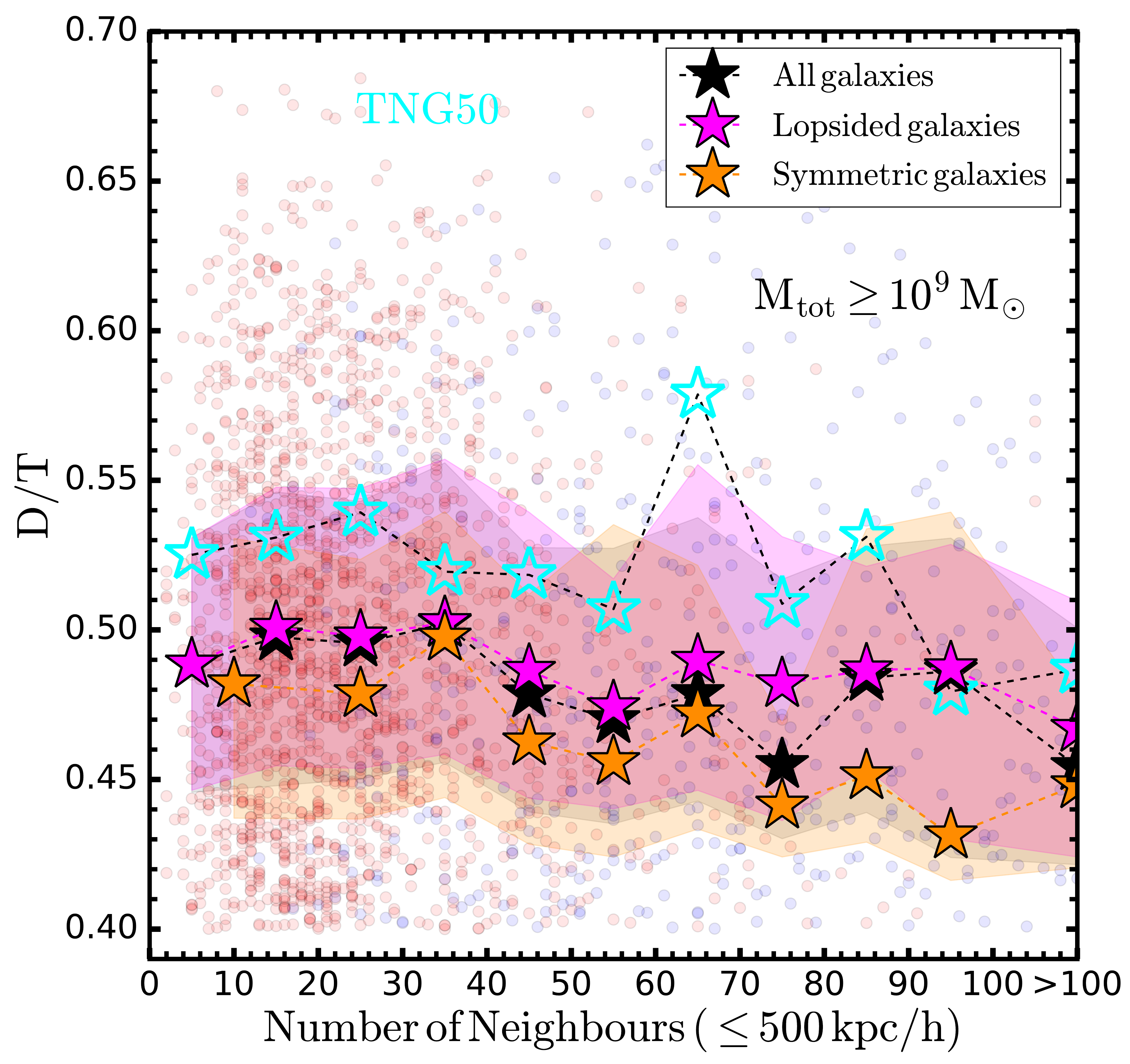}
    \includegraphics[width=0.3\textwidth]{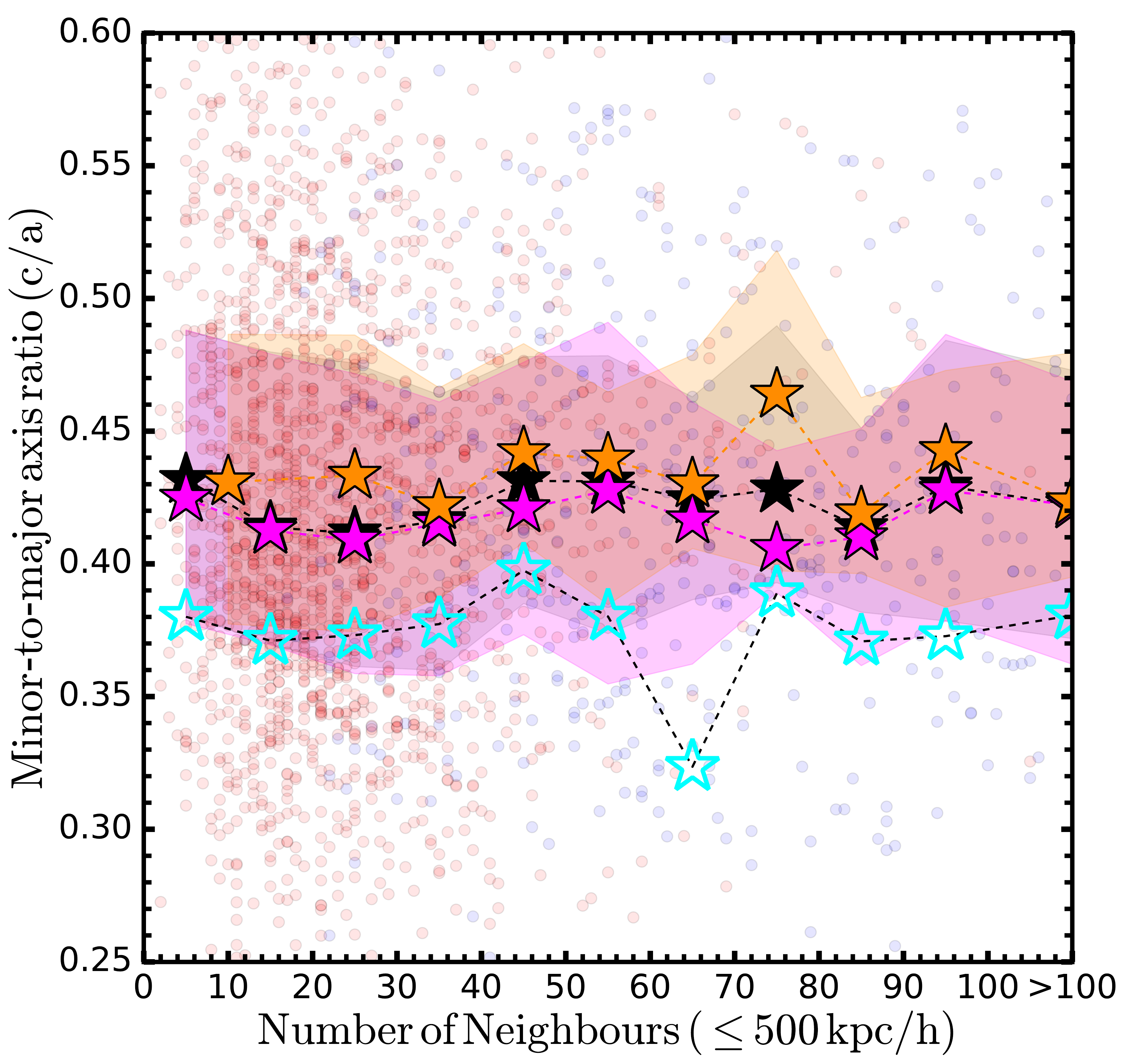}
    \includegraphics[width=0.3\textwidth]{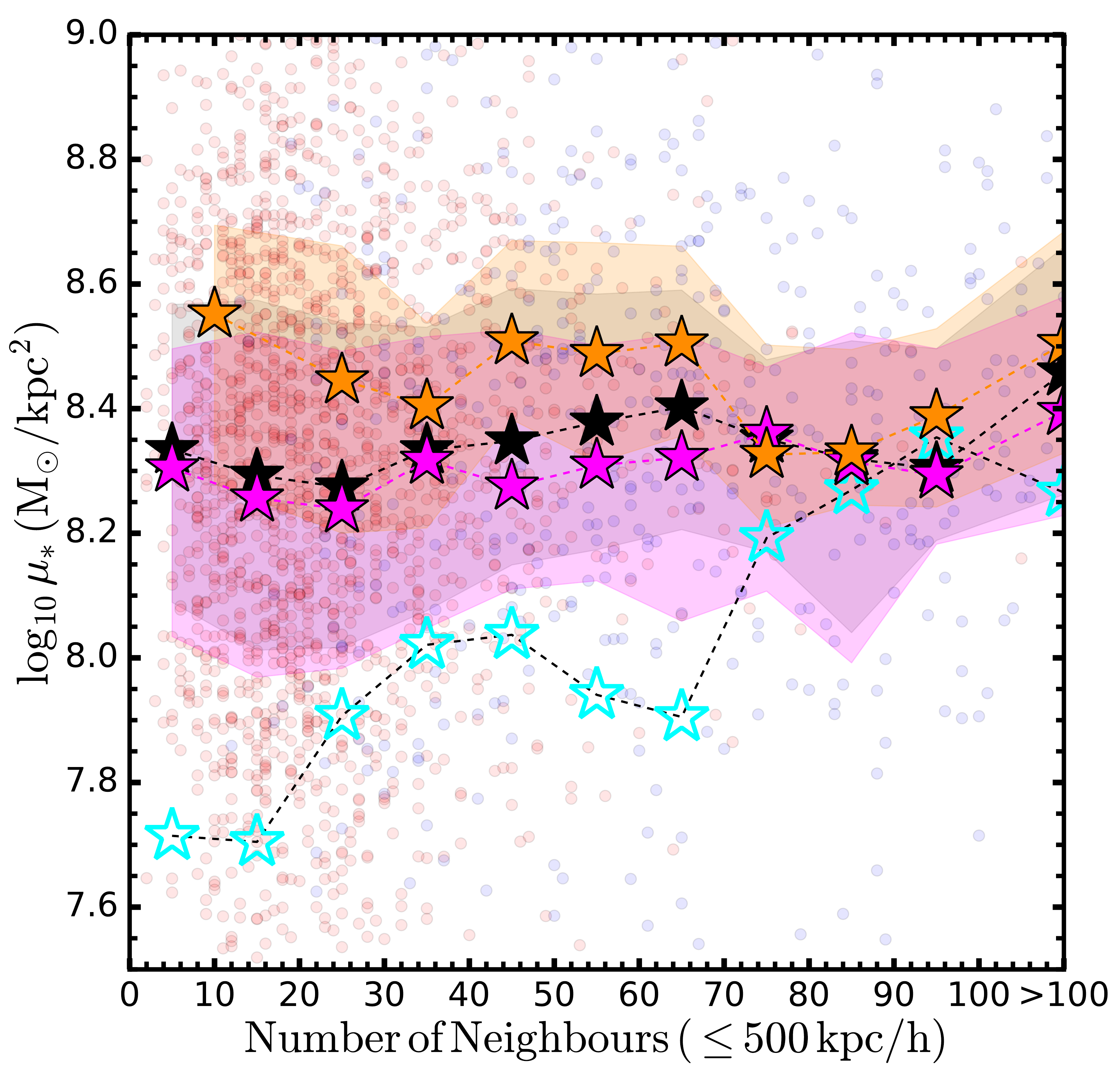}

    \includegraphics[width=0.3\textwidth]{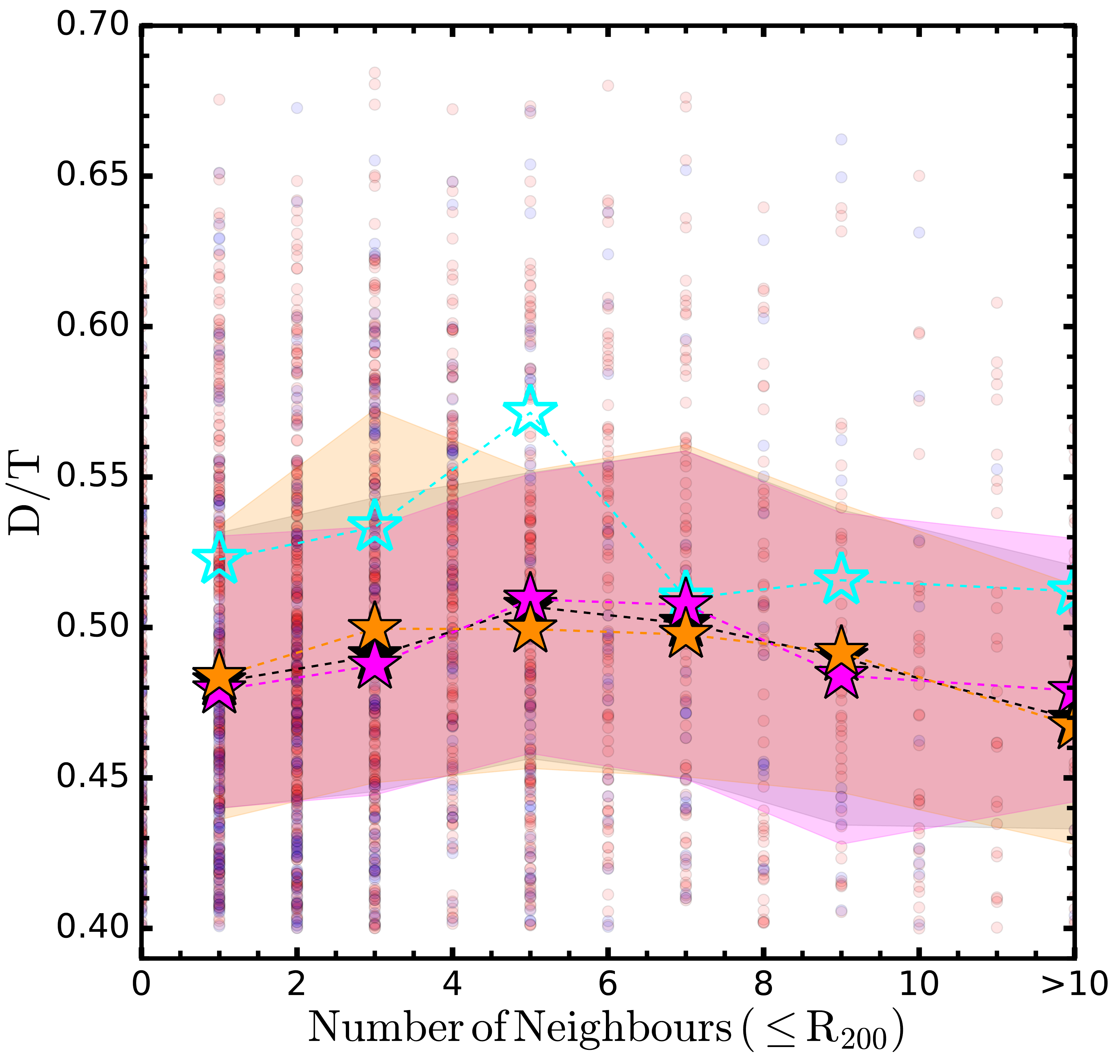}
    \includegraphics[width=0.3\textwidth]{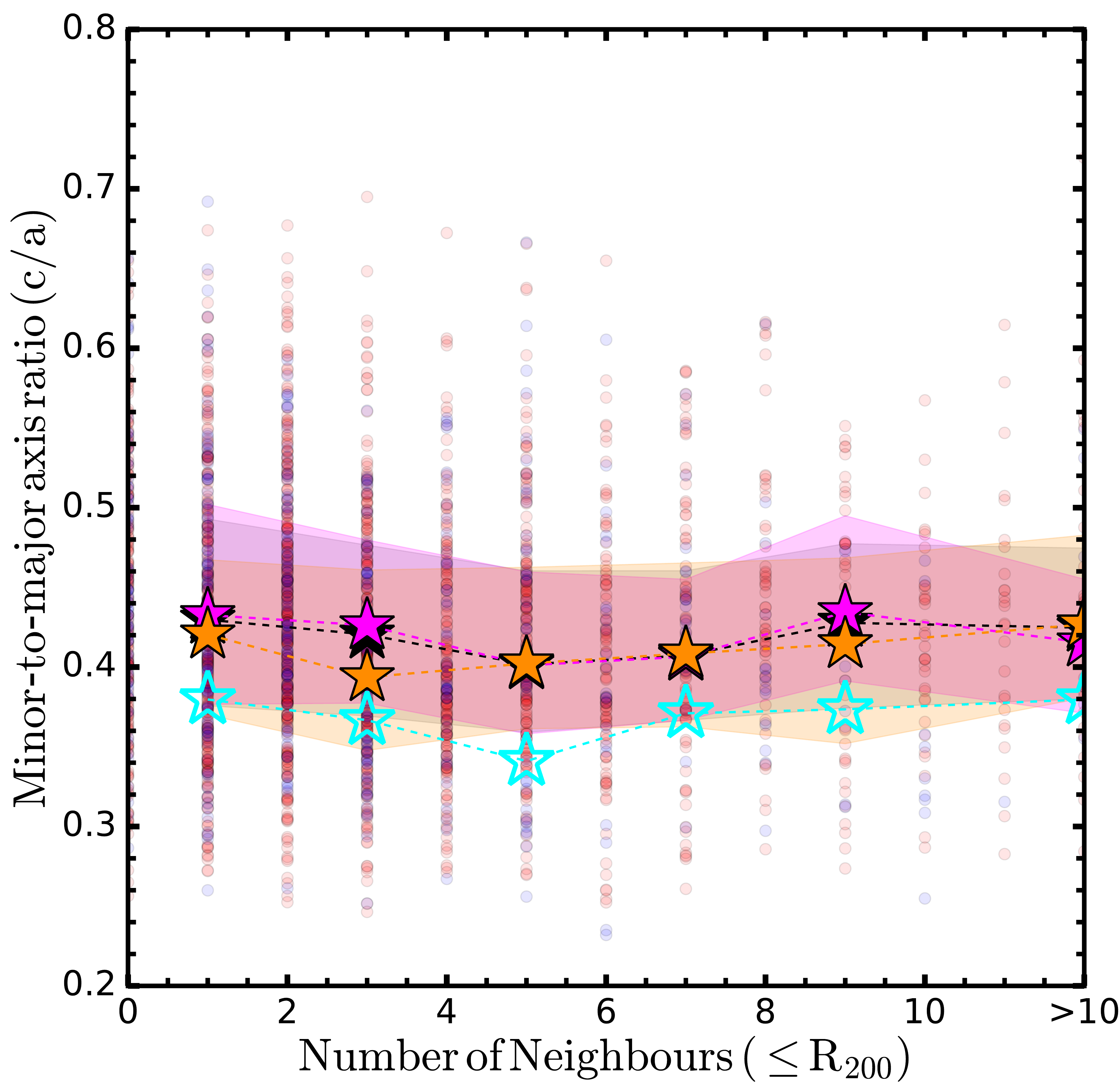}
    \includegraphics[width=0.3\textwidth]{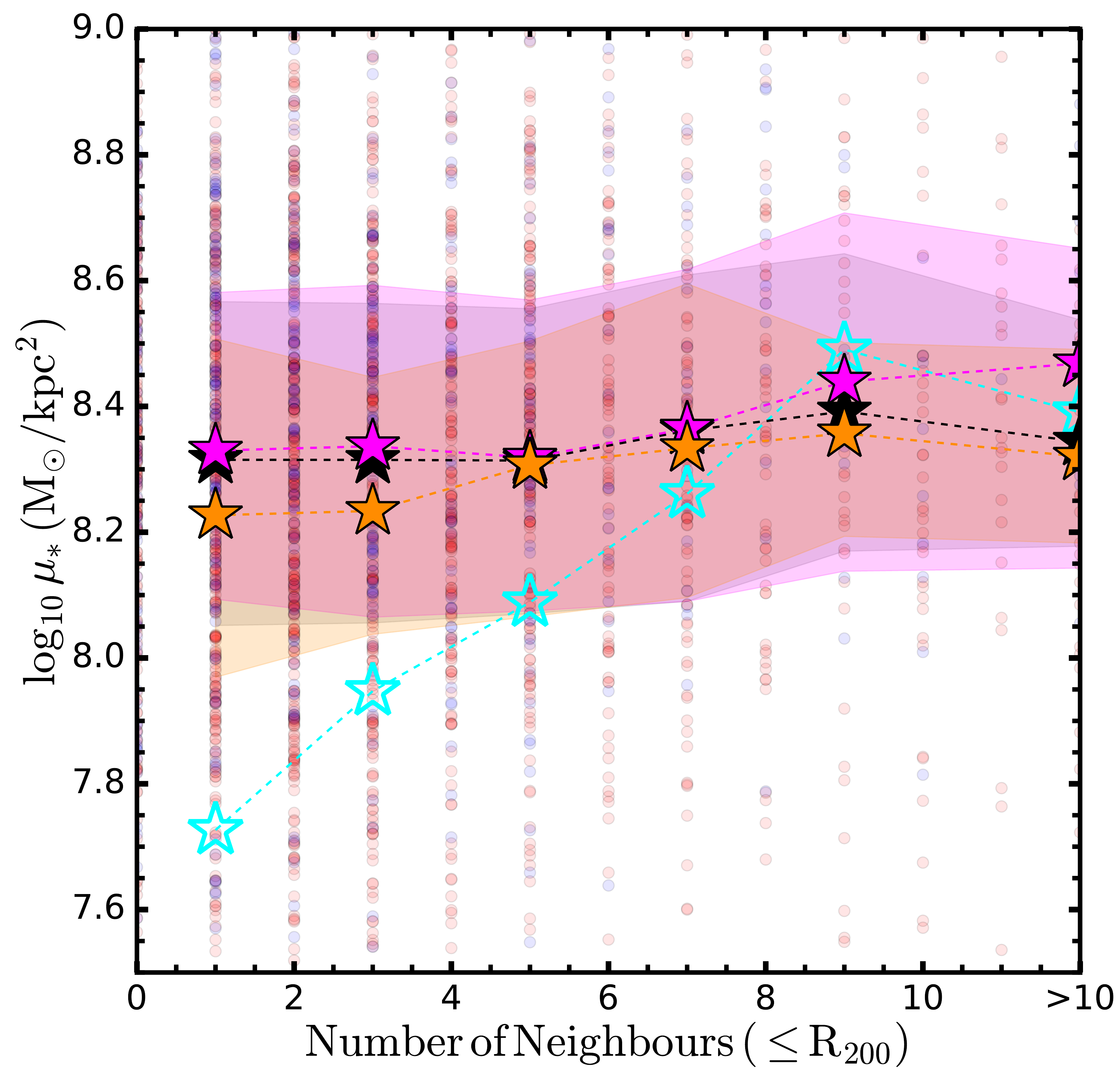}
    
    \includegraphics[width=0.3\textwidth]{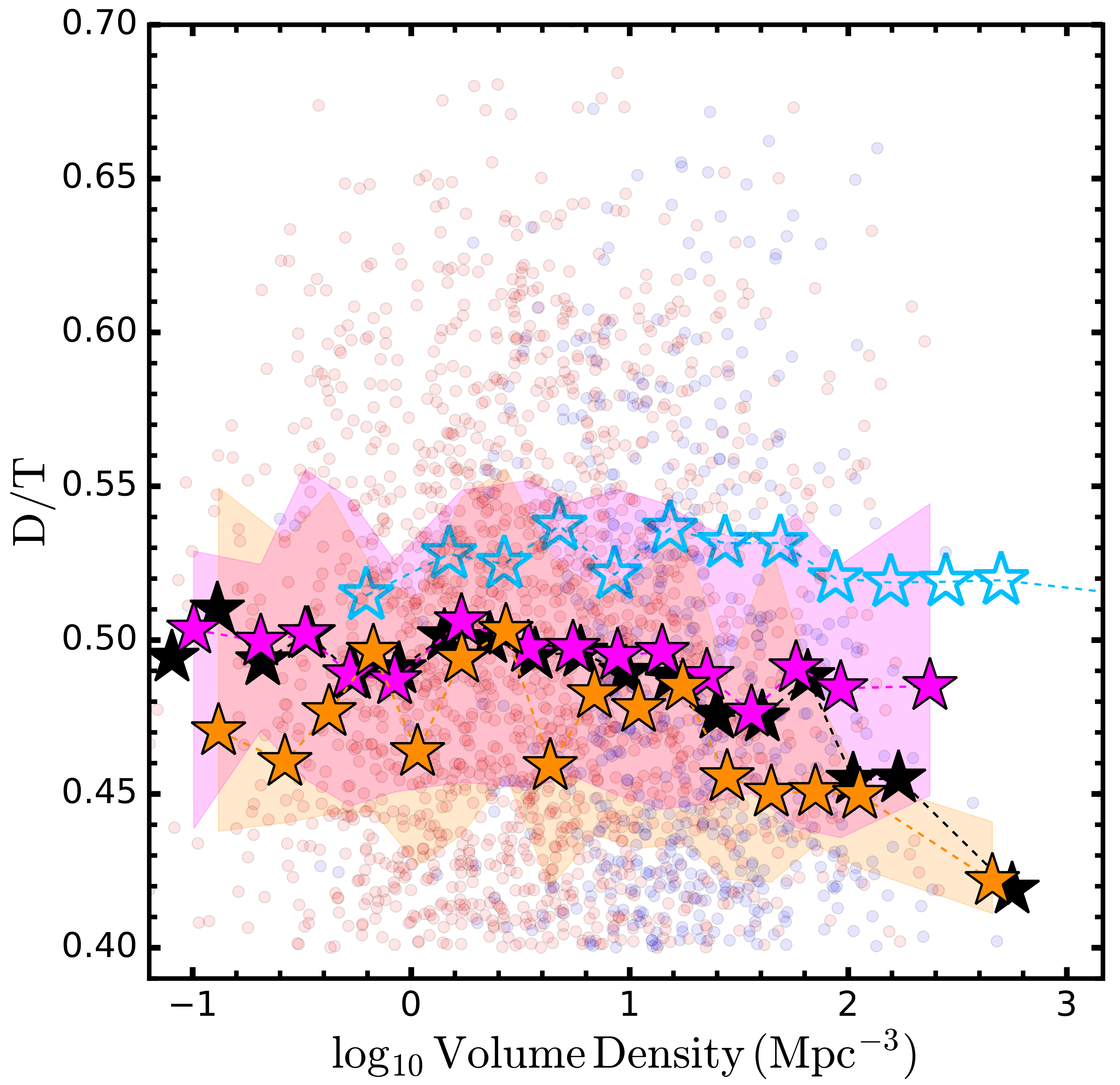}
    \includegraphics[width=0.3\textwidth]{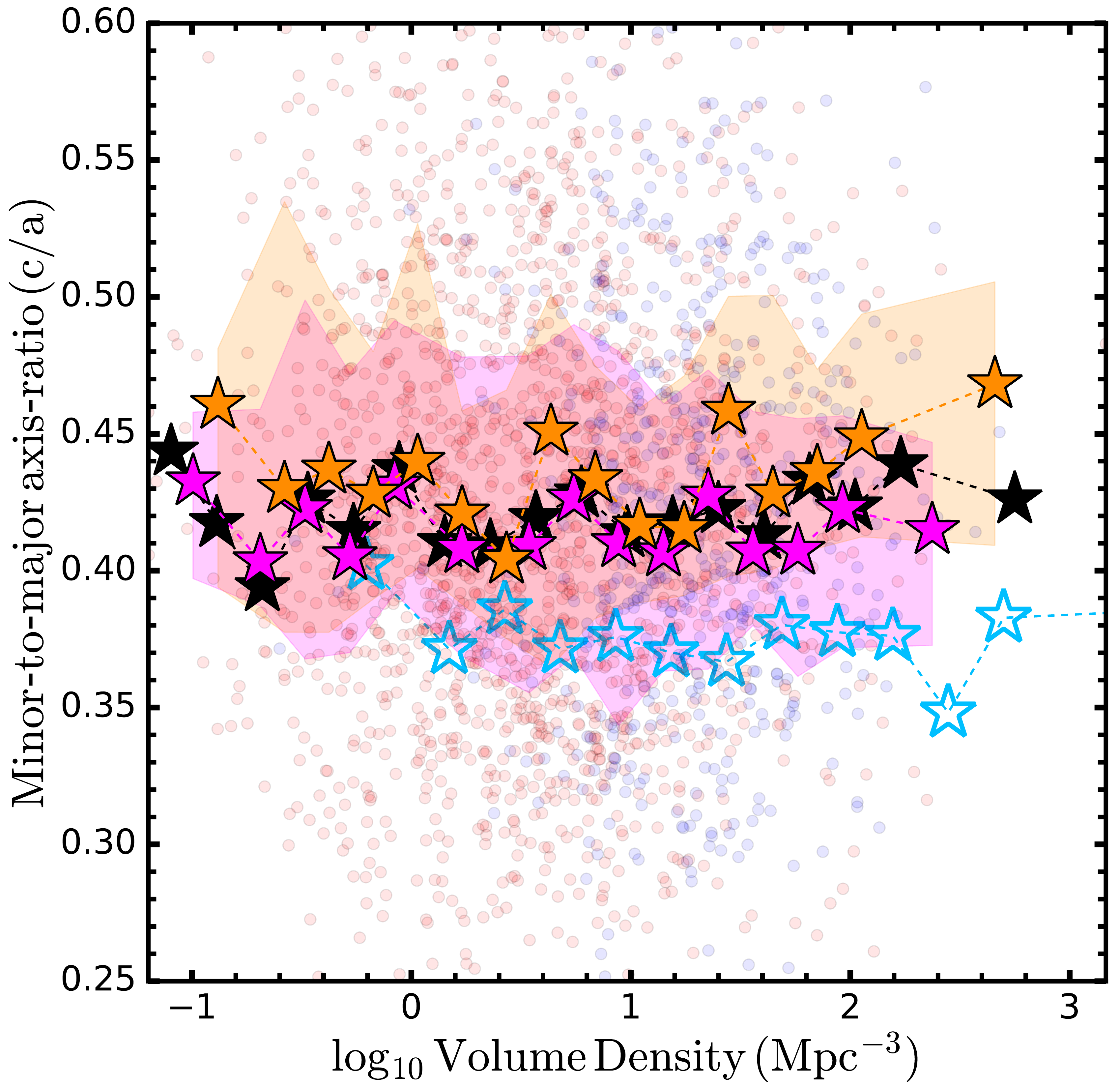}
    \includegraphics[width=0.3\textwidth]{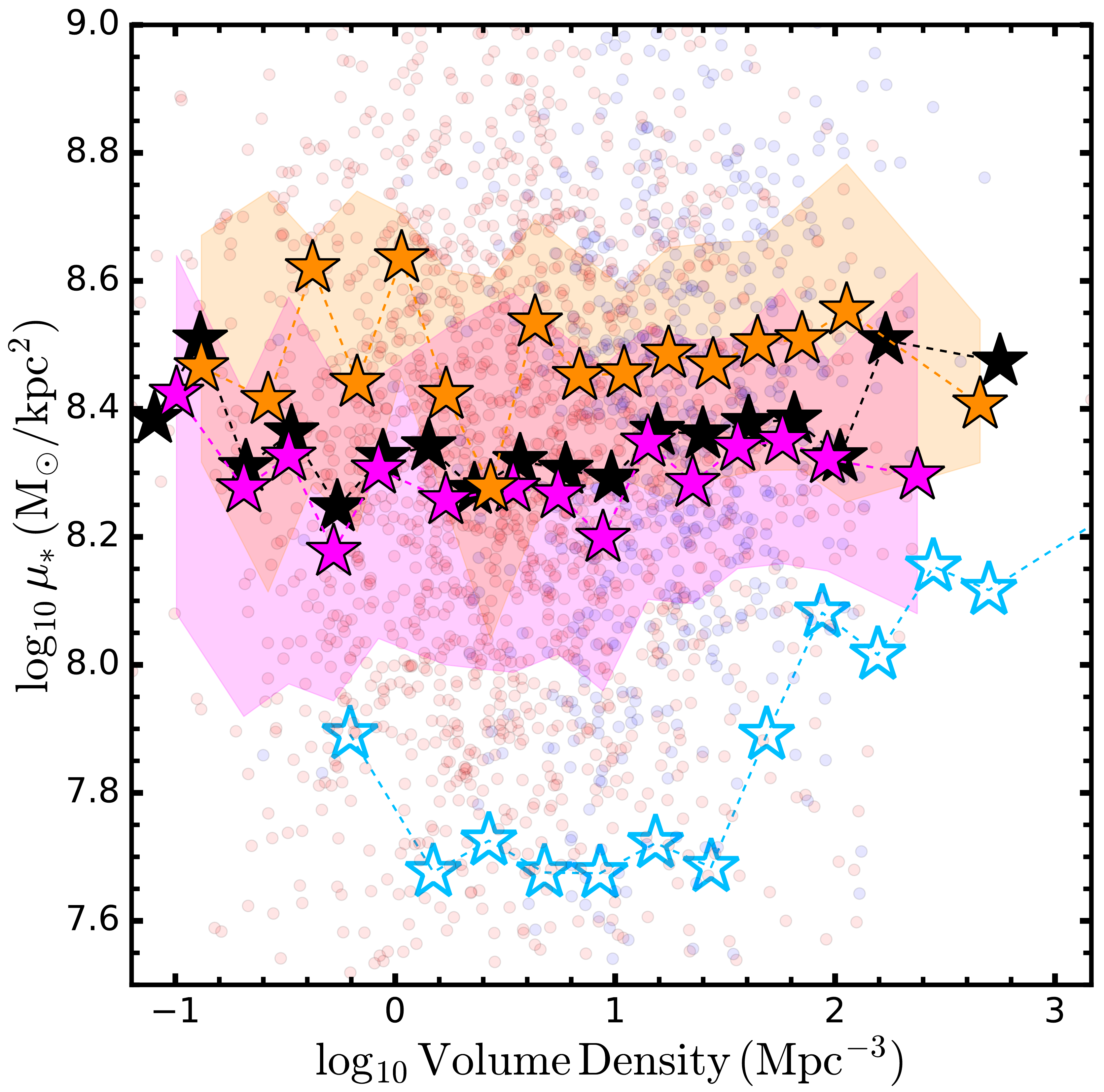}
    \caption{From the left to the right panel, the median disk-to-total ratio ($\mathrm{D/T}$), minor-to-major axis ratio ($\mathrm{c/a}$) and central stellar mass density ($\mu_{*}$) as a function of the local environment at $z=0$ defined in i)-iii), respectively, for all (black stars), only the lopsided (magenta stars) and only the symmetric (orange stars) galaxies in our selected galaxy sample in the TNG100 simulation. The light blue stars represent the represent the result from the TNG50 simulation shown in Fig. \ref{fig:environment_dependence_neighbours_galaxy_properties}. The description is as in Fig. \ref{fig:environment_dependence_neighbours_galaxy_properties}. We see that the galaxies tend to be characterized by on average smaller $\mathrm{D/T}$ and larger $c/a$ and $\mu_{*}$ in TNG100 than in TNG50, possibly suggesting resolution effects, and that the lopsided and symmetric galaxies show on average distinct internal properties as seen in Fig. \ref{fig:environment_dependence_m200_tng100} as a function of $\mathrm{M_{200}}$, even though these differences between the internal properties of the lopsided and symmetric galaxies are less evident in TNG100 than in TNG50 (see Fig. \ref{fig:environment_dependence_m200} and \ref{fig:environment_dependence_neighbours_galaxy_properties}).} 
    \label{fig:environment_dependence_neighbours_galaxy_properties_tng100}
\end{figure*}

Overall, the study of the lopsidedness carried out in this Appendix \ref{sec:role_environment_tng100} using the TNG100 simulation, which provides a larger statistical galaxy sample in high-density environments, seems to confirm the results from the TNG50 simulation in Sec. \ref{sec:role_environment} that the lopsidedness amplitude does not significantly depend on the environment but rather it correlates more strongly with the distinct internal properties of the galaxies, such as their disk fraction, disk thickness and central stellar mass density, overall consistent also with the findings from other previous works (e.g. \citealt{Reichard2008,Jog2009,Varela-Lavin2022}).

\section{Stellar half-mass and optical radii distribution}
\label{sec:rhalf}

\begin{figure*}
    \centering
    \includegraphics[width=0.45\textwidth]{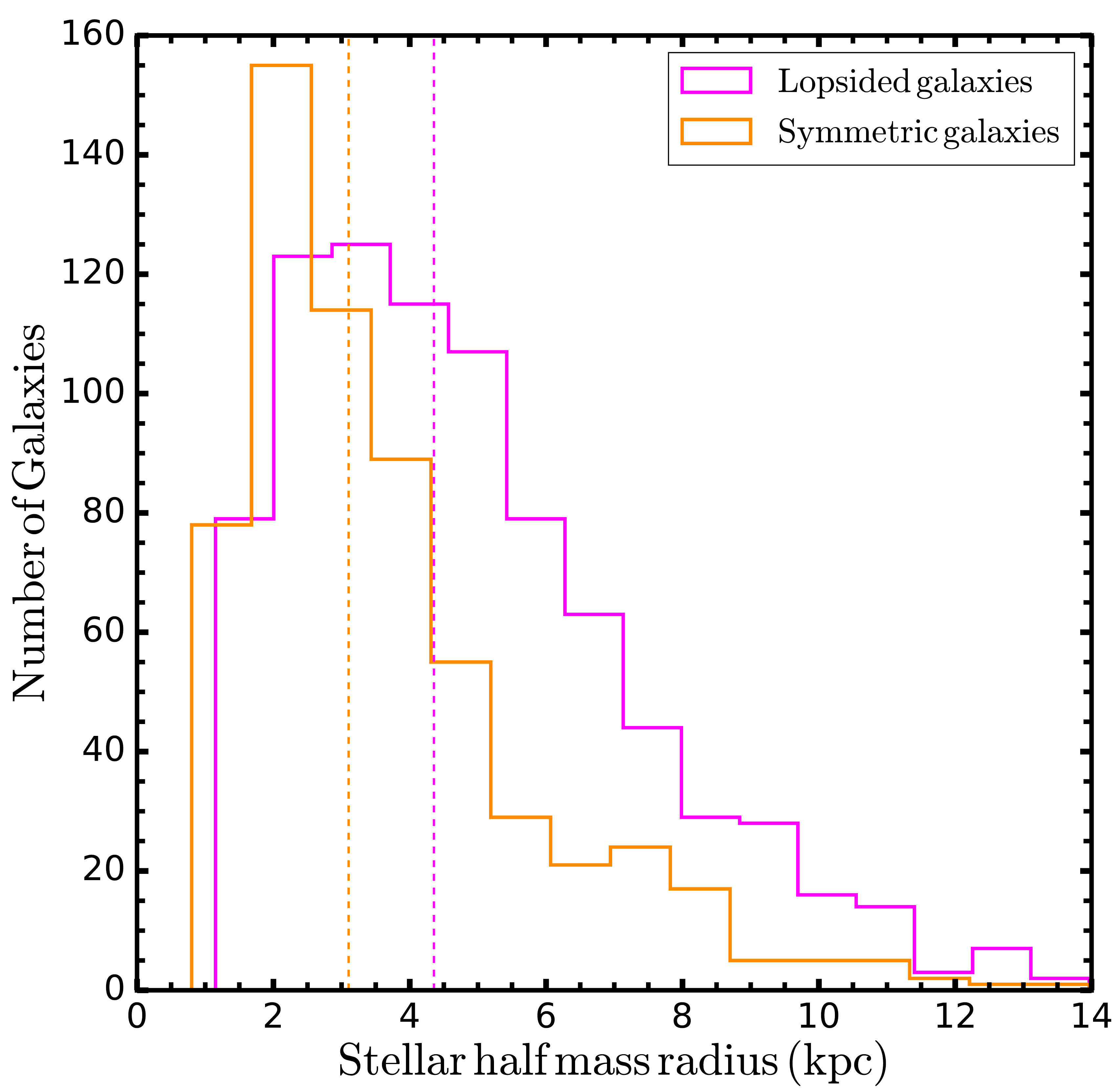}
    \includegraphics[width=0.45\textwidth]{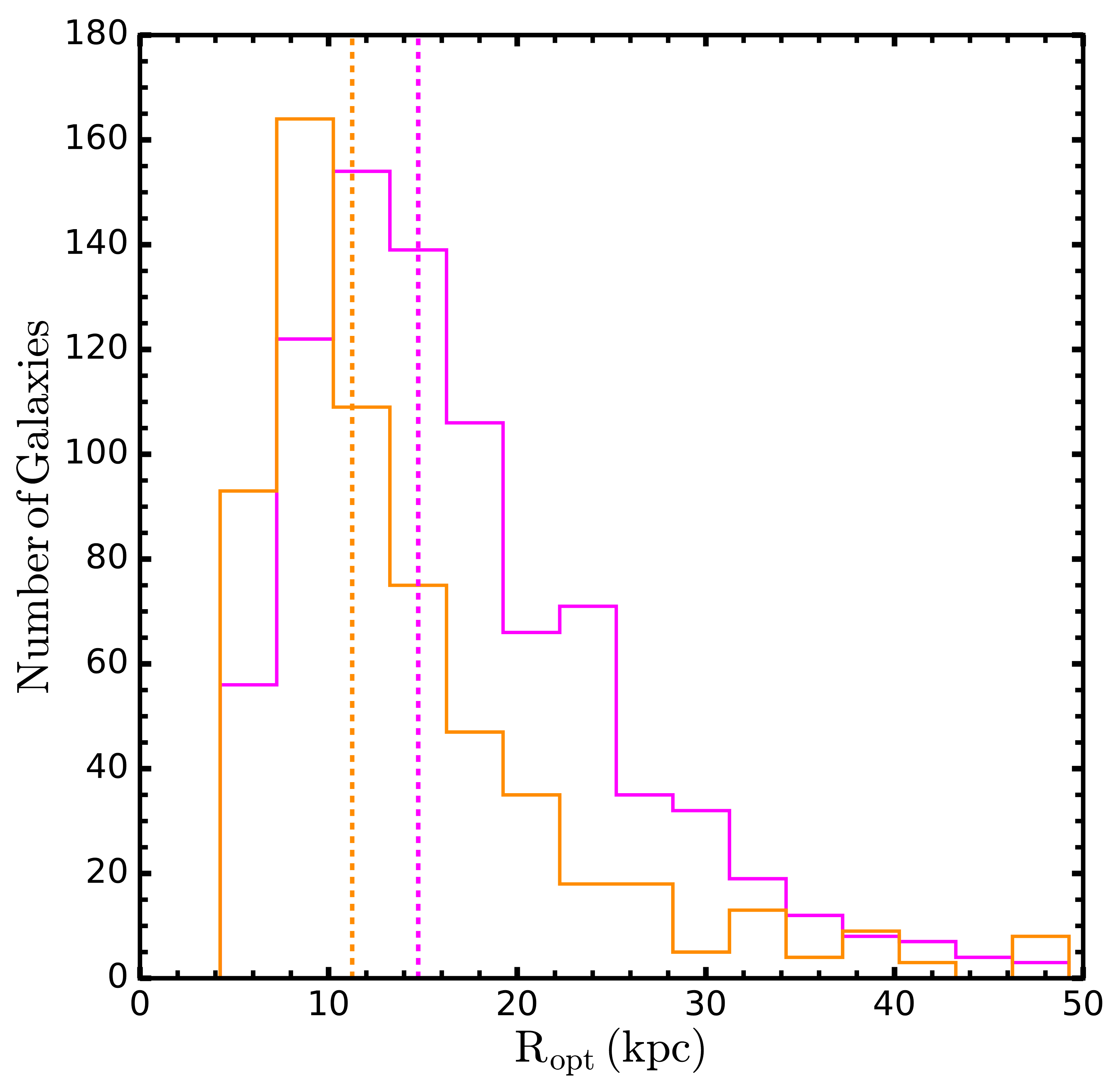}
    \caption{{\bf Left:} The distribution of the stellar half-mass radius, $R_{\mathrm{half}}$, of the lopsided and symmetric galaxies in our sample, which is calculated from the face-on projection. {\bf Right:} The distribution of the optical radius, $R_{\mathrm{opt}}$, of the lopsided and symmetric galaxies, which is calculated as described in Sec. \ref{sec:lopsidedness}. The vertical dashed lines represent the corresponding median values of the distributions. We see that the lopsided galaxies typically have larger $R_{\mathrm{half}}$ and $R_{\mathrm{opt}}$ than the symmetric counterpart, suggesting that lopsided galaxies are characterized by a more extended mass distribution.}
    \label{fig:rhalf}
\end{figure*}


\bsp	
\label{lastpage}
\end{document}